\newcommand{\beq}{\begin{equation}}
\newcommand{\eeq}{\end{equation}}

\newcommand{\bear}{\begin{eqnarray}}
\newcommand{\eear}{\end{eqnarray}}

\newcommand{\tn}{\textnormal}

\newcommand\numu{\nu_\mu}

\newcommand\barparen[1]{\overset{(-)}{#1}}

\RequirePackage[switch]{lineno}

\documentclass[review]{elsart5p}

\usepackage{graphicx}
\usepackage{lscape}
\usepackage{enumerate}
\usepackage{pifont}
\usepackage{esvect}
\usepackage{amssymb}
\usepackage{bbding}
\usepackage{rotating}
\usepackage{footnote}
\usepackage{tablefootnote}
\usepackage[flushleft]{threeparttable}
\usepackage{amsmath}
\usepackage{color}
\usepackage[dvipsnames]{xcolor}
\usepackage{amsfonts}
\usepackage{dsfont}

\usepackage{url}
\usepackage{tabularx}

\usepackage{multicol}
\usepackage{float}
\usepackage{booktabs}
\usepackage{xspace}
\usepackage{pbox}
\usepackage{verbatim}
\usepackage[tocflat]{tocstyle}
\usetocstyle{standard}
\usepackage{color}\definecolor{darkblue}{rgb}{0,0,1}

\makesavenoteenv{tabular}
\makesavenoteenv{table}

\begin{document}

\begin{frontmatter}

\title{Gaseous and dual-phase time projection chambers for imaging rare processes}
\bigskip
\author[1]{Diego Gonz\'alez-D\'iaz\corauthref{aut1}},
\corauth[aut1]{Corresponding author.}\ead{Diego.Gonzalez.Diaz@usc.es}
\author[2]{Francesc Monrabal},
\author[3]{Sebastien Murphy}
\address[1]{Instituto Galego de F\'isica de Altas Enerx\' ias (IGFAE), Universidade de Santiago de Compostela, Spain}
\address[2]{Department of Physics, University of Texas Arlington, USA}
\address[3]{ETH Zurich, Institute for Particle Physics, Switzerland}

\begin{abstract}
Modern approaches to the detection and imaging of rare particle interactions through gaseous and dual-phase
time projection chambers are discussed. We introduce and examine their basic working principles and
enabling technological assets.
\end{abstract}

\begin{keyword}
Time Projection Chambers \sep TPCs \sep drift chambers \sep imaging chambers

\PACS 29.40 \sep Cs
\end{keyword}
\end{frontmatter}


\tableofcontents

\section{Introduction}
\label{intro}

The introduction of the time projection chamber (TPC) by David Nygren in 1974 \cite{Dave1st} has exerted a perdurable influence in particle and nuclear physics, casting its shadow over much of today's instrumentation. TPCs revolutionized experimentation at colliders with the introduction of a novel scheme for reconstructing particle trajectories, in which electric and magnetic fields would be set parallel to each other \cite{Dave2nd}: with the passage of charged particles, ionization electrons are locally released in the detector medium and then collected after meter-long drift distances, their spread reduced through the convenient orientation of the $\vec{E}$ and $\vec{B}$ fields. Once collected at an $x$,$y$-sensitive image plane, their arrival times are back-converted to longitudinal positions ($z$) through their average drift velocity, a technique borrowed from drift chambers (e.g., \cite{Walenta1,ISIS}).

By identifying a preferred alignment relative to the spectrometer's magnetic field, Nygren posited a detector configuration that not only avoided the problems associated to the traditionally undesirable `$\vec{E}\times\vec{B}$ effect', it incorporated it as a defining feature of the new device. Through the 80's and 90's, TPCs offered thus the closest representation of the long sought idea of \emph{imaging particle tracks through space}, however in an uniform magnetic field, with ultimate spatial precision and at a speed greatly exceeding that of cloud or bubble chambers. The first TPC was integrated in the PEP4 spectrometer and helped at studying electron-positron collisions at a center of mass energy of 29\,GeV, in the Stanford Linear Accelerator Center (SLAC). This pioneering development embraced the concepts of continuous readout, particle identification through energy loss and diffusion suppression in a way that made the effort technologically unprecedented \cite{PEP}. By showing what could be done, it paved the way to increasingly sophisticated devices.\footnote{An excellent review covering the technological aspects of TPCs, mainly from a collider standpoint, can be found in a recent work by the late H. J. Hilke \cite{Hilke}, while for a historical account the reader is referred to \cite{Galison_TPC}.}

But what could be the relevance of the TPC technique for imaging rare particle/nuclear interactions, the subject of this review?, and is it really meaningful to introduce such a notion?. First, we are referring here to situations where reaction products can arrange themselves in a wealth of topologically complex multi-track patterns (and even particle showers) as it often happens with neutrino interactions \cite{Antonello:2012hu, Badertscher:2013wm}. At the same time, however, TPCs in this category may have to deal with nearly point-like \cite{CAST, Phan} or tortuous \cite{NEXT1} tracks, too. These broad topological characteristics, together with the need of huge (and, in some cases, flexible) active volumes, can be expected to reduce the feasibility and benefits of a magnet, and indeed the effect of the magnetic field is considerably subtle in some instances (e.g. \cite{JoshB, Galan}). Therefore, and contrary to TPCs at colliders, operation under magnetic field will be seldom found throughout this text. But even in rcases where the event topology is dull and a handful of fairly straight tracks must be detected, the ionization profiles may need to be reconstructed in modern TPCs with a very fine pixelization (mm and sub-mm scale), in order to identify the reaction products \cite{ZimmermanPhD,ACTAR,pp2}, the direction and sense of their momentum vector \cite{Phan,NEXT1} or the polarization of an incoming particle \cite{Pol1, XrayPol}. Despite recent progress (e.g. \cite{CASTMichel}), such a fine sampling is still out of reach for contemporary TPCs used at colliders, given the large dimensions involved. The position resolution on the other hand (that is, the precision with which the barycenter of a portion of the track can be reconstructed), can reach similar levels in both types of TPCs, meaning that they will recover each track's geometrical information with a comparable performance. Importantly, in TPCs used for the study of rare interactions, global information like the particle's range (e.g., \cite{Phan}) and/or the total energy transferred to ionization and scintillation (\cite{XENON,LZ,Cao:2014jsa,JoshHPNeutrons}) can become crucial particle discriminants. Such assets require, unlike in collider TPCs, full event containment. Finally, these new imaging chambers do not always display some of the other features that are found invariably at colliders, namely i) the existence of a start time of the event (hereafter $T_0$) or some a priori knowledge of the interaction vertex, ii) the presence of positive-ion space charge capable of distorting the drift field and iii) the necessity of developing techniques to mitigate the associated aging effects. A striking paradigm shift is that the selection of suitable materials becomes essential in many cases, their main figure of merit being however a low nuclear activity, rather than a low chemical one \cite{XENONcampaign, NEXTcampaign}. In summary, a text describing TPCs used at colliders and another describing TPCs for the study of rare processes may well look like a recipe for making a good omelette (e.g. \cite{Blum-Rolandi,Knoll}) and a description of all its various types and tastes (e.g., \cite{Aprile,Chepel}).\footnote{While perhaps not very relevant technologically, it is possible to identify one single feature that determines the situation with very little ambiguity: the topology of the TPC itself. Since TPCs aimed at rare processes need to maximize the collection of information, the active volume is seamless (homeomorphic to a sphere), whereas it is hollow and thus fully penetrated by the unreacted beam particles, in collider and in most high energy physics applications (i.e., homeomorphic to a torus). The containment (or not) of the reaction vertex (and usually of the full event, indeed) is another way to identify the situation.}

\begin{figure}[h!!!]
\centering
\includegraphics*[width=\linewidth]{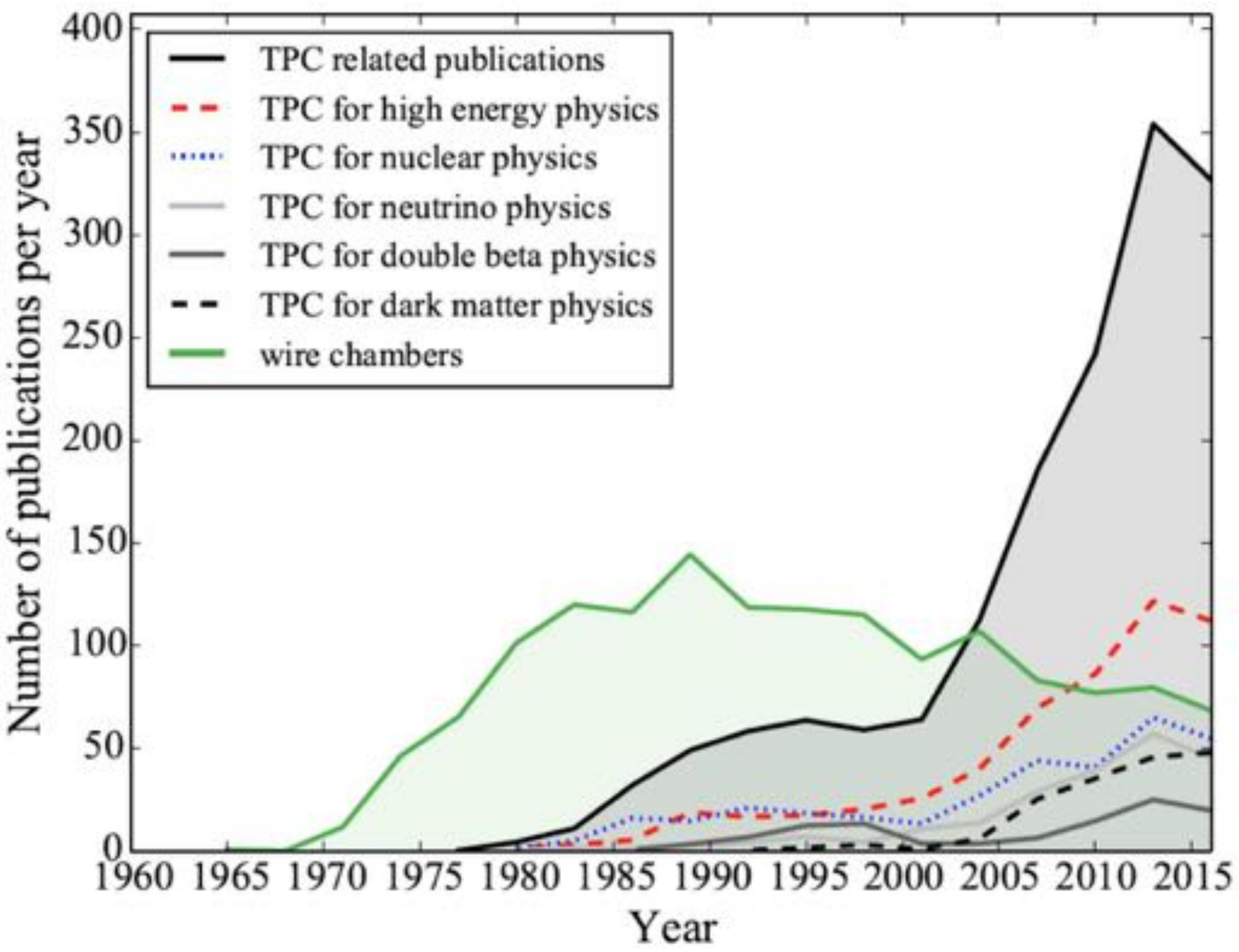}
\caption{Number of peer-reviewed publications per year (averaged over 3 years) containing the words `time projection chamber' or `TPC' in the area of physics, for different sub-fields \cite{scopus}. (The purity and efficiency of these selections is higher than $80\%$)}
\label{TPCpubs}
\end{figure}

Despite the differences, there is little doubt that `TPC' has been generally adopted as the trademark name for a number of detectors that aim at the detailed imaging of particle interactions in gas, liquid or solid, through the time-to-space projection of the drift coordinate \cite{TPCconferences}. This can be seen from the publication trend in Fig. \ref{TPCpubs}, were three revealing regions are highlighted (inception: (1975-1990), consolidation: (1990-2001), expansion: (2001-today)).

\begin{figure}[h!!!]
\centering
\includegraphics*[width=\linewidth]{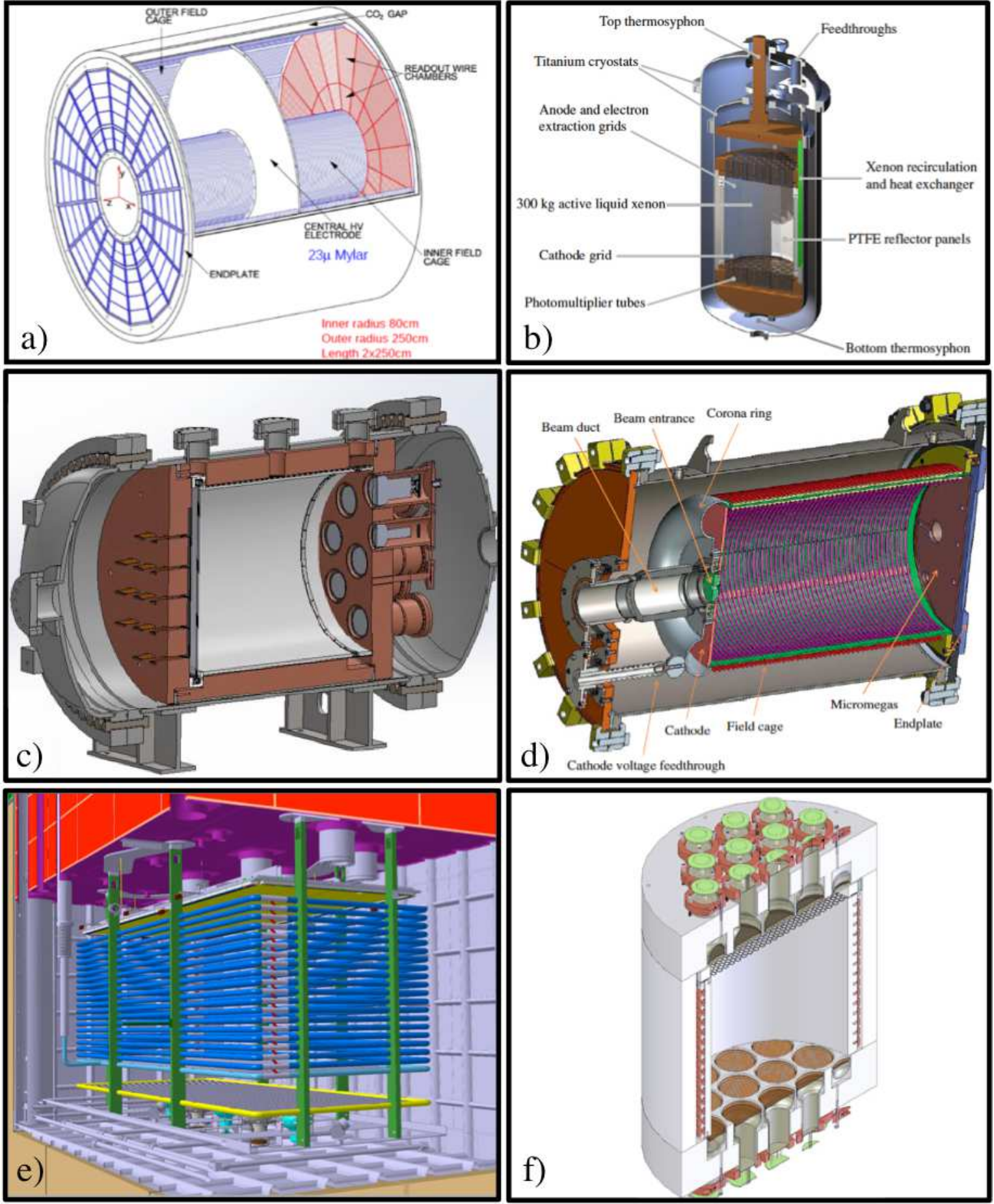}
\caption{Some TPCs discussed in text: a) the state of the art (\mbox{MWPC}-based) collider TPC from ALICE, at the Large Hadron Collider (LHC)  \cite{ALICE}; b) LUX, the \mbox{dual}-phase (optically read) xenon TPC for dark matter searches at the Sanford Underground Research Facility (SURF) \cite{LUX}; c) NEW, the high pressure xenon TPC (optically read) developed for phase-I of the $\beta\beta0\nu$ experiment NEXT, at Laboratorio Subterr\'aneo de Canfranc (LSC) \cite{NEXT1,NEW}; d) phase-I of the (Micromegas-based) Active Target TPC (AT-TPC), at the Facility for Rare Isotope Beams (FRIB) in Michigan \cite{AT-TPC}; e) $3\times1\times1$\,m$^3$ \mbox{dual}-phase argon prototype based on LEMs/thick GEMs for the far detector of the neutrino oscillation experiment DUNE \cite{Murphy:2016ged}; f) The DarkSide-50 dual-phase argon detector for dark matter searches at Laboratori Nazionali del Gran Sasso (LNGS) \cite{DarkSide}.}
\label{4TPCs}
\end{figure}

Modern TPCs comprise a category of imaging chambers with vastly different designs (Fig. \ref{4TPCs}),
making a technological classification increasingly impractical: (classical) charge-readout \cite{ACTAR, AT-TPC, Gotthard, NEXTMM}, optical \cite{NEXT1,ZimmermanPhD, OTPC2pp}, negative-ion \cite{Martoff, DRIFTlast}, liquid \cite{EXO, ICARUS, WA104}, dual-phase \cite{Badertscher:2013wm, XENON, LZ, Cao:2014jsa, DarkSide}, solid \cite{Filipenko}, Penning-fluorescent \cite{MePenning}, Cherenkov \cite{Woody}, radial \cite{cyl1,cyl2}, spherical \cite{spherical, spherical2}... In the case of gaseous and dual-phase chambers (the topic of this review), their operational range of pressures goes from tens of mbar \cite{DDMpeople} to above 10\,bar \cite{NEXT1}, their temperatures from 87\,K \cite{Badertscher:2013wm} to 300\,K, their readouts presently include wires \cite{DRIFTlast}, GEMs \cite{Phan}, thick GEMs/LEMs \cite{Badertscher:2013wm}, Micromegas \cite{NEXTMM}, $\mu$-dot/$\mu$-PIC \cite{NEWAGE}, InGrid \cite{CAST}, photomultipliers (PMs) \cite{XENON}, silicon PMs \cite{NEXT1}, and CCD or CMOS cameras \cite{Leyton,Orange1}. Besides imaging the primary ionization with high accuracy they can, in some configurations, retrieve essential information from the gas scintillation, too. As we will see throughout the text, and in particular in section \ref{classification6}, there are numerous hints that future TPCs will be able to do more than `just' that (e.g., \cite{MePenning, BaTa1, BaTa2, ColReco, PositiveIon, CO2Henriques, Xenon bubble chamber}.

This review is organized in such a way so as to first introduce the main characteristics of the images obtained with this new generation of imaging chambers (section \ref{image}), proceeding then to discuss their enabling assets (section \ref{assets}), and the most successful designs used in fundamental science (sections \ref{classification1}-\ref{D-P}). Section \ref{classification6} is dedicated to a discussion of some of the most recent and intriguing ideas. For completeness, TPC designs that do not fit well to the classification scheme chosen in text are summarized in this section, too. In general, our aim has been to provide a technological overview of gas and dual-phase TPCs used in the detection of rare processes, with a particular focus on its imaging characteristics and enabling assets. Our choice to emphasize one development or another may not be totally free of bias, but we believe that the end result is representative of the whole.

\section{Images, patterns and their building blocks} \label{image}

Some of the events that must be imaged using the techniques described in this article involve radioactive decays with lifetimes that are at least one quadrillion times the age of the Universe \cite{Kamland}, or super-symmetric particles that, were they to exist, would easily cross 1 million Earths without interacting \cite{1TonLimit}, and for sure neutrinos in any of their flavours. The energy released can be as small as that needed to create a few primary ionizations, producing for instance distinct images of low energy nuclei that would otherwise extend just 10's to 100's of nm in a solid \cite{emulsions}. The processes are so rare in some cases, that interactions caused by solar and supernova neutrinos \cite{DDMpeople} or other extremely rare processes like double beta decay \cite{reviewbb0, LauraNeutrinos}, constitute very real irreducible backgrounds that ultimately limit the detector sensitivity. In other cases, the events described can be considered rare simply because they don't appear naturally on Earth, but they can be produced in dedicated facilities at practical reaction and decay rates like, e.g., double proton decay or the resonant photo-production of excited Hoyle states \cite{ZimmermanPhD, pp2}.

Unfortunately, such `rare' physics processes do not always result in measurable patterns that are sufficiently unconventional, and therefore backgrounds can abound. In some cases the presence of a well defined interaction vertex (e.g., in accelerator-driven nuclear reactions), or the detector capability of pointing on and off-source (eg., solar neutrinos, axions and axion-like particles, or even weakly interacting massive particles gravitationally trapped in the galaxy -as we will see) can provide a high event purity. In others, the absence of those handles coupled to the need of very long measurement times, make background sources conspire in unpredictable ways, eventually mimicking the process under investigation. Defeated by the burden of irreproducible claims \cite{KK,DAMA}, some experimentalists are turning to the search of defining event features to beat possible uncontrolled systematic errors.\footnote{It is important to note that this is not the only solution to this riddle: KIMS \cite{Kims} and ANAIS \cite{Anais} collaborations are currently pursuing a verification of earlier claims of dark matter observation by the DAMA/LIBRA collaboration in NaI/CsI(Tl) crystals, whereas the GERDA experiment has recently excluded a previous $\beta\beta0\nu$ observation in $^{76}$Ge \cite{GerdaExcl}. The potential importance of finding additional event features for the confirmation of any future positive claim seems hardly in question, though.}

The 3-dimensional images produced by the ionization trails left by end-state particles are amongst the most powerful patterns that nature provide us with (Fig. \ref{Images}).
Indeed, good imaging capabilities do not only defeat systematics, they can increase statistical significance as well: in the absence of backgrounds the sensitivity of a typical discovery experiment improves linearly with the system's exposure ($Mt$) but only with its square root otherwise \cite{JJSense}.
Therefore, when aimed at the reconstruction of rare processes, the notion of imaging can be associated to the ability
of recognizing one, or several, event patterns that may ultimately lead to unambiguous (background-free) identification. Roughly speaking, they can be
 classified in three types:

\begin{enumerate}
\item Track `topology'. The topology/shape of the event is usually its most defining feature: if there is a reaction line, time, or a vertex that is a priori known and can be reconstructed, it provides a highly unambiguous signature (Fig. \ref{Images}-a,c,i). The tracks' direction can be correlated with the direction of impact of an incoming particle (eg. Fig. \ref{Images}-b,f,g,i), the relative orientation of the reaction plane with its polarization (Fig. \ref{Images}-g). A straight track suggests high momentum (Fig. \ref{Images}-a,f,g,h), a wavy track a low one (Fig. \ref{Images}-b,d,e), while a curved trajectory in a magnetic field provides its value, directly (Fig. \ref{Images}-i). The ability to separate two closely approaching tracks is a handle in many practical situations (Fig. \ref{Images}-f,g,i). The main enemies here are multiple scattering (Fig. \ref{Images}-b,d) and diffusion (Fig. \ref{Images}-b,g), as well as an insufficient segmentation and/or an inadequately broad `point spread function' ($\mathcal{PSF}$) at the image plane.
\item Differential energy loss (or `$d{\varepsilon}/dx$'). If track topology refers to patterns in space, differential energy loss refers to how those patterns are drawn. The presence of high ionization density `blobs' towards the track's end-point (Bragg's peak) combined with enhanced multiple scattering is one of the most pursued features (Fig. \ref{Images}-d). When the products are highly ionizing nuclei, however, the entire energy loss profile should be used to reconstruct and identify the emerging species (Fig. \ref{Images}-a,b,c,i), and its direction of motion (Fig. \ref{Images}-b). This latter feature is known in the field of direct dark matter detection as the `head-tail' signature.\footnote{Perhaps anti-intuitively, the highest energy is deposited at the beginning of the track and the lowest at its end. Part of the effect is due to the fact that the event's energy is in this case to the left of the Bethe-Bloch maximum.} Characteristic x-ray emission can lead to a displaced ionization cluster and it also points to the origin of the event. The main enemy here is noise, and the main figure the signal to noise ratio ($\mathcal{S}/\mathcal{N}$) per sensor, at the image plane.

\begin{figure}[ht!!!]
\centering
\includegraphics*[width=\linewidth]{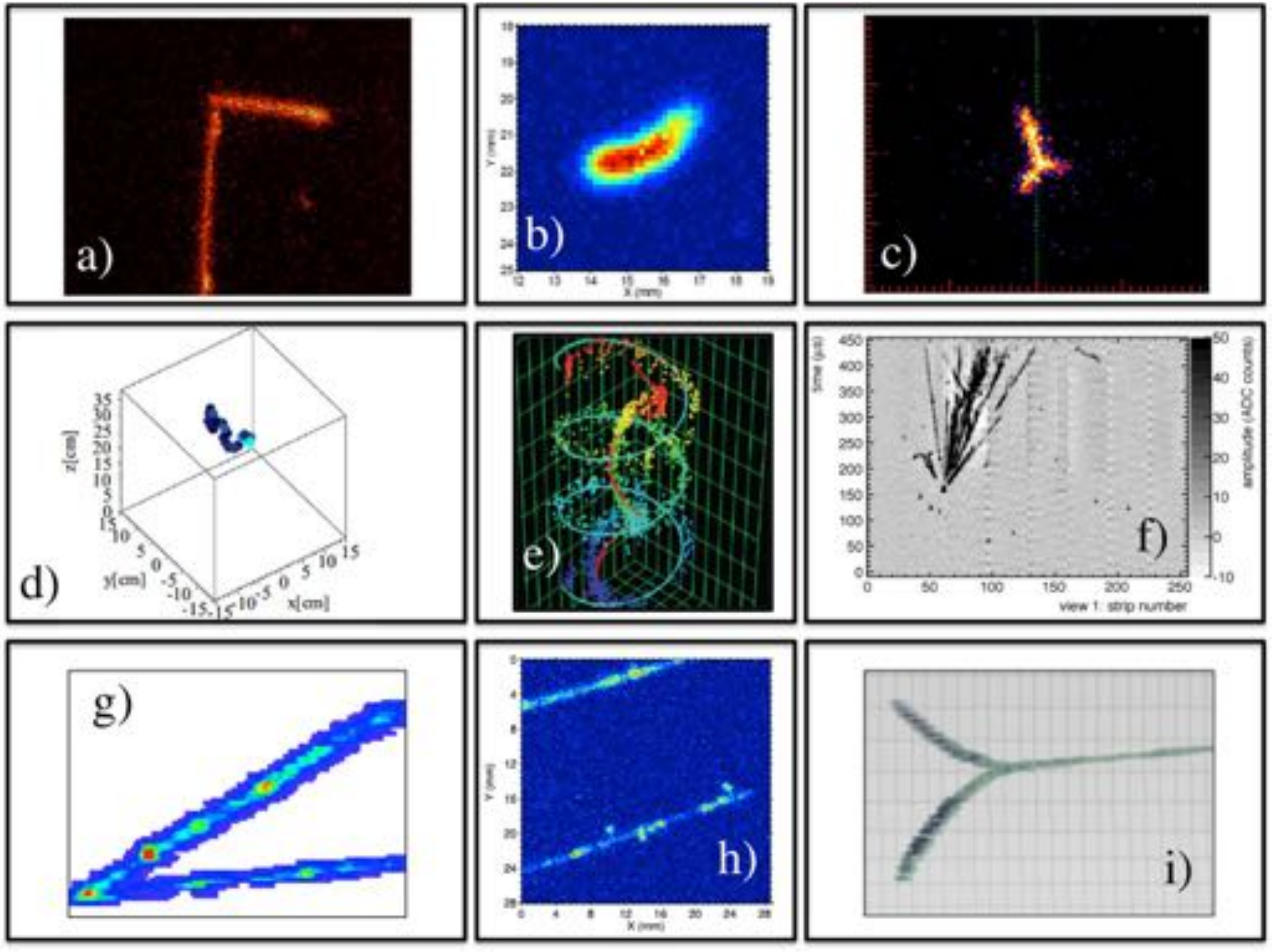}
\caption{Some representative images obtained with state of the art TPCs employed outside collider physics:
a) $\beta$-delayed proton emission from $^{46}$Fe \cite{OTPClast}; b) a low energy C or F nucleus ($\varepsilon=214$\,keV) recoiling against a neutron \cite{Phan}; c) a triple $\alpha$ event produced after the reaction $^{12}$C($\gamma$,$\alpha$)$^8$Be in a 150\,mbar CO$_2$/N$_2$ active target \cite{ZimmermanPhD}; d) a 1.275\,MeV photoelectron from a $^{22}$Na source in a 10\,bar xenon/TMA admixture \cite{NEXTMM}; e) a low energy electron spiraling in a magnetic field, reconstructed in a $\sim1$\,cm$^3$ mini-TPC with an InGrid device \cite{CAST}; f) a cosmic ray shower obtained with the dual-phase argon TPC of the WA105 collaboration \cite{Badertscher:2013wm}; g) pair production of a 74.3\,GeV photon in the HARPO polarimeter, based on pressurized argon \cite{HARPO}; h) electrons with energies above 30\,keV, reconstructed in 50\,mbar of CF$_4$; i) elastic scattering between two $\alpha$-particles at around 1\,bar, reconstructed with the AT-TPC \cite{AT-TPC2}.}
\label{Images}
\end{figure}

\item Global event variables. The relation between range and ionization provides a direct way to separate low energy (10's of keV) electrons from nuclei at low gas density (Fig. \ref{Images}-b,h), or $\alpha$-particles from MeV electrons (\cite{NEXTrecoFirst}, for instance). This simple and powerful procedure will allow, in general, to identify different particle and nuclear species in a variety of experimental situations, whenever the reaction products can be contained inside the chamber. But even for point-like events, the combination of ionization and scintillation signals can provide excellent particle identification capabilities both in gas and liquid \cite{LZ, JoshHPNeutrons}. With enough precision, a calorimetric measurement based on the ionization signal can become an essential handle by itself, like in the reconstruction of $\beta\beta0\nu$ decays \cite{NEXT1}. Lastly, the start time of the event ($T_0$) allows the determination of the absolute position along the drift coordinate, and it is, to date, the only practical way to eliminate backgrounds emanating from the electrodes.
\end{enumerate}

Given the difficulty of reconstructing and unambiguously identifying events like the ones shown in Fig. \ref{Images}, it is not surprising that deep neural nets have joined the effort \cite{Baldi, Aurisano, CERNseminar}, yielding (in some particularly difficult cases) a considerable advantage over human-based analysis \cite{JoshNN}. Particle flow algorithms typically used at colliders \cite{Pandora1} have been successfully applied to the reconstruction of the complex topologies arising in neutrino interactions, too. Indeed, for an experiment that needs to approach zero-background conditions, every bit of information going into the reconstructed image can be significant. This is discussed in detail in the next two subsections.

\subsection{Generation of primary ionization}

Assuming a fully contained particle or nucleus of energy $\varepsilon$, ionization is produced on average as $\bar{n}_e=\varepsilon/W_I$, with an intrinsic spread $\sigma_{{n}_e}/\bar{n}_e = \sqrt{F_e/\bar{n}_e}$.\footnote{We will often use in the following $\Delta\varepsilon$ instead of $\varepsilon$ in order to describe simultaneously the case of fully contained events ($\Delta\varepsilon\equiv\varepsilon$) as well as partially contained ones, for which $\Delta\varepsilon$ refers to the average energy lost before exiting the gas volume under consideration. A discussion about the ionization fluctuations in this latter case can be found later in text. Since full containment is the preferred situation in the TPCs discussed hereafter, it is used here as a reference case.} $W_I$ is the average energy to create an electron-ion pair and $F_e$ is the Fano factor. For either gas or liquid, $W_I$ stays commonly in the range 15-40\,eV (i.e., 25-65\,e$^-$/keV), and $F_e$ around 0.1-0.25 \cite{Aprile, ICRU}. $W_I$ ultimately derives from the ionization cross sections and so, to first order, simple additivity rules apply to gas mixtures \cite{Sauli77}:
\beq
Q_y \equiv \frac{1}{W_I} = \sum_i \frac{f_i}{W_{I\!,i}} \label{FirstEq}
\eeq
where $Q_y$ is the charge yield and $f_i$ refers to the relative (molar) fraction of each species.

\subsubsection{Penning transfers and the Jesse effect}
In eq. \ref{FirstEq} it has been assumed that the $W_{I\!,i}$ values describe well the ionization response for pure gases, but the charge yield, $Q_y$, has been introduced for the admixture instead of $1/W_I$, in order to consider more general situations. $Q_y $ can be increased for instance if energy transfers between species are important, in particular from excited states ($N_{ex}$) into ionization, a.k.a. as Penning transfer:
\beq
Q_y = \sum_i \frac{f_i}{W_{I\!,i}} + \frac{1}{\Delta\varepsilon}\sum_i\sum_j N_{ex\!, ij} \times r_{ij}(\vec{P}) \label{Jesse}
\eeq
with $r_{ij}$ referring to the transfer probability for each excited state ($j$) of each species ($i$), and $\vec{P}$ is the array of partial pressures of the components in the admixture. Historically, this increase in ionization stemming from Penning-transfers is named
`Jesse effect' \cite{Jesse}.\footnote{In addition to the terms in eq. \ref{Jesse}, the ionization of the additive by `sub-excitation electrons' (i.e, with energy below the energy of the first excited state of the noble gas) was pointed out by Platzman in \cite{Platzman}. Except for high additive concentrations, this latter effect will represent a minor contribution in general.} By analogy with $W_I$, the average energy needed to promote a gas species to an excited level $j$ can be approximated through:
\beq
\frac{N_{ex,ij}}{\Delta\varepsilon} = \frac{f_i}{W_{ex,ij}}
\eeq
and so eq. \ref{FirstEq} reads:
\beq
Q_y = \sum_i \frac{f_i}{W_{I\!,i}} + \sum_i \sum_j \frac{f_i}{W_{ex,ij}} ~ r_{ij}(\vec{P})
\eeq

In the practice of gaseous detectors, it is notably difficult to isolate the transfer probabilities for individual states, and effective coefficients per species are used instead:
\beq
Q_y \simeq \sum_i f_i \left( \frac{1}{W_{I\!,i}} + \frac{1}{W_{ex\!,i}} ~ r_{i}(\vec{P}) \right)
\eeq

\begin{figure}[h!!!]
\centering
\includegraphics*[width=\linewidth]{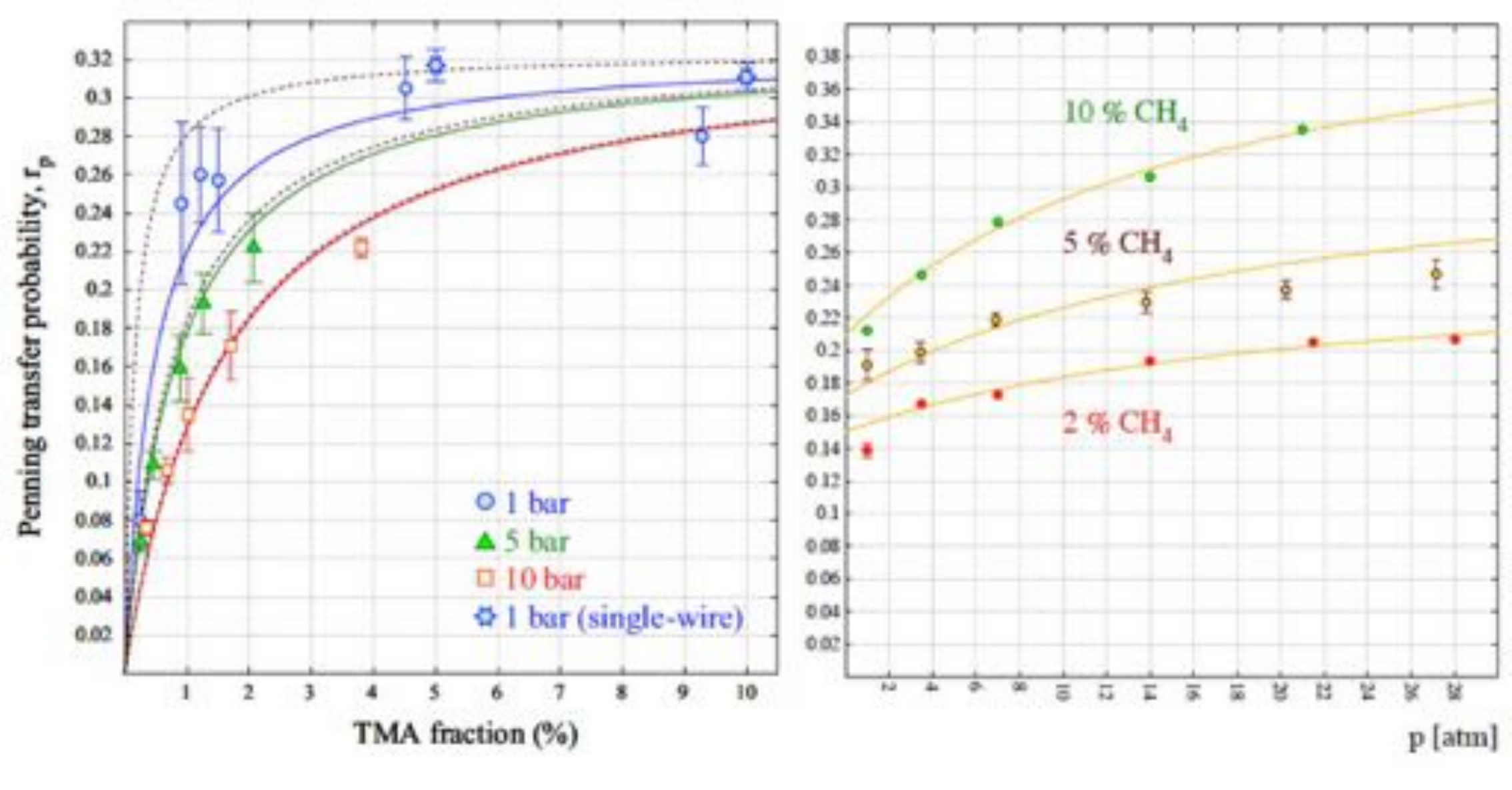}
\caption{Effective Penning transfer probability obtained for two typical Xe \cite{XePen} (left) and Ar \cite{ArPen} (right) -based binary mixtures (the additive is TMA and CH$_4$, respectively) under avalanche conditions. The trends represent microphysical models that include transfer rates as well as the competing channels. Note the authors' convention for the effective transfer coefficient ($r_p$) instead of the one chosen in text ($r$).}
\label{Penning}
\end{figure}

Penning transfer probabilities in mixtures of interest to TPCs have been the subject of extensive research in recent years, through the study of the avalanche process. As a result, they are nowadays available for binary mixtures ($r_{i}(\vec{P})\equiv r(f_i,P))$ based on Ne \cite{NePen}, Ar \cite{ArPen} and Xe \cite{XePen} (e.g., Fig. \ref{Penning}).\footnote{By assuming a single transfer coefficient $r$ we are neglecting homonuclear associative ionization of highly excited noble gas states \cite{Molnar}: $X^{**}+X \rightarrow X_2^+ + e^-$. For small additive concentrations we may assume that this process is little affected, therefore being implicitly included in the corresponding $W_{I\!,i}$.} According to the analysis in \cite{ArPen}, for instance, Penning transfer probability in Ar/Xe and Ar/C$_2$H$_2$ mixtures can approach 100\%. On the other hand, and despite the lower transfer probability observed in Xe/TMA admixtures, an approximate agreement was found in that case with a direct measurement of the Jesse effect \cite{TMAYasu}. Given the considerable interest in lowering the detection threshold in many of the TPCs discussed here, and the long history of studies around these phenomena,\footnote{Nearly a factor of two increase in $Q_y$ was observed for helium mixtures in the presence of sub-\% additives, as far back as 1954 \cite{Jesse}.} it is perhaps surprising that Jesse effect has not found application in TPCs, yet. Part of the difficulty is related to the phenomenon discussed next.

\subsubsection{Photo-ionization}

A separate treatment of Jesse effect and photo-ionization is documented in literature since very early on (e.g. \cite{Person}). Having just slightly different energy thresholds, they can be difficult to distinguish through a calorimetric measurement \cite{TMATEAinliquid}, although the negative impact of this latter effect on image reconstruction can be very substantial \cite{BreskinTeaSeminal}. Its inclusion leads, with approximate character, to the following expression:
\bear
& Q_y \simeq & \sum_i f_i \Big(\frac{1}{W_{I\!,i}} + \frac{1}{W_{ex\!,i}} \Big[ ~ r_{i}(\vec{P}) +... \nonumber \\
&& \mathcal{P}_{scin,i}(\vec{P}) \times (1-e^{-L^*/\Lambda_I}) \Big]\Big) \label{eq+photoI}
\eear
This new equation introduces the scintillation probability ($\mathcal{P}_{scin,i}$) as the probability that, after an excited state has been populated, emission of a photon ensues: $\mathcal{P}_{scin,i}=N_{\gamma,i}/N_{ex,i}$.\footnote{In the field of gaseous detectors it is common to refer to the quantity ($1-P_{scin}$) as the `quenching' probability, that is an useful concept provided `well quenched' gases exhibit higher working gains. In order to avoid confusion with the definition of quenching preferred in text (eqs. \ref{Q}, \ref{L}) we will favour the use of the magnitude ${P}_{scin}$, or will refer to this effect as `excimer/VUV-quenching'. A situation that is rarely found under practical detector operation is that in which $\mathcal{P}_{scin}>1$, and that can occur in the special case of a broad range of emission wavelengths being considered.} Although this expression again does not consider all excited states explicitly, it approximates well the situation in a noble gas for typical TPC conditions, since virtually all noble gas' excited states are precursors to scintillation \cite{Filomena, oliveira}. The emission of a photon leads to the de-localization of information (in time and space), depending on the mean free path for photo-ionization, $\Lambda_I$, the characteristic system dimensions $L^*$, and the time constant of the scintillation process. Photo-ionization has been used to explain the large ionization enhancement observed in Xe-TMA and Xe-TEA mixtures in liquid phase \cite{TMATEAinliquid}, for instance. Compared to photo-ionization, Penning transfers between species are much faster (and localized) and do not lead to image blurring.

\subsubsection{Charge recombination}

For high ionization densities and/or low diffusion conditions, the recombination of the electrons and ions released can modify the situation further \cite{Ramsey, Diana, Icarus, BoxModel}. The effect depends strongly on the electric field, and partly on the track topology and orientation \cite{Kanne, Jaffe}. Assuming that all processes in eq. \ref{eq+photoI} suffer the effect in a similar way, and defining $\mathcal{R}$ as the fraction of charge that undergoes recombination, one obtains:
\beq
Q_y \simeq (1-\mathcal{R}) \Big( \sum_i \frac{f_i}{W_{I,i}} + \frac{r}{W_{ex,p}} + \frac{\mathcal{P}_{scin}}{W_{ex,\gamma}} \times  \mathcal{P}_{p.i.}\Big) \label{Reco}
\eeq
that is the final formula that we will use. We have made in eq. \ref{Reco} a set of additional simplifying assumptions by effectively re-defining $r$ as the Penning transfer probability, $\Delta\varepsilon/W_{ex,p}$ as the number of excited states that can undergo Penning transfer and $\Delta\varepsilon/W_{ex,\gamma}$ as the number of excited states that are precursors to scintillation. $\mathcal{P}_{p.i.}$ is the short-hand notation for the gas photo-ionization probability. The magnitudes without sub-index refer to effective values for the mixture and need to be obtained by a direct measurement or simulation. As it is self-evident from eq. \ref{Reco}, they are not expected to follow general additivity rules, contrary to the $W_{I,i}$ values.

\begin{figure}[h!!!]
\centering
\includegraphics*[width=8cm]{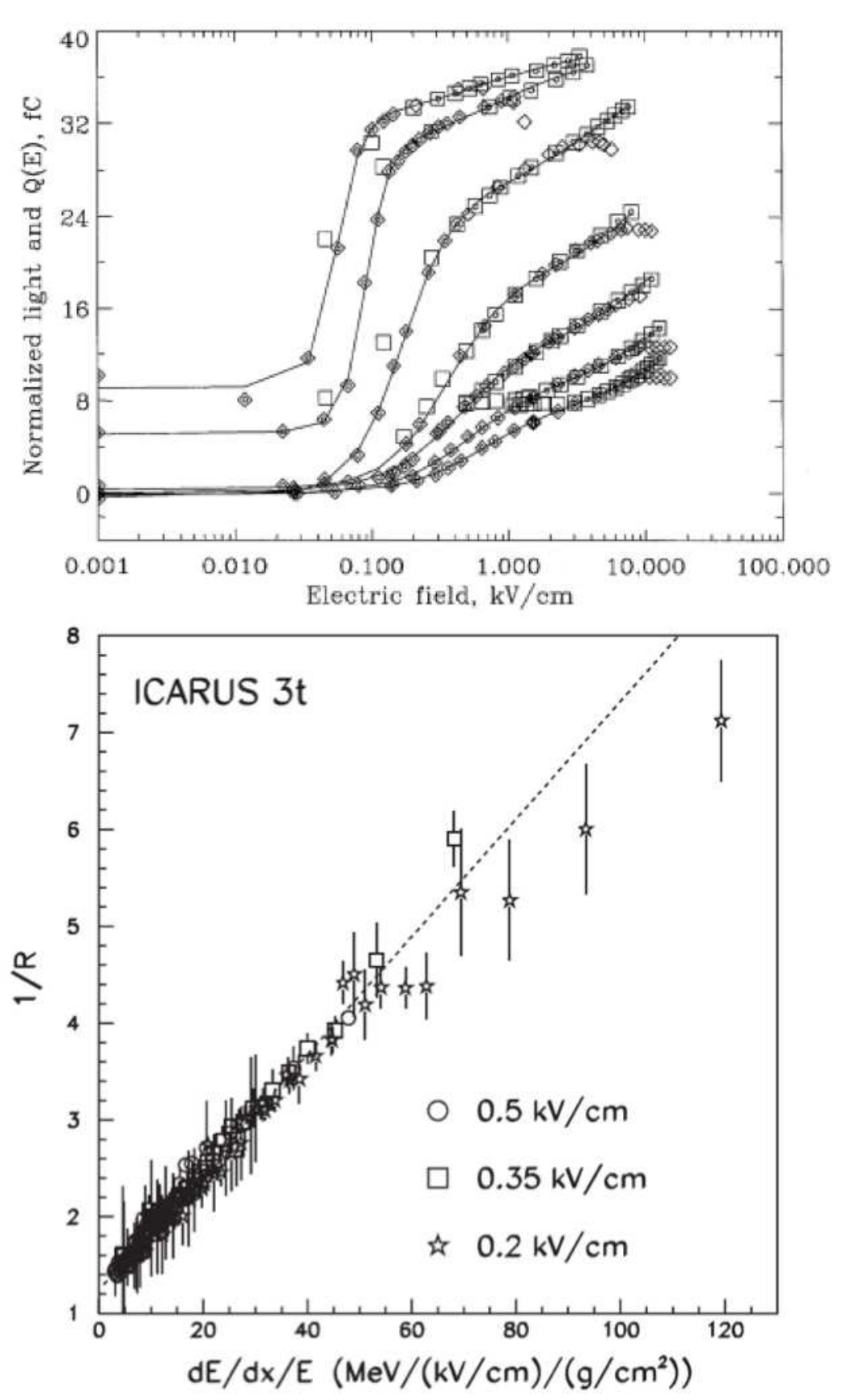}
\caption{Top: charge collected for $\alpha$-particles as a function of the electric field (diamonds), for xenon densities of 0.06, 0.08, 0.21, 0.33, 0.5, 0.63, 0.74 g/cm$^3$ (series are read from top to bottom), \cite{Ramsey}. For reference, 0.06\,g/cm$^3$ corresponds approximately to the density of 10\,bar of xenon at 20\,$^\circ$C. Bottom: recombination factor ($R\equiv (1-\mathcal{R})$ using the notation in text) as a function of the theoretical energy loss (divided by field), obtained by the ICARUS collaboration \cite{Icarus}. The dashed line in the lower figure is a fit to eq. \ref{BirksLaw}, and the continuous lines in the upper figure represent the result of an iterative fitting procedure based on a similar relation.}
\label{RecoFig}
\end{figure}

To first order, charge recombination can be analytically approximated by the high (gas) or low (liquid) diffusion limits of the hydrodynamic equations \cite{Jaffe, Kramers, Chabod1},\footnote{These solutions assume the presence of an additional recombination term in the hydrodynamic equations (found later in eq. \ref{Eq_hydro}) as $-k \!\cdot\! \mathcal{N}_e \!\cdot\! \mathcal{N}_i$, with $\mathcal{N}_{e(i)}$ being the number density of electrons (ions). Given the high medium density, the solution in liquid may be approximated by neglecting diffusion. The solution in gas, on the other hand, can be obtained by assuming that recombination is small and $\mathcal{N}_{e(i)}$ can be approximated in the hydrodynamic equation through the solutions obtained when neglecting it. This is the main assumption of what has become known as the `Jaff\'e model' \cite{Jaffe}.} or through numerical evaluation \cite{Largon}. Most of the available information is experimental, though, given the difficulty of including additional effects like the recombination produced in the inmediate vicinity of the parent ions \cite{Onsager} or $\delta$-rays \cite{AprileReco}. As a result, small modifications of the following phenomenological formula are frequently used in practice, provided it can approximate well measurements performed both in liquid and gas phase \cite{XePen,Ramsey,Icarus,AprileReco}:
\beq
(1-\mathcal{R}) \sim \frac{1}{1+\frac{k}{E_d} \cdot d\varepsilon/dx} \label{BirksLaw}
\eeq
with $k$ being a fitting constant. The choice of formula \ref{BirksLaw} is not completely coincidental: it corresponds to the high drift field ($E_d$) approximation of the Jaff\'e solution for charge recombination within a column of ionized gas, obtained as early as 1912.

In the limit of low diffusion (e.g., liquid phase), on the other hand, a particular solution of the hydrodynamic equations can be found under the so called `Thomas and Immel' model \cite{Cao}, that results in the functional relation:\footnote{Eq. \ref{BoxModel} is also referred to as the `box model' or the `Boag empirical formula' \cite{BoxModel}.}
\beq
(1-\mathcal{R}) \sim \frac{E_d}{k'}\ln(1+\frac{k'}{E_d}) \label{BoxModel}
\eeq
and $k'$ is again a fitting constant. The reader should consider the above formulas as a mere guidance, as it is frequent to find them with additional terms. It can be readily noted, however, that both expressions agree again at high fields!. Some arguments explaining this fairly (albeit approximate) universal behaviour can be found in a recent theoretical work \cite{BoxModel}.

Illustratively, Fig. \ref{RecoFig}-up shows the effect of charge recombination as a function of the electric field for $\alpha$-particles in high pressure xenon, after \cite{Ramsey}. In the lower figure, the complementary of the recombination probability (note the authors' convention: $R\equiv (1-\mathcal{R}))$ measured in liquid argon is shown, as a function of the differential energy loss divided by drift field. A fit to eq. \ref{BirksLaw} has been overlaid (dashed line).

Despite being usually an undesirable feature, charge recombination plays an important technological role in some modern TPCs, in conjunction with the effect discussed next.

\subsubsection{Ionization quenching}

Experimentally, an important pattern is observed during the evaluation of $Q_y$: the $W_I$-value obtained in pure gases for low ionizing particles, e.g., `minimum ionizing particles' (mips), keV-electrons and x-rays is a rather well defined magnitude, irrespective from the particle energy and the gas density \cite{ICRU}. Additionally, charge recombination is usually very small in those cases (\cite{Bolot},\cite{Balan}) and Penning transfers as well as photo-ionization can be kept within a few \% except if the admixed gases are purposely chosen to maximize the effect \cite{Agra}. This set of facts is used regularly for the calibration of gaseous detectors as well as for the experimental determination of the $W_I$ values, and has a large historical background \cite{Sauli77}. Therefore, deviations for a particle `$X$' relative to the `standard' situation for x-rays/electrons are commonly described through an `ionization quenching factor' $\mathcal{Q}$:
\beq
\mathcal{Q} = \frac{Q_{y}|_X}{Q_{y}|_{el}} \label{Q}
\eeq

If $\mathcal{Q}$ is experimentally evaluated in recombination-free conditions (i.e., $E_d\rightarrow \infty$), it can be readily interpreted as the ratio $W_I/W_I|_{X}$. In the 60's, Lindhard, Scharff and Schi${\o}$tt developed an unified theory (hereafter LSS \cite{Lindhard}) aimed at the accurate evaluation of the two main contributions to the energy loss by a charged particle, i.e.: i) losses to the electrons of the surrounding atoms ($\mathcal{S}_e = d\varepsilon/dx|_e$) and ii) losses into translatory movement of the atom as a whole ($\mathcal{S}_n = d\varepsilon/dx|_n$). This allows defining $\mathcal{Q}$ as:
\beq
\mathcal{Q} \simeq \frac{d\varepsilon/dx|_e}{d\varepsilon/dx|_e + ~ d\varepsilon/dx|_n} \label{LSSeq}
\eeq

At low energies, the electronic stopping power $\mathcal{S}_e$ is proportional to the particle velocity and the so called `nuclear stopping power' $\mathcal{S}_n$ may be approximated by a constant, leading thus to $\mathcal{Q} \sim \sqrt{\varepsilon}/(k \sqrt{\varepsilon}+1)$ \cite{Lindhard}. Consequently, ionization transfers become unlikely for slow particles, and the ensuing cascade only enhances the effect.\footnote{Both the energized nuclei and the released electrons will continue the ionization process, hence eq. \ref{LSSeq} is only approximate. Additionally, a more rigourous treatment requires identifying the fraction of energy that goes into excitation and ionization in the term $d\varepsilon/dx|_e$. The associated scintillation quenching factor is introduced later in text.} In practice, $\mathcal{Q}$ can be particularly small for heavy nuclei with energies $\lesssim 1$\,MeV: values of $\mathcal{Q}=0.17$ have been estimated for instance in high pressure xenon for Xe nuclei up to 150\,keV \cite{JoshHPNeutrons}, and they vary in the range $\mathcal{Q}=0.2$-$0.5$ for C, F nuclei with energies between $5$ and $55$\,keV, in CF$_4$ gas at a pressure around 50\,mbar \cite{Santos}.

While the simulation code SRIM/TRIM \cite{SRIM} includes a parameterization of the LSS model for nuclei, and it is often used to that aim, an approximate expression is sometimes found, for the case where the atomic number of the moving nuclei and that of the target are similar \cite{Lindhard}:
\beq
\mathcal{Q} = \frac{k \cdot g(\varepsilon)}{1 + k \cdot g(\varepsilon)} \label{LSSeq2}
\eeq
where $k$ is  a constant, and $g(\varepsilon)$ a slowly increasing function of the particle's energy.

\begin{figure}[h!!!]
\centering
\includegraphics*[width=\linewidth]{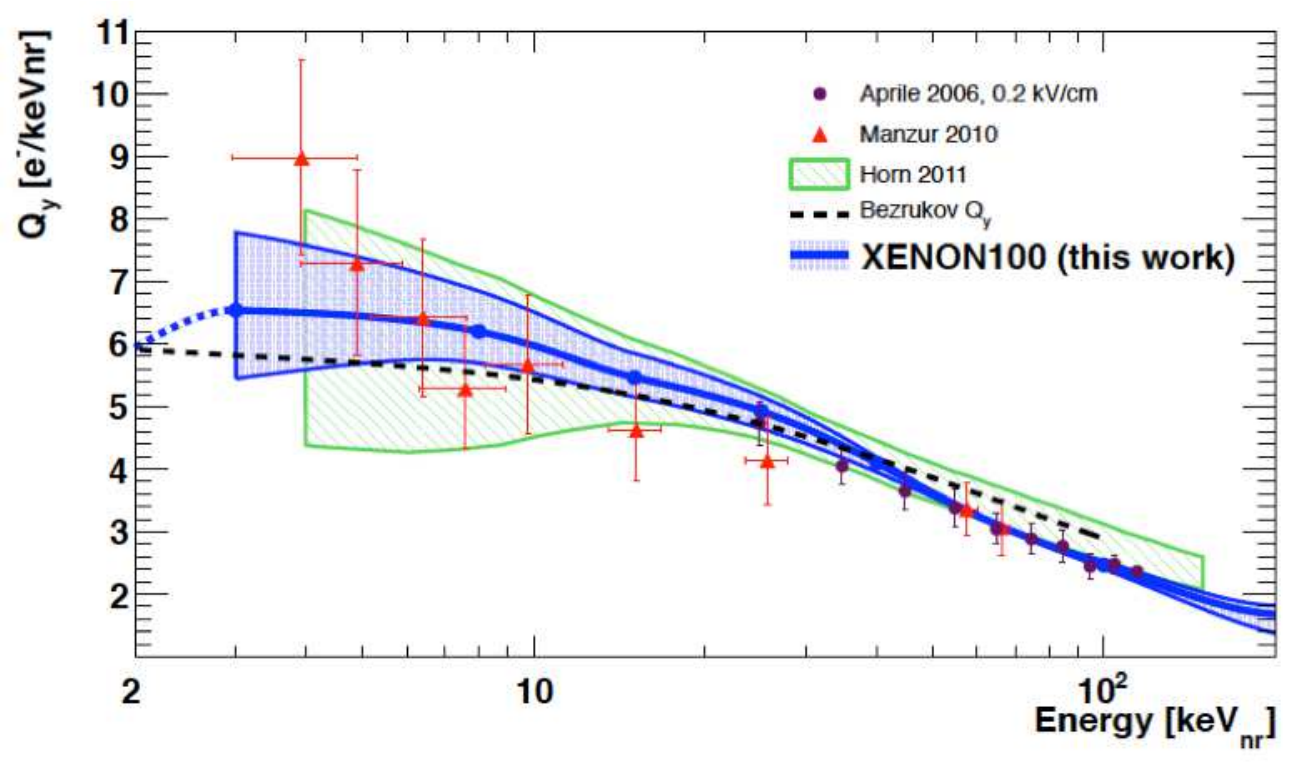}
\caption{Figure illustrating the combined effect of ionization quenching and charge recombination, as measured by the XENON100 dark matter experiment (reprinted from \cite{XENON100}): shown is the charge yield ($Q_y$) for Xe nuclei in liquid xenon (at a drift field $E_d=530$\,V/cm) as a function of their energy in keV (dubbed `keVnr'). As a reference, the charge yield for electrons is within $Q_y=60$-$70$\,e$^-$/keV in the same energy range \cite{NEST}.}
\label{Qy_XENON}
\end{figure}

Ionization (and scintillation, see later) quenching are the main handles for distinguishing recoiling nuclei from electrons in direct-detection dark matter experiments (Fig. \ref{Qy_XENON}). Its impact in one of the most pressing problems of particle physics has led some experimentalists to wonder whether the presence of additional terms in eq. \ref{Reco} can be used favourably to enhance their TPC response for other applications, too. This has given birth to proposals for using columnar recombination as a handle for determining nuclear recoil directionality \cite{ColReco, Diana, Li}, the possibility of sub-Fano mixtures in the search of double-beta decay \cite{DaveSub}, or photo-ionizable mixtures that would allow grabbing the primary scintillation in the TPC itself \cite{SubPenning}.

The magnitudes introduced in text in order to characterize the primary ionization depend (for a given mixture) on field, pressure and particle type, qualitatively as summarized in table \ref{TableQy}.

\begin{table}[h]
  \centering
  \begin{tabular}{|c|c|c|c|c|}
     \hline
                & ~ particle type ~ & ~ energy ~ & ~ field ~ & ~ density ~\\
     \hline
     ~ $\mathcal{R}$                & ~ yes ~       & ~ yes ~   & ~ yes ~   & ~ yes ~        \\
     ~ $W_{I} ^{~*a}$               & ~ small ~     & ~ small ~    & ~ no ~    & ~ no$^{*b}$ ~  \\
     ~ $r$                          & ~ no ~        & ~ no ~    & ~ small ~ & ~ yes ~        \\
     ~ $W_{ex} ^{~~*a}$             & ~ small ~     & ~ small ~    & ~ no ~    & ~ no$^{*b}$ ~  \\
     ~ $\mathcal{P}_{scin}$         &  ~ no ~       & ~ no ~    & ~ no ~    & ~ yes ~        \\
     ~ $\mathcal{P}_{p.i.(p.a.)}$   & ~ no ~        & ~ no ~    & ~ no ~    & ~ yes ~        \\
     ~ $F_e$                        & ~ small ~     & ~ small ~ & ~ small ~ & ~ no$^{*b}$ ~  \\
     ~ $F_\gamma$                   & ~ small ~     & ~ small ~ & ~ small ~ & ~ no$^{*b}$ ~  \\
     ~ $\mathcal{Q}$                & ~ yes ~       & ~ yes ~   & ~ no ~    & ~ yes ~        \\
     ~ $\mathcal{L}$                & ~ yes ~       & ~ yes ~   & ~ no ~    & ~ yes ~        \\
     \hline
   \end{tabular}

  \caption{Qualitative dependencies of the main variables ruling the ionization and scintillation yields in TPCs, under typical operating conditions. From top to bottom: recombination probability $\mathcal{R}$; average energy to create an electron-ion pair $W_{I}$; Penning transfer probability $r$; average energy to create an excited state (or an exciton) $W_{ex}$; scintillation probability $\mathcal{P}_{scin}$; photo-ionization (absorption) probability $\mathcal{P}_{p.i.(p.a.)}$; ionization Fano factor $F_e$; scintillation Fano factor $F_\gamma$; ionization quenching factor $\mathcal{Q}$; scintillation quenching factor $\mathcal{L}$.}\label{TableQy}
     \begin{tablenotes}
    \item[1] $^{*a}$ Defined here by convention for low-ionizing radiation. Deviations from this situation should be included through $\mathcal{Q}$, $\mathcal{L}$, according to definitions in text.
    \item[1] $^{*b}$ Except close to the phase transition or in condensed phase.
  \end{tablenotes}
\end{table}

\subsubsection{Fluctuations in the ionization process}

\begin{figure}[h!!!]
\centering
\includegraphics*[width=7.5cm]{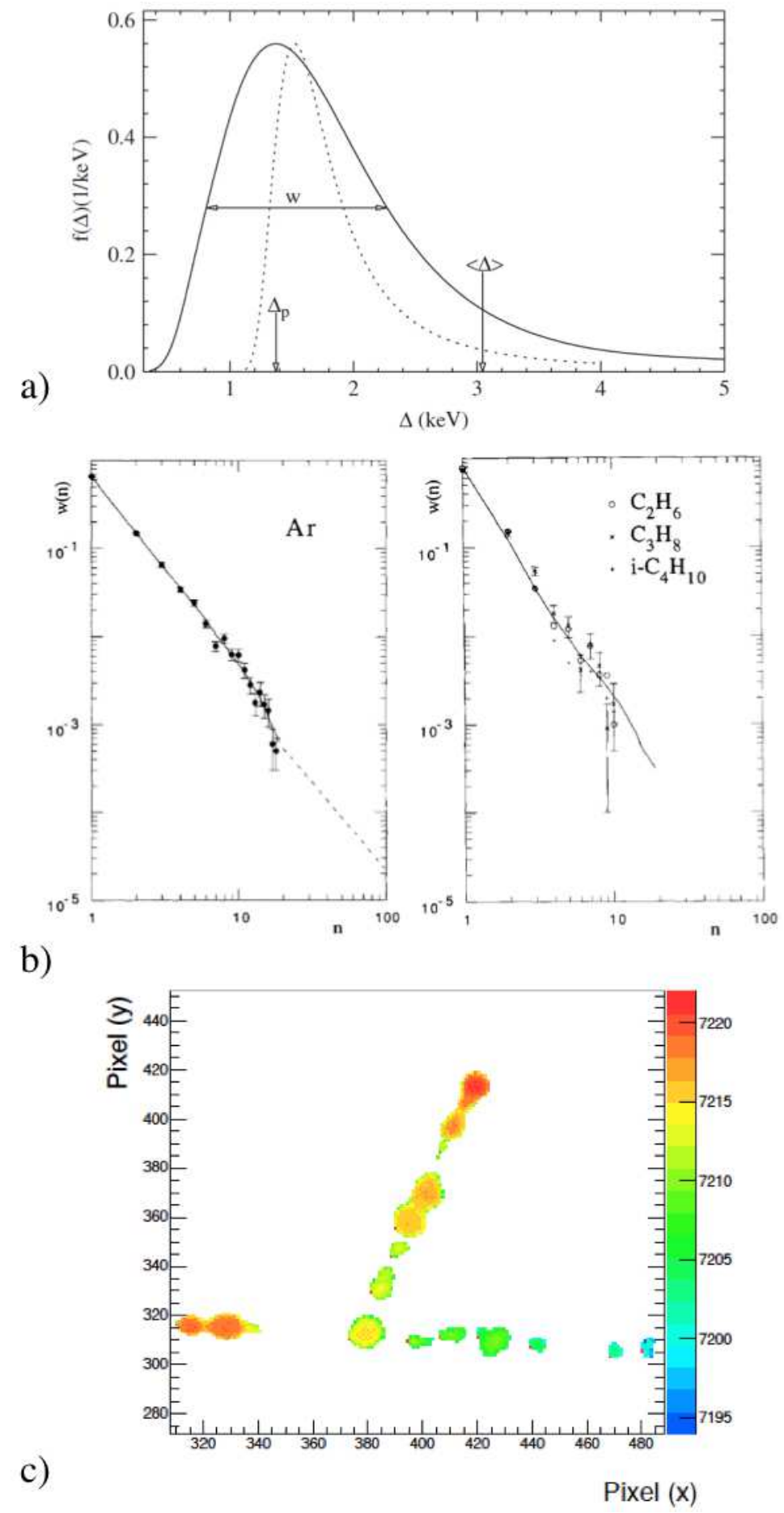}
\caption{Three fundamental (related) aspects of the energy loss process by charged particles, with focus on minimum ionizing ones (mips): a) energy loss distribution of singly charged particles with $\gamma\beta=3.6$, when traversing $\Delta{x}=1.2\,$cm of argon gas at around 1\,bar \cite{Bichsel}. The mean value of the distribution is denoted by $<\!\Delta\!>$ and $\Delta_p$ is the most probable value (using the notation in text $<\!\Delta\!> \equiv \Delta \varepsilon$ and $\bar{n}_e = Q_y \Delta \varepsilon$). The value $\Delta\varepsilon/\Delta{x}$ corresponds to the average energy loss, estimated by the Bethe-Bloch formula. The function obtained with the original Landau formalism is shown by a dashed line, mostly to illustrate its approximate character; b) cluster size distribution $w(n)$ (probability of producing a given number of ionization electrons $n$ upon a single ionizing encounter), obtained for fast electrons ($\beta=0.8$-$0.97$) with a setup specially commissioned at Heidelberg University in the early 90's \cite{Fischle}. Results for argon (left) and some typical hydrocarbons (right) are shown. The continuous line is a fit and the dashed line represents an extrapolation proposed by the authors; c) a candidate $\delta$-electron emitted by a 120\,GeV proton, as reconstructed with a 3-GEM detector coupled to a Timepix chip, in a gas mixture based on Ar/CO$_2$/CF$_4$ \cite{GEMPIX}. The image displays several ionization clusters along the particle trajectory (coming from the left). Although a plausible track model can be easily drawn in this image, the task will be considerably hardened if more conventional $1$-$10$\,mm $\times$ $1$-$10$\,mm pixelated sensors would be used (instead of the $50\,\mu$m$\times 50\,\mu$m employed for this measurement).}
\label{LandauFig}
\end{figure}

It must be noted that $Q_y$ accounts for the ionization produced directly by the impinging particle, as well as for that produced by the released electrons themselves. It is mainly through this second process that large ionization `clusters' can be created (e.g. \cite{BreskinIonization, Fischle}).\footnote{For small cluster sizes, multiple ionization of a single atom/molecule through direct interaction with the passing particle represents an important contribution too.} Eventual distortions of the ionization trail caused by energetic $\delta$-electrons as well as by the sheer point-to-point fluctuations along the trail can severely impair the reconstruction accuracy, and even the calorimetric response if the particle track is not fully contained inside the chamber. The effect, illustrated in Figs. \ref{LandauFig}a-c, is particularly important for minimum ionizing particles and can not be neglected in TPCs when used as particle trackers.

Throughout this text we will refer to these fluctuations of the primary ionization as `Landau fluctuations', and the reader is referred to the recent review work \cite{Bichsel} for a detailed treatment of the underlying physics processes. A discussion about the impact of these fluctuations for track reconstruction is given later in section \ref{PosResSec}, while their role in the calorimetric measurement of non-contained particle tracks is discussed in section \ref{LandauSec}.

Naturally, the unavoidable presence of such secondary ionization processes renders the additive relation in eq. \ref{FirstEq}, even if neglecting additional contributions, just a (good) approximation. Illustratively, we can write for pure gases \cite{Fischle}:
\bear
&\bar{n}_e      &=  Q_y \Delta\varepsilon = \bar{n}_{cl} \cdot \bar{n}_{e/cl} \\
&\bar{n}_{cl}   &= \sum^\infty_{k=0} k \cdot \mathcal{P}(k) \\
&\bar{n}_{e/cl} &= \sum^\infty_{i=1} i \cdot w(i) \\\label{PrimSec}
\eear
that separates the average number of ionization encounters, or clusters ($\bar{n}_{cl}$), and the average number of electrons produced per cluster ($\bar{n}_{e/cl}$). $\mathcal{P}(k)$ is the Poisson probability for $k$ ionizing encounters and $w(i)$ the probability
that $i$ electrons are produced for each of them (see, e.g., Fig. \ref{LandauFig}b). For pure gases, $\bar{n}_{e/cl}$ can be found in the range 1.5-7, however only $\bar{n}_{cl}$ will follow, strictly, additivity rules for the case of admixtures.\footnote{It is common to find alternative designations, chiefly $N_{T} \equiv \bar{n}_e$ and $N_{P} = \bar{n}_{cl}$, that are tabulated for instance in \cite{Amsler}.}

\begin{figure}[h!!!]
\centering
\includegraphics*[width=\linewidth]{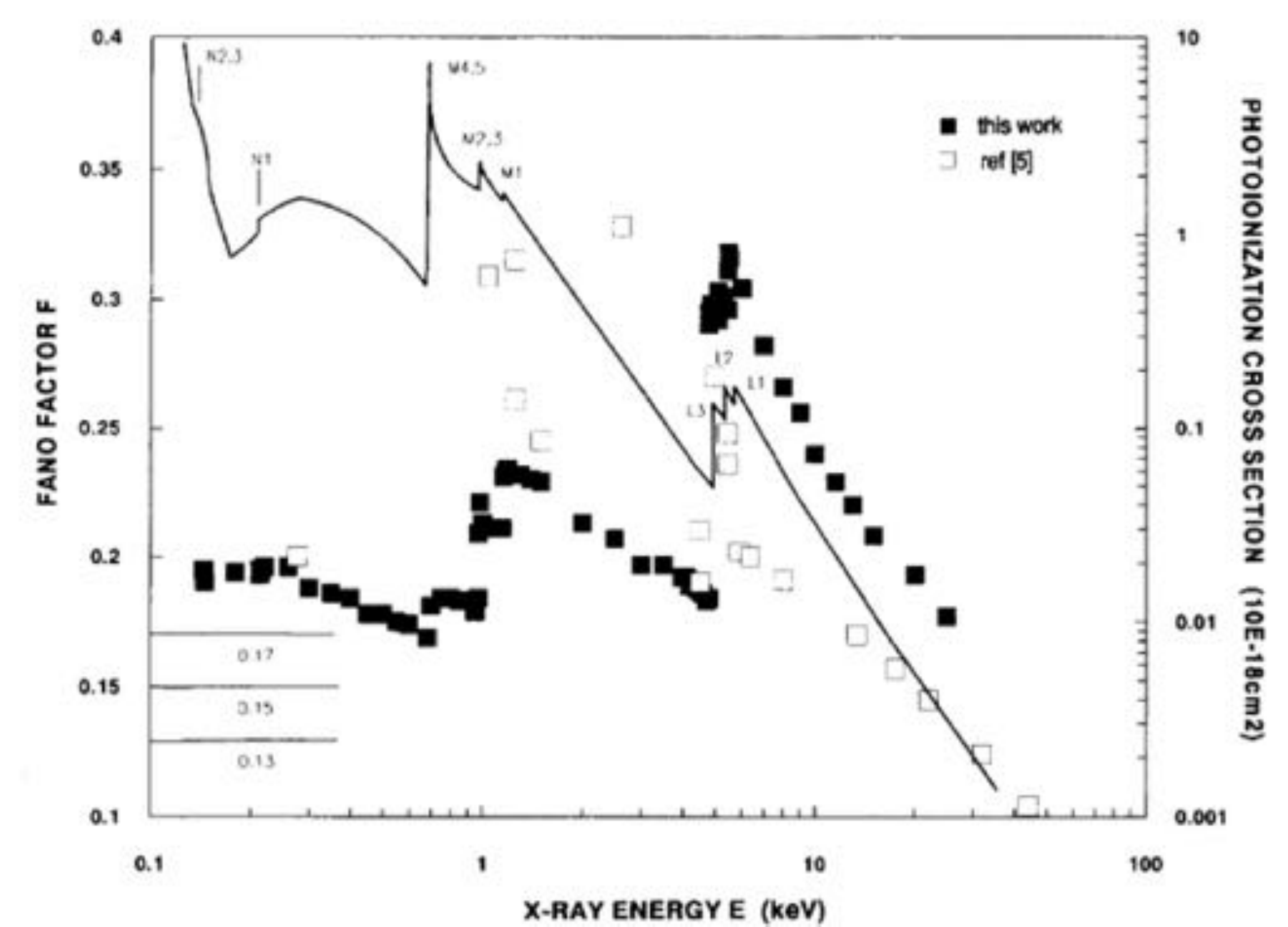}
\caption{Measured Fano factor (open squares) from \cite{Kowalski}, compared to a numerical calculation (filled squares) from \cite{Filomena}, for x-ray interactions in xenon. Lines indicate the photo-ionization cross section, with axis on the right.}
\label{FanoFig}
\end{figure}

TPCs used for the detection of rare processes are very often designed with the aim of fully containing the reaction under study, conditions under which the relevant measure of the ionization fluctuations is the Fano factor ($F_e$) and Landau fluctuations along the trail play a secondary role. Similar to the latter, $F_e$ is highly elusive to analytical treatment and often evaluated numerically through microscopic transport \cite{Filomena, Alkhazov, Magboltz}.\footnote{Some modern open software packages like Degrad can do this type of calculation (section \ref{Micro}).} In the case of Penning mixtures, however, a handy lower bound to the resulting Fano factor ($F_{e,p}$) can be derived as a function of the effective Penning transfer probability ($r$):
\beq
F_{e,p} \geq (1 - r) F_e
\eeq
where $F_e$ refers to the Fano factor in the absence of the Penning additive. As an example, a value for $F_{e,p} = (1 - 0.7 r) F_e$ was calculated for Ar/C$_2$H$_2$ mixtures in \cite{Alkhazov}. Taking $F_e=0.17$ (pure argon) and a Penning transfer probability of 70\%, the obtained value ($F_{e,p}=0.09$) was found to be in close agreement with measurements. In general, $F_e$ depends only mildly on the particle energy (Fig. \ref{FanoFig}), but it can achieve relatively large values near the binding energy of closed shells \cite{Kowalski}.

Additional fluctuations will appear in the presence of charge recombination, that may be approximated by a binomial probability distribution to first order. Under this latter assumption, the intrinsic calorimetric response of a TPC obeys the formula:
\beq
\frac{\sigma_{{n}_e}}{\bar{n}_e} \simeq \sqrt{\frac{F_e +\mathcal{R}}{\bar{n}_e}}
\eeq

\subsection{Generation of scintillation}

Assuming a fully contained particle or nucleus, scintillation photons are produced on average as $\bar{n}_\gamma=\varepsilon/W_{sc}$ (or $\varepsilon/W_{ph}$ in some works), with an intrinsic spread given by $\sigma_{{n}_\gamma}/\bar{n}_\gamma = \sqrt{F_\gamma/\bar{n}_\gamma}$. Since light production is isotropic, only a small fraction of the produced photons, $\Omega$, (typically of the order of few \%) can be collected. As an example, a 30\,keV x-ray produced in xenon gas under a (not unreasonable) 1\% photo-detector coverage at 30\% quantum efficiency (QE) results in approximately one photon recorded, on average (see, e.g. \cite{Mua}). This is one thousand times less than the information conveyed in the ionization sector, thereby making a calorimetric measurement specially challenging. Present measurements and simulations indicate that $W_{sc}$ (defined as the average energy that it takes to produce a scintillation photon) is fairly constant for primary electrons/x-rays \cite{Mua,AprileLiq}. Compared to $W_I$, their values span over a broader range, from 15 to 70\,eV, with experimental uncertainties being larger too.

Clearly, the faint nature and isotropic characteristics of primary scintillation are not optimal for imaging on large volumes. However, despite its intrinsic limitations, it provides some valuable assets too: i) outside accelerator-based experiments, detecting the primary scintillation represents the most common way to obtain the reference (`start') time of the event ($T_0$), a standard technique to remove background events that accumulate at the electrodes \cite{XENON}, and to correct for eventual losses occurring during charge drift; ii) as illustrated below, the sheer scintillation yields can contain information about the particle type, too.

\subsubsection{Scintillation mechanisms}

\begin{figure}[h!!!]
\centering
\includegraphics*[width=\linewidth]{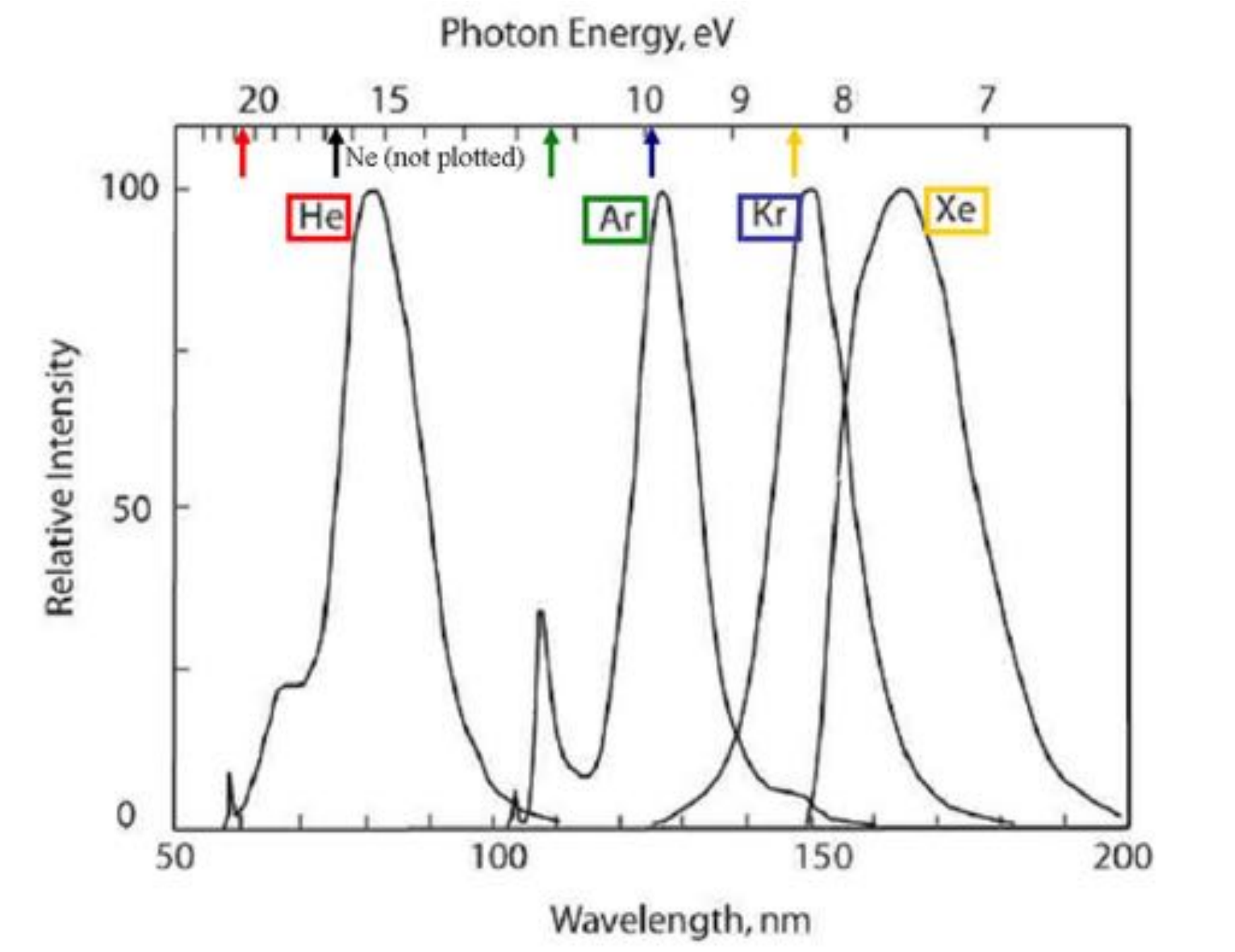}
\includegraphics*[width=8cm]{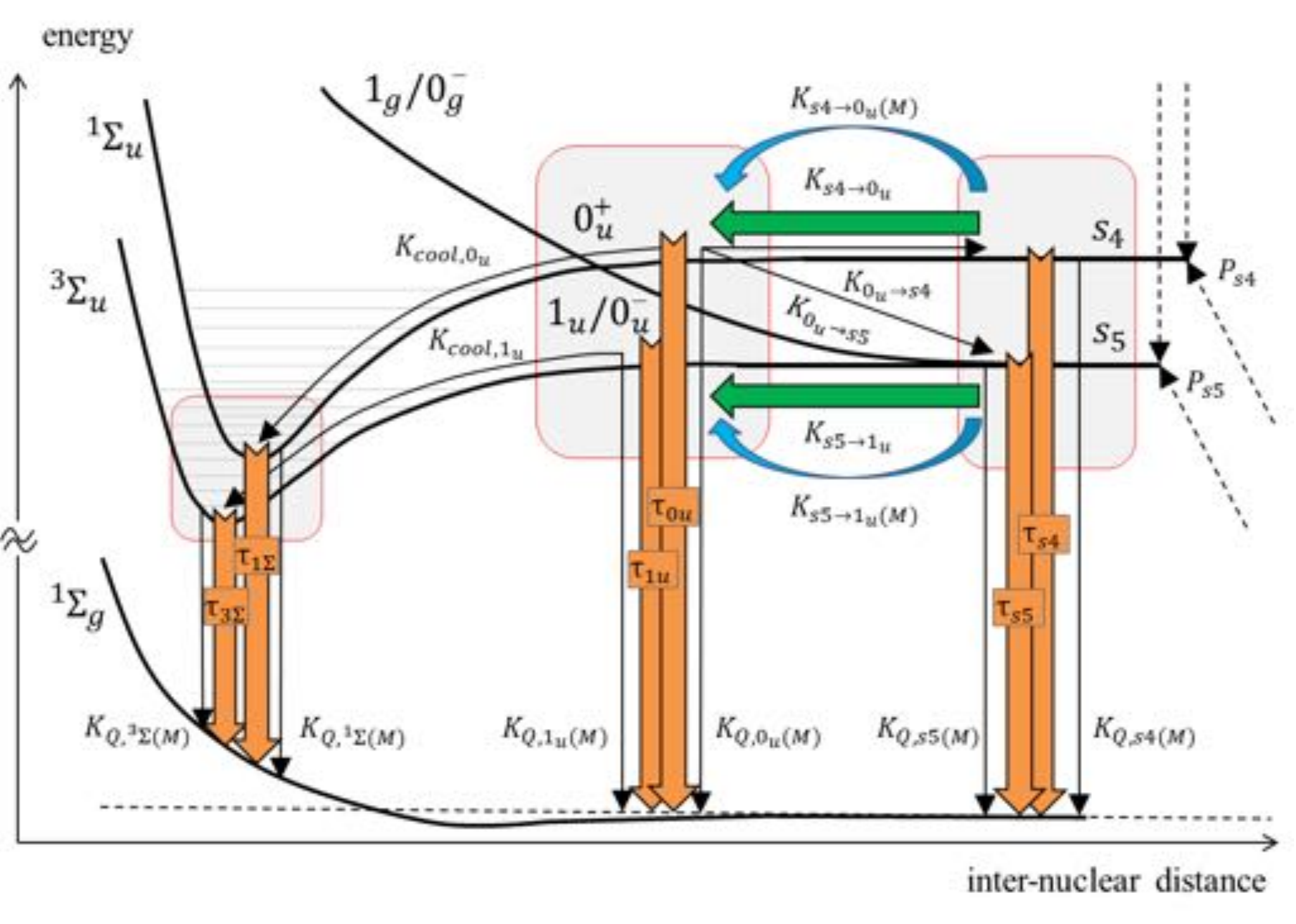}
\caption{Top: the so called 2$^{\tn{nd}}$ continuum emission of noble gases, a dominant feature above few 100's of mbar and up to condensed phase, and that is particularly well understood. Arrows indicate the approximate position of the two lowest lying atomic states that are precursors to the scintillation process: the first metastable (with alternative notations: 1s$_5$//6s$[3/2]_2$//$^3$P$_2$) and resonant (1s$_4$//6s$[3/2]_1$//$^3$P$_1$) states. The shift of the vertical arrows with respect to the maximum of the continua appears due to the excimer nature of the emission: at the above pressures, excimers are formed in 3-body collisions and afterwards cooled collisionally prior to emission, hence leading to self-transparency. (Figure adapted and reprinted from \cite{Aprile}). Bottom: path diagram for xenon scintillation (from \cite{Mua}). This partly artistic diagram indicates the three type of states involved, that appear associated to different radii (from right to left: unbound $\equiv$ atom, loosely bound, and strongly bound or `excimer'), each identified with boxes. Quenching pathways have been labeled with subindex $Q$. States responsible for scintillation in the 2$^{\tn{nd}}$ continuum are $^1\Sigma_u$ (singlet) and $^3\Sigma_u$ (triplet).}
\label{NobleGasScint}
\end{figure}

Scintillation in noble gases is strongly dependent on the concentration of impurities, that can easily quench the triplet and singlet precursors as well as leading to photo-absorption \cite{BenTransp, ArN2Suzuki}. Other scintillating gases like CF$_4$ seem to be more robust in this respect \cite{Margato}. The scintillation process is, comparatively, much harder to describe with approximate formulas than ionization: since it often requires energy transfers between several levels and species \cite{TMAYasu, ArN2Suzuki, Moutard, ArCF4Fraga} not even approximate additivity rules can be expected to hold. It commonly involves molecular states, leading to broad spectra (illustratively, the reader is referred to Fig. \ref{NobleGasScint}).

Scintillation time constants can be generally found in the range 1-100\,ns, although they can exceptionally exceed the $\mu$s landmark (the argon triplet state is a remarkable exception, with a lifetime of around $1\,\mu$s and $3\,\mu$s in liquid and gas phase, respectively)\footnote{Lighter noble gases display increasingly longer lifetimes \cite{NeTriplet, HeTriplet}.}.
In low pressure gases (around 100\,mbar and below), the build-up time constants can become important, too \cite{Moutard}. By analogy with $Q_y$, the scintillation yield may be expressed as:
\beq
L_y = \frac{1}{W_{sc}} = \frac{\mathcal{P}_{scin}}{W_{ex,\gamma}} \label{Ly}
\eeq
using the definitions of previous subsection.\footnote{In liquid, $\Delta\varepsilon/W_{ex}$  must be interpreted as the number of excitons, the formal treatment being analogous to that in gas phase.} Provided the scintillation probability approaches 1 in pure noble gases, $W_{sc} \simeq W_{ex,\gamma}$ (i.e., the number of emitted photons is very similar to the number of excited species). A useful (but not general) description in the presence of additives is the `triplet dominance model' (\cite{Mua},\cite{ArN2Suzuki}), under which:
\beq
\mathcal{P}_{scin} \simeq \frac{1}{1 + f\!\cdot\! \cdot\!\tau_{_{^3\Sigma}}\!\cdot\! K_{Q,^3\Sigma}} \label{TDM}
\eeq
with $f$ being the additive concentration (binary mixture assumed), $\tau_{_{^3\Sigma}}$ the lifetime of the triplet state and $K_{Q,^3\Sigma}$ its quenching rate (in units of $[T^{-1}]$).\footnote{Throughout the text $K$ is defined as the reaction rate in the limit where the additive approaches 100\,\% of the detection medium, since this value can be readily obtained from the tabulated reaction rate constants (e.g., \cite{Mua}).} 

\begin{figure}[h!!!]
\centering
\includegraphics*[width=7 cm]{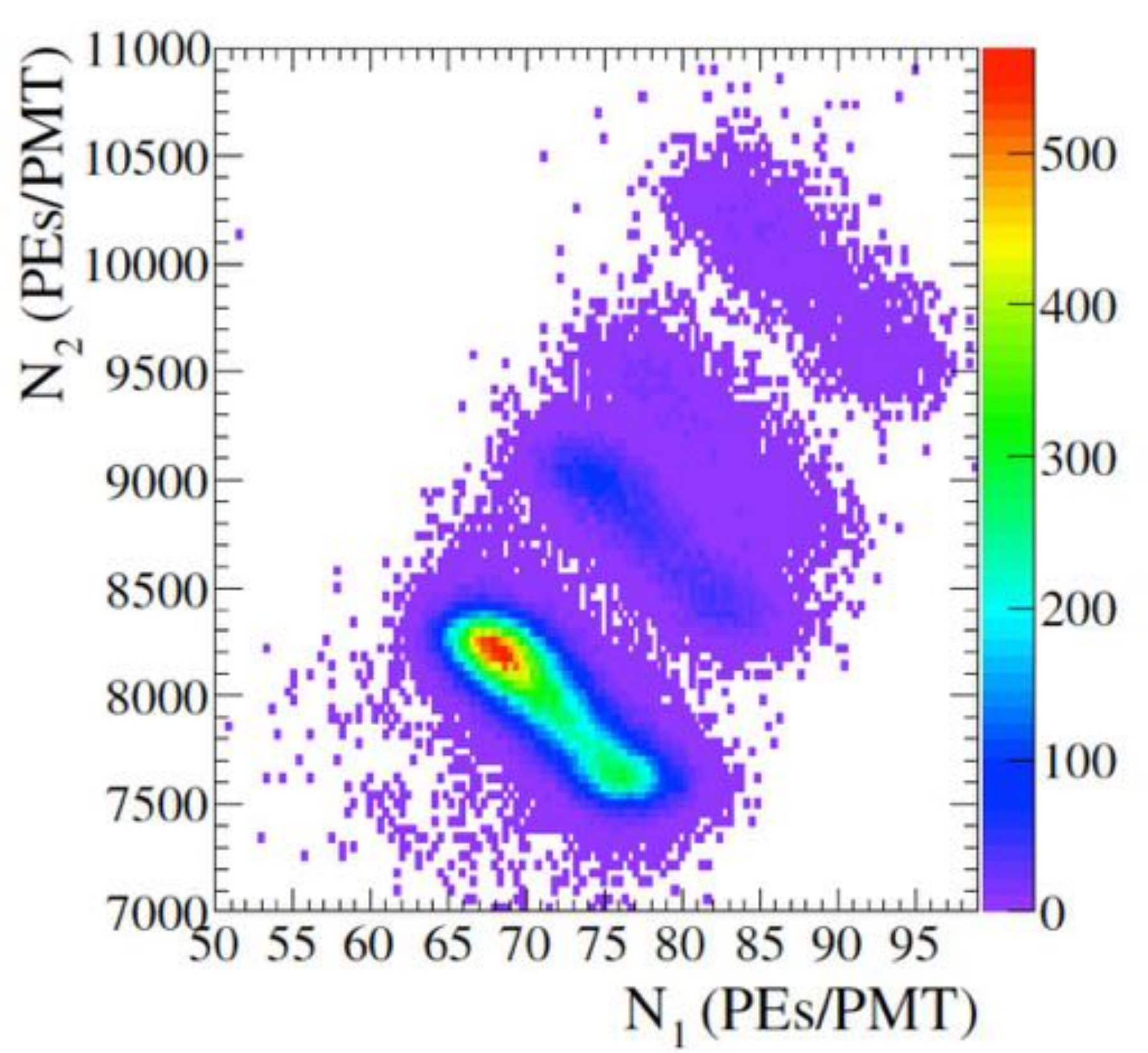}
\caption{Anti-correlation between primary ionization ($N_2$) and scintillation ($N_1$) in high pressure xenon, obtained by the NEXT experiment. Bands correspond to $\alpha$-particles emitted in the decay of the various nuclear species belonging to Rn progeny (stemming from a Rn source diffused in
the detector volume), at 10\,bar and $E_d=300$\,V/cm \cite{SerraLast}.}
\label{RecoCorr}
\end{figure}

In noble gases, recombined electrons populate the high-lying p-states. 
After cascading very fast ($\lesssim$ 1ns) to the singlet an triplet states, a strong anti-correlation between ionization and scintillation arises, as shown in Fig. \ref{RecoCorr}. Equation \ref{Ly} can be easily generalized in order to account for this possibility, after making use of the expression for $Q_y$ from eq. \ref{Reco}:
\beq
L_y = \Big(\frac{1}{W_{ex,\gamma}} + \frac{\mathcal{R}}{1-\mathcal{R}} Q_y \Big)\mathcal{P}_{scin} \label{LyReco}
\eeq
that is the final formula that we will use. Charge recombination can be very disturbing in condensed (or high pressure) phases, especially when a precise calorimetric measurement is required. With the ancillary measurement of the scintillation, however, the situation can be partially restored \cite{EXO}.

\begin{figure}[h!!!]
\centering
\includegraphics*[width=8cm]{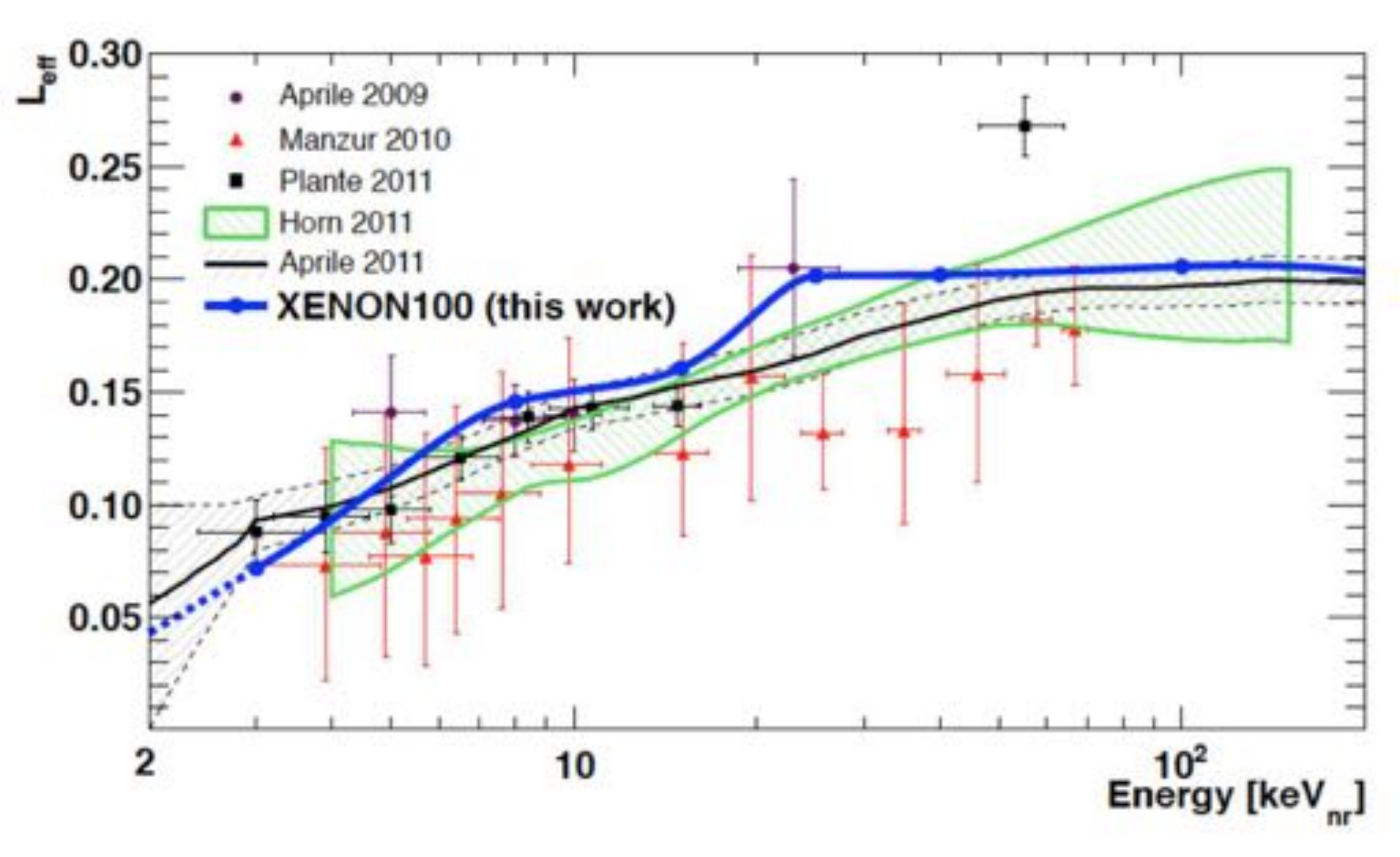}
\includegraphics*[width=8cm]{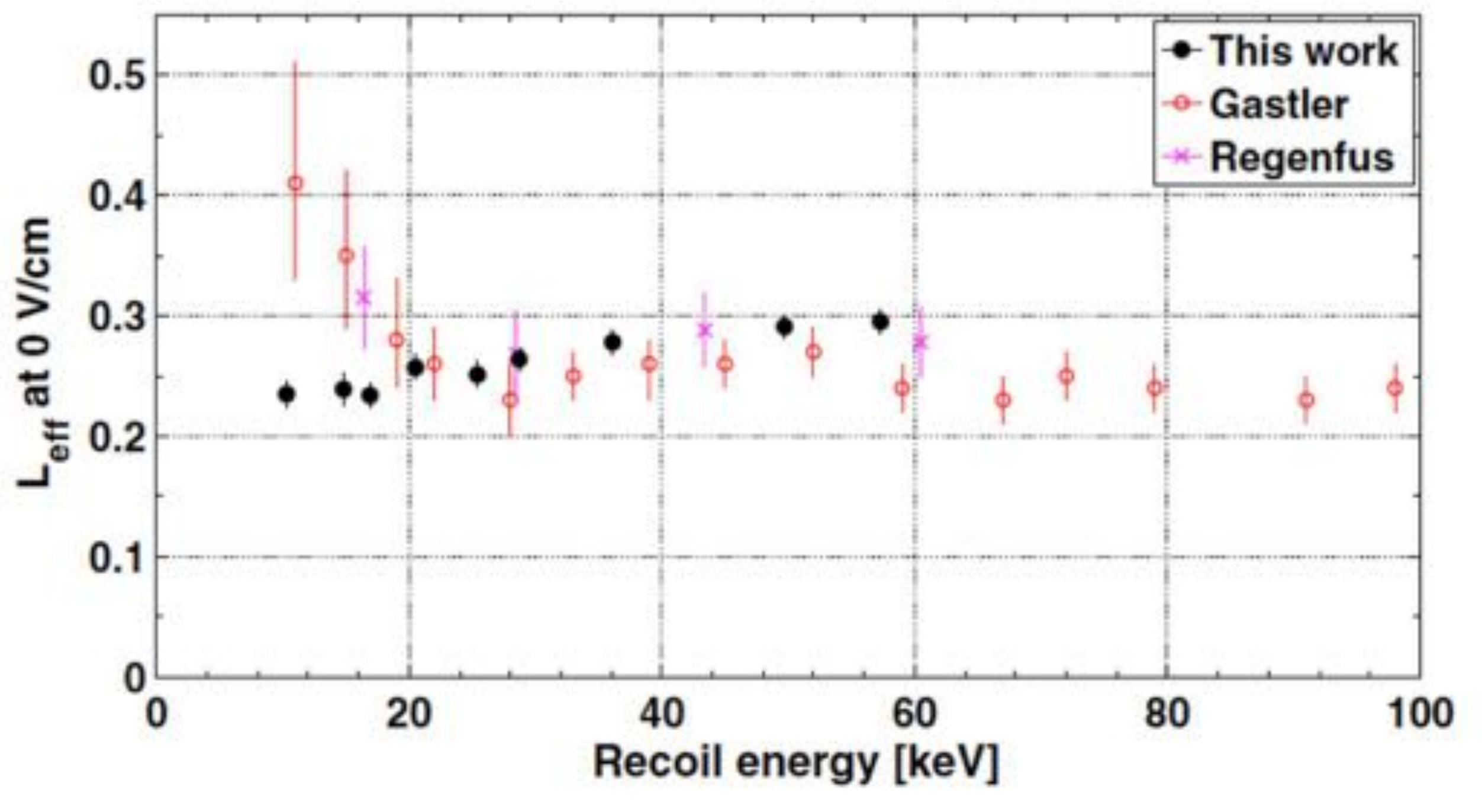}
\caption{Top: scintillation quenching factor, $\mathcal{L}$, for liquid xenon at at around 500\,V/cm electric field (note the authors' convention: $L_{eff} \equiv \mathcal{L}$). Bottom: scintillation quenching factor for liquid argon at zero drift field. (Reprinted from \cite{Cao, XENON100})}
\label{L_ARGON}
\end{figure}

Similarly to the ionization quenching factor, a scintillation quenching factor can be defined as:
\beq
\mathcal{L} = \frac{L_y|_X}{L_y|_{el}} \label{L}
\eeq
The scintillation quenching factor for low energy nuclei has been measured to be around 0.5 in high pressure xenon \cite{JoshHPNeutrons} and it can be as small as 0.1 in liquid xenon (Fig. \ref{L_ARGON}-top), and around 0.25 in liquid argon (Fig. \ref{L_ARGON}-bottom). Again, as for the case of $\mathcal{Q}$,
it is frequent that measurements are not free from recombination effects.\footnote{In addition to the LSS scintillation quenching, some authors speculated about the presence of other processes in condensed media, resulting from the high density of excitons. This would allow them to interact with each other and eventually dissipate their energy collisionally \cite{Mei}. Hence, by resorting to Birk's phenomenological formula for the saturation of scintillation in organic scintillators \cite{Birks}, the measured $\mathcal{L}$ in liquid argon could be reproduced. The reader is referred to recent works \cite{NEST,Cao} for a discussion on this topic.}

\begin{figure}[ht!!!]
\centering
\includegraphics*[width=\linewidth]{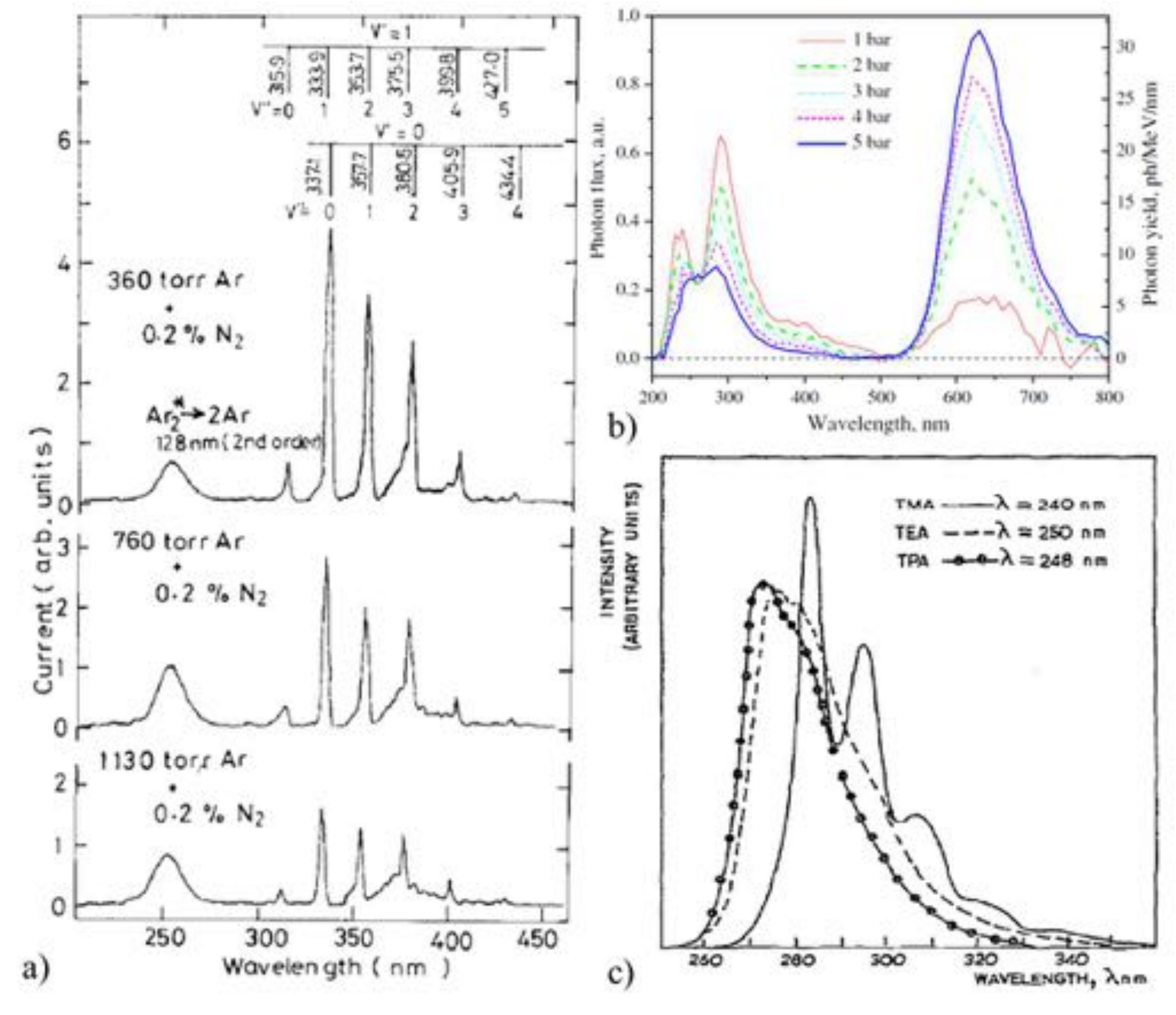}
\caption{a) Secondary scintillation in Ar-N$_2$ mixtures \cite{ArN2Suzuki}, b) primary scintillation by $\alpha$-particles in pure CF$_4$ as a function of pressure \cite{Morozov}, c) scintillation of TMA, TEA and TPA upon photon excitation with different wavelengths, given in inset \cite{Cureton}.}
\label{AdditiveScintillation}
\end{figure}

It is important to note that, through the use of eq. \ref{LyReco}, it is not made explicit whether the number of scintillation precursors ($N_{ex,\gamma} = \Delta\varepsilon/W_{ex,\gamma}$) is produced by primary interaction or through transfers between species. This is because a handy general expression does not exist. As an example (that hardly represents an exception), at around atmospheric pressure both xenon and argon will display scintillation in the same band, if the latter is admixed with as little as 3\% xenon \cite{ArN2Suzuki}. Scintillation in noble elements is by far the most pursued scintillation source, but its use is not exclusive. TPCs relying on the scintillation of TEA (triethylamine), TMA (trimethylamine), CF$_4$ and N$_2$ have been built and are described later. Their scintillation spectra are given in Fig. \ref{AdditiveScintillation} and for additional information the reader is referred to \cite{Takeshi}, published in this volume.

Throughout this text we will discuss as well the case of \emph{secondary} scintillation: a process that relies on the ionization electrons being driven to a high field region, in which they can excite the medium and produce scintillation.\footnote{Note that secondary \emph{ionization} has been  defined as that produced by high energy electrons released during the ionization process. Such use is not widespread in the case of scintillation, though, and therefore it should not lead to confusion in the present context.} Very often, processes leading to primary scintillation (by the impinging particles) and secondary scintillation (by the ionization electrons) are, in fact, rather similar, in particular for noble gases. This may be counter-intuitive, since scintillation by a primary charged particle is largely a top-down process, that relies on the cascade of a relatively broad distribution of excited states; field-assisted scintillation by ionization electrons, on the other hand, is of the bottom-up type, leading to an overwhelming majority of low-lying excited states. Nevertheless, if the cascade proceeds fast (e.g., via collisions, that is the case under typical operating pressures in TPCs) and the low-lying states are responsible for scintillation (e.g., for noble gases), the resulting spectra and time constants will be very similar in both cases. In practice, primary scintillation may include, additionally, the processes of recombination light and scintillation quenching (and Cherenkov light, if conditions allow), while secondary scintillation will depend strongly on the electron transport under the external field. Modern simulation frameworks allow to obtain both primary and secondary scintillation in cases of interest, and for a detailed discussion the reader is referred to \cite{Mua}.

\subsection{Other information bits}

A strong continuum around 1250\,nm in Ar/Xe and Xe mixtures, with yields close to those observed in the VUV region, has been recently reported \cite{Borghe, Belo}. Technically, IR detection can be accomplished at near 100\% quantum efficiency, and the scintillation time constants will be presumably faster than the ones of the VUV precursors \cite{Mua}. Cherenkov light is also fast, and it may be available in some conditions \cite{Woody,YannisChe}. Authors in \cite{Xenon bubble chamber} have suggested to use, as additional particle discriminant in a TPC, the local temperature increase resulting from the kinetic energy of the atoms/molecules; in fact, low-energy recoiling nuclei produce detectable bubbles along their trajectories in liquid xenon \cite{arXiv:1702.08861v1}. For nuclear decays, the ultimate milestone would be to be able to identify the daughter nucleus, something that is actively pursued in liquid as well as in gas for the case of $^{136}$Xe \cite{BaTa1,BaTa2}. Lastly, a bold proposal for the detection of positive ions by resorting to Auger emission upon neutralization at the cathode has been recently made in \cite{PositiveIon}, despite it lacks experimental verification yet. It could be used, in its simplest version, for $T_0$-determination in the absence of primary scintillation. Some of these ideas are expanded later in section \ref{classification6}.

\section{Technical problems, solutions, and enabling technological assets} \label{assets}

\subsection{Collection of event information}


\subsubsection{Collection of primary ionization} \label{ChargeColl}

The charge collection process can be conveniently described for typical TPC conditions through the hydrodynamic approximation of the Boltzmann equations \cite{Huxley}:\footnote{We assume that charge transport starts after all types of charge recombination, that may be considered separately. Although the effect can be included in a hydrodynamic framework (and a solution found in some cases, as discussed earlier in text), a rigorous treatment is outside the scope of this work.}
\bear
&\frac{\partial \mathcal{N}_e}{\partial t} &+ v_d \frac{\partial \mathcal{N}_e}{\partial z'} - D_T\bigg(\frac{\partial^2 \mathcal{N}_e}{\partial x'^2} + \frac{\partial^2 \mathcal{N}_e}{\partial y'^2}\bigg) - D_L\frac{\partial^2 \mathcal{N}_e}{\partial z'^2} \nonumber \\
&&= -\eta v_d \mathcal{N}_e \label{Eq_hydro}
\eear
Here $\mathcal{N}_e$ is the density of electrons per unit volume, $D_{L(T)}$ the longitudinal (transverse) diffusion coefficient, $v_d$ the drift velocity and $\eta$ the attachment coefficient.

\begin{figure}[h!!!]
\centering
\includegraphics*[width=7 cm]{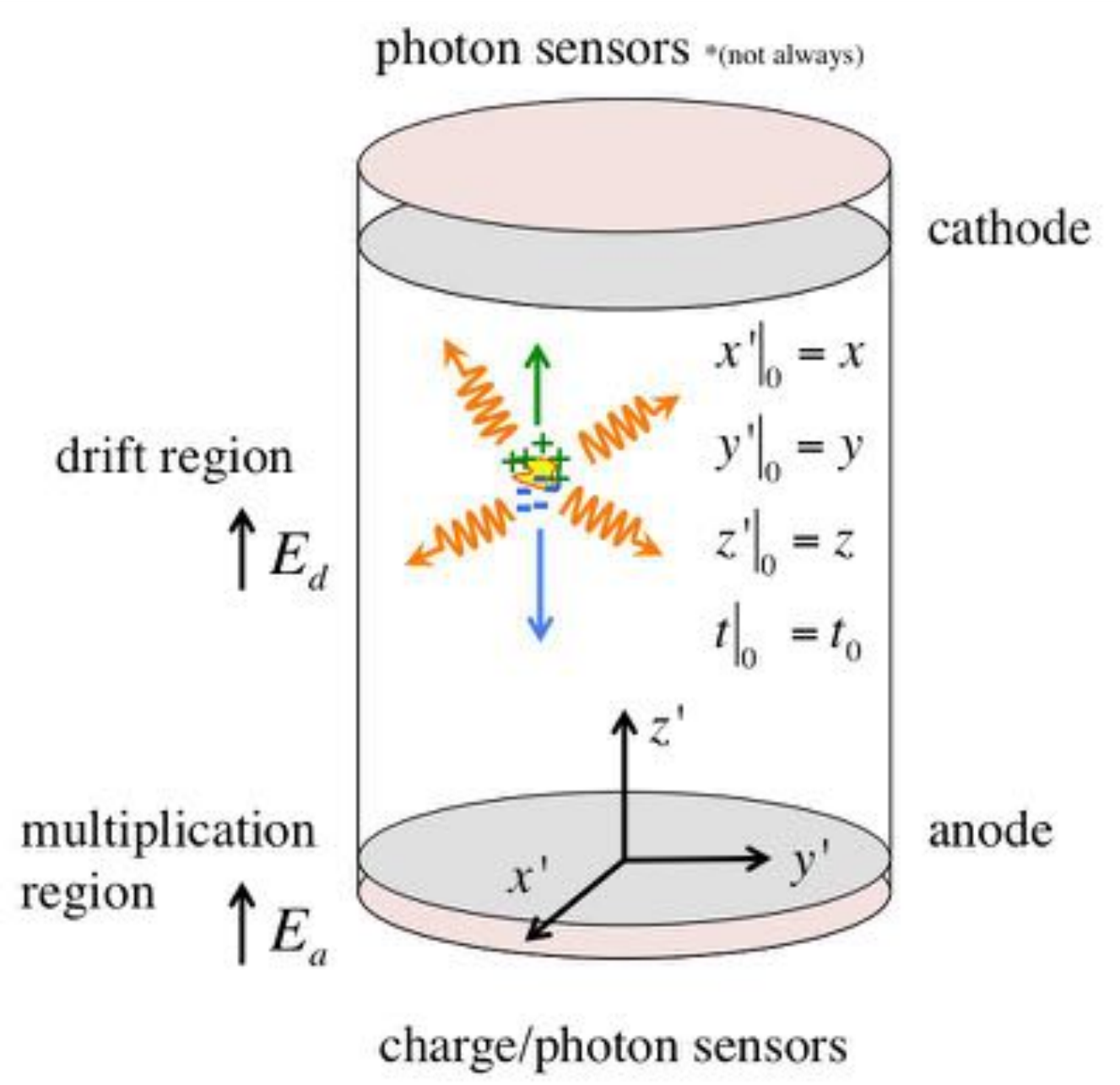}
\caption{Sketch of a generic TPC devised for imaging a rare process, together with some of the conventions used in text. Although the orientation and the chosen cylindrical shape is largely arbitrary, it is common for gaseous TPCs to find the drift direction perpendicular to gravity, and parallel to it for dual-phase (obviously there is no freedom of choice in this latter case). Vessels compatible with vacuum/pressurization tend to be cylindrical. The shape of the active volume (represented in this figure) can be actually chosen at will, as long as the two main electrodes are set parallel to each other. In practice, cylindrical, rectangular parallelepiped or other higher order right prisms can be found.}
\label{TPCdrawing}
\end{figure}

When far from the TPC boundaries, eq. \ref{Eq_hydro} can be readily solved, leading to:\footnote{Since diffusion for chambers described here is at the mm's-scale even for full cathode-anode propagation, clearly this `infinite-volume' assumption is well suited. A similar argument applies to the final derivation leading to eq. \ref{THESOL3}.}
\bear
&\mathcal{N}_e(x',y',z',t)\! & = \! \frac{e^{-\frac{(x'-x)^2+(y'-y)^2}{4D_T(t-t_0)}} e^{-\frac{((z'-z)+v_d(t-t_0))^2}{4D_L(t-t_0)}}}{(4\pi D_T (t-t_0))(4\pi D_L (t-t_0))^{1/2}} \times \nonumber\\
&& \bar{n}_e \cdot e^{-\eta v_d (t-t_0)}\label{THESOL}
\eear
We indicate by $(x',y',z',t)$ the variables measurable by an observer and $(x,y,z,t_0)$ is the initial position and time of the ionization cloud (Fig. \ref{TPCdrawing}), assumed to be point-like and containing $\bar{n}_e = Q_y\Delta{\varepsilon}$ electrons (eq. \ref{Reco}). The solution in eq. \ref{THESOL} is an asymmetric gaussian cloud that broadens and loses carriers as it goes on. Arbitrary track topologies can be propagated directly by superposition of such solutions, and analogous (or related) expressions are available for the coupled equations describing the movement of positive and negative ions.

In TPCs, the field configuration is strongly distorted in the anode region (at the mm or sub-mm scale) in order to provide some form of multiplication. This small scale, together with the very short transit times associated to it ($\lesssim$ ns), suggest that `drift' and `multiplication' regions can be treated separately. In that case, the charge distribution arriving at a virtual multiplication plane placed at $z'=0$ can be obtained exactly as:
\bear
& \mathcal{N}_e(x',y',t) & = \frac{e^{-\frac{(x'-x)^2+(y'-y)^2}{4D_T(t-t_0)}} e^{-\frac{(t-t_0- z/v_d)^2v_d^2}{4D_L(t-t_0)}}}{(4\pi D_T (t-t_0))(4\pi D_L(t-t_0))^{1/2}} \times \nonumber\\
&& \bar{n}_e \cdot  e^{-\eta v_d (t-t_0)}\label{THESOL2}
\eear
We may approximate eq. \ref{THESOL} under the assumption that all charges arrive at a fixed time $t$-$t_0 \simeq z/v_d$, instead:
\bear
&\mathcal{N}_e(x',y',z') &=  \frac{e^{-\frac{1}{2}(\frac{x'-x}{D_T^*\sqrt{z}})^2} e^{-\frac{1}{2}(\frac{y'-y}{D_T^*\sqrt{z}})^2} e^{-\frac{1}{2}(\frac{z'}{D_L^*\sqrt{z}})^2}}{(2\pi D_T^{*,2} z)(2\pi D_L^{*,2} z)^{1/2}}\times \nonumber\\
&& \bar{n}_e \cdot e^{-\eta z}\label{THESOL3}
\eear
with the definition $D_{L,T}^* = \sqrt{2D_{L,T}/v_d}$.\footnote{Note that the natural units of $D_{L,T}$ are [$L^2\, T^{-1}$] and the ones of $D_{L,T}^*$ are [$L^{1/2}$]. This apparently odd choice stems from the fact that the latter can be expressed in, e.g., mm/$\sqrt{\tn{m}}$ units (as done throughout the text), thus being directly interpreted as the spread in mm for 1\,m drift.}  As a result of the condition $t$-$t_0 \simeq z/v_d$, the $z'$ position appears to jitter around $z'=0$, and so the reconstructed drift distance $z$-$z'$ does as well. Eq. \ref{THESOL3} is our final expression and it represents, almost invariably, the starting point of any Monte Carlo simulation designed to model the response of a TPC, since it is numerically easier to handle than eq. \ref{THESOL2}, and sufficiently accurate.
Clearly, in order to extract the maximum information of the obtained images, all parameters in eq. \ref{THESOL3} need to be optimized and experimentally understood to great detail, something that is frequently done with dedicated setups. We discuss below some related results and their implications.

\paragraph{Attachment.}

The ability to drift the primary electrons to the anode region can be conveniently described through the `electron lifetime':
\beq
\tau_e = (\eta ~ v_d)^{-1}
\eeq
Assuming that charge is transferred from drift to multiplication region with a probability $\mathcal{T}_e$, and integrating eq. \ref{THESOL3}, the total collected charge will be:
\bear
&& \bar{n}_{e,c}(z) = \mathcal{T}_e \cdot (1-\mathcal{A}(z)) \cdot \bar{n}_e \label{Ncoll} \\
&& \mathcal{A}(z) = 1 - \exp(-(t-t_0)/\tau_e) = 1 - \exp(-\eta z)
\eear

\begin{figure}[h!!!]
\centering
\includegraphics*[width=8 cm]{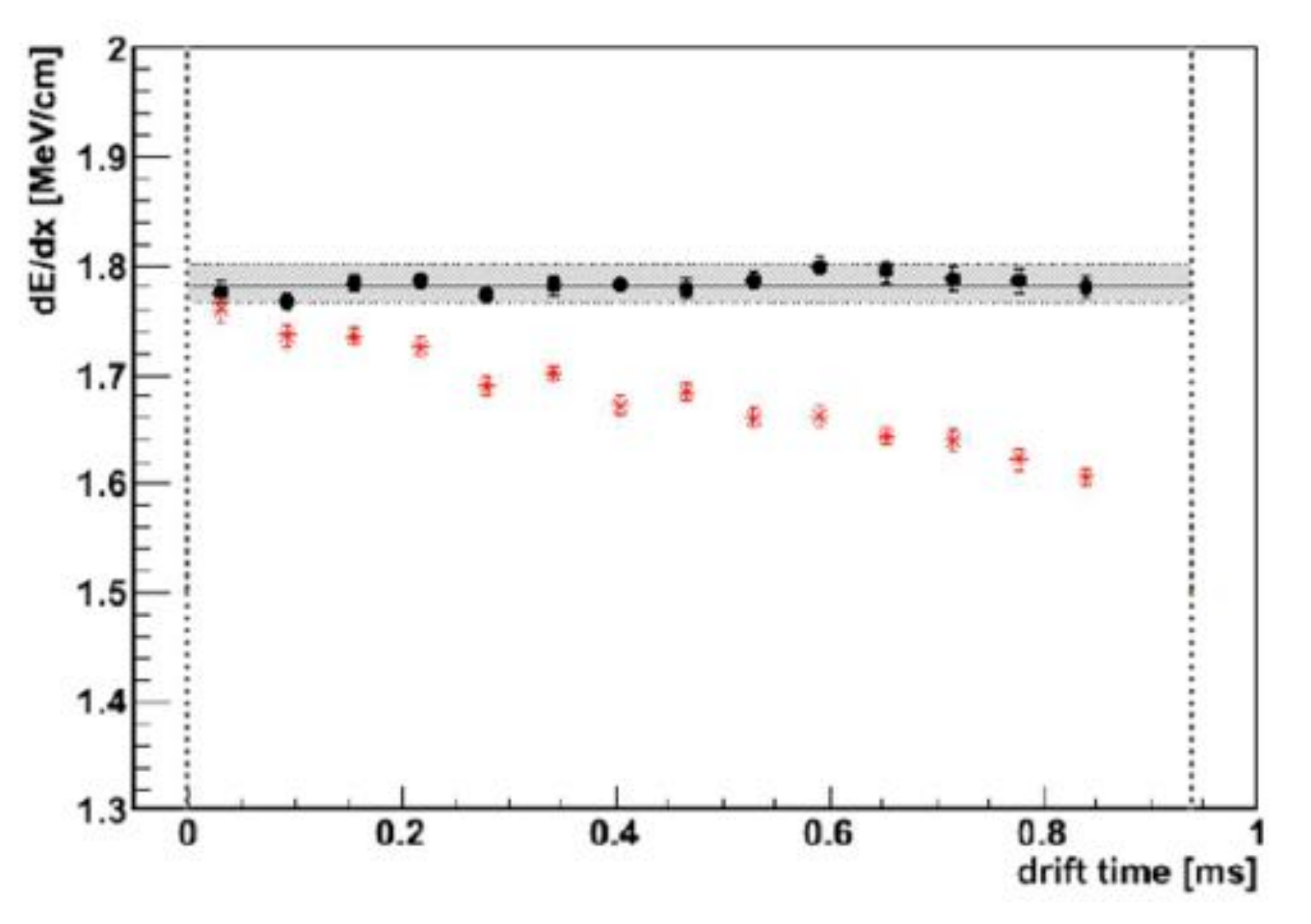}
\caption{Mean energy loss ($d\varepsilon/dx$) for minimum ionizing particles in liquid argon as a function of drift time (red asterisks), corresponding to an electron lifetime of $\tau_e=9$\,ms (obtained by the ICARUS T600 collaboration in \cite{ICARUSlifetime}). From the evaluation of eq. \ref{lifetimeDUNE}, the presence of O$_2$ contamination at around 30\,ppt can be inferred. (Black circles indicate the corrected energy loss after an exponential fit)}
\label{ICAlife}
\end{figure}

The main cause of attachment in large TPCs is related to the presence of O$_2$, a very effective scavenger of low energy electrons under typical density conditions.\footnote{Similar to CO$_2$ or CF$_4$, and unlike SF$_6$ (all mentioned later in text), attachment for isolated O$_2$ molecules is mainly of the dissociative type, meaning that a minimum amount of energy needs to be furnished to the electron so that it can break the molecule and be captured by one of the resulting oxygen atoms. Either direct (SF$_6$) or dissociative (O$_2$, CO$_2$, CF$_4$) attachment proceed through 2-body reactions, but 3-body reactions can open additional channels, becoming important at pressures of interest for TPCs (as discussed in text).}  A possible reaction scheme from medium to high pressures (dubbed BBH) is attributed to Bloch, Bradbury and Herzenberg \cite{Bloch,Herzenberg}, and proceeds in two steps:
\bear
& \tn{e}^- + \tn{O}_2  &\rightarrow \tn{O}_2^{-*} \\
&\tn{O}_2^{-*} + \tn{X} &\rightarrow \tn{O}_2^{-} + \tn{X}^*
\eear
with the main gas constituents (X) acting as stabilizers against the competing channels:
\bear
&\tn{O}_2^{-*} &\rightarrow \tn{O}_2 + \tn{e}^- \\
&\tn{O}_2^{-*} + \tn{X} &\rightarrow \tn{O}_2^{-} + \tn{X} + \tn{e}^-
\eear
Such a 2-step process yields a $P^2$ dependence, making attachment especially problematic at high pressures.
Indeed, after the extensive experimental survey in \cite{Huk}, the electron lifetime of Ar-based TPCs has been shown to follow the expression:
\beq
\tau_e = \frac{1}{P^2[\tn{bar}^2]}\frac{\tau_{e0}[\tn{ms}]}{f_{O_2}[\tn{ppm}]} \label{P2atta}
\eeq
as a function of the O$_2$ concentration, $f_{O_2}$ (and similar expressions can be expected in other gases). Here $\tau_{e0}=13.8\pm5$ ms, for instance, in case of Ar/CH$_4$/i-C$_4$H$_{10}$ (88/10/2) at a reduced field of $E_d^*=100$\,V/cm/bar \cite{Huk}.\footnote{We introduce for the first time the reduced field $E_d^*=E_d/P$ explicitly. It will become clear in section \ref{Prole} the relevance of this choice for gases.} For liquid phase the dependence is even more dramatic, and the following relation has been reported for argon at $E_d=500$\,V/cm \cite{IcarusMonitor}:
\beq
\tau_e = \frac{300\,\tn{ms}}{f_{O_2}[\tn{ppt}]} \label{lifetimeDUNE}
\eeq

\begin{figure}[h!!!]
\centering
\includegraphics*[width=6 cm]{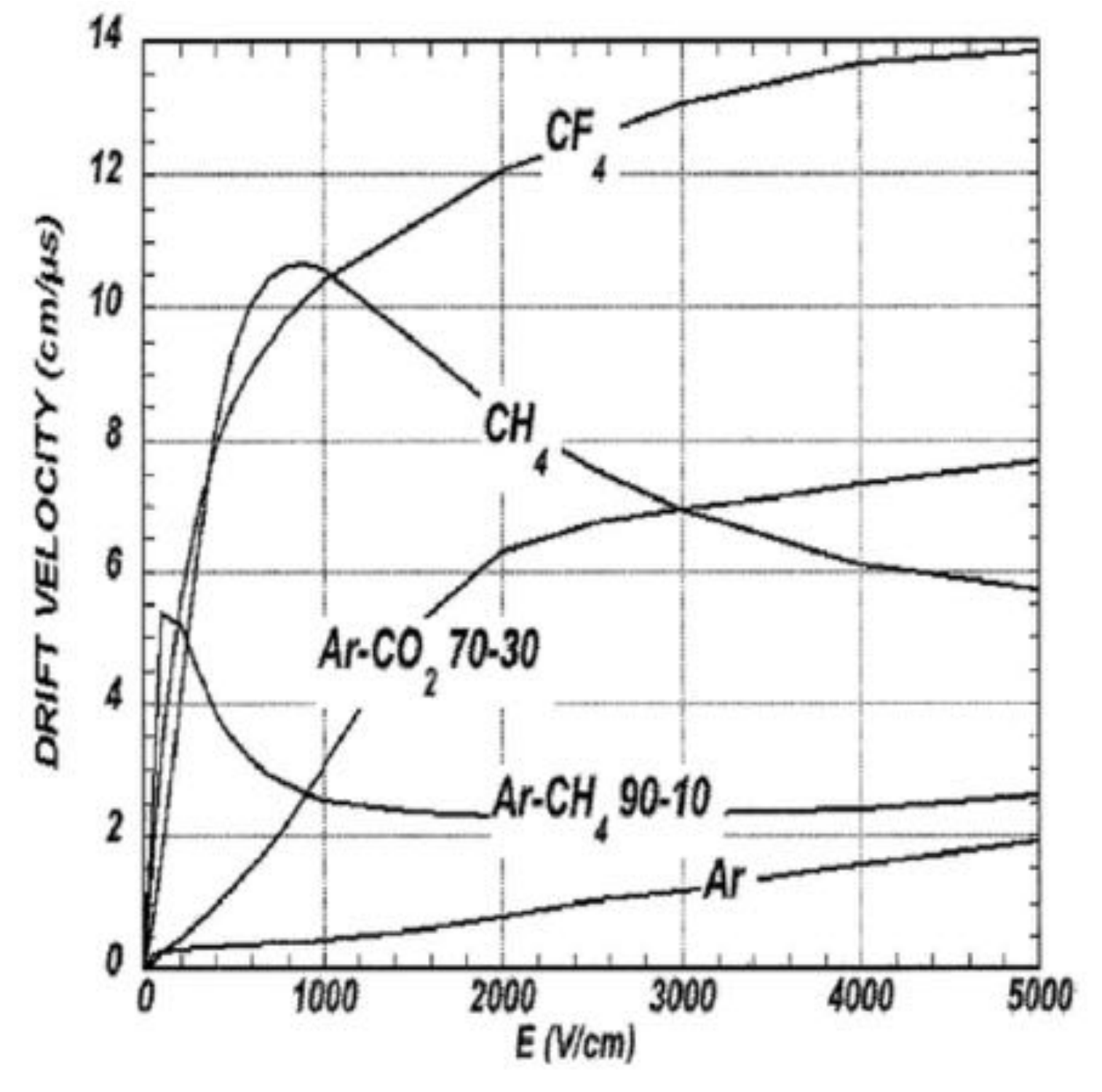}
\caption{Compilation of drift velocities for different gases and admixtures around atmospheric pressure, illustrating the increase attainable through the addition of molecular species \cite{Amsler}.}
\label{vdSeveral}
\end{figure}

At a drift velocity that can be as low as $v_d = 1\tn{\,mm}/\mu{\tn{s}}$ for operation under pure noble gases, some experiments need to approach lifetimes of several ms in order to efficiently collect charge along m-long drifts \cite{NEXT_TDR, DUNE_CDR}. The only realistic way to achieve the necessary purity levels (Fig. \ref{ICAlife}) is through the selection of low-outgassing materials, continuous recirculation and purification, and monitoring with dedicated devices \cite{IcarusMonitor}. The situation becomes much more comfortable for operation in gas at around atmospheric pressure (due to the lower attachment), and especially in the presence of molecular additives (due to the increased drift velocity). The latter allow electron drift times as small as $\Delta{T}=20\,\mu$s per m, for reduced fields around $E_d^* = 200$\,V/cm/bar (Fig. \ref{vdSeveral}).

\paragraph{Fluctuations in charge collection.} \label{LandauSec}

Electron attachment can be corrected on average if the $t_0$ is known, and thus the absolute $z$ position (Fig. \ref{ICAlife}), but it introduces an uncorrectable source of fluctuations in the reconstructed charge. In case of fully contained events it reads, approximately:\footnote{The expression is valid for small charge losses, but it has a simple generalization otherwise (e.g. \cite{XePen}).}
\beq
\frac{\sigma_{_{\Delta\varepsilon}}}{\Delta\varepsilon} \equiv \frac{\sigma_{n_{e,c}}}{\bar{n}_{e,c}}(z) \simeq \sqrt{\frac{F_e + \mathcal{R} + \mathcal{A}(z) + 1-\mathcal{T}_e}{\bar{n}_{e,c}(z)}} \label{EresColl}
\eeq

Besides attachment, ionization may be also lost during the collection process due to fringe fields at the TPC boundaries. These losses (as well as eventual distortions of the electron cloud caused by modifications of the parameters of eq. \ref{THESOL3} in that region) are mitigated with the help of a structure that allows grading the field between anode and cathode (usually referred to as the `field cage'). There are numerous satisfactory ways to build well-behaved field cages and the reader is referred to \cite{Blum-Rolandi} for details. In practice, a field cage reduces fringe fields down to the characteristic scale of the grading structure ($\sim1$\,cm) that, for the case of experiments designed to fully contain the physical process under study, can be easily `fiducialized'-off during analysis.

\begin{figure}[h!!!]
\centering
\includegraphics*[width=8 cm]{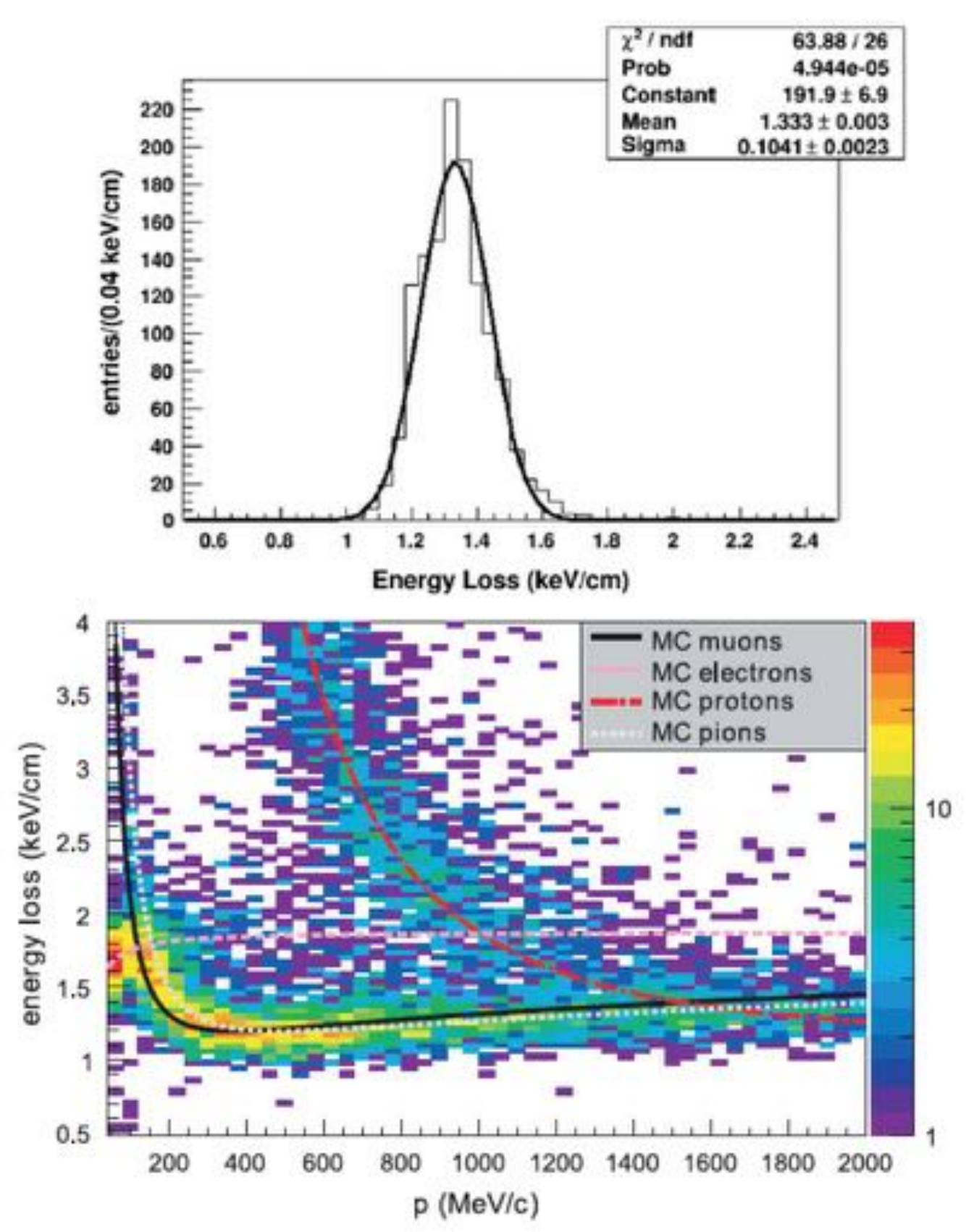}
\caption{Some results obtained by the T2K TPC (from \cite{T2KNIM}). Top: energy loss distribution for negatively charged particles with momenta in the range 400-500\,MeV/c. It corresponds to a $d\varepsilon/{dx}$-resolution ($\sigma$) of $7.8\pm0.2\%$ (the design goal was 10\%). Bottom: measured $d\varepsilon/{dx}$ vs momentum together with the theoretical expectations for various species.}
\label{ElossT2K}
\end{figure}

As compared to expression \ref{EresColl}, fluctuations in charge collection are easily over-shadowed by Landau fluctuations when particles can not be fully contained (Fig. \ref{LandauFig}-a). An approximate expression for mips in Ar/CH$_4$ at around 90/10 has been derived very early on in \cite{Walenta2,AllisonCobb}, and is still in use \cite{Kalweit, Yamamoto}:
\beq
\frac{\sigma_{_{\Delta\varepsilon}}}{\Delta\varepsilon} \equiv \frac{\sigma_{n_{e,c}}}{\bar{n}_{e,c}} \simeq \frac{0.41}{(n\! \cdot \!\Delta{x})^{0.32} \! \cdot \! N_{samples}^{0.43\tn{-}0.46}} \label{EresColl2}
\eeq
with similar expressions for other noble gases \cite{AllisonCobb}. Here $n=N/N_o$ is the gas density relative to standard conditions and $\Delta{x(y)}$ the readout segmentation in the $x$($y$) dimension (evaluated in cm). For argon at atmospheric pressure one expects on average $\Delta\varepsilon=d\varepsilon/dx \!\cdot \!\Delta{x}= 2.5$\,keV in 1\,cm pad, that would correspond to an energy resolution ($\sigma$) around 41\% according to eq. \ref{EresColl2}. Additional losses during drift will enter eq. \ref{EresColl2} approximately under a cube root (the product $n\! \cdot \!\Delta{x}$ conveys the information about the ionization collected per pad), and additional measurements ($N_{samples}$) approximately under a square root. Evaluation of formula \ref{EresColl} for a fully contained event of $\varepsilon=2.5$\,keV on the other hand (e.g., an x-ray) would barely exceed 5\% at this energy, even if assuming 10\% losses for all terms. As a reference, Poisson fluctuations would yield 9\%. In view of this, additional fluctuations due to the charge multiplication process at the anode are usually neglected in the evaluation of formula \ref{EresColl2} (e.g., \cite{AllisonCobb}) provided they do not exceed the Poisson-limit by much, even under extreme circumstances (see next subsection). Despite the large fluctuations, the contribution of multiple samples per track (easily 10's or even 100's of pads) allows `$d\varepsilon/{dx}$-resolutions' down to $\simeq5\%$ \cite{ALICE, Yamamoto}, (e.g., Fig. \ref{ElossT2K}-top). In the presence of a magnetic field, $d\varepsilon/{dx}$ information can be combined with momentum information to perform particle identification (Fig. \ref{ElossT2K}-bottom).

\paragraph{Charge spread in space.}

Even if primary ionization can be efficiently collected at the multiplication plane, the electrons' random motion can distort the ionization trails to the extent that essential topological information may be lost. Following eq. \ref{THESOL3}, the characteristic $z$-dependent spread is given by:
\beq
\sigma_{z,xy} = D_{L,T}^*\sqrt{z}
\eeq

\begin{figure}[h!!!]
\centering
\includegraphics*[width=\linewidth]{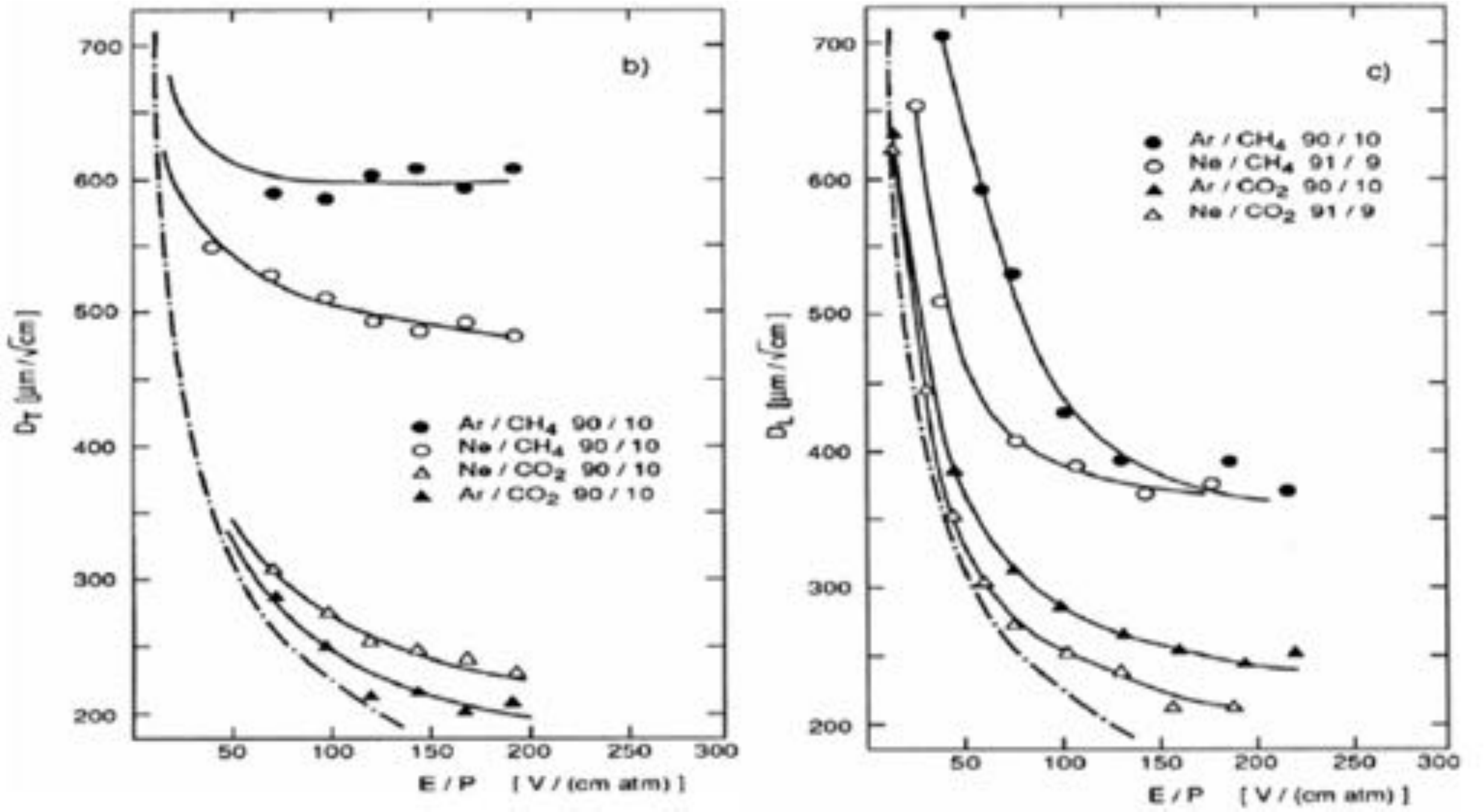}
\caption{Transverse (left) and longitudinal (right) diffusion coefficients for argon and neon mixtures as a function of drift field, at near-atmospheric pressure (from \cite{Hilke}). The dot-dashed line shows the thermal diffusion limit. (Using the convention in text, coefficients should carry a star)}
\label{Diff}
\end{figure}

Since transverse and longitudinal diffusion coefficients can be very different in practice (Fig. \ref{Diff}), a point-like charge cloud moving inside a TPC is in general deformed like an ellipsoid. This is not just a mere technical subtlety: for operation in pure noble gases at low fields ($\sim 10$'s of V/cm/bar), the difference can exceed a factor of four (e.g. \cite{NEXT_TDR}). As noted when the TPC was first introduced in 1974, a magnetic field parallel to the electric one can suppress transverse diffusion as:
\beq
\frac{D^*_T(B)}{D^*_T(0)} = \sqrt{1/(1 + \omega^2 \tau^2)} \label{TransDiff}
\eeq
Here $\omega = (q_e/m_e) |B|$ and $\tau$ is the mean time between collisions ($q_e$ and $m_e$ are the charge and mass of the electron, respectively). Under strong pressurization some existing TPCs can achieve $D_{L(T)}^*\simeq 1\tn{\,mm}/\sqrt\tn{m}$ \cite{NEXTMM}, however values as small as $D_T^* = 0.19, 0.56$\,mm/$\sqrt{\tn{m}} $ have been measured at atmospheric or sub-atmospheric pressure under strong magnetic fields \cite{T2KDixit}. As shown later in section \ref{Prole}, $D^*_T(0)$ displays a $1/\sqrt{P}$ scaling and $\tau$ an inverse linear one. This explains why the lowest (transverse) diffusions are achieved under a strong magnetic field and relatively low pressure: by evaluating eq. \ref{TransDiff} in the low pressure limit we obtain the relation $D^*_T\propto \sqrt{P}/|B|$, whereas $D^*_T\propto 1/\sqrt{P}$ is obtained in the high pressure one (note that the latter becomes independent from $B$).

While mm-diffusion over m-drift may sound like a manageable number, many experiments cannot afford high pressure conditions. In very competitive fields where the system size needs to be flexible, compatibility with a magnet is not easy, either. An elegant solution to the diffusion problem in low pressure chambers (as the ones needed for imaging low energy sub-MeV nuclei) was introduced by Martoff in \cite{Martoff}, who suggested the use of negative ions. Contrary to electrons, ions are rapidly thermalized through elastic collisions and acquire a Maxwellian energy distribution, that corresponds to the minimum diffusion achievable in a medium (Fig. \ref{Diff}, dot-dashed line). Such a thermal limit can be readily obtained through the Einstein relation:
\beq
\frac{D_{L,T}}{\mu} = \frac{k_{_B}T}{q_e}, ~~ (v_d = \mu E_d)
\eeq
where $\mu$ is defined as the mobility and $k_{_B}$ is the Boltzmann constant. By recalling the definition of $D_{L,T}^*$ one obtains:
\beq
D_{L,T}^* = \sqrt{\frac{2D_{L,T}}{v_d}} =  \sqrt{\frac{2 k_{_B}T}{q_e}\cdot \frac{1}{E_d}} = \sqrt{\frac{2 k_{_B}}{q_e}\frac{T}{P}\cdot \frac{1}{E_d^*}} \label{thermalLimit}
\eeq
that provides a ball-park value of $D_{L,T}^* = 1\tn{\,mm}/\sqrt{\tn{m}}$ at a field of $E_d = 500$\,V/cm.

In strongly electronegative mixtures based on CS$_2$ or even SF$_6$, the mean attachment distance $1/\eta$ can be reduced to a minute fraction of the total propagation distance, such that the drift-diffusion process is dominated by the ionic behavior. Once the negative ions reach the multiplication region, they will be stripped off their excess electron if sufficiently high values of $E/P$ can be reached. At the same time, since the ratio multiplication/attachment increases at high fields, an avalanche can develop. This technique is nowadays used in some experiments and has reached a particularly high level of sophistication in \cite{DRIFTlast}.

\paragraph{Charge arrival at the multiplication plane.}

Once the electrons/ions arrive at the anode region a multiplication process starts, after which they become detectable by customized electronics. The multiplication process alters little, however, the basic relation between the $z$-position where the ionization was produced and the recorded time, through $v_d$. The drift velocity is therefore a crucial magnitude in the context of the readout electronics: once a given voxel size in the $z$ dimension is targeted by design, the choice of $v_d$ determines the electronics buffer size and sampling frequency. Conversely, an inadequate buffer can limit the voxel size, the total recorded time per event, or both.

\subsubsection{Collection of scintillation} \label{ScinColl}

Scintillation is notably difficult to handle, too, and full collection+detection is not even close to be a realistic possibility for modern TPCs, yet. First there is the obvious problem of coverage (due to the $4\pi$ characteristics of the emission) and, second, the limited quantum efficiency of conventional PMs ($20$-$40\%$). A greatly enhanced collection efficiency $\Omega$ can be achieved with the help of reflectors: as an example, the LUX experiment approaches a striking $50$\% value, benefiting from the excellent VUV-reflectivity ($\gtrsim 95\%$) of teflon submersed in liquid xenon \cite{Yamashita2004692}. When a fine position sampling is demanded, on the other hand, it is possible to use lenses (eventually coupled to image intensifiers), albeit with a much higher penalty factor for $\Omega$. Exemplarily, if choosing a fast lens\footnote{I.e., with a lens $f$-number around 1 or, in other words, with an aperture close to the physical limit.} and assuming a camera sensor of area $A_s$, the light collection efficiency for an object placed at a distance $L$ (when requiring that it is perfectly focused) reads (e.g. \cite{DDMpeople}):
\beq
\Omega_{_{CCD}} \simeq \frac{A_s}{4\pi L^2} \simeq \frac{1}{16(1+1/M)^2} \label{OmegaCCD}
\eeq
where $M$ is the magnification of the optical system.\footnote{$M$: ratio of the object size at the image plane divided by its size at the object plane.} Hence, a configuration consisting of a typical sensor of $1$\,cm$^2$ imaging a $\sim 1$\,m$^2$ plane will display $\Omega \sim 10^{-5}$. Perhaps surprisingly, there are indeed ways to overcome this limitation in some applications, with the help of avalanche scintillation (see next subsection).

\begin{figure}[h!!!]
\centering
\includegraphics*[width=7 cm]{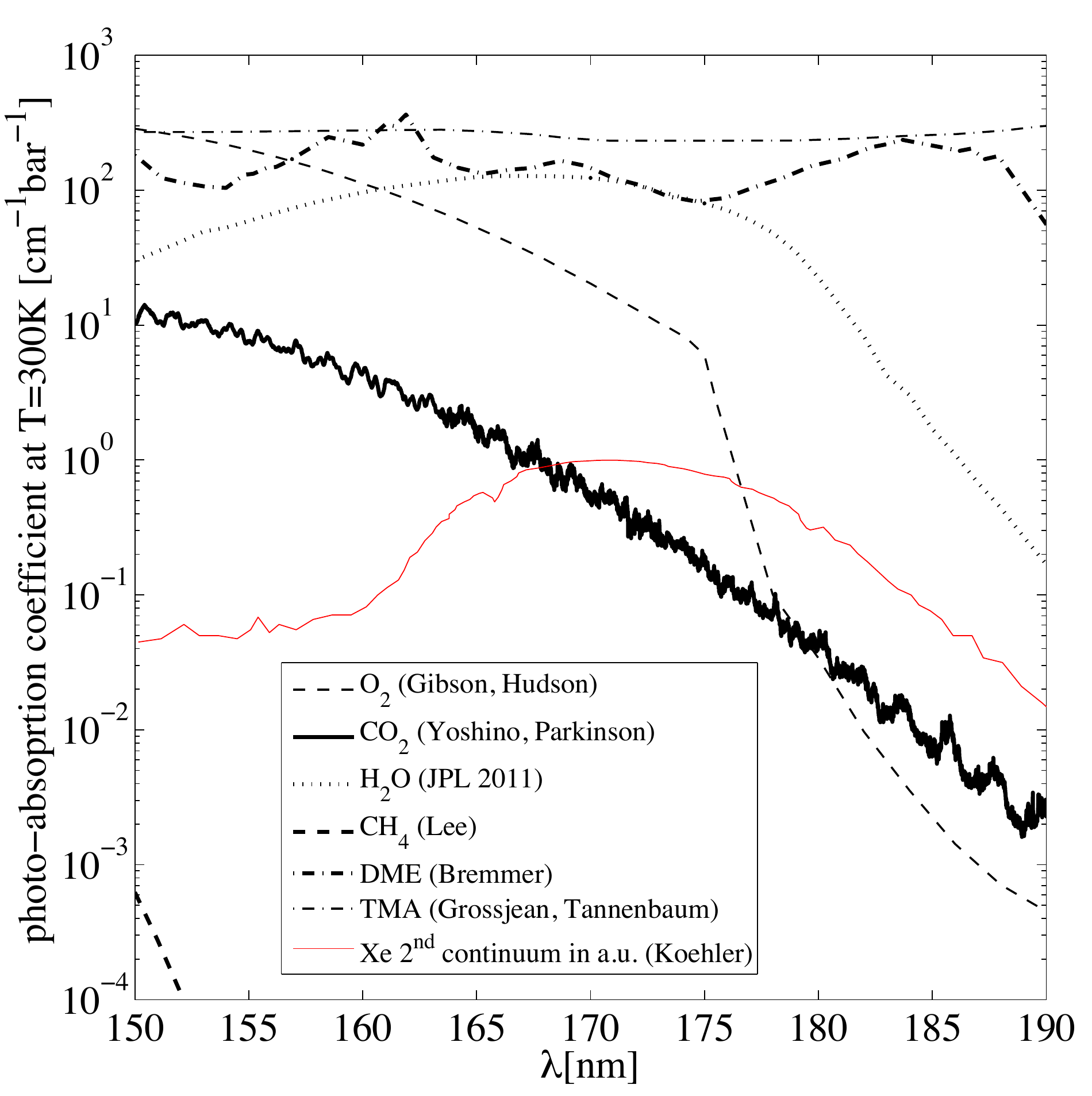}
\caption{Reduced photo-absorption coefficient ($\Pi_a/P$) in xenon as a function of the photon wavelength, after Gibson \cite{Gibson} and Hudson \cite{Hudson1, Hudson2} (for O$_2$), Yoshino \cite{Yoshino} and Parkinson \cite{Parkinson} (for CO$_2$), JPL \cite{JPL} (for H$_2$O), Lee \cite{Lee} (for CH$_4$), Bremmer \cite{Bremmer} (for DME), Grossjean \cite{Grossjean}, and Tannembaum \cite{Tannembaum} (for TMA), together with the $2^{\tn{nd}}$ xenon continuum from Koehler \cite{Koehler}. Information retrieved largely from the Mainz database \cite{Mainz}.}
\label{Xsec}
\end{figure}

\begin{figure}[h!!!]
\centering
\includegraphics*[width=\linewidth]{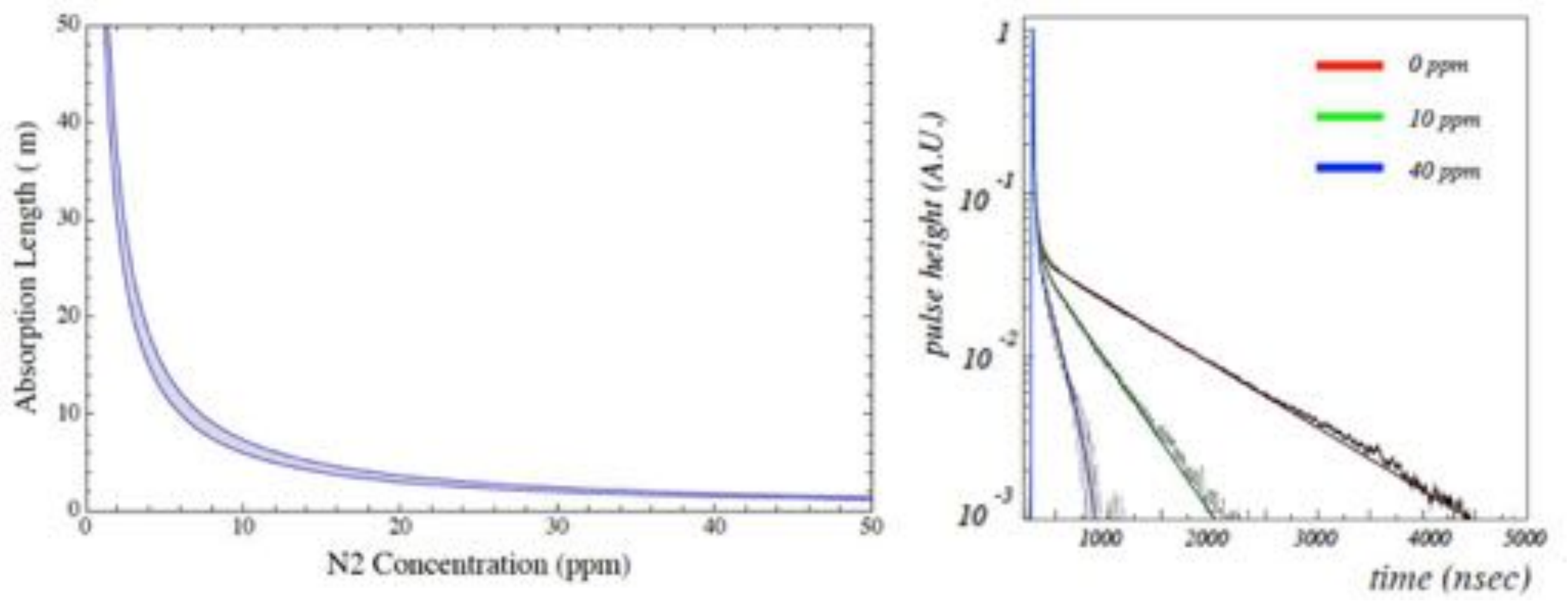}
\caption{Some scintillation properties of argon VUV-light in liquid argon, for different concentrations of N$_2$. Left: light attenuation \cite{BenTransp}. Right: excimer quenching (adapted from \cite{WarpQuenching}).}
\label{FigLambda}
\end{figure}

To the previous geometrical difficulties one must add the loss of scintillation to impurities, especially in big systems. Eq. \ref{LyReco} can be easily modified to include the effect in the number of collected photons:
\bear
&& n_{\gamma, c}= \mathcal{T} \cdot L_y \cdot \Delta\varepsilon \\
&& \mathcal{T}= 1-P_{p.a.} \sim \exp(-\Pi_a L^* ) \label{photoabs}
\eear
with $L^*$ defined again as the characteristic system size and $\Pi_a$ as the photo-absorption coefficient. If $\Pi_a$ does not change sizeably over the scintillation spectrum (not always the case, see for instance Fig. \ref{Xsec} for O$_2$ or CO$_2$ under xenon VUV-light), the `attenuation length' $\Lambda_{a}$ may be defined experimentally from the transparency $\mathcal{T}$, measured over a suitable length. Further, if the absorbed light does not lead to re-emission inside the photo-sensor band, then $\Lambda_{a}=\Pi_a^{-1}$.\footnote{As already noted, for non-monochromatic light or non-constant cross sections, eq. \ref{photoabs} should be weighted over the scintillation spectrum.} This fact has been used for instance to obtain the attenuation length for argon scintillation in liquid argon doped with N$_2$ \cite{BenTransp}, leading to:
\beq
\Lambda_{a}[\tn{m}] = -\frac{1}{100 \log{(1-p \cdot f_{N_2}[\tn{ppm}])}} \label{attLambda}
\eeq
with $p = 1.51\pm0.15 \times 10^{-4}$\,ppm$^{-1}$. The evaluation of eq. \ref{attLambda}
produces the curve in Fig. \ref{FigLambda}-left, from which a value for $\Lambda_{a}\simeq6$\,m at a N$_2$ concentration of $f_{N_2}=10$\,ppm can be extracted.
N$_2$ is a very common (but also particularly benign) contaminant. Most complex molecules can absorb even the very penetrating xenon light with ease (Fig. \ref{Xsec}) and so 10\,ppm of TMA in 10\,bar of xenon result in much lower values of around 0.3\,m for $\Lambda_{a}$.

It must be taken into account that the experimental determination of $\Lambda_{a}$ can be affected in liquid phase by Rayleigh scattering, whose interaction length $\Lambda_{s}$ is in the range 30-50\,cm in liquid xenon and 66-90\,cm in liquid argon (see \cite{Chepel} and references therein). Measurements for different system sizes, with and without reflectors are needed in order to disentangle effects.

Besides the problem of light attenuation, the scintillation probability ($P_{scin}$) is rapidly reduced in the presence of additives, too. Under the triplet dominance model, the scintillation probability can be experimentally obtained from the Stern-Volmer relations \cite{Stern-Volmer} for either the yield (eq. \ref{TDM}) or the modified lifetime, as:
\beq
\frac{1}{\tau'_{_{^3\Sigma}}} = \frac{1}{\tau_{_{^3\Sigma}}} + f \!\cdot \!K_{Q,^3\Sigma} \label{S-V}
\eeq

An example of this later experimental procedure is shown in Fig. \ref{FigLambda}-right, yielding $K_{Q,^3\Sigma}=K_{Q,^1\Sigma}$ = 110 ns$^{-1}$ for N$_2$ in pure liquid argon ($\tau_{_{^3\Sigma}}=1.26$\,$\mu$s), implying that ppm-level N$_2$ concentrations will keep the scintillation probability within a 90\% \cite{WarpQuenching}. An analysis performed over xenon mixtures by working directly on the yields, led to $\tau_{_{^3\Sigma}}=100$\,ns and $K_{Q,^3\Sigma} = 11$\,ns$^{-1}$ (at 1\,bar) for CO$_2$ as an additive, thus a $\times 150$ higher resilience of the scintillation throughput \cite{Mua}.

\subsection{Detection of charge and light}

\subsubsection{Photon detection}

\begin{figure}[h!!!]
\centering
\includegraphics*[width=\linewidth]{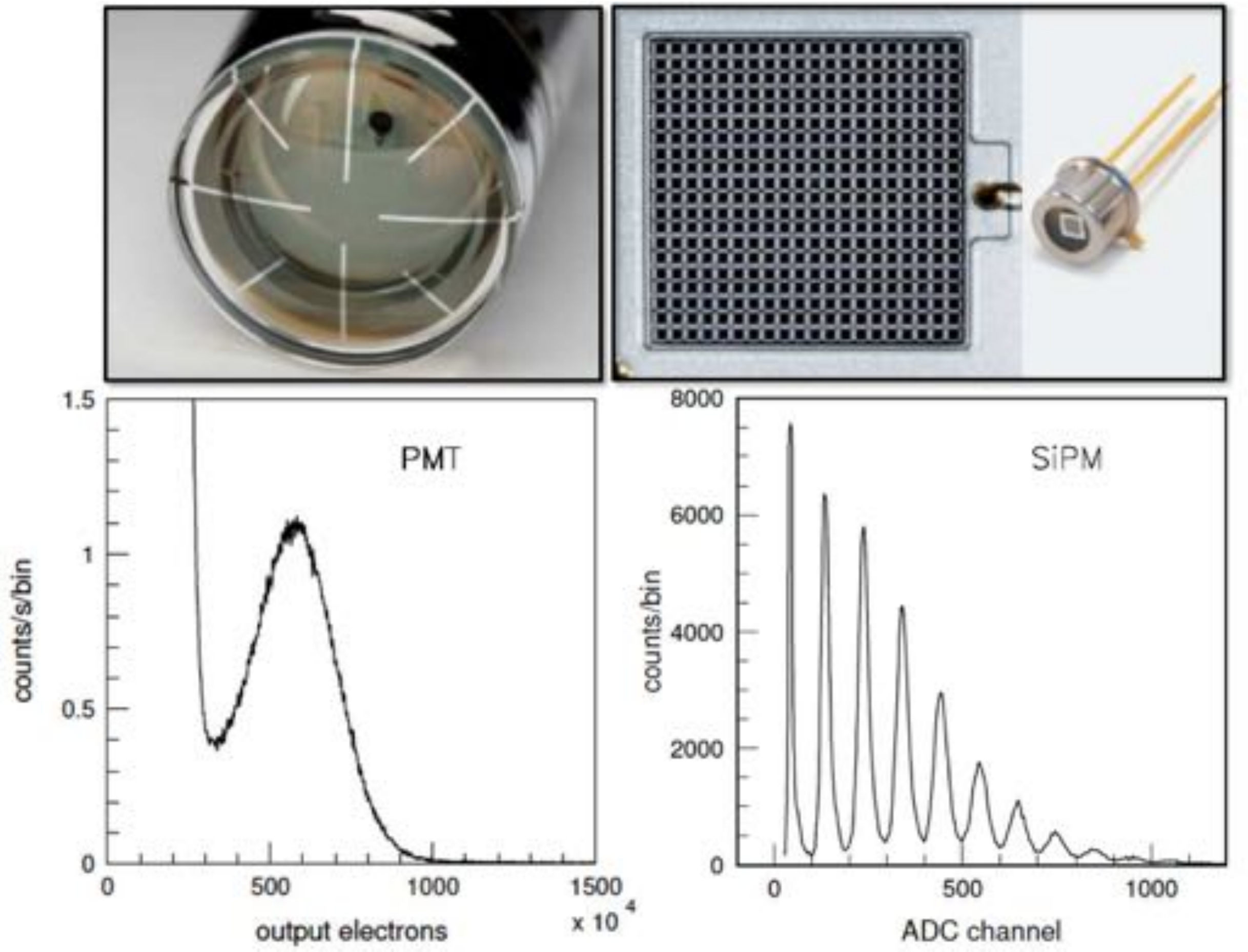}
\caption{Two representative photo-sensors for (left) large area coverage at high sensitivity, featuring a photomultiplier tube (PM), model R7724Q-MOD, and (right) fine position sampling with excellent photon counting capabilities, featuring a MPPC/silicon photomultiplier (SiPM). (Figures from \cite{Chepel})}
\label{Fig:Photosensors}
\end{figure}

\begin{figure}[h!!!]
\centering
\includegraphics*[width=\linewidth]{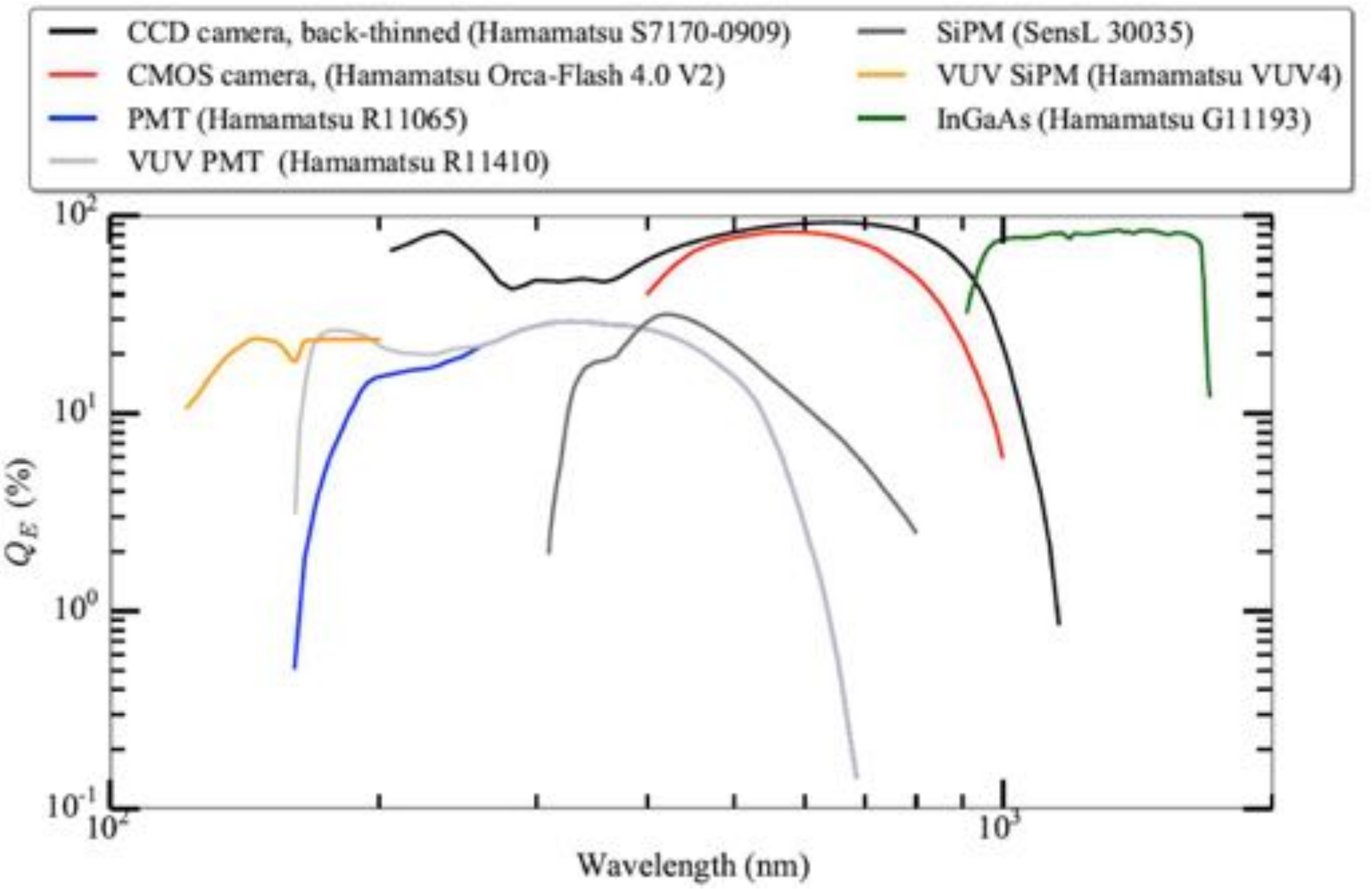}
\caption{A compilation of some devices typically used for recording scintillation from 120\,nm to nearly 2000\,nm in TPCs.}
\label{photocathd}
\end{figure}

Photon detection has been discussed at length in a recent review by Chepel and Araujo \cite{Chepel} and we refer the reader to it for additional extensive information. The problem of photo-detection in TPCs is ultimately related to sensitivity and spatial sampling. The first aspect encompasses: i) obtaining a good `peak-to-valley' ratio (equivalently: good single photon to noise pedestal separation),\footnote{A more rigorous figure of merit for the evaluation of the single photon detection capability is given later in section \ref{Single-det}.} that can be usually achieved with conventional vacuum-based photomultipliers (PMs) or silicon-based ones (SiPMs) as shown in Fig. \ref{Fig:Photosensors}, but also a number of other devices, like hybrid photo-detectors (HPD, \cite{HPD}), QUPIDs\footnote{QUartz Photon Intensifying Detector.} \cite{QUPID}, micro-channel plates (MCPs, \cite{MCP}) or gaseous photomultipliers \cite{MicromegasCsI}...; ii) minimizing thermionic and field emission (this is especially problematic for silicon-based devices in virtue of the small band gap of 1.12\,eV, but can be alleviated under cryogenic operation); iii) adjusting the photocathode/sensor characteristics to enhance the sensor response from the VUV to the near-IR region (Fig. \ref{photocathd}) and down to cryogenic conditions; and iv) affordability to cover large areas.

PMs are widely used for large area coverage in TPCs, since they can provide a good single-photon response (e.g., Fig. \ref{Fig:Photosensors}-left) and an acceptable QE in the range 20-40\%. The use of silicon-based devices like SiPMs (Fig. \ref{Fig:Photosensors}-right) or CCD/CMOS cameras unavoidably sacrifices coverage, but is indispensable for producing accurate images. It is thus not surprising that some optical TPCs decouple position reconstruction from other functions like calorimetry or detection of primary scintillation \cite{JJreview}. At the moment, SiPMs, PMs and CCD/CMOS cameras (coupled to solid wavelength-shifters or image intensifiers when needed) dominate the optical readout of gas and dual-phase TPCs. They are commercially available and researchers have been able to establish close collaborations with industry to tune the existing developments to their needs.

\begin{figure}[h!!!]
\centering
\includegraphics*[width=7 cm]{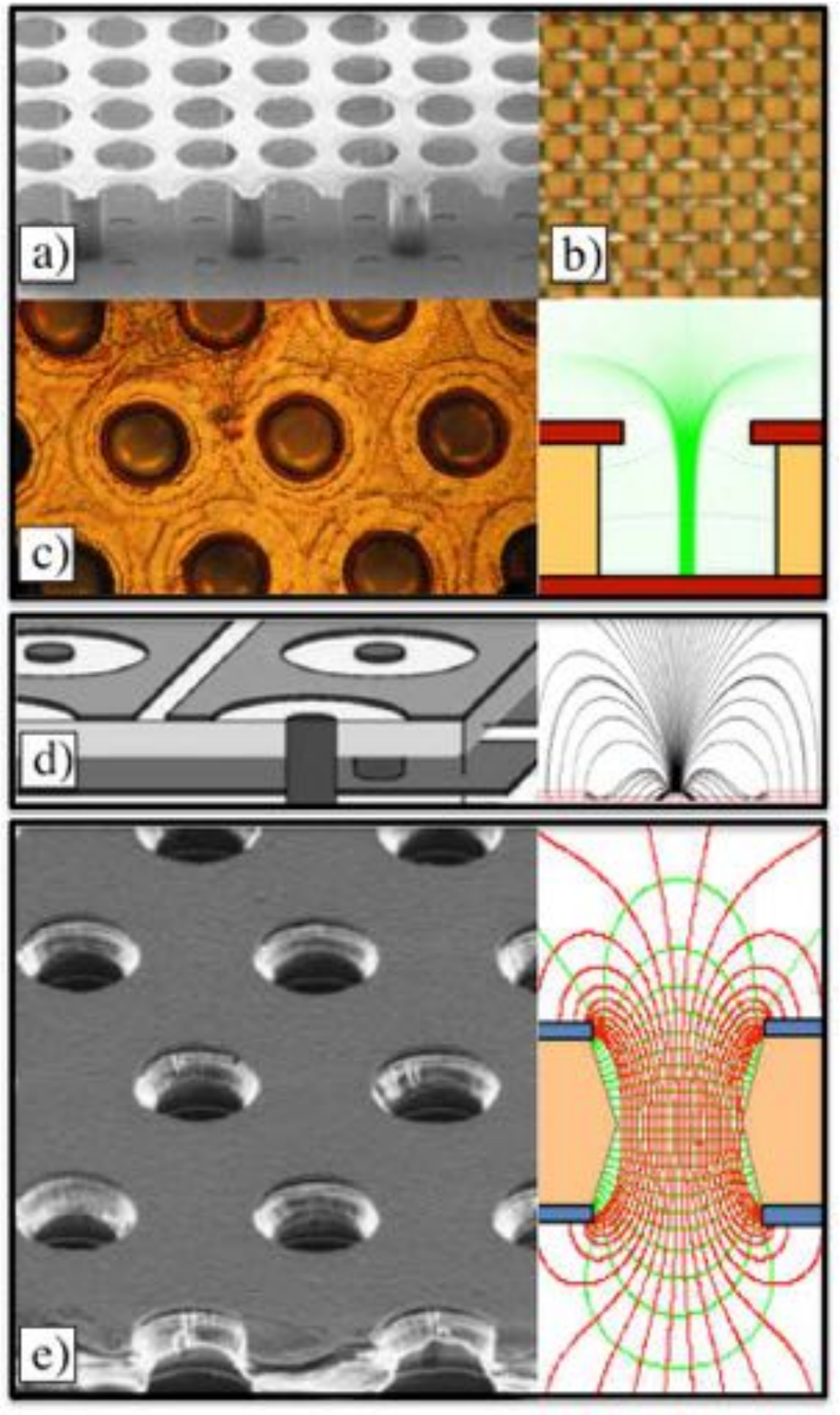}
\caption{Various representative micro-pattern gas detectors (MPGDs) used for charge multiplication inside TPCs: a) an InGrid device \cite{CAST}; b) top view of a woven mesh in a bulk Micromegas; 
c) top view of a microbulk Micromegas (resembling a GEM with an anode lid), courtesy of D. C. Herrera; d) $\mu$-PIC structure from \cite{MicroDotFirst}; e) a gas electron multiplier (GEM) \cite{SauliBook}. The diameter of the holes and anode-cathode distance in Figs. a), c), e) is $\simeq 50$\,$\mu$m. Also shown, for each set of structures, the electric field lines during typical operation conditions (right).}
\label{MPGDFigs}
\end{figure}

\subsubsection{Charge detection (I. avalanche multiplication)} \label{ChargeMultSec}
Comparatively, the detection of primary ionization in a TPC is much less standardized than photon detection: modern chambers must image particles over a broad range of ionization densities, gases and pressures and, unlike for photon detection, the multiplication structure must work necessarily under those very same conditions, and in a stable manner. Additionally, for low-background experiments in which radiopurity is a concern the (otherwise ubiquitous) glass fiber/epoxy composites have to be avoided and material selection enforced.

Almost invariably, the boundary between the drift and multiplication regions is physically realized at one of the TPC equipotential surfaces (very much akin to a Frisch grid \cite{Frisch}), and the field configuration suitably chosen in that region in order to optimize charge transmission (e.g., Fig. \ref{MPGDFigs}-right). This transfer should be done through structures with conveniently small feature sizes, not to distort the original ionization trail. At that point, the experiment requirements for sensitivity and spatial precision determine the multiplication factor and readout segmentation. Given these and the aforementioned constraints, and the high stakes of most experiments described here, the development of customized multiplication structures is a desirable feature.

\begin{figure}[h!!!]
\centering
\includegraphics*[width=7 cm]{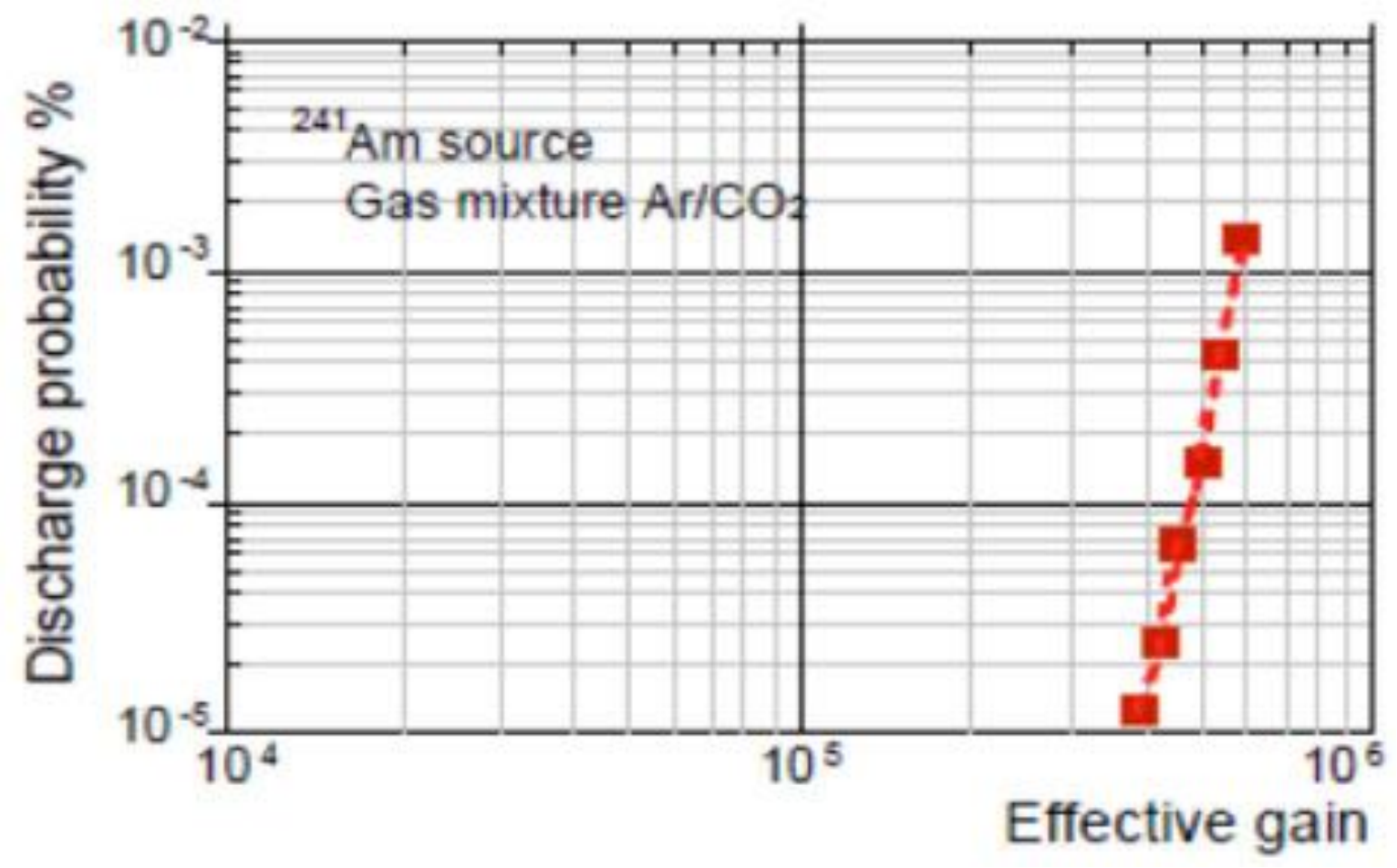}
\caption{Discharge probability as a function of the effective charge gain, obtained for a 3-GEM structure operating in an Ar/CO$_2$ mixture during the CMS R$\&$D phase \cite{JeremyPhD}. Irradiation done with $\alpha$'s from $^{241}$Am.}
\label{Jeremy}
\end{figure}

While wire-based multiplication (at some mm's pitch) was used at collider TPCs for decades, nowadays either hole, point, strip or gap-based multiplication can be easily achieved, in structures having their feature size below $100$'s of $\mu$m. Furthermore, such a precision is available over m$^2$-areas, benefiting from industrial mass-scale procedures such as computerized drilling, photo-lithography and chemical etching. Structures of this kind are commonly referred to as micro-pattern gas detectors, or MPGDs, and have been subject to an active development over the past two decades (for details the reader is referred to \cite{RopeTitov}, and to the comprehensive textbook on gaseous detectors \cite{SauliBook}). MPGD architectures are versatile (Fig. \ref{MPGDFigs}) and, in addition, they can achieve stable charge gains nearing $10^6$ through genuine 3D manufacturing techniques like field-correcting electrodes, embedded resistors
or stacking \cite{RopeTitov, Max, JeremyPhD}, to name a few. Signal induction and routing benefit, similarly to charge multiplication, from established techniques for the micro-fabrication of the readout electrodes (e.g., for applications demanding high spatial precision), as well as embedded vias or multi-layer technology (to deal with high channel densities). Conveniently, this allows in some applications for compact `all-in-one' designs with the charge multiplication, induction and routing planes bonded into the same structure (e.g, 
, \cite{uBulkMM}) and even the front-end electronics can be integrated too (e.g., \cite{CAST}).

\begin{figure}[h!!!]
\centering
\includegraphics*[width=6.5 cm]{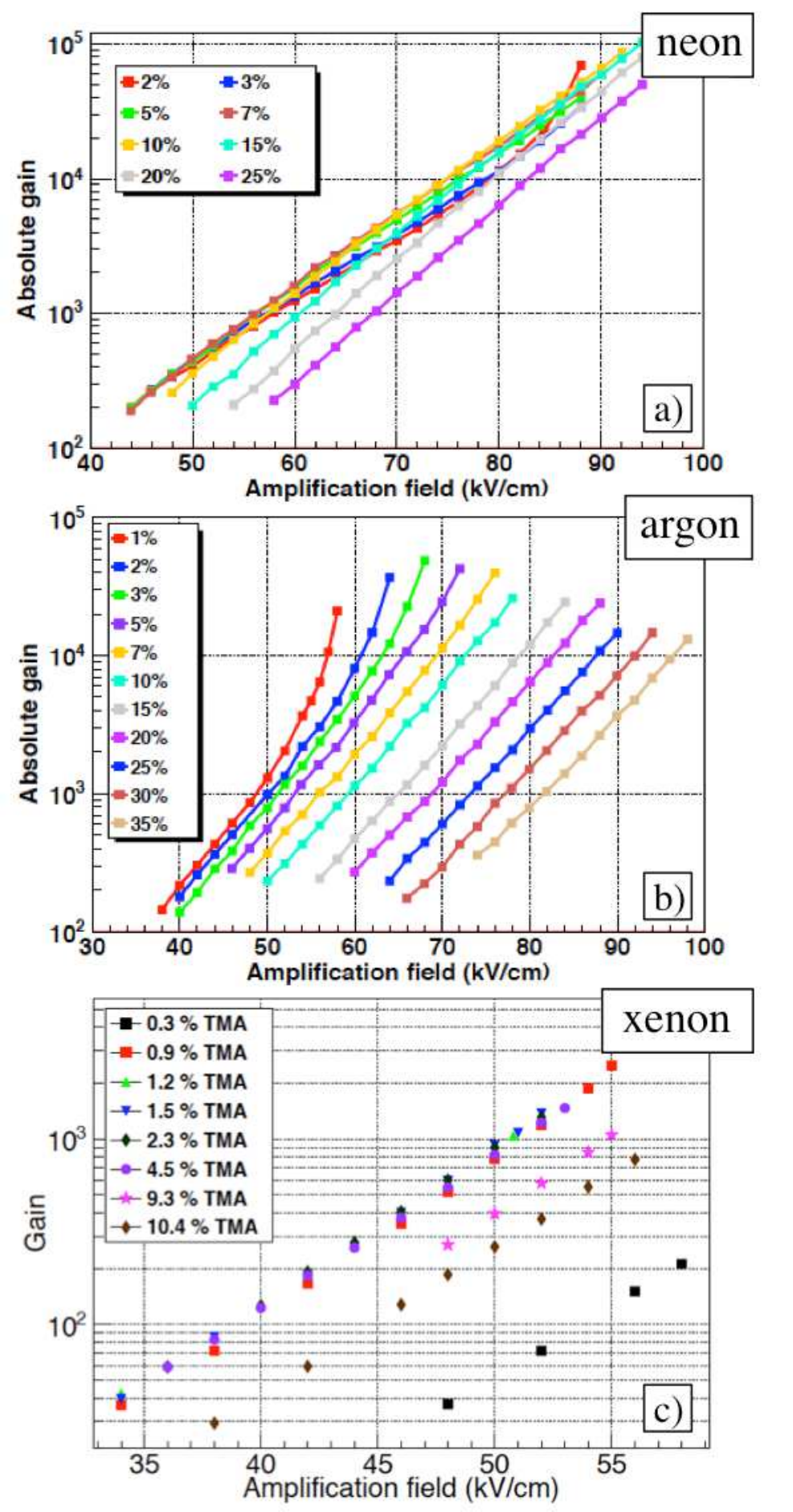}
\caption{Gain curves obtained for $\sim 20$\,keV x-rays in noble gases at around atmospheric pressure, by using small area (10\,cm$^2$) ${50\,\mu}$m-thick microbulk Micromegas, after \cite{PacoNeAr} and \cite{DianaTMA2}. a) Ne/i-C$_4$H$_{10}$; b) Ar/i-C$_4$H$_{10}$; c) Xe/TMA. }
\label{GainHierarchy}
\end{figure}

\paragraph{Operation in quenched gases.}

\begin{figure}[h!!!]
\centering
\includegraphics*[width=\linewidth]{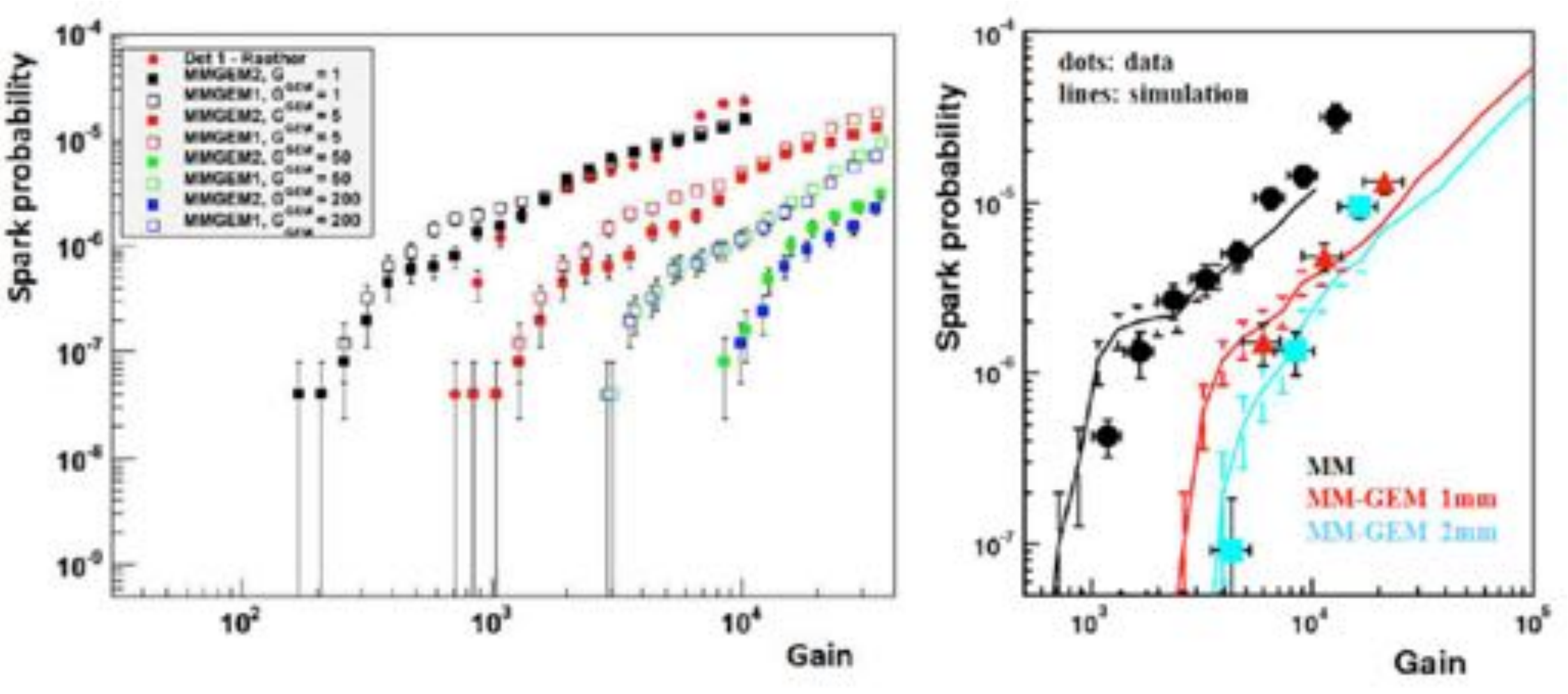}
\caption{Left: simulations of the spark probability for a 2-stage amplification structure (GEM+Micromegas) in Ar/i-C$_{4}$H$_{10}$(90/10) when assuming a charge density -based breakdown criteria ($d_{s}^{lim} = 2\times10^9\,$e$^-/$mm$^2$) and comparison with a classical breakdown criteria (Raether \cite{Raether}) based on the total charge (circles). Right: comparison with experiment for three different configurations (in decreasing order of charge density at the Micromegas plane: black, red, and blue). (After \cite{ProcureurGain})}
\label{ProcuFantastic}
\end{figure}

With MPGDs, an \emph{effective} charge gain of $m^*=10^4$-$10^6$ is regularly achieved for ionizing particles under well quenched gases, especially for cascaded amplification (Fig. \ref{Jeremy}).\footnote{The presence of several nearby electrodes in the case of MPGDs can lead to small charge induction losses, resulting in the concept of `effective gain' (e.g. \cite{bellazini}). It will be denoted through letter $m$ (standing for multiplication factor) with a star.} An important hierarchy has been established recently through a series of theoretical and experimental works indicating that (except perhaps for helium) the multiplication process shows systematically higher fluctuations the heavier the noble gas \cite{XePen, Rob, Zerguerras}. Although hardly the only explanation, as pointed out in \cite{Zerguerras}, this fact certainly mirrors the hierarchy observed in the maximum achievable gains in single-stage devices, that can differ by more than one order of magnitude between neon and xenon -based mixtures (see for instance works leading to Fig.\ref{GainHierarchy}, as well as \cite{Thers}). This is of course a very important practical observation. Authors in \cite{Procureur} argued that the difference in the $W_I$-values as well as nuclear interaction rates is insufficient to explain the observed effect, hinting strongly at the role of the ionization density, an effect identified later in \cite{ProcureurGain}. For illustration, Fig. \ref{ProcuFantastic}-right shows the discharge probability for three different situations, corresponding to different charge densities at the entrance of the main amplification structure (a Micromegas detector): if assuming a critical charge density $d_{s}^{lim} = 2\times10^9\,$e$^-/$mm$^2$, a good description of the measured spark probability could be achieved.\footnote{A similar work has recently appeared for the case of single GEMs, pointing to a critical charge value of $4.7(7.3) \times 10^6$\,e$^-$/hole in case of argon(neon)-based mixtures \cite{Piotr}. It translates (if crudely taking the 50\,$\mu$m inner diameter of the GEM holes as a reference) into $d_{s}^{lim} = 2.4(3.7)\times10^9\,$e$^-/$mm$^2$.} In fact, much more comfortable gains could be achieved for xenon mixtures, approaching $2\times10^4$, in a 3-GEM detector operated in Xe/CH$_4$ (98/2) under $\sim 30$\,keV x-rays \cite{BondarXeCH4}.

\paragraph{Operation in noble gases.}

\begin{figure}[h!!!]
\centering
\includegraphics*[width=8cm]{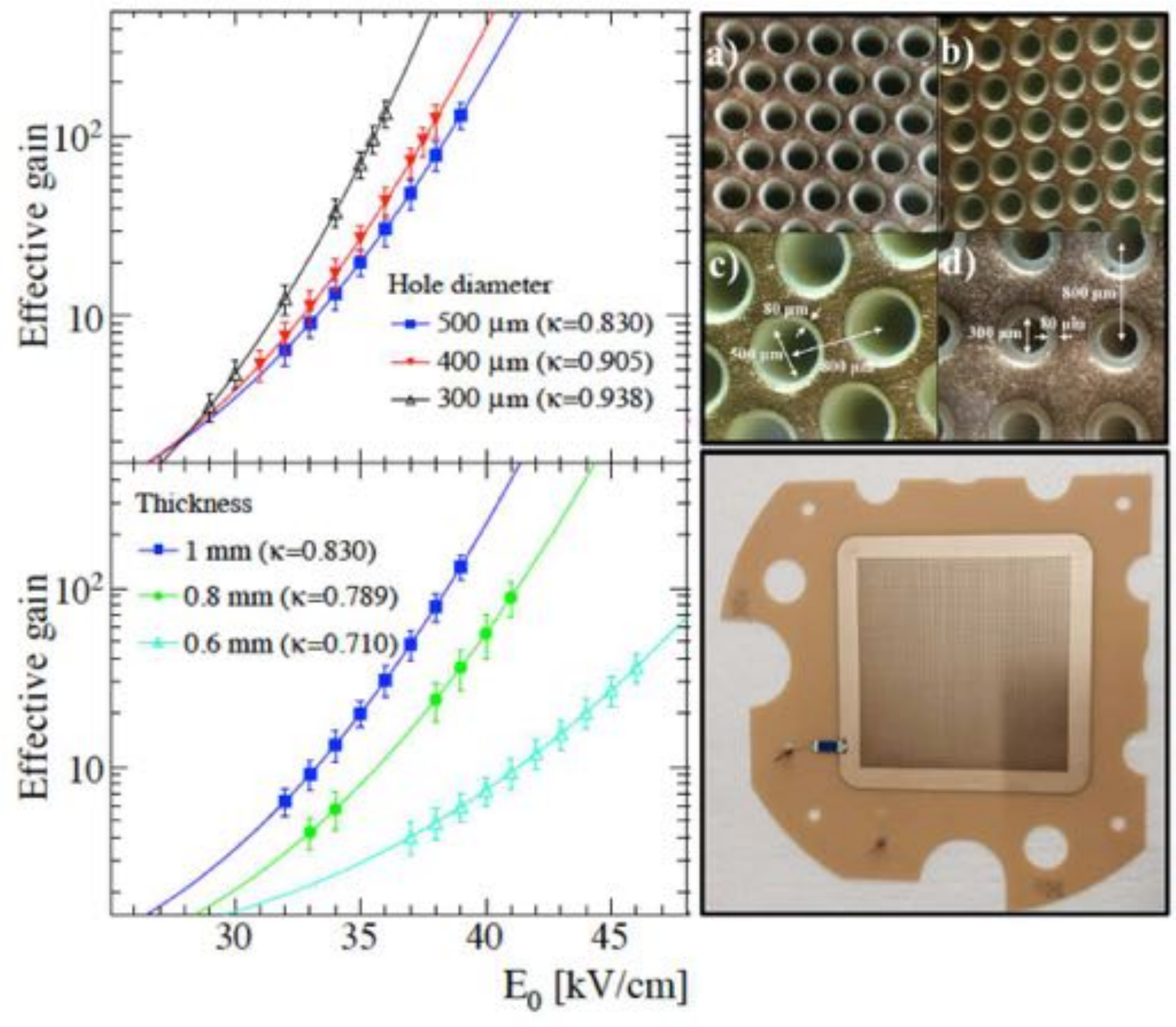}
\caption{Left: gain curves obtained for cosmic ray muons in several LEM structures, operated in the gas phase of a dual-phase argon chamber purified down to ppm levels \cite{Cantini:2014xza}. Several of the architectures studied appear photographed to the right. The final working gains were situated in the range $m^*=20$-$40$, after accounting for charging-up (see later in text).}
\label{LEMgain}
\end{figure}

MPGDs can operate reliably in ultra-pure noble gases, too (essential for experiments exploiting scintillation), however a gain of just a few 10's becomes challenging, in general \cite{PeskovGain}. Extensive prototyping has established the possibility of working at stable gains in the range 20-40 for single-stage LEMs \cite{Cantini:2014xza}, in ultra-pure argon under cryogenic conditions ($T=87$\,K, $P\sim1$\,bar), and is shown in Fig. \ref{LEMgain}. It is often argued (although unproven) that such gains are possible by virtue of the closed geometry of MPGDs, that confines scintillation and minimizes photoelectric effect at metallic electrodes. Moved by this driving force, a vast research has taken place over the last decade, and values much higher than 10 and even 100 can be found in literature for several MPGD geometries operated in pure noble gases (for an extensive review work, see \cite{Buzu}). As noted in \cite{PeskovGain}, however, the maximum stable gain can drop by a big factor depending on the actual gas purity conditions, and plausibly the available scintillation too. The interested reader is advised to look carefully at the purity conditions under which each measurement has been realized.

\begin{figure}[h!!!]
\centering
\includegraphics*[width=8cm]{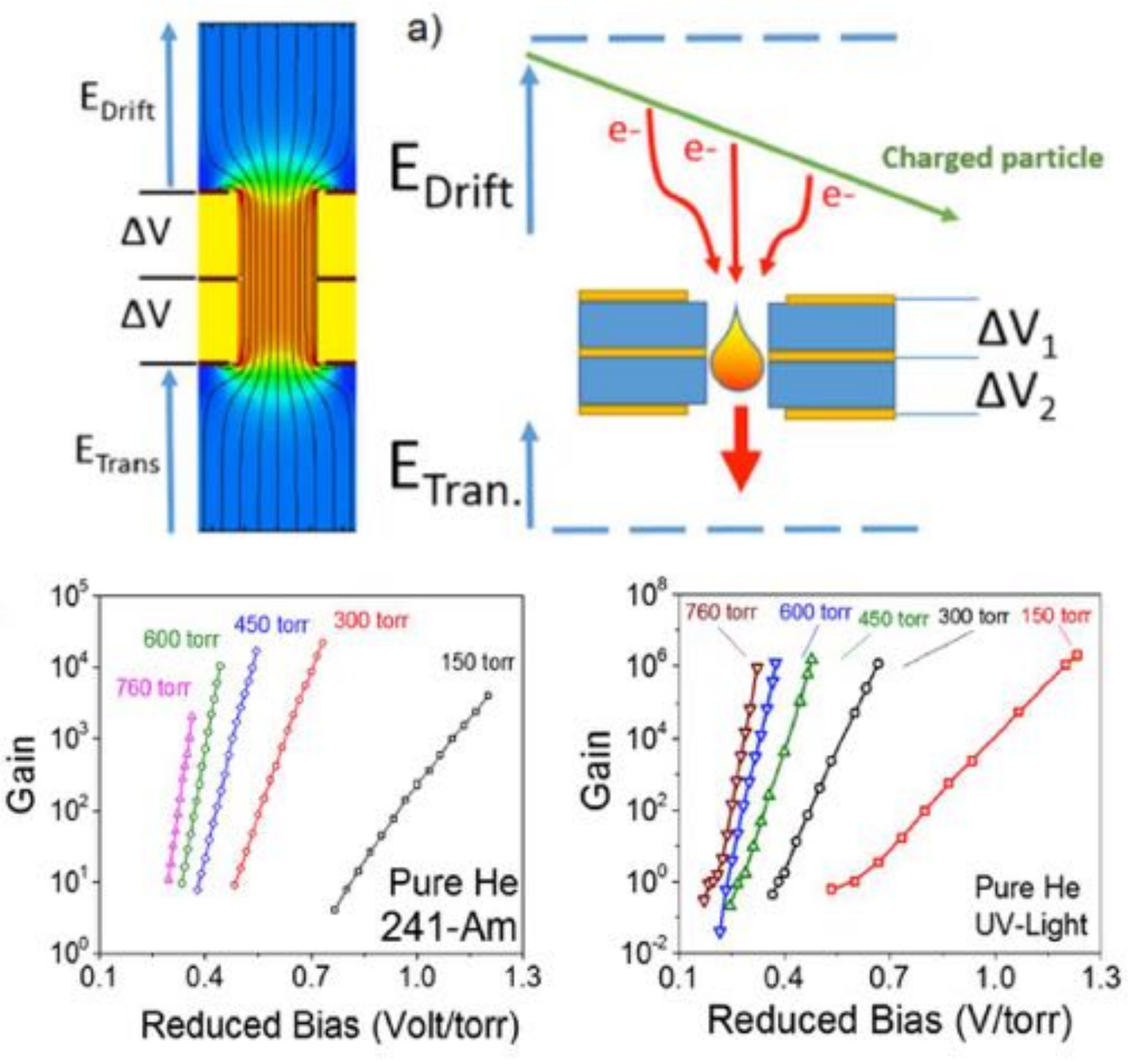}
\caption{Top: field configuration and layout of the recently introduced multi-layer thick GEM, aimed at operation in pure helium under variable pressure conditions. Bottom-left: gain curves for $\alpha$'s from a $^{241}$Am source. Bottom-right: gain under single-electron conditions. (Figures from \cite{Cortesi})}
\label{FigCortesi}
\end{figure}

There are situations, however, where trace-amount impurities are likely to have much less practical importance, e.g. for experiments involving nuclear reactions on light nuclei, that exploit the possibility of performing complete event reconstruction in inverse kinematics. In this context, a novel amplification structure has been recently introduced and operated in pure helium \cite{Cortesi}. The structure (dubbed multi-layer thick-GEM) works like an extended thick-GEM with intermediate field grading (Fig. \ref{FigCortesi}). The authors achieved an encouraging $m^*=10^4$ landmark value for $\alpha$ particles, although purity conditions were not reported.

\paragraph{Operation in electronegative gases.}

\begin{figure}[h!!!]
\centering
\includegraphics*[width=7cm]{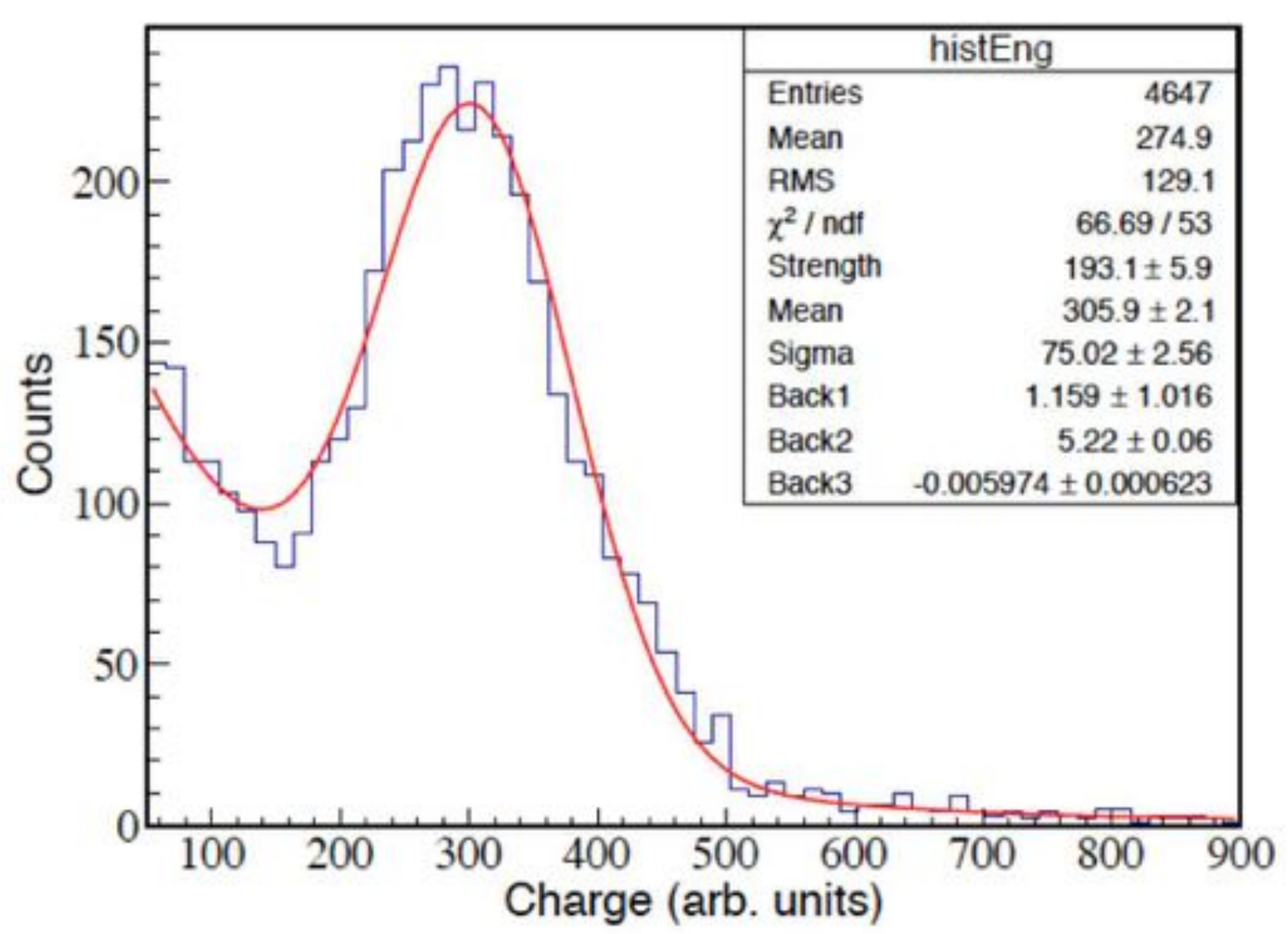}
\caption{Energy spectrum from $^{55}$Fe x-rays ($\varepsilon=5.9$\,keV) in pure SF$_6$ at around 40\,mbar, recorded with a 0.4\,mm-thick GEM with 0.4\,mm diameter holes \cite{PhanSF6}. The authors claim SF$_6$ to be the ideal gas for directional dark matter detection given its high cross section, low diffusion and availability of minority-carriers for fiducialization.}
\label{FigPhanSF6}
\end{figure}

A remarkable, and yet more unconventional, operation has been demonstrated for electronegative gases. Either GEMs \cite{MiyaCS2} or thick-GEMs \cite{PhanSF6} have been shown to allow an efficient negative ion stripping at pressures of technological interest, even for the extremely electronegative SF$_6$ ion (electron affinity = 1.06\,eV). The reduced electric field applied in \cite{PhanSF6} to obtain the $^{55}$Fe x-ray spectrum shown in Fig. \ref{FigPhanSF6} (500\,kV/cm/bar) dwarfs the ($\sim 100$\,kV/cm/bar) fields reached in the fastest SF$_6$-doped timing RPCs to date \cite{DiegoTiming}. An effective gain of 3000 was reported in those conditions.

\paragraph{Some technical details about the operating gain in MPGDs.}

It has been already mentioned that charge density is an important parameter for the stability of the multiplication process in MPGDs, and therefore primary ionization density and particle type need to be taken into account when interpreting the maximum stable gain achievable as a function of the bias voltage. Besides this important fact, and provided electrical breakdown is a stochastic process, the maximum gain achieved is bound to depend on the sensor area (through the probability of having a defect) and particle rate (through the probability for a particle to `hit' that defect). Those have, however, a minor impact in the conditions discussed in this paper, due to the relatively low particle rates characteristic of TPCs, and the high mechanical accuracy of MPGDs.\footnote{HV-segmentation may be additionally implemented in order to limit the larger spark energy ensuing when dealing with large areas, e.g., for GEM detectors.} As an example, the 3-GEM detector in \cite{PatrickRate} could be operated up to 10\,MHz/cm$^2$ of 5.9\,keV x-rays without any sign of instability. At much lower rates corresponding to around 100\,Hz, the maximum gain achieved for a microbulk Micromegas detector in \cite{NEXTMM} decreased by only a factor of two when going from 10\,cm$^2$ to 700\,cm$^2$ sensors. In the same conditions, the energy resolution ($\sigma$) for x-rays deteriorated from  $17.7\%/\sqrt{\varepsilon[\tn{keV}]}$ to $19.8\%/\sqrt{\varepsilon[\tn{keV}]}$ at 1\,bar, and from $21\%/\sqrt{\varepsilon[\tn{keV}]}$ to $34\%/\sqrt{\varepsilon[\tn{keV}]}$ at 10\,bar. Similar figures have been reported for LEMs, gain-wise.

\begin{figure}[h!!!]
\centering
\includegraphics*[width=\linewidth]{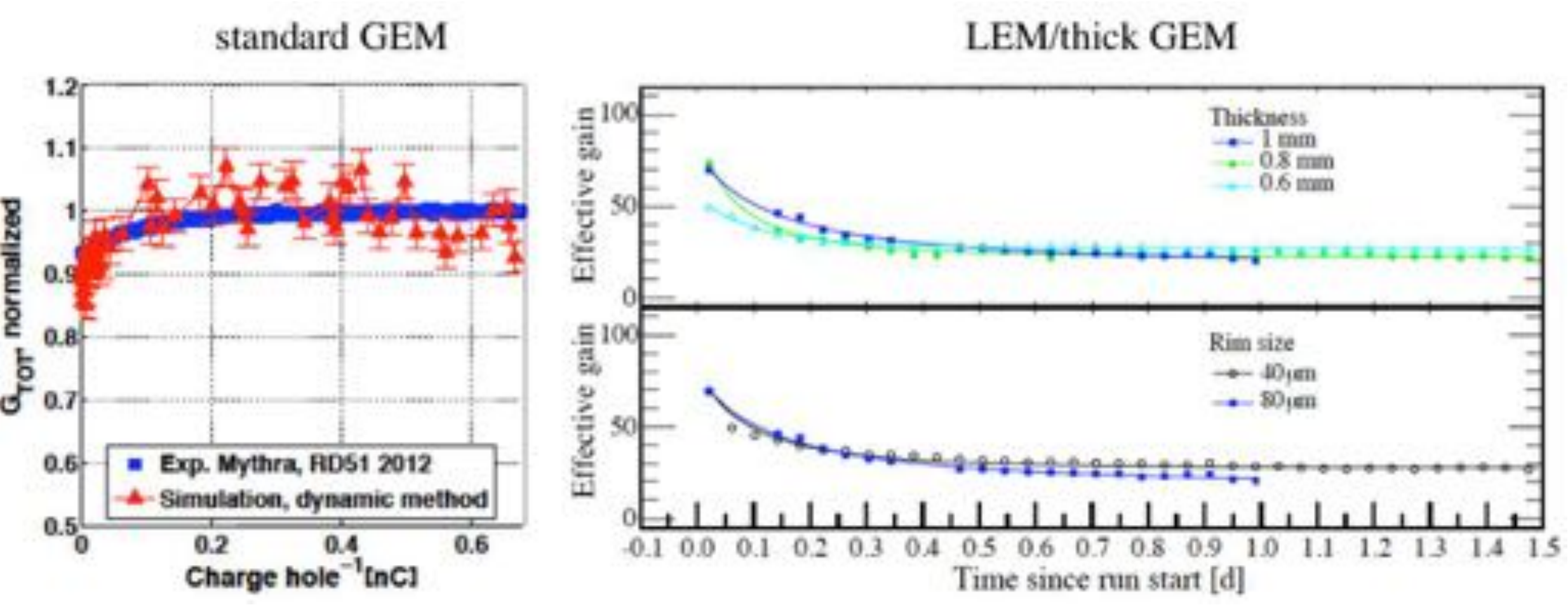}
\caption{Left: measurements of the transient gain behaviour as a function of the charge rate per hole in a GEM detector, and comparison with a Garfield++ simulation (from \cite{Correia}). Right: transient gain behaviour observed for several thick-GEM architectures, after \cite{Cantini:2014xza}. The different behaviour for each architecture is related to the geometry in the hole region.}
\label{TransientFig}
\end{figure}

A notable characteristic of MPGDs when compared to wire-based amplification is that, due to the use of insulating materials, they can introduce additional time constants that ultimately lead to field reconfiguration in the multiplication region. The effect is important for structures in which the insulator is exposed to the avalanche. This behaviour leads for instance to a $\sim10$\% gain increase in standard GEMs, but can easily cause a 50\% decrease in thick-GEMs (Fig. \ref{TransientFig}). The charging-up process depends on the incoming particle rate, and can be particularly slow for low rate experiments. After the charging-up process ends, the detector remains stable.

\paragraph{Gain fluctuations.}

Unavoidably, the charge multiplication process introduces an additional source of fluctuations in the number of recorded electrons ($\bar{n}_{e,r}$) that translates, for contained events, into an energy resolution:
\bear
& \frac{\sigma_{_{\Delta\varepsilon}}}{\Delta\varepsilon} \equiv \frac{\sigma_{n_{e,r}}}{\bar{n}_{e,r}} & \simeq \sqrt\frac{F_e + \mathcal{R} + \mathcal{A} + 1-\mathcal{T}_e + f_m}{\bar{n}_{e,c}} \label{f-factor} \\
& \bar{n}_{e,r} & = m^* \cdot \bar{n}_{e,c} \label{Nread}
\eear
(and analogously for ions). In single multiplication stages, $f_m$ can be interpreted as the relative standard deviation of the multiplication process, squared
($\sigma_{m^*}/{\bar{m}^*})^2$. Versions of eq. \ref{f-factor} can be obtained for cascaded architectures \cite{Varga} and in the presence of small construction imperfections \cite{XePen}.

\begin{figure}[h!!!]
\centering
\includegraphics*[width=\linewidth]{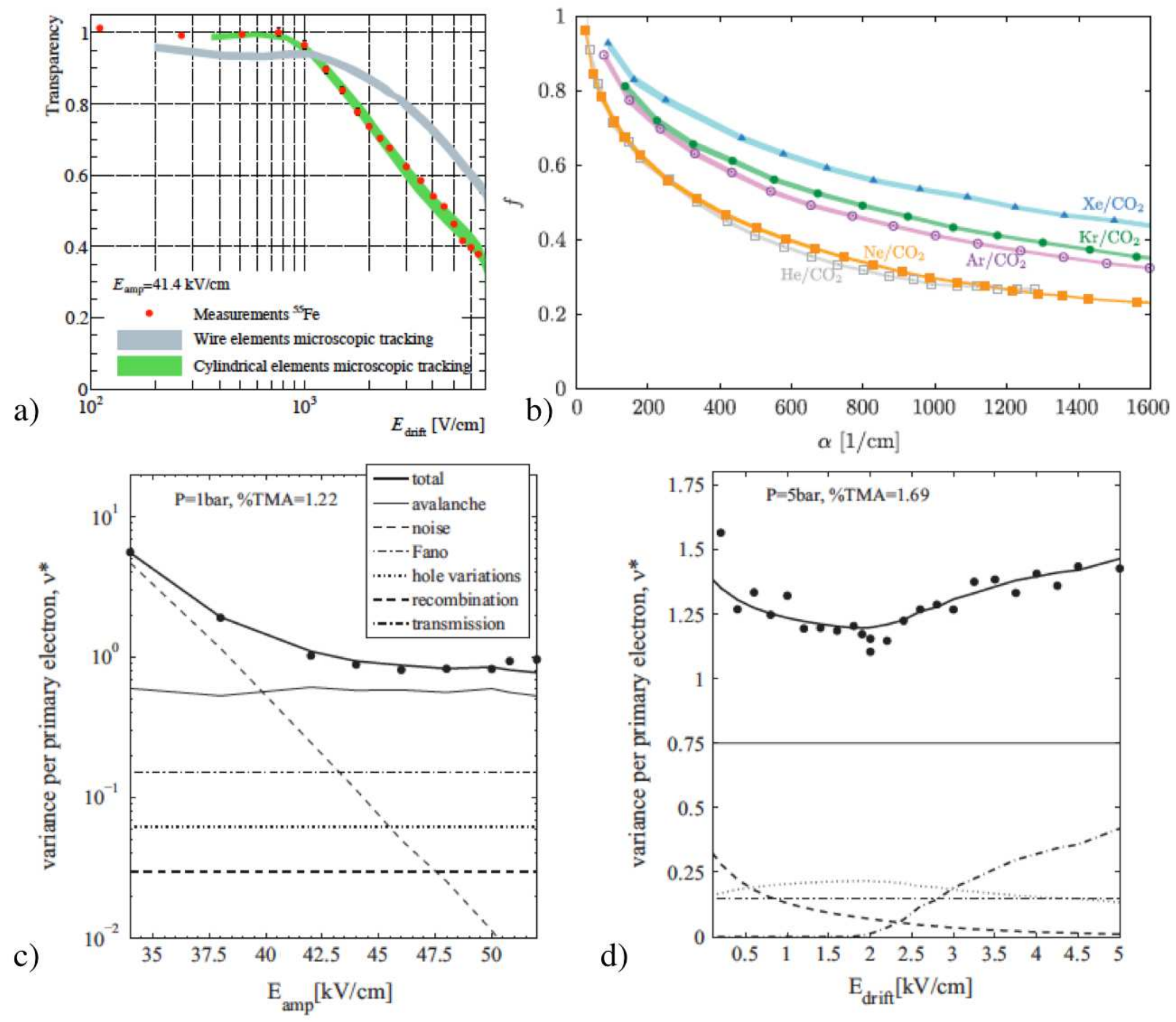}
\caption{Measurements and comparison with Garfield++ simulations \cite{Garfield++} of some important aspects for the calorimetric response achievable with MPGDs. a) Electron transmission, $\mathcal{T}_e$, (dubbed `transparency' by the authors) for a woven mesh as the ones typically employed in bulk Micromegas \cite{refTranspRob} (results for holes can be found for instance in \cite{XePen}). b) Simulated $f_m$-value (relative spread of the multiplication process, squared) obtained in noble gas/CO$_2$ mixtures under uniform fields, as a function of the multiplication coefficient $\alpha$ \cite{Rob} (note the authors' convention $f\equiv f_m$). This behaviour, with the multiplication spread being larger the heavier the noble gas, can be found experimentally too \cite{XePen, Zerguerras}. c), d) Variance per collected electron, $v^* = \bar{n}_{e,c}\cdot(\sigma_{\Delta\varepsilon}/\Delta\varepsilon)^2$, obtained for microbulk Micromegas under Xe/TMA admixtures for two different pressures \cite{XePen}, together with the different contributions in eq. \ref{f-factor}. Note the increase observed at 5\,bar, relative to 1\,bar.}
\label{TranspAndF}
\end{figure}

While electron transmission ($\mathcal{T}_e$) can be made nearly 100\% in practical operating conditions for conventional architectures based on either meshes or holes (e.g. Fig. \ref{TranspAndF}-a) and thus it has a small impact on the detector performance, $f_m$ can deteriorate the intrinsic energy resolution of a TPC by a factor of two or three, depending on gas and pressure (Fig. \ref{TranspAndF}-b,c,d).

\subsubsection{Charge detection (II. secondary scintillation)} \label{SecScint}

\begin{figure}[h!!!]
\centering
\includegraphics*[width=\linewidth]{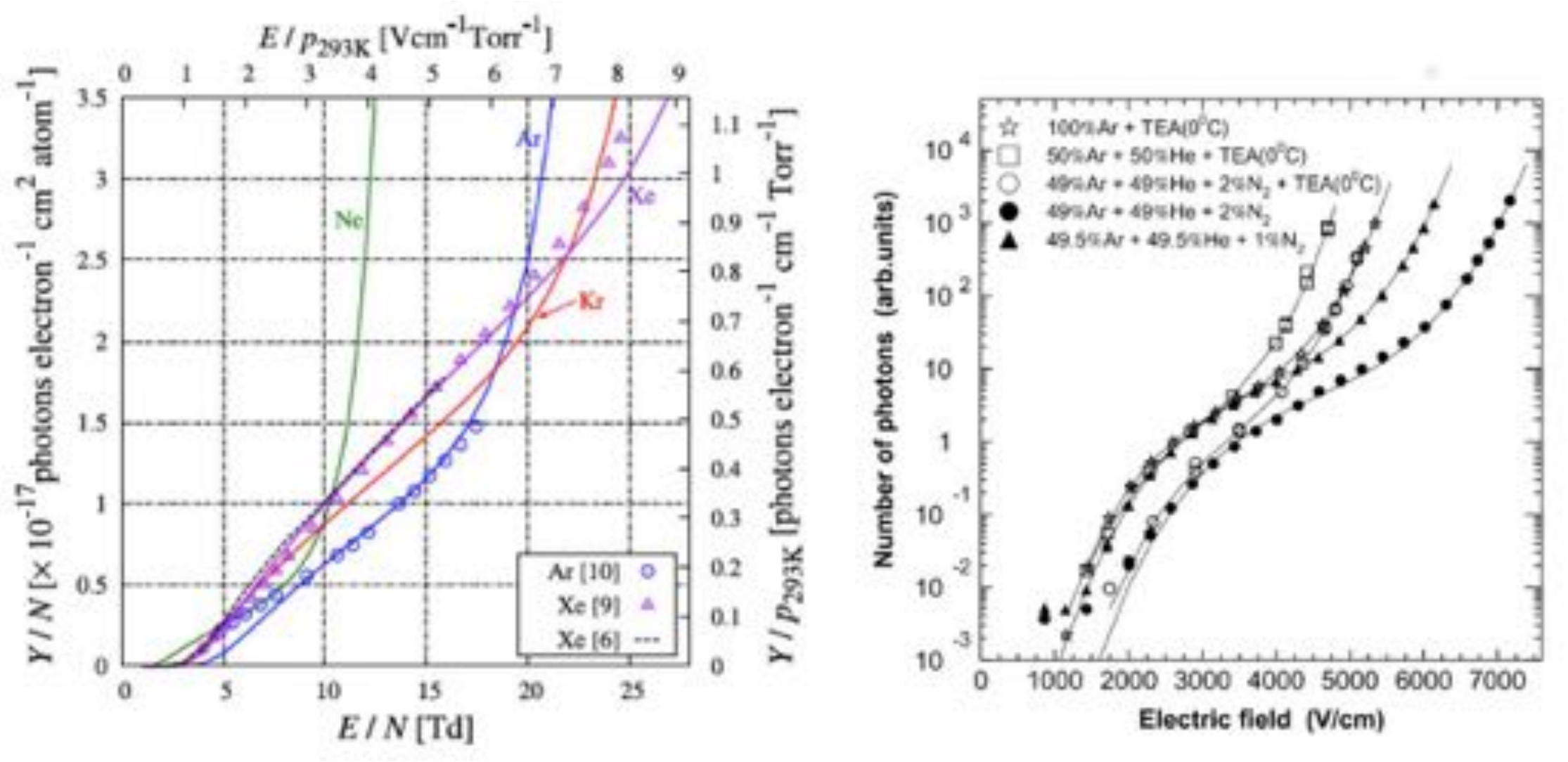}
\caption{Left: measured electroluminescence yield ($Y$) in xenon (triangles) and argon (circles) as a function of the electric field, and comparison with simulations for all gases, after \cite{oliveira} (recent results from krypton can be found in \cite{CrisKr}). Right: scintillation process in the presence of molecular additives, from \cite{OTPCfirst}. A proportional scintillation regime followed by an exponential (avalanche-driven) one can be appreciated in both cases.}
\label{ELOTPC}
\end{figure}

The large sensitivity of optical devices allows operation of imaging chambers in pure scintillation mode, a mode known as
`proportional scintillation' or `electroluminescence' (EL) \cite{Policarpo}. It naturally occurs for electrons at intermediate fields,
under which they can promote the gas species to their various excited states, while avalanche multiplication is still negligible.
Electroluminescence is a very energy-efficient scintillation process particularly in noble
gases, provided i) only elastic interaction can compete with excitation, ii) virtually any excited state is a precursor to VUV-scintillation,
iii) noble gases are self-transparent for typical TPC pressure conditions and, contrary to
molecular species (e.g., \cite{TMAYasu,Cureton}), their scintillating states cannot self-quench. These facts result,
above a certain threshold, in an approximate linear behaviour of the optical gain (number of photons per electron)
as a function of the total energy gained by the electron:
\beq
m_{\gamma} \simeq K(V-V_{th}) \label{eqEL}
\eeq
It is customary to introduce the scintillation yield per unit length and per electron ($Y$), after which eq. \ref{eqEL} can be rewritten as:
\beq
\frac{Y}{N} = K(E^*-E^*_{th}) \label{EL2}
\eeq
with $N$ being the number density, i.e. the number of atoms (molecules) per unit volume, and $E^*=E/N$ the reduced field. The threshold field for electroluminescence is in the neighborhood of $E^*_{th}\simeq1$\,kV/cm/bar ($\sim 4$\,Td),\footnote{1 Townsend $\equiv 1$\,Td = $10^{-17}$\,Vcm$^2$ are
reduced field ($E/N$) units. Since at $P=1$\,bar and $T=20\,^\circ{\tn{C}}$ the number density is $N_0=2.5\cdot 10^{19}$\,cm$^{-3}$ for an ideal gas, 1\,Td roughly corresponds to a reduced field ($E/P$) of 250\,V/cm/bar at $T=20 ^\circ{\tn{C}}$. Either Td, V/cm/bar or V/cm/Torr are units found often in the practice of gaseous electronics.} slightly depending on the noble gas (Fig. \ref{ELOTPC}-left).

Contrary to the $f_m$-value for charge multiplication, the high efficiency
of the electroluminescence process results in very small fluctuations ($Q \ll f_m$), at the level of $10^{-3}$ or less \cite{oliveira}. After including charge
losses in the drift, it is possible to write an approximate formula for the energy
resolution of fully contained events, as \cite{oliveiraRes}:
\bear
& \frac{\sigma_{_{\Delta\varepsilon}}}{\Delta\varepsilon} \equiv \frac{\sigma_{n_{pe}}}{\bar{n}_{pe}} & \simeq \sqrt{\frac{F_e + \mathcal{R} + \mathcal{A} + 1-\mathcal{T}_e + Q}{\bar{n}_{e,c}} + \frac{2}{\bar{n}_{pe}}} \label{Q-factor} \\
& \bar{n}_{pe} & = Y \cdot h \cdot \bar{n}_{e,c} \cdot \Omega_{EL} \cdot QE \\
& Q &= \left(\frac{\sigma_{m_\gamma}}{m_\gamma}\right)^2
\eear
with $\bar{n}_{e,c}$ the total number of electrons collected in the multiplication region (eq. \ref{Ncoll}), $\bar{n}_{pe}$ the total number of detected photoelectrons, $\Omega_{EL}$ the solid angle covered for a point-like source crossing the EL region and $h$ its size.
Due to its smallness, the experimental determination of $Q$ is extremely difficult, and it has been performed to date only for electronegative mixtures \cite{CO2Henriques}.

\begin{figure}[h!!!]
\centering
\includegraphics*[width=7cm]{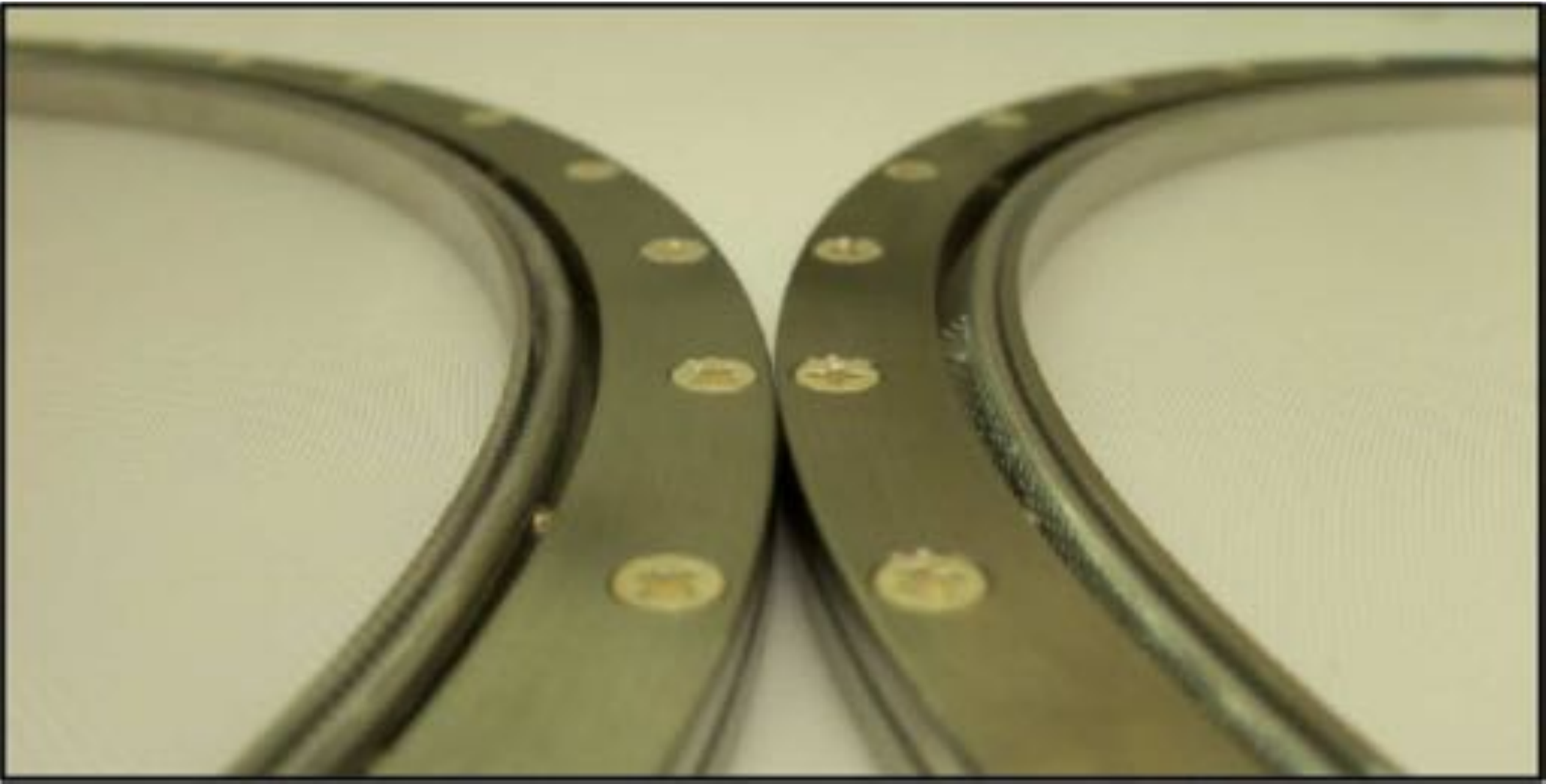}
\includegraphics*[width=7cm]{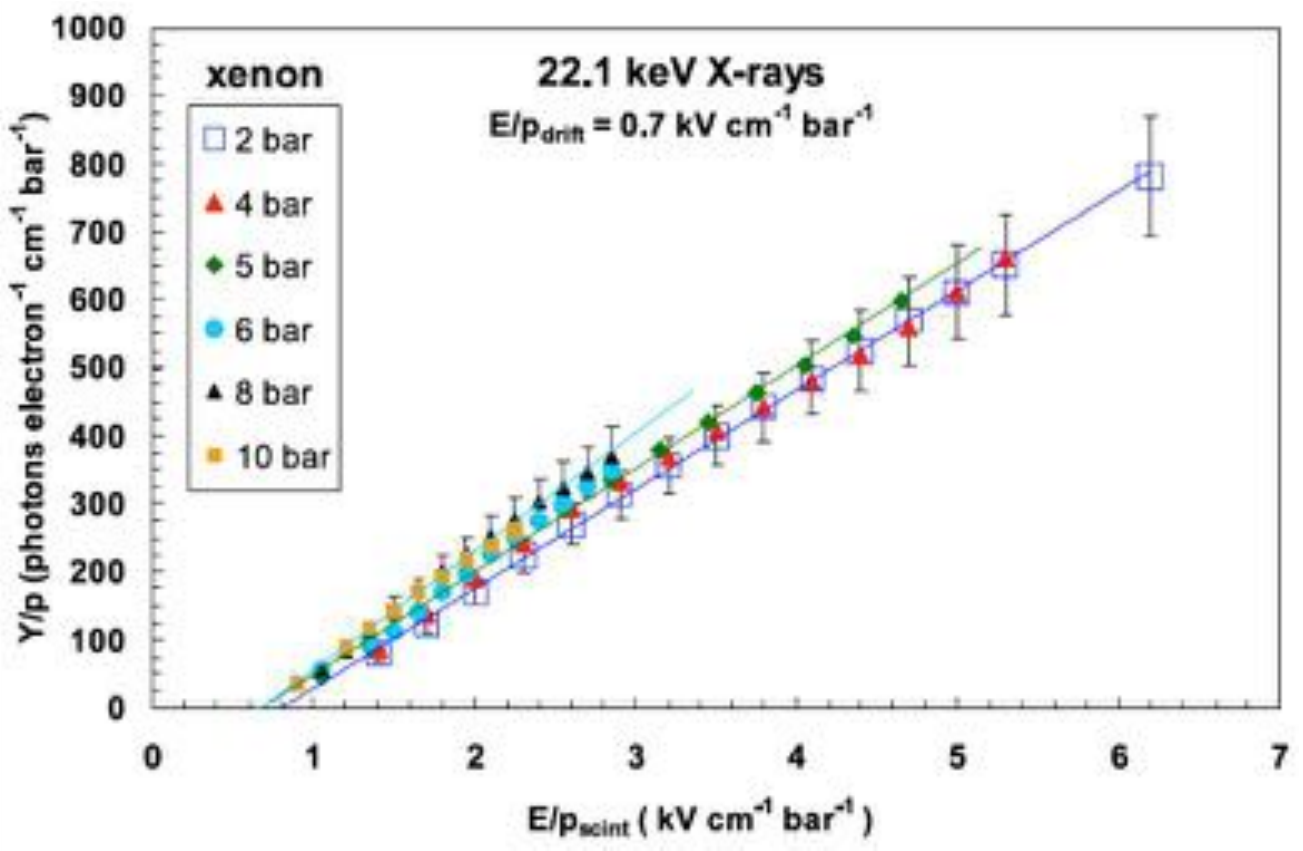}
\caption{Top: meshes used to define the electroluminescence gap in \cite{DEMO}. Bottom: photons per incoming electron, bar and cm ($Y/P$) as a function of the reduced field $E/P$ at different pressures (the gas is xenon and the events are 22.1\,keV x-rays from a $^{109}$Cd source), for different pressures. The maximum optical gain ($m_\gamma = Y/P \cdot h \cdot P$ in this case) nearly doubles at 10\,bar compared to 2\,bar.}
\label{ELallP}
\end{figure}

Electroluminescence is the leading technique when energy resolution close to the Fano factor \cite{NEXT_TDR} and/or sensitivity down to a few primary electrons \cite{Aprile:2013blg} are essential. Since the optical gain depends on the applied voltage through eq. \ref{eqEL}, higher values are obtained in practice for large gaps and pressures (Fig.\ref{ELallP}-bottom), provided the breakdown voltage increases in that situation.

\begin{figure}[h!!!]
\centering
\includegraphics*[width=7cm]{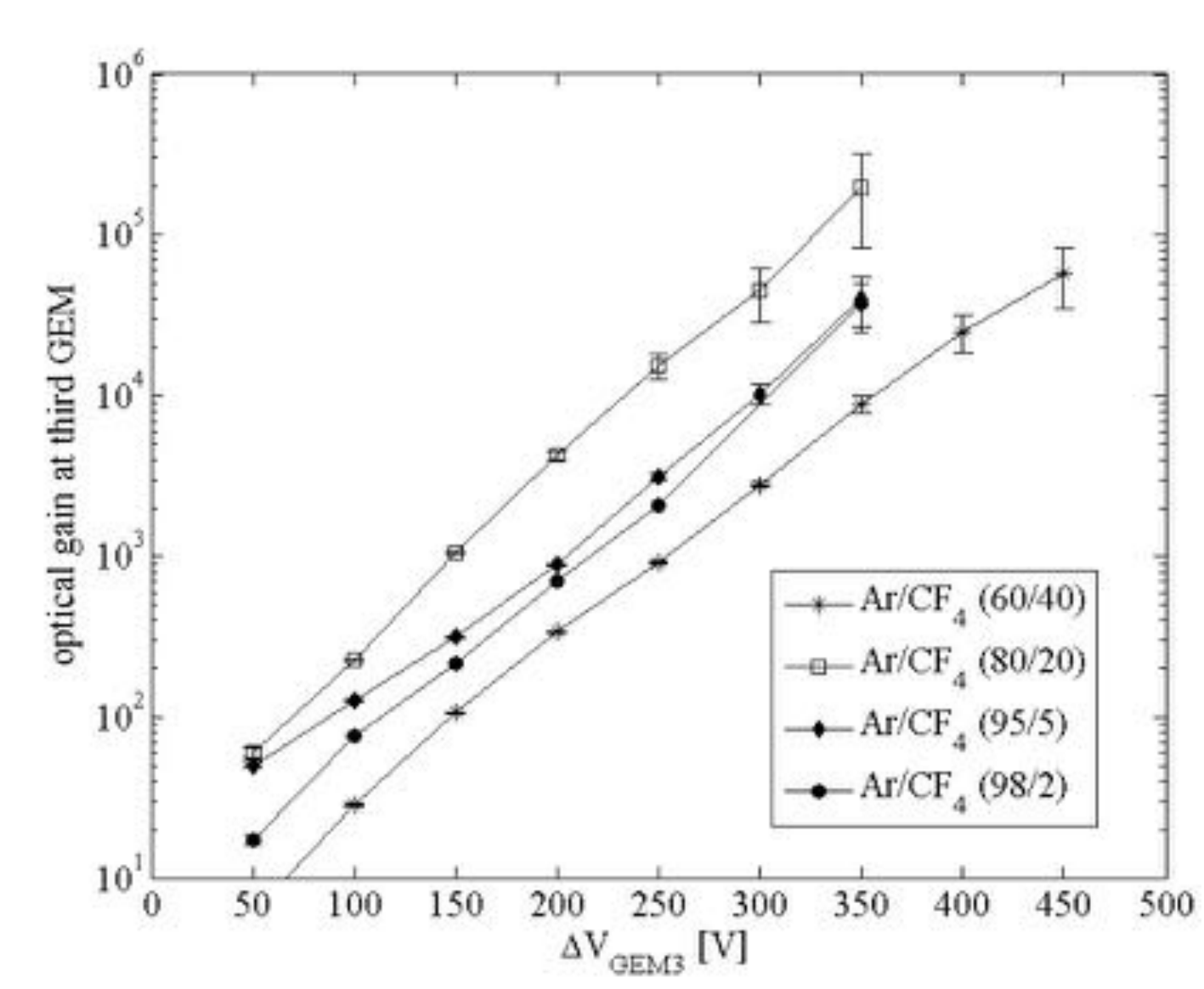}
\caption{Optical gain $m_\gamma$ (number of photons emitted per primary electron) achieved for Ar/CF$_4$ mixtures in a 3-GEM structure at around atmospheric pressure. The exponential behaviour follows the one observed for avalanche multiplication. (After \cite{MeVienna})}
\label{ArCF4}
\end{figure}

There are strong hints of electroluminescence by molecular additives as well (\cite{OTPCfirst, DominikTMA1, DominikTMA2}, Fig. \ref{ELOTPC}-right), however the light yields are generally too low to be usable as such at the moment (\cite{BreskinTeaSeminal}, for instance). The favoured scintillation mechanism in this case involves some type of avalanche multiplication, under which $Q \simeq f_m$ and $m_{\gamma}\sim m$. Secondary scintillation in any of the two forms (electroluminescence or avalanche scintillation) can be easily achieved through relatively large mm's-scale gaps delimited by fine meshes (e.g., Fig.\ref{ELallP}-top). Lately, optical gains as high as $m_{\gamma}=10^5$ have been achieved with multi-GEM structures operated in He/CF$_4$ and Ar/CF$_4$ mixtures (Fig. \ref{ArCF4}) and similar values have been reported for thick GEM, too \cite{Leyton, Orange1, MeVienna, TakeshiJap}.

\subsection{Formation of images}


\begin{figure}[h!!!]
\centering
\includegraphics*[width=\linewidth]{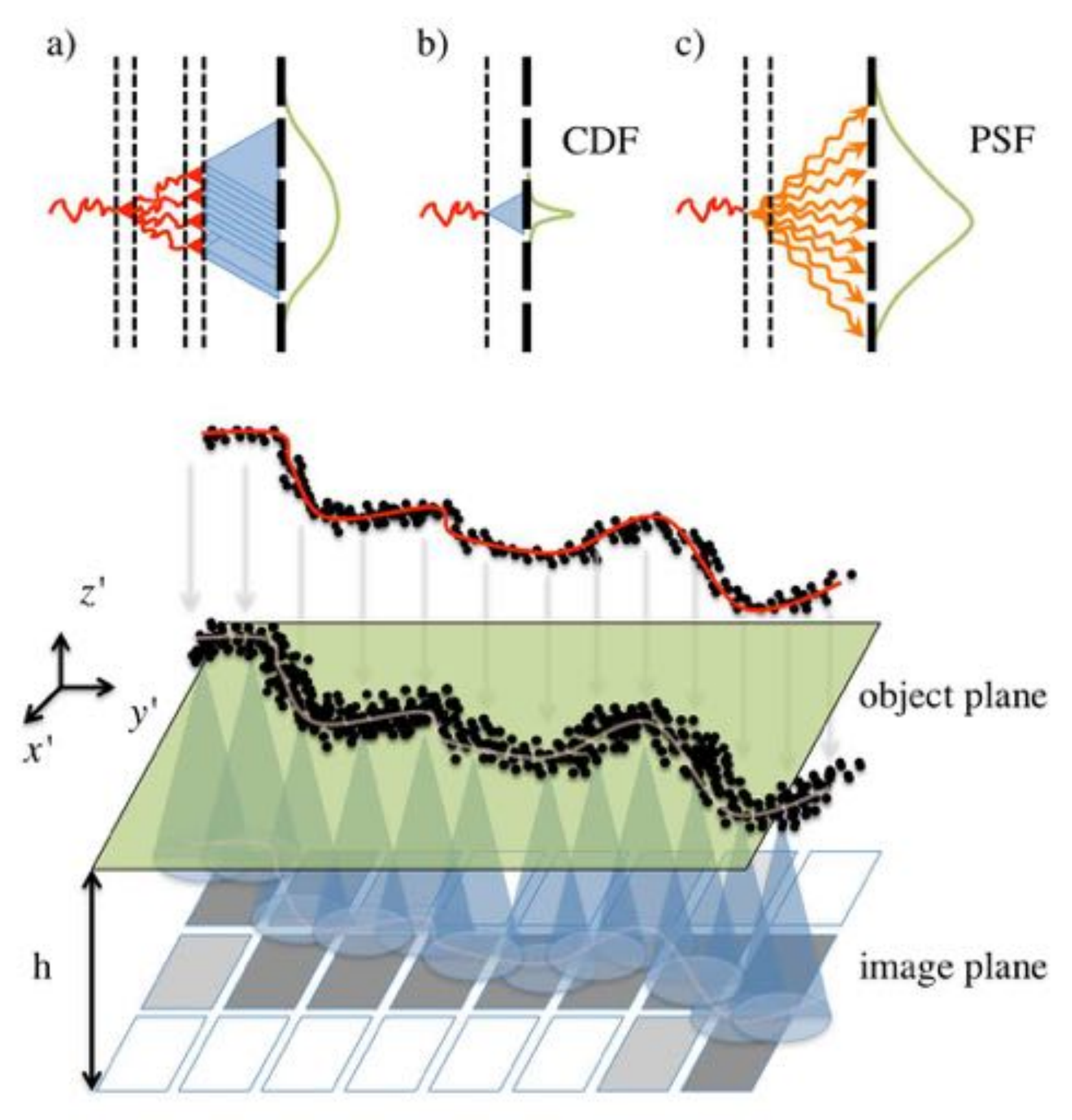}
\caption{Top: illustration of various physical processes in the multiplication region, producing different point spread functions (green curve). Bottom: representation of the imaging process in a TPC. The entrance plane of the multiplication region is labelled as `object plane' and the sensor plane as `image plane'.}
\label{ImagFig}
\end{figure}
\subsubsection{Sensor granularity and point spread function}

The following discussion is simplified if we unify two concepts that are widely used for the readout of charge multiplication -based systems and for optical ones, namely: the charge distribution function (CDF) and the point spread function (PSF), respectively. The CDF is the (normalized) charge profile induced by a point-like avalanche \cite{Gati,Mathieson}. It is a continuous function that represents the probability of finding a fraction of the total induced charge at a given distance from the source (Fig. \ref{ImagFig}-b). Similarly, for an optically-read TPC we can see the PSF as the analogous probability distribution, but for the emitted photons. For sufficiently high numbers the PSF can be regarded as a continuous function too (Fig. \ref{ImagFig}-c). Both the CDF and PSF are 2D probability distributions defined at the sensor/image plane, for a point-like charge that enters the multiplication region at position $x,y$:
\beq
\delta(x'-x,y'-y) \rightarrow CDF,PSF(x'-x,y'-y)
\eeq
We will globally refer to these as the point spread function in the $xy$ plane ($\mathcal{P}\mathcal{S}\mathcal{F}_{xy}$), defined by convenience with the following normalization:
\beq
\int_{-\infty}^{\infty}\int_{-\infty}^{\infty} \mathcal{P}\mathcal{S}\mathcal{F}_{xy}(x'-x,y'-y) dy' dx' = 1
\eeq

Several analytical $\mathcal{P}\mathcal{S}\mathcal{F}$'s can be found in literature, and some illustrative examples are listed in table \ref{tableRF}. Clearly, in the absence of a focusing lens, the widths of these distributions are bound to be dominated by geometrical effects, in particular by the distance between the object and the image plane, $h$. As shown below, this can be beneficially adjusted in practice to some degree.

Unfortunately, the analytical solutions in table \ref{tableRF} do not contain a number of relevant effects, in particular the spread of the charges across the multiplication region or the dependence of the induction process with the distance to the image plane (Fig. \ref{ImagFig}-a). For optical systems, the overall transmission up to the sensor through meshes and windows needs to be included, too. For that, numerical simulations combining induction and transport might be the only way to compute the system point spread function with enough accuracy.\footnote{In the presence of resistive electrodes, the temporal behaviour of the point spread function can become important (and even beneficial) too. Although a very significant technique, it has yet to be applied to TPCs. For a detailed discussion the reader is referred to \cite{DIXITchargeDispersion} and to the recent exhaustive work in \cite{RieglerLast}.}

\begin{table}[h]
  \centering
  \begin{tabular}{|c|c|c|c|}
     \hline
     type                  & point spread function & width & Ref. \\
    \hline
     MWPC            & ~$ \!\frac{K_1}{h} \frac{1\!-\!\tanh^2(K_2 \!\cdot\! (x'\!-\!x)/h)}{1\!+\!K_3\tanh^2(K_2\!\cdot\! (x'\!-\!x)/h)} $ ~ & ~ $\frac{4 \tn{arctanh}(1-\sqrt{2+K_3})}{\pi(1-0.5\sqrt{K_3})} h$~ & \cite{Gati} \\
    single wire              & $\frac{1}{4 h} \cdot \tn{sech} \left( \frac{\pi}{2}(x'-x)/h  \right)$ & $\frac{4\sqrt{2}}{\pi} h$ & \cite{Endo} \\
    PPC      & $\frac{\pi}{8 h} \cdot \tn{sech}^2 \left( \frac{\pi}{2}(x'-x)/h  \right)$ & $\frac{4}{\pi} h$        & \cite{LeeCDF} \\
    light point      & $\frac{1}{1+((x'-x)/h)^2}$ & $2 h$ & - \\
     \hline
   \end{tabular}

  \caption{Some classical 1-dimensional $\mathcal{P}\mathcal{S}\mathcal{F}$'s found in literature, represented as a function of $x'$ and assuming a point source at position $x$. Their width at half maximum is also given (for a definition of the $K_1$, $K_2$, $K_3$ constants see for instance \cite{Mathieson}).}\label{tableRF}
\end{table}

\paragraph{Precision in position reconstruction (`position resolution').}\label{PosResSec}

\begin{figure}[h!!!]
\centering
\includegraphics*[width=\linewidth]{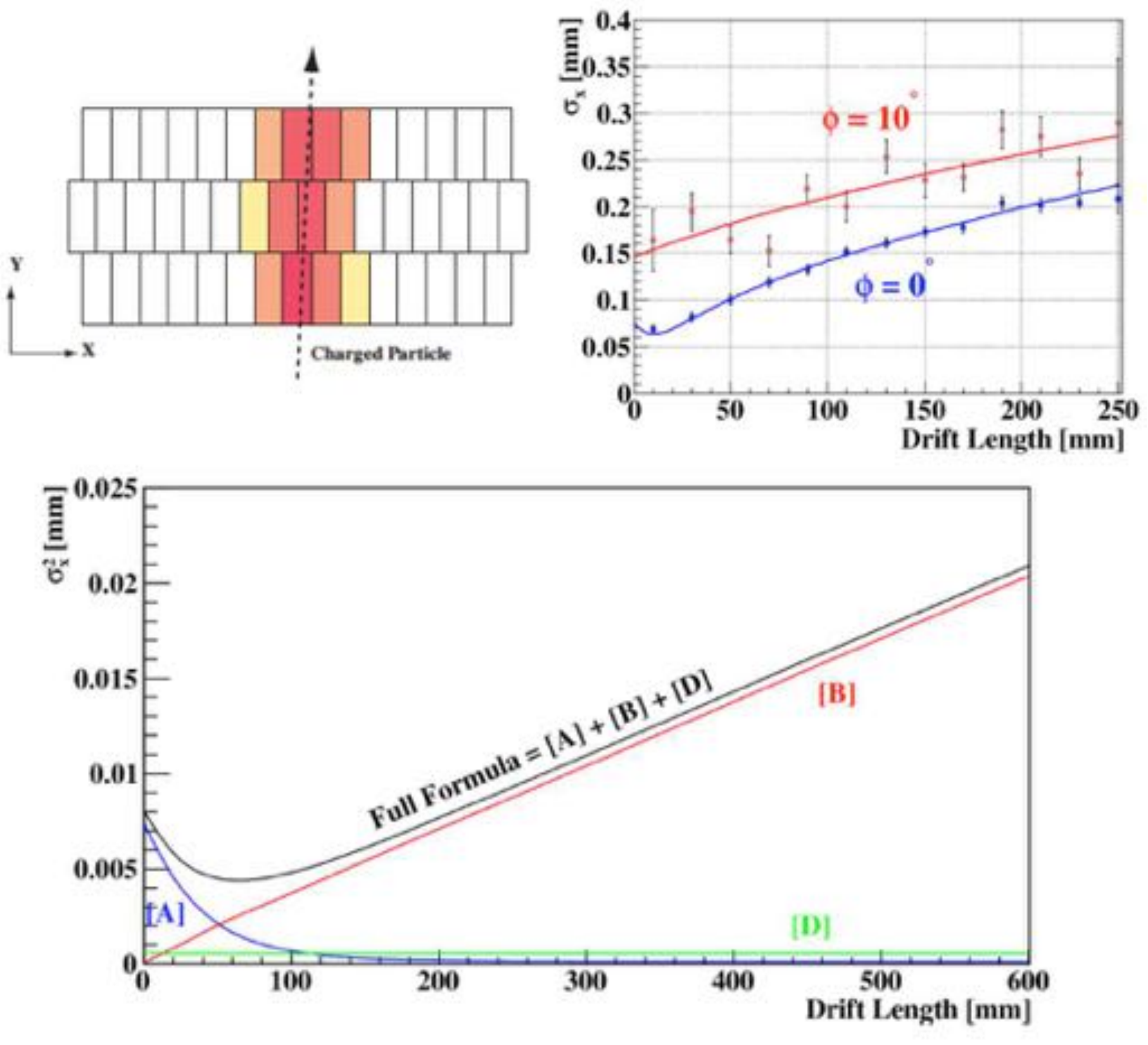}
\caption{Top: simulation (lines) and measurements (points) of the position resolution in the $x$-direction as a function of distance $z$ and angle $\phi$ (relative to the $y$ axis) in an Ar/CH$_4$/CO$_2$ (93/5/2) mixture \cite{JapaneseZaragozaPosRes}. The lower plot (simulation) illustrates how, close to the multiplication region, the precision in the reconstruction (i.e, the position resolution) improves slightly when increasing $z$, due to the larger effective spread function. Far from that region, any additional spread does not bring an advantage and the position resolution follows the expectation from diffusion. The terms $A$, $B$, $D$ are discussed in text.}
\label{FigPosRes}
\end{figure}

If considering the additional spread during the transit along the drift region, and a sufficiently large number of primary charges, the width of the point spread function of
a TPC can be expressed as:
\beq
\sigma_{\mathcal{P}\mathcal{S}\mathcal{F}_{xy}}^{*,2} \simeq \sigma_{\mathcal{P}\mathcal{S}\mathcal{F}_{xy}}^{2} + D_T^{*,2}z \label{PSF*}
\eeq
For extended tracks, eq. \ref{PSF*} mirrors the achievable precision in the position reconstruction of each of its segments, that can be experimentally approximated by:
\beq
\sigma_{x,y}^2 \simeq \sigma_0^2 + \sigma_D^2 z + D(\phi) \label{PosRes}
\eeq
where $\phi$ represents an angle (or angles) defined at the image plane, $\sigma_0$ and $\sigma_D$ are parameters, and $D$ is a system-dependent function \cite{PEP4,ALEPH}.
Clearly, if the $\mathcal{P}\mathcal{S}\mathcal{F}$ is sufficiently finely sampled at the readout, and $\mathcal{S}/\mathcal{N}$ allows, the initial $x$,$y$ position can be reconstructed to almost arbitrary precision, and $\sigma_0\rightarrow0$ (an excellent example of this is \cite{DIXITchargeDispersion}). A $\mathcal{P}\mathcal{S}\mathcal{F}$ that spreads over 2-4 pads guarantees a comfortable situation in this regard, in practice. For low segmentation on the other hand $\sigma_0\rightarrow\Delta{x(y)}/\sqrt{12}$ (with $\Delta{x(y)}$ referring to the segmentation along the $x$($y$) plane).
For illustration, a recent calculation of the position resolution for MPGD-based TPCs is shown in Fig. \ref{FigPosRes}, \cite{JapaneseZaragozaPosRes}.

Naturally, in the above competition between $\mathcal{P}\mathcal{S}\mathcal{F}$ and segmentation, the transverse spread due to diffusion represents an additional (and irreducible) contribution. For the purpose of further illustration, it is interesting to use the analytical limit of eq. \ref{PosRes} derived for high $z$ ($D_T^*\sqrt{z}\gg\Delta{x}$) and $\phi=0$. The authors in \cite{ColasPos} made the additional assumption that the $\mathcal{P}\mathcal{S}\mathcal{F}$ is much narrower than the layout segmentation ($\Delta{x} \gg \sigma_{\mathcal{P}\mathcal{S}\mathcal{F}_{xy}}$, a standard situation for Micromegas detectors), so that it can be neglected (i.e., assimilated to a $\delta$-function). For the sake of simplicity we may disregard noise, too. In that situation, the following expressions for the parameters of eq. \ref{PosRes} were obtained by the authors:
\bear
&\sigma_0  & = \frac{1}{\sqrt{\bar{n}_{e,c}^*}} ~ \frac{\Delta{x}}{\sqrt{12}} \label{sigma0}\\
&\sigma_D  & = \frac{1}{\sqrt{\bar{n}_{e,c}^*}} ~ D_T^*\\
&D(\phi=0)      & = 0\\
&\bar{n}_{e,c}^* & = \Bigg( \overline{\Big(\frac{1}{n_{e,c}}\Big)} \frac{m^2 + \sigma_m^2}{m^2} \Bigg)^{-1} =  \overline{\Bigg(\frac{1 + f_m}{n_{e,c}} \Bigg)}^{ ~-1}\label{eqTricky}
\eear
In eq. \ref{eqTricky}, the average over the inverse of the number of electrons collected per pad ($n_{e,c}$) must be performed, in order to obtain a magnitude that represents an \emph{effective} number of electrons collected per pad: $\bar{n}_{e,c}^*$. The above expressions convey an intuitive result: when diffusion dominates (and in the absence of noise), the centroid estimate deteriorates as expected for a normal distribution of width $D_T^*\sqrt{z}$ (this is the term $B$ in Fig. \ref{FigPosRes}). Since $f_m$ is bounded to a maximum value of 1, the multiplication process will reduce $\bar{n}_{e,c}^*$ in less than a factor of two relative to $\bar{n}_{e,c}$. The skewed distribution associated to the energy loss process of minimum ionizing particles (Figs. \ref{LandauFig}-a,b) can imply a much stronger suppression on $\bar{n}_{e,c}^*$, on the other hand. Therefore, even if Landau fluctuations may not impact a calorimetric measurement for a fully contained event, they can still dominate image reconstruction in some cases. In practice, an accurate evaluation of $\overline{1/n_{e,c}}$ should be probably done numerically, despite authors in \cite{JapaneseZaragozaPosRes} pursued an analytical treatment.

In order to understand the other major contribution to the position resolution, it is necessary to introduce the term $A$ in Fig. \ref{FigPosRes}, that dominates in the limit $z \rightarrow 0$. This term is responsible for an improvement of the position resolution beyond the geometrical limit dictated by $\sigma_0$ in eq. \ref{sigma0}, as diffusion spreads the charge over the neighboring pads at increasing values of $z$. The improvement comes from a $\mathcal{P}\mathcal{S}\mathcal{F}$ that becomes effectively broader, except that in this case diffusion involves an additional random behaviour that builds up $B$, and that eventually takes over.

At last, the term $D$ in Fig. \ref{FigPosRes} is found to depend on the track angle $\phi$ with respect to the $y$-axis (Fig. \ref{FigPosRes} up-left) according to \cite{JapaneseZaragozaPosRes}:
\beq
D(\phi) = \frac{1}{\sqrt{\bar{n}_{e,c}^*}} ~ \frac{\Delta{y}}{\sqrt{12}} \tan{\phi}
\eeq

\paragraph{Sensitivity to primary ionization.}

Irrespective of how intuitive $\mathcal{P}\mathcal{S}\mathcal{F}$'s might be, calculating the charge induced on a given pad is comparatively simpler through the reciprocity theorem, and the several weighting field techniques that result from it \cite{Blum-Rolandi}. This allows to directly retrieve the charge induced on a given pad or strip, something that is commonly dubbed the `pad response function'. Aiming at a higher generality, we will refer to it as the sensor response function, $\mathcal{S}\mathcal{R}\mathcal{F}$. $\mathcal{S}\mathcal{R}\mathcal{F}$ and $\mathcal{P}\mathcal{S}\mathcal{F}$ are related through the integral over the sensor area as:
\beq
\mathcal{S}\mathcal{R}\mathcal{F}(x,y) = \int_{A_s} \mathcal{P}\mathcal{S}\mathcal{F}_{xy}(x'-x,y'-y) dy' dx' \label{SRF}
\eeq
The number of electrons recorded in a given sensor as part of a track image can be hence re-written as:\footnote{Diffusion has been neglected under the assumption that the number of electrons diffused out of a pad or strip is comparable to the number of those diffusing in, from neighboring pads/strips.}
\beq
\bar{n}_{e,r} \equiv \bar{n}_{e,r}(x,y) \cdot \mathcal{S}\mathcal{R}\mathcal{F}(x,y)
\eeq
that can be expanded by using eq. \ref{Nread} as:
\beq
\bar{n}_{e,r} \approx \frac{\Delta\varepsilon}{W_I} \cdot  \mathcal{Q}  \cdot  (1-\mathcal{R})  \cdot  (1-\mathcal{A})  \cdot \mathcal{T}_e \cdot \mathcal{G} \cdot \mathcal{S}\mathcal{R}\mathcal{F} \label{neFINAL!}
\eeq
Reading from left to right, eq. \ref{neFINAL!} includes the energy deposited by the to-be-detected particle divided by the average energy needed to create an electron-ion pair (defined here after including Jesse effect and photo-ionization, if pertinent), times the ionization quenching factor, times the probability that ionization survives recombination, times the probability that ionization survives attachment, times the probability that ionization enters the multiplication region, times the multiplication factor $\mathcal{G}$, times the fraction of signal that reaches the sensor under consideration. $\mathcal{G}$ equals $m^*$ for charge multiplication and $m_{\gamma} \cdot QE \cdot m_{_{PM}}$ for secondary scintillation, with $m_{_{PM}}$ being the gain of the light sensor. For tracks, $\Delta\varepsilon \simeq d\varepsilon/dx \cdot \Delta{x}$.

The system sensitivity can be characterized by how much the average signal in electrons (from eq. \ref{neFINAL!}) is above the system's electrical noise. The latter is customary described by its standard deviation, the so-called equivalent noise charge ($ENC$, given in electrons hereafter):
\beq
\mathcal{S}/\mathcal{N}   \equiv \frac{\bar{n}_{e,r}}{ENC} \label{StoN}
\eeq
In this way, the above definition of the signal to noise ratio ($\mathcal{S}/\mathcal{N}$) represents the number of $\sigma$'s that the average signal is above the noise level. Such a figure of merit is particularly important in order to understand the lowest energy deposit on which the device can trigger. Given that electrical noise is (quite frequently) not merely determined by the thermal limit expectation, a high value of $\mathcal{S}/\mathcal{N}$ minimizes the presence of noise-related systematic effects and is therefore of high practical importance.

For an assessment of the image quality on the other hand, and especially given the large signal fluctuations produced by minimum ionizing particles, it is frequent to find a related quantity (unfortunately bearing the same name), that we name here as the \textit{effective} $\mathcal{S}/\mathcal{N}$ (e.g. \cite{ChineseCCD}):
\beq
\mathcal{S}/\mathcal{N}^* \equiv \frac{\tn{average signal}}{\tn{signal spread}} = \frac{\bar{n}_{e,r}}{\sqrt{\sigma_{n_{e,r}}^2 + ENC^2}} \label{StoN*}
\eeq
As an example, mip particles tracked in 1\,cm-steps at around $\sim 1$\,bar will display a $\mathcal{S}/\mathcal{N}^*$ in the region of two to three per pad. This is a standard figure at collider TPCs, further because it improves only with the cube root of the product $P\cdot\Delta{x}$, following eq. \ref{EresColl2}. For imaging in liquid phase, however, or for reactions involving highly ionizing particles, it can be expected that $\mathcal{S}/\mathcal{N}^*\gtrsim10$. The image quality will thus improve accordingly.

\paragraph{Layout of the image plane.}

\begin{figure}[h!!!]
\centering
\includegraphics*[width=\linewidth]{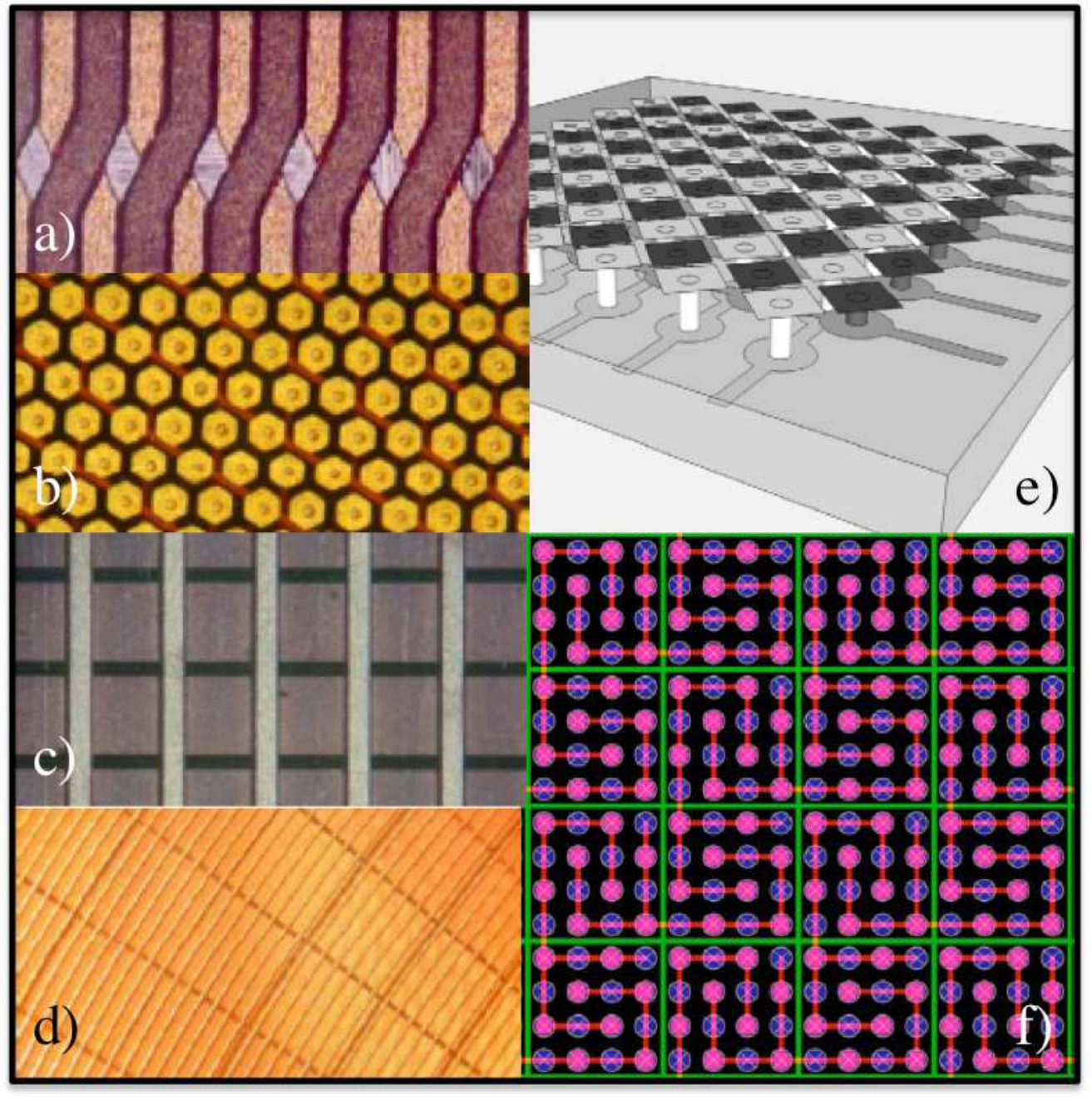}
\caption{Some 2D+2D layouts, capable of reconstructing $xz$ and $yz$ projections in TPCs. Left column : a) small angle, b) `hexaboard', c) cartesian, d) radial (figures a-d from \cite{LeszekTalk}), e) checkers \cite{Checkers}, f) low capacitance \cite{Cantini:2014xza}.}
\label{2Dlayouts}
\end{figure}

The segmentation of the image plane can be performed with previous considerations in mind, both targeting a certain position resolution ($\sigma_{x,y}$) and a spatial sampling such that a sufficient signal to noise ratio ${S}/\mathcal{N}^*$ can be obtained. However there is still a large degree of freedom concerning the sensor layout. We may distinguish three main layout types, that can be found throughout the text:
\begin{enumerate}
\item `3D'. Based on pads/pixels, this is by far the most pursued type for high multiplicity environments, since it provides 3D voxels that can be unambiguously assigned to tracks, except at the few places where tracks cross. \label{enum1}
\item `2D+2D'. For low multiplicities and relatively straight tracks, an induction plane segmented in $x$ and $y$ strips is often used. Placing several readout strips at an angle, (e.g., `2D+2D+2D') has been tried too, in order to reduce ambiguities compared to a simple 2D+2D layout (Fig. \ref{2Dlayouts}-b). Considerable ingenuity has been historically devoted to this problem, as illustrated by Fig. \ref{2Dlayouts}. \label{enum2}
\item `2D+1D'. In conditions similar to those above, optical readouts based on CCD cameras can represent a good practical solution, too, allowing a high reconstruction accuracy. Since CCD cameras are not sufficiently fast yet to allow 3D reconstruction, auxiliary PMs are used to reconstruct the $z$-profile in order to minimize ambiguities. \label{enum3}
\end{enumerate}

Layouts based on (\ref{enum2}) or (\ref{enum3}) are aimed at achieving a very fine image sampling (down to 100\,$\mu$m-200\,$\mu$m) that would be prohibitively expensive or too impractical in a large system otherwise (e.g., if resorting to layout (\ref{enum1})). In some applications the combination of (\ref{enum2}) and (\ref{enum3}) might be advantageous, as recently suggested in \cite{FlorianLast}, in order to improve the 3D-reconstruction of complex topologies.

\subsubsection{Front-end Electronics}

To allow reconstruction of the $z$-coordinate from the time information, the electronics used for registering the primary ionization must work in continuous mode. This is particularly simple for the 2D+1D readout discussed earlier, provided the time evolution is encoded in a few PM signals, that allow a high-performing `ad hoc' solution. Digitization of a large number (10$^3$-10$^5$) of signal waveforms is by far a more frequent problem, that requires custom-designed electronics. Its main characteristics are briefly discussed here on the
light of a particularly successful development, the AFTER chip \cite{AFTER}. Other illustrative developments can be found in \cite{Badertscher:2013wm, STARFEE, ALTRO,aGET,JaviSiPM}. AFTER was originally developed for the T2K TPC and has been adopted, since, for the charge readout of a number of TPCs. It is a rugged and versatile ASIC, whose input is protected against the micro-discharges that are characteristic of charge multiplication in gases, and at the same time it can adapt to a number of TPC conditions in terms of dynamic range, sensitivity and sampling time. Its good inter-channel isolation ensures cross-talk levels of less than 0.1\%.

\begin{figure}[h!!!]
\centering
\includegraphics*[width=\linewidth]{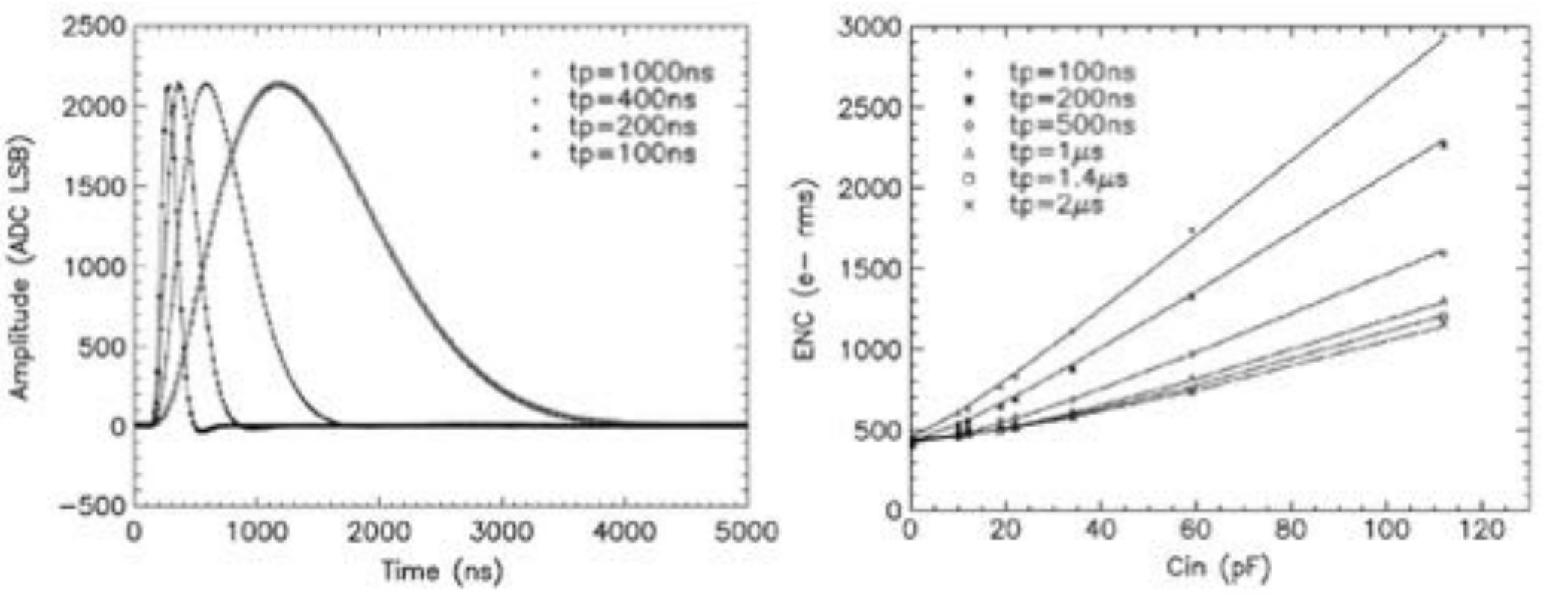}
\caption{Left: impulse response function ($\mathcal{IRF}$) of the AFTER electronics, for several shaping times $\tau_{_s}$ (dubbed `$tp$' in inset). Right: equivalent noise charge ($ENC$)
as a function of the input capacitance, for different shaping times (after \cite{AFTER}).}
\label{FigT2KFEE}
\end{figure}

Generally speaking, minimizing white noise as well as high frequency pickup suggests the use of a certain shaping time ($\tau_{s}$). AFTER performs signal shaping through three RC-CR stages, resulting in a quasi-gaussian impulse response function as (Fig. \ref{FigT2KFEE}-left):\footnote{The impulse response function, or $\mathcal{IRF}$, is mathematically defined as a system's response to an input function that can be assimilated to a $\delta$-function. Experimentally, for an electronic system, it can be obtained by injecting an input current whose fastest frequency component (defined at 3\,dB drop) is much higher than the electronics cutoff frequency, $\sim 1/\tau_s$.}
\beq
\mathcal{IRF} = A e^{-3t/\tau_{_s}} \left(\frac{t}{\tau_{_s}}\right)^3 \sin \left(\frac{t}{\tau_{_s}} \right) \label{IRF}
\eeq

Similarly to the analysis in the $xy$ plane (eq. \ref{PSF*}), the spread of the $z$ coordinate can be approximated for a point-like ionization cloud as:
\beq
\sigma^{*,2}_{\mathcal{PSF}_z} \simeq v_d^2 \cdot \sigma_{_{IRF}}^2 + D_L^{*,2} z
\eeq
where $\sigma_{_{IRF}}$ is the standard deviation of the $\mathcal{IRF}$. In principle, similar arguments to those followed in previous section apply to the precision achievable in the reconstruction of the $z$ coordinate. In practice, however, it is customary to chose the $\sigma_{_{IRF}}$ as small as practically possible, and in particular smaller than the desirable voxel size in $z$, so as not to lose $\mathcal{S}/\mathcal{N}$. Values for $\sigma_{_{IRF}}= 80$\,ns, $\Delta{t}=100$\,ns ($\Delta{z}=2.8$\,mm) have been chosen in \cite{ALICE}, for instance. AFTER allows a range of shaping times $\tau_s=[100$-$2000]$\,ns (with $\sigma_{_{IRF}}\simeq 0.65 \tau_{_s}$). Through switched capacitor arrays, it can record drift times from 10\,$\mu$s to 511\,$\mu$s with 9\,bits in time and 12\,bits in amplitude. Effectively, for a typical drift velocity around $\sim 1\tn{cm}/{\mu}\tn{s}$ it can voxelize in $z$ from $\Delta{z}=200$\,$\mu$m for a 10\,cm drift to $\Delta{z}=1$\,cm for a 5\,m drift.

The equivalent noise charge in AFTER allows a simple parameterization as a function of the input capacitance ($C_{inp}$) and shaping time as (Fig. \ref{FigT2KFEE}-right):
\beq
ENC \!=\! \frac{{\alpha}^2\!(C_o \!+ \!C_{inp})}{\tau_s}\! + \!\gamma^2\! (C_o \!+ \!C_{inp})^2\! +\! \beta \tau_s^2\! +\! D^2
\eeq
with $C_o, \gamma, \beta, D$ being fitting constants. For the T2K TPC the pad capacitance is around 20\,pF, on top of which a typical 100\,pF cable would situate the noise in the range $1000$-$3000$\,e$^-$. The experiment design avoids the use of cabling (not always possible), allowing an overall system value of $ENC=800\pm 100$e$^-$.

In general, if considering a typical number of time bins per waveform around 1000, the number of noise entries at $4\times ENC$ can be expected to be around 5\%. Due to that, it is desirable that a threshold can be set per channel at $\times 5$ above the $ENC$ level at least, and the $\mathcal{S}/\mathcal{N}$ should be much higher than that. Practical $\mathcal{S}/\mathcal{N}$ are commonly found at the level of 10 or higher.

\subsubsection{Sensitivity limits} \label{Single-det}

Single photon and electron detection represent the ultimate sensitivity limits in TPCs. Although there are nowadays several systems with technical requirements for performing one if not both, practical operation in those conditions can suffer from the lack of stability of the multiplication structures, or the sheer background levels, and needs to be demonstrated beyond the nominal technical specifications. The ability to detect one photon or one electron has different implications as well: in case of a scintillation signal, especially due to the lack of perfect coverage, the underlying statistics will be largely Poisson-like; for the ionization signal, on the other hand, much lower fluctuations (Fano-like) are expected.

\paragraph{Single-photon detection.}

For CCD/CMOS cameras, the optical analogous to the $ENC$ (i.e., the $1\sigma$-spread of the distribution of the number of photons in the absence of any signal) is often used for evaluating the response to single-photons. Denoted here as $\sigma_{\gamma}|_{b}$, it can be expressed as the sum of three contributions: dark rate from the device itself ($r_{dark}$), external ambient light ($r_{amb}$) and read noise ($\sigma_r$). The resulting expression for a given exposure time ($t_{exp}$) is:
\beq
\sigma_{\gamma}|_{b} = \sqrt{(r_{dark}+r_{amb})\cdot t_{exp} + \sigma_r^2} \label{PhNoise}
\eeq
As an example, latest CMOS cameras can achieve a $\sigma_r$ with a median around 1\,$\gamma$ and 1.5\,$\gamma$ pixel-to-pixel spread. In such conditions, values
for the effective signal to noise (eq. \ref{StoN*}), i.e. after including the Poisson fluctuations in the signal itself, will approach the physical limit $\mathcal{S}/{N}^*|_\gamma=1$ for an average signal of one photon \cite{ChineseCCD}.

For photon sensors based on charge multiplication, however, (e.g., PMs, SiPMs, APDs, MCPs...) it is more convenient to define a figure of merit for $\mathcal{S}/{N}^*$ based on the actual electrical response of the device. In such a case the $ENC$ and the width of the single-photon multiplication distribution can be used for the evaluation, with background sources considered separately.\footnote{For the sake of clarity we recall the definition of $\mathcal{S}/{N}^*$ given earlier in eq. \ref{StoN*}, applied to the case of single electron (or photon) detection: $\mathcal{S}/{N}^* = m_{_{(PM)}}/\sqrt{\sigma_{e(\gamma)}^2 + ENC^2}$. Here $\sigma_{e(\gamma)}$ refers to the width of the single electron (or photon) charge distribution (in electrons) and $m_{_{(PM)}}$ to the average multiplication factor in the gas (or PM).} Illustratively, values of $\mathcal{S}/{N}^*\gtrsim 2$ can be typically achieved in single-photon conditions for PMs operated at standard temperature, approaching $\mathcal{S}/{N}^*=10$ for SiPMs. Using this definition for the case of CCD/CMOS cameras, a more modest value around $\mathcal{S}/{N}^* \simeq 1/\sigma_r = 0.6$-$1$ is obtained. Similarly to the figure of merit defined from eq. \ref{PhNoise}, the actual ability to detect individual photons emitted inside the chamber will depend on the background rates, and on the statistics of the collection process.

\paragraph{Single-electron detection.}

Based on the gains of several times $10^4$ shown in section \ref{ChargeMultSec} for gaseous multiplication (at least for a given structure and a certain set of gases and pressure conditions), and taking noise values around $ENC=1000\,$e$^-$ as reported in previous section, a $\mathcal{S}/\mathcal{N}$ as high as 10 can be anticipated even for individual electrons. Moreover, some specific designs like InGrid (with the multiplication stage coupled to a finely pixelated ASIC for the readout, - TimePix \cite{TimePix} in this case) allow approaching single-electron identification for a gas multiplication of several 1000's \cite{CASTMichel}. Recently, a new ASIC (TopMetal), with a (nominal) noise of 15\,e$^-$ per $83\,\mu$m pixel, has been introduced in \cite{TopMetal} and is discussed in the last section.

On the basis of the above discussion one may naively expect a sensitivity to the primary ionization down to an energy equivalent to the $W_I$ value of the gas. A critical analysis shows some technical caveats, however, like the difficulty of triggering in those conditions under a presumably large background stemming from field-emission at the cathode. Additionally, the width of the multiplication distribution ($\sqrt{f_m}$) is typically 0.7-1 (Fig. \ref{TranspAndF}), the resulting $\mathcal{S}/{N}^*$ being $= 1/\sqrt{f_m + (\mathcal{N}/\mathcal{S})^2} < 1.2$ in this case. Thereby, compared to photon detection, the application of a charge threshold results in a relatively inefficient detection process. The statistics of the primary ionization (and specially the efficiency of the collection process) is expected to be much more benign than the one of the primary scintillation, though. Photon detection via electroluminescence is the most successful solution to this conundrum, enjoying both the good statistical behaviour of the photo-detection process and the one of the primary ionization, and is the most mature technique to assure (nominally) the ability to detect single electrons in TPCs.

\subsubsection{Energy resolution limits} \label{Single-det}

Energy resolution may be limited in a TPC by a number of practical considerations, besides those discussed earlier in eq. \ref{f-factor}. They may include: the presence of damaged/low-performing sensors, sensor-to-sensor gain equalization, (intra-)sensor uniformity of response, cross-talk and noise (for a detailed account see \cite{NEXTMM,XePen}). The procedures leading to the ultimate resolution achievable in a TPC are laborious and strongly system-dependent (e.g. \cite{Lorca}). We include here for completeness the effect of electrical noise for the case of charge readouts, that leads to the following formula:
\beq
\frac{\sigma_{_{\Delta\varepsilon}}}{\Delta\varepsilon}\! \equiv \! \frac{\sigma_{n_{e,r}}}{\bar{n}_{e,r}} \! \simeq \! \sqrt{\!\frac{F_e \!+\! \mathcal{R} \!+\! \mathcal{A} \!+\! 1\!-\!\!\mathcal{T}_e \!+\! f_m \!+\!\! \Big(\!\frac{ENC}{m^*}\!\Big)^2\!\!\!\frac{1}{\bar{n}_{e,c}}}{\bar{n}_{e,c}}}
\eeq
if the signal is registered by a single sensor (or if there is a dominant common-mode noise and all sensors registering the signal have similar gain).

\subsection{The role of pressure} \label{Prole}

Gaseous TPCs provide an approximate lensing effect, in which density largely determines magnification (or demagnification, depending on the experimental needs). Hence, in the range of one to ten mbar, ionization clusters extending along an equivalent size of few tens of nm in tissue can be resolved in tissue-equivalent gases \cite{Tea, Fabi}; at 20-130\,mbar, the direction of travel of mm and sub-mm long nuclei down to 40\,keV kinetic energy become distinctly clear in CF$_4$ \cite{Phan, DRIFTlast}; light p/He nuclei with MeV energies extend over several cm in either 150\,mbar of CO$_2$ \cite{TUNL} or Ar/He ($\sim50/50$) at 1\,bar \cite{OTPClast}; at 10\,bar, a 2.45\,MeV $\beta\beta$-event in xenon conveniently expands over a $\sim20$\,cm length \cite{JJreview}... Fortunately, handling the electric fields inside a TPC through this four order of magnitude pressure range is facilitated by the existence of number density ($N$) and reduced field ($E_d/N$) scalings.

\begin{table}[h]
  \centering
  \begin{tabular}{|c|c|}
     \hline
     magnitude             & ~ scaling ($n=N/N_0$) ~ \\
     \hline
     ~ electron, ion drift velocity $v_d$     & ~ $v_d(E/n)$ ~      \\
     ~ electron, ion diffusion coefficients $D^*_{L,T}$        & ~ $\frac{1}{\sqrt{n}}D^*_{L,T}(E/n)$ ~      \\
     ~ attachment coefficient $\eta$ ~         & ~ $n\cdot \eta(E/n)$ $^{*a}$ ~      \\
     ~ Light transparency $\mathcal{T}$ ~   & ~ $\exp{(-n \Pi_a L^*)}$ ~        \\
     ~ scintillation probability $P_{scin}$ ~   & ~ $\frac{1}{1+ n \tau k}$ ~        \\
     ~ particle range $R$ ~  & ~ $R/n$ ~     \\
     ~ Fano factor $F_e$, $W_{I}$, $W_{ex}$ ~  & ~ $\sim\tn{constant}$ ~     \\
     \hline
     \hline
     ~ charge multiplication coefficient $\alpha$~           & ~ $n \cdot \alpha(E/n)$ $^{*b}$ ~      \\
     ~ secondary scintillation coefficient $Y$  ~     & ~ $n \cdot Y(E/n)$ $^{*b}$ ~    \\
     \hline
   \end{tabular}

  \caption{Approximate number density ($N$) and reduced electric field ($E/N$) scalings commonly found in the practice of gaseous electronics.}\label{TableN}
       \begin{tablenotes}
    \item[1] $^{*a}$ For 3-body attachment (e.g., O$_2$) an additional factor $n$ is needed.
    \item[2] $^{*b}$ In the presence of Penning (or wavelength-shifting) transfers, deviations from these scalings will appear.
  \end{tablenotes}
\end{table}

In table \ref{TableN} we show $N$ and $E_d/N$-scalings that are known to be highly accurate in the practice of gaseous electronics such as those of the attachment, charge and light multiplication coefficients, drift-diffusion parameters \cite{Huxley, Urquijo} and particle range \cite{AzevedoP}. We also give some approximate scalings for the transparency and scintillation probability according to formulas derived in previous sections. In practice, $N$ is controlled in gaseous TPCs through $P$ adjustment, so both quantities can be used interchangeably in the present context.\footnote{A relevant exception is the gas phase of dual-phase chambers, where both $P$ and $T$ can deviate significantly from NTP conditions. This will be explicitly stated when needed.} For small chambers (scale $30\times30\times30$\,cm$^3$) some elements and vessels developed for ultra-high vacuum applications are often compatible with pressurization but, outside those dimensions, custom-designed chambers are needed. Excluding deviations from the above scalings in extreme conditions \cite{Bolot,Japa}, some technical problems associated to their implementation in a practical situation should be noted:

\begin{enumerate}
\item High voltage in the drift region. In general-purpose chambers, minimizing the pressure dependence of $v_d$, $D_L$ and $\eta$, requires that the drift voltages follow $P$ so that $E_d/P \sim \tn{constant}$. Therefore, especially at high pressure, insulation problems may develop (section \ref{HVsec}). The necessary voltage increase is much milder for charge multiplication (since $\alpha\sim P \cdot \alpha(E_a/P)$) so an additional factor of two in bias voltage can easily cope with a $\times10$ pressure increase (Figs. \ref{FigCortesi}, \ref{FigLowP}, \ref{FigHighP}). As shown in \cite{XePen}, however, energy resolution suffers at high $P$ from the unavoidable $E_a/P$ reduction and the associated increase of the avalanche fluctuations (Fig. \ref{FigHighP}).

\begin{figure}[h!!!]
\centering
\includegraphics*[width=\linewidth]{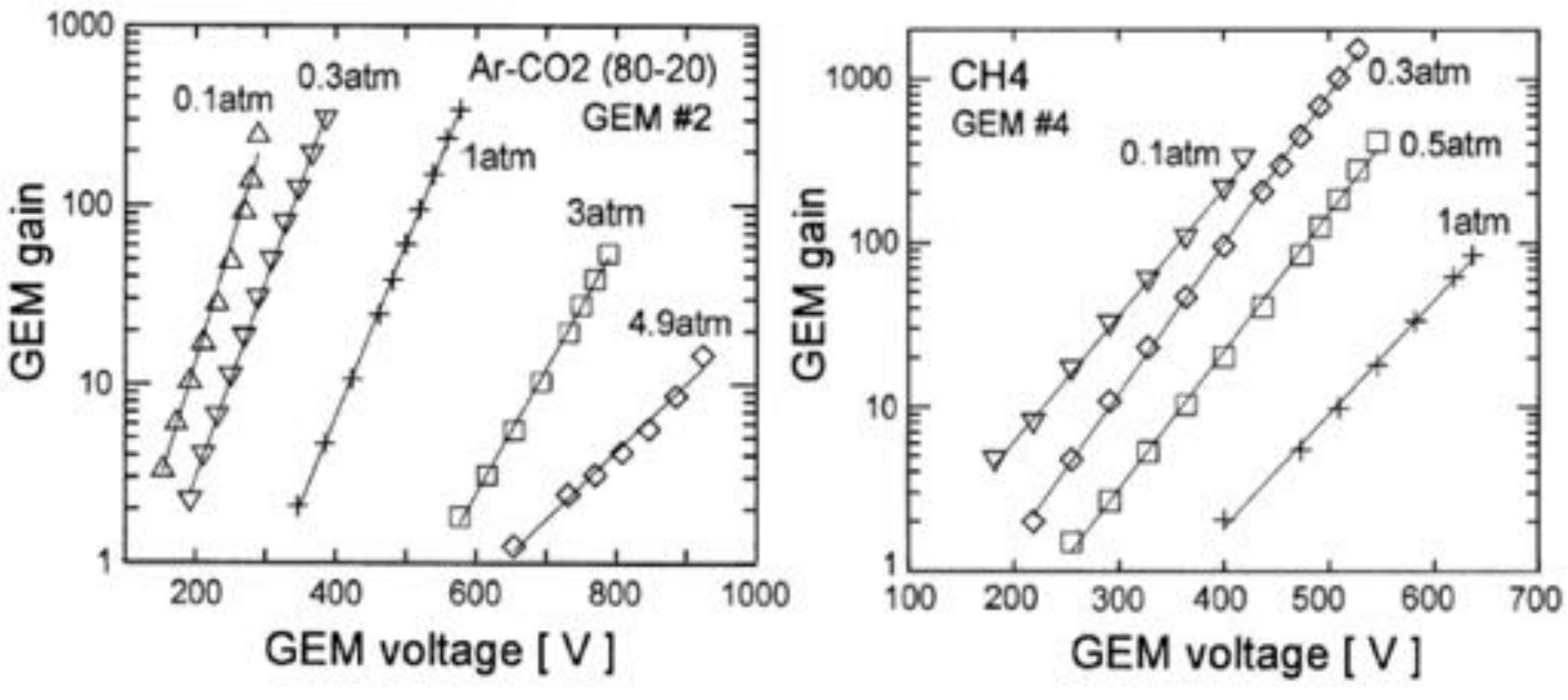}
\caption{Gain curves at different pressures obtained in early GEMs. Left: Ar/CO$_2$ at 80/20 concentration. Right: pure CH$_4$. (Adapted from \cite{BondarLowP})}
\label{FigLowP}
\end{figure}

\begin{figure}[h!!!]
\centering
\includegraphics*[width=\linewidth]{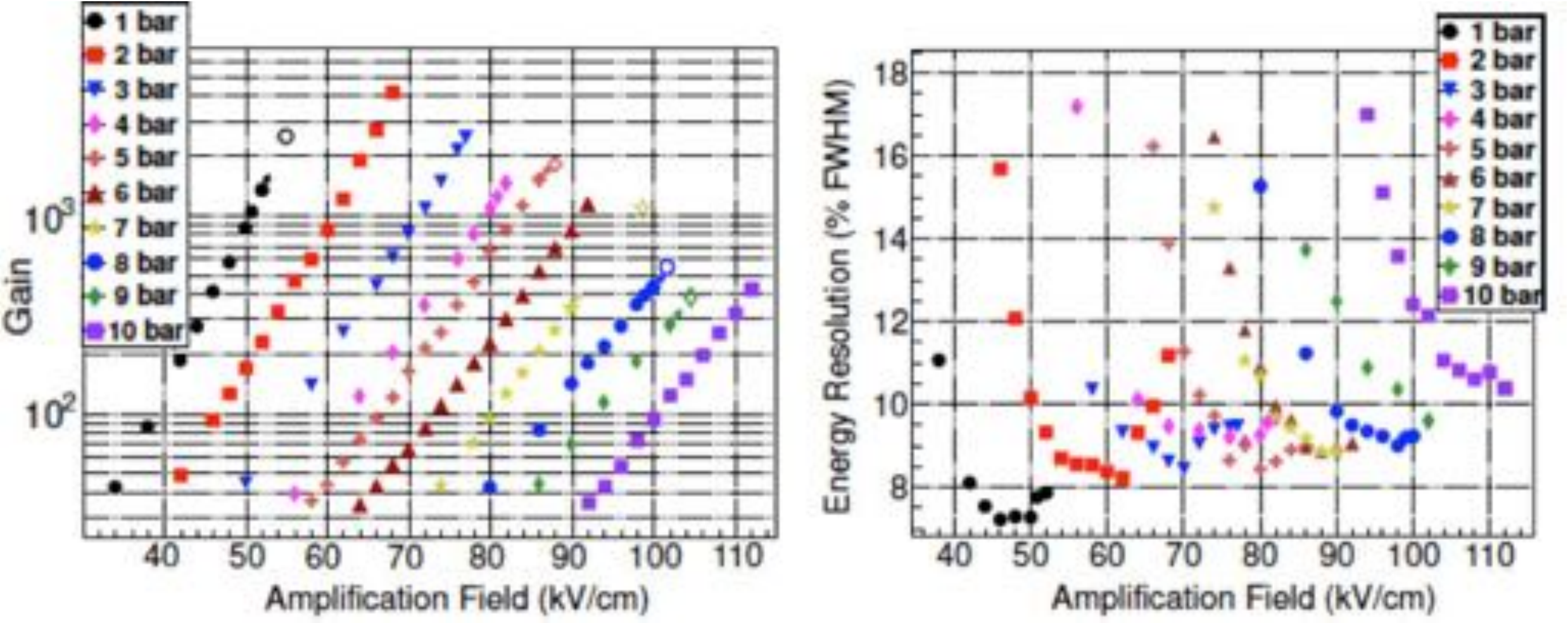}
\caption{Left: gain curves in microbulk Micromegas at different pressures for Xe/TMA admixtures at around 99/1. Right: energy resolution under the same conditions. (Figs. from \cite{DianaTMA})}
\label{FigHighP}
\end{figure}

\item Maximum (stable) charge multiplication. While comfortable charge multiplication has been achieved in a number of structures from the mbar regime \cite{Phan, Ref2BreskinFromPhan} up to around 10\,bar \cite{TREX-DM_last, SmithFirst, DianaTMA}, most works underline the fact that, for a given multiplication structure and gas, performance rapidly deteriorates outside a certain pressure regime (for example, Figs. \ref{FigCortesi}, \ref{FigLowP}, \ref{FigHighP}). The existence of an optimum gap (and pressure) displaying a higher resilience against mesh sagging was shown for instance in \cite{GiomatarisMech} for the case of classical Micromegas, and a similar effect has been reported for microbulk Micromegas, concerning variations on the holes size \cite{XePen}. It is very unlikely that these observations alone can explain such a universal behaviour for structures of vastly different geometries. A general description of the observed trends will plausibly require of additional considerations related to the behaviour of diffusion, ionization density, formation of excited states, photon quenching and transparency as a function of pressure \cite{RD51meetingOnThis}. Understanding and mitigating this pressure-dependence of the maximum gain will be clearly a major milestone towards next-generation TPCs.
\end{enumerate}

\subsection{The role of gas mixture}\label{GasMix}

The choice of gas mixture involves a number of considerations: i) its interaction with
the primary particle (e.g., in active target experiments), ii) its interaction with the secondary particles
produced in the reaction (mainly through multiple scattering and $d\varepsilon/dx$), iii) its influence on the collection of the scintillation and ionization released by those particles (sections \ref{ChargeColl}, \ref{ScinColl}); iv) its chemistry (namely, transfer reactions mediated by photons as well as atomic and
molecular reactions). We will not refer in this latter case to detector ageing, considering that this is an unusual situation in TPCs devoted to imaging rare processes.

\begin{figure}[h!!!]
\centering
\includegraphics*[width=\linewidth]{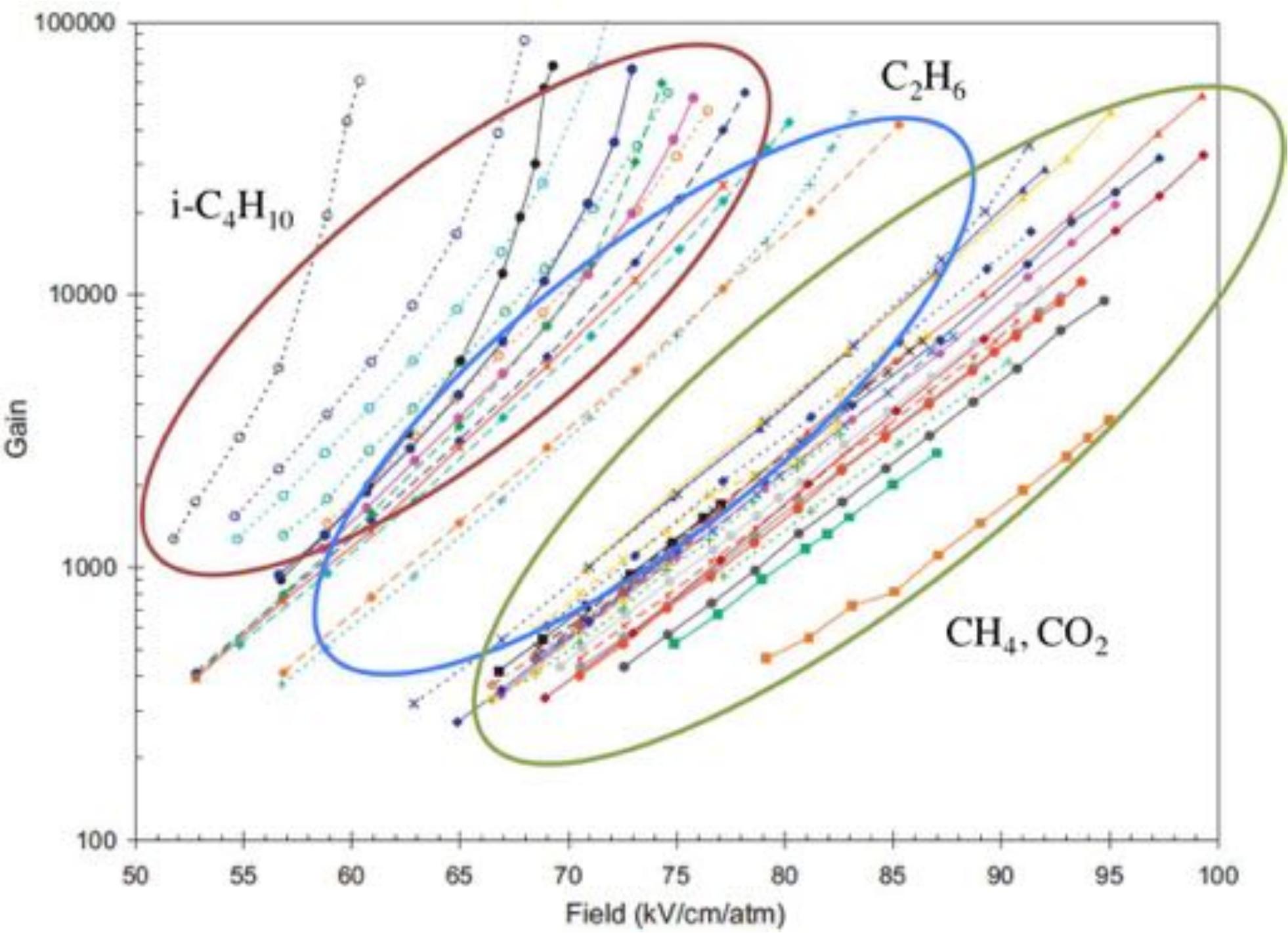}
\caption{Measured gain in microbulk Micromegas structures ($50\,\mu$m gap) under x-ray illumination, obtained for argon gas admixed with common quenchers (i-C$_4$H$_{10}$, C$_2$H$_6$, CH$_4$, CO$_2$), at around atmospheric pressure. The figure has been adapted from \cite{AttieReview}, where the exact concentrations for each series can be found. The bands shown are approximate. It must be noted that the author in \cite{AttieReview} has essayed (and included in this compilation) ternary mixtures based on CF$_4$ as well, in order to fine tune the TPC response in the drift region. Besides the observation that the operating voltage is reduced in inverse relation with the reactivity of the additive, the maximum achievable gain can be generally found in the range of several times 10$^4$, fairly independently from the additive(s) employed.}
\label{MixVarious}
\end{figure}

A particularly convenient situation for detector design is that where the primary particle or nuclei is implanted and then let decay, since the TPC can be optimized in such a case with a large freedom of gas choices. Active target experiments on the other hand enforce the use of a given gas, and the TPC should function under those given conditions too. TPCs based on H$_2$, D$_2$, He, Ne, Ar, Xe, CF$_4$ or SF$_6$ for instance, are described in this text in the context of nuclear reactions in inverse kinematics (mainly colliding heavy nuclei against H and He targets), muon capture, neutrino interactions or dark matter searches. However, even in those cases, it is common that the physics process under study is not perturbed much by the use of \%-additives.
Those levels can be already sufficient for processes like electron cooling and VUV-quenching, but also Penning \cite{ArPen} and wavelength-shifting transfers \cite{ArN2Suzuki} to fundamentally alter the chamber response, and in a beneficial way. We refer in the following to the two most common processes to bear in mind during the choice of gas mixture for optimizing the TPC behaviour:

\begin{enumerate}
\item Electron cooling: molecular additives possess a number of low-energy rotational and vibrational degrees of freedom, starting at around 100\,meV (Fig. \ref{Xsections}-up), that cause electrons to lose their energy very efficiently as compared to elastic interactions. In this latter case, the fraction of energy transferred by an ionization electron in each encounter can be expressed as $\Delta\varepsilon/\varepsilon \simeq 2 m_e/M$, with $M$ being the atom (or molecule) mass. Therefore, while an electron with energy around 100\,meV will be able to transfer most of it to an additive in any inelastic encounter, it will only transfer a fraction $\sim10^{-4}$ elastically. Hence, despite the generally smaller cross sections for inelastic encounters relative to elastic (by up to a factor $10$-$100$), a mere 1\% of a molecular additive can be already sufficient to dominate the energy loss process. Additional cooling will result in a gradual reduction of the longitudinal and transverse diffusion, approaching the thermal limit (eq. \ref{thermalLimit}). The drift velocity, on the other hand, can experience a maximum as a function of the electric field (Fig. \ref{vdSeveral}), when the electrons' energy distribution is brought into the neighborhood of the so-called `Ramsauer minimum' of the elastic cross section (typically at some 100's of meV).

\begin{figure}[h!!!]
\centering
\includegraphics*[width=7cm]{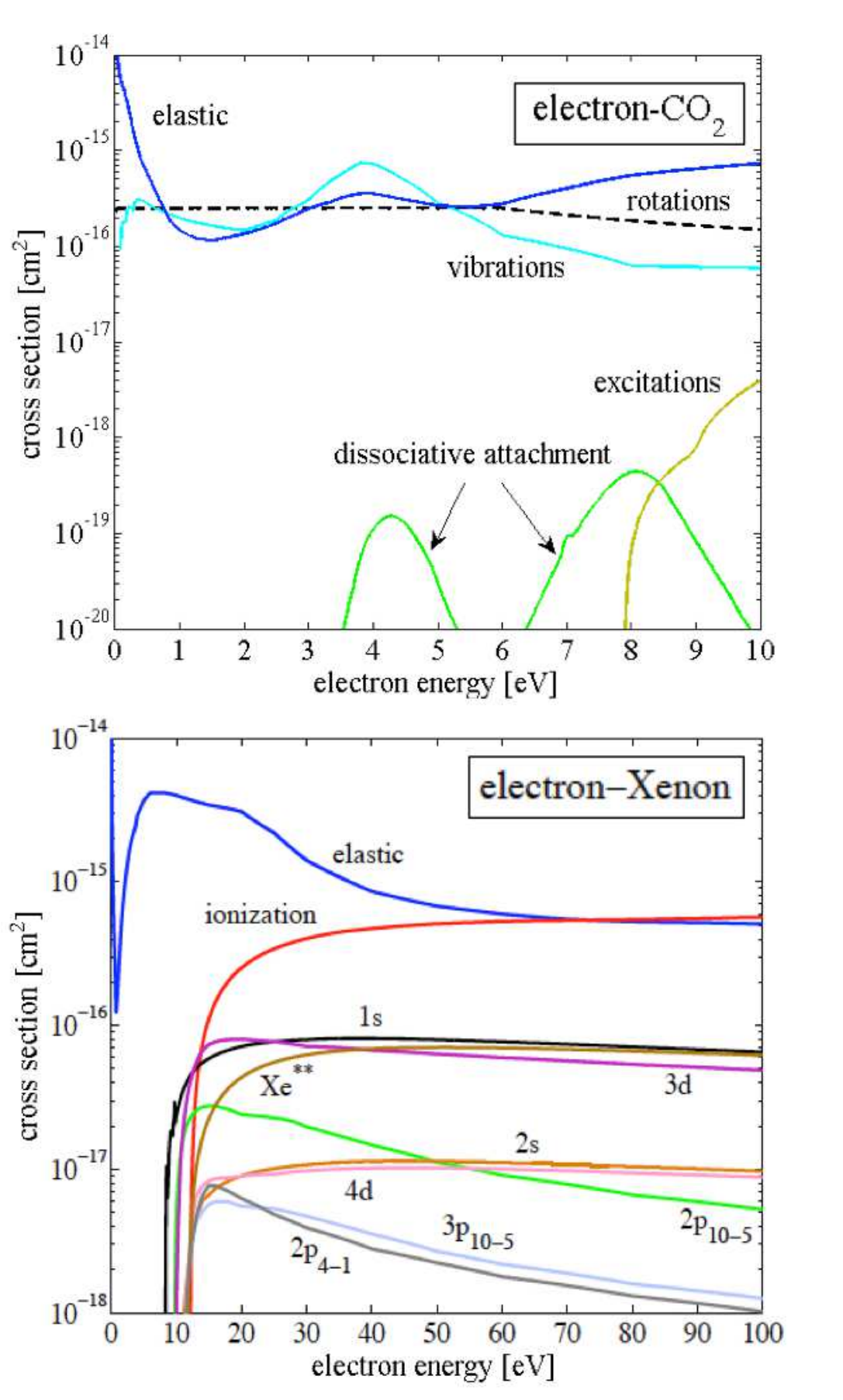}
\caption{Most common electron cross section types for two representative gases. Top (CO$_2$): indicating elastic, vibrational, rotational, excitation and attachment terms in the low energy region, below 10\,eV. Bottom (xenon): featuring only elastic, ionization and excitation terms, the later separated in multiplets (taken from \cite{Mua}). The Ramsauer minimum of the elastic cross section can be appreciated to the left of the plot. (Both figures obtained from the Magboltz database)}
\label{Xsections}
\end{figure}

Hydrodynamic (swarm) parameters of common gases, $v_d$, $D^*_{L(T)}$, $\eta$, $\alpha$ (eq. \ref{Eq_hydro}, table \ref{TableN}) have been extensively studied for pure gases, usually at low pressures (e.g. \cite{Raju}), and measured values are routinely extrapolated to atmospheric and high pressure conditions based on $E/N$ and $N$-scalings. The problem appears in the case of mixtures, since these parameters do not follow additivity rules. Presently, one of the most reliable ways around this difficulty is to resort to the simulation code Magboltz \cite{Magboltz} (by S. F. Biagi). First introduced nearly 30 years ago \cite{Biagi}, it performs electron transport in Monte Carlo fashion by using elementary cross sections, and computes the parameters by analogy to the solutions of the hydrodynamic equation. Its cross section database covers at the moment around 50 common gases, and is regularly updated. Some recent results are shown for illustration in Fig. \ref{DiffAndVmag}.

\begin{figure}[h!!!]
\centering
\includegraphics*[width=7.5cm]{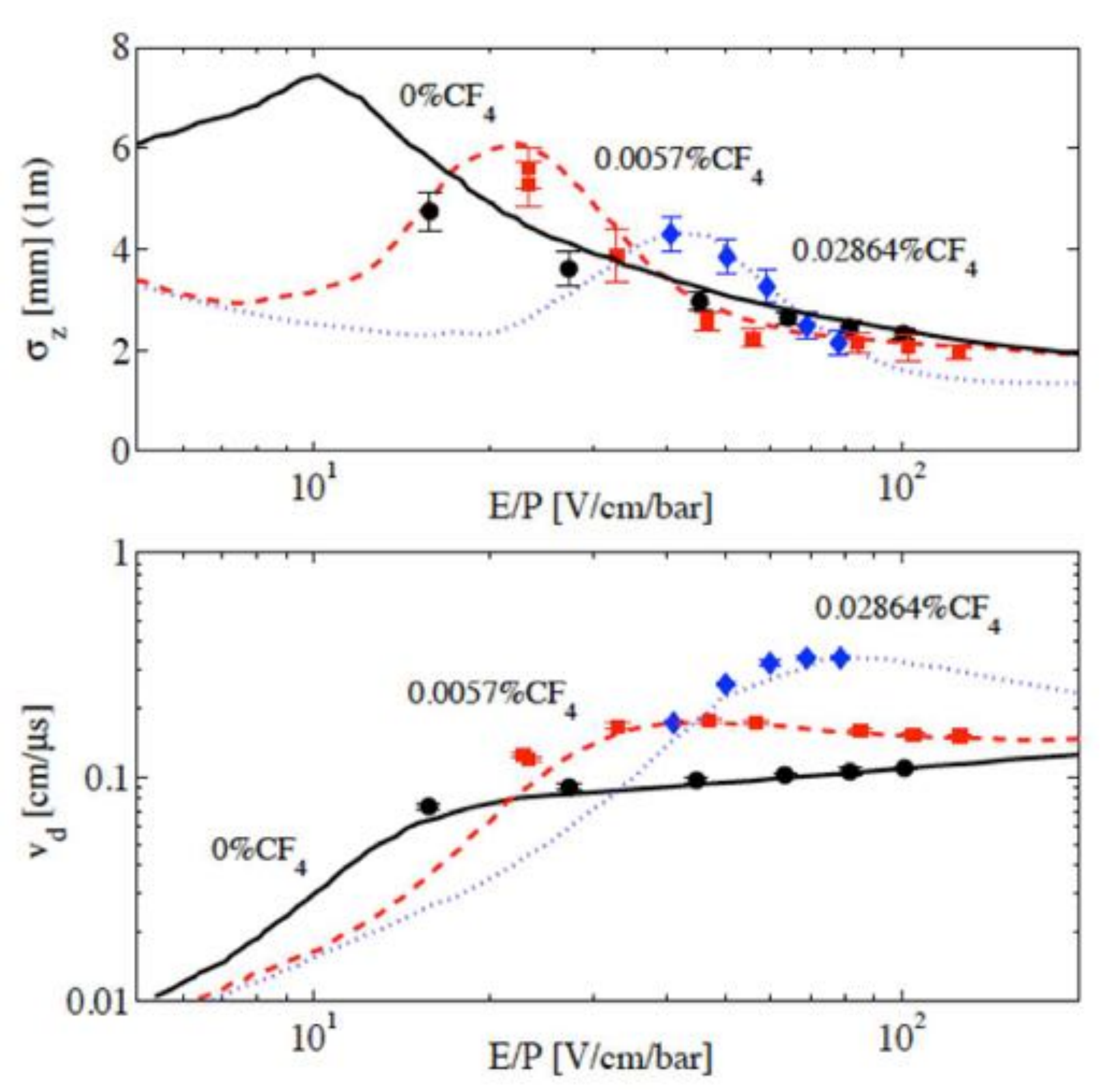}
\caption{Comparison between experiment and Magboltz simulation for Xe/CF$_4$ mixtures at 10\,bar. Top: longitudinal spread in 1\,m. Bottom: drift velocity. (Figures from \cite{Mua})}
\label{DiffAndVmag}
\end{figure}

\item Excimer/VUV-quenching: the use of additives has an extra benefit for chambers whose readout is based on charge multiplication: it both quenches (eq. \ref{S-V}) and absorbs (eq. \ref{attLambda}) the dangerous VUV light emitted from the noble gas, minimizing photoelectric effect and allowing comfortable working gains. Alkanes are very common quenchers in the practice of gaseous detectors, but CO$_2$ is a popular choice too. Experimentally, a gas mixture is sufficiently well quenched above a few \% for the purpose of charge multiplication, or at least the maximum gain depends only weakly on it (an overwhelming compilation by D. Attie can be found in Fig. \ref{MixVarious}, \cite{AttieReview}).
\end{enumerate}
The relation between electron cooling and VUV-quenching is so intertwined in TPCs that it has taken a long time to realize and demonstrate that they can be adjusted independently to some extent, for instance leading to a low diffusion and weakly VUV-quenched mixture \cite{CO2Henriques}, an important asset for chambers based on electroluminescence.

\begin{figure}[h!!!]
\centering
\includegraphics*[width=7cm]{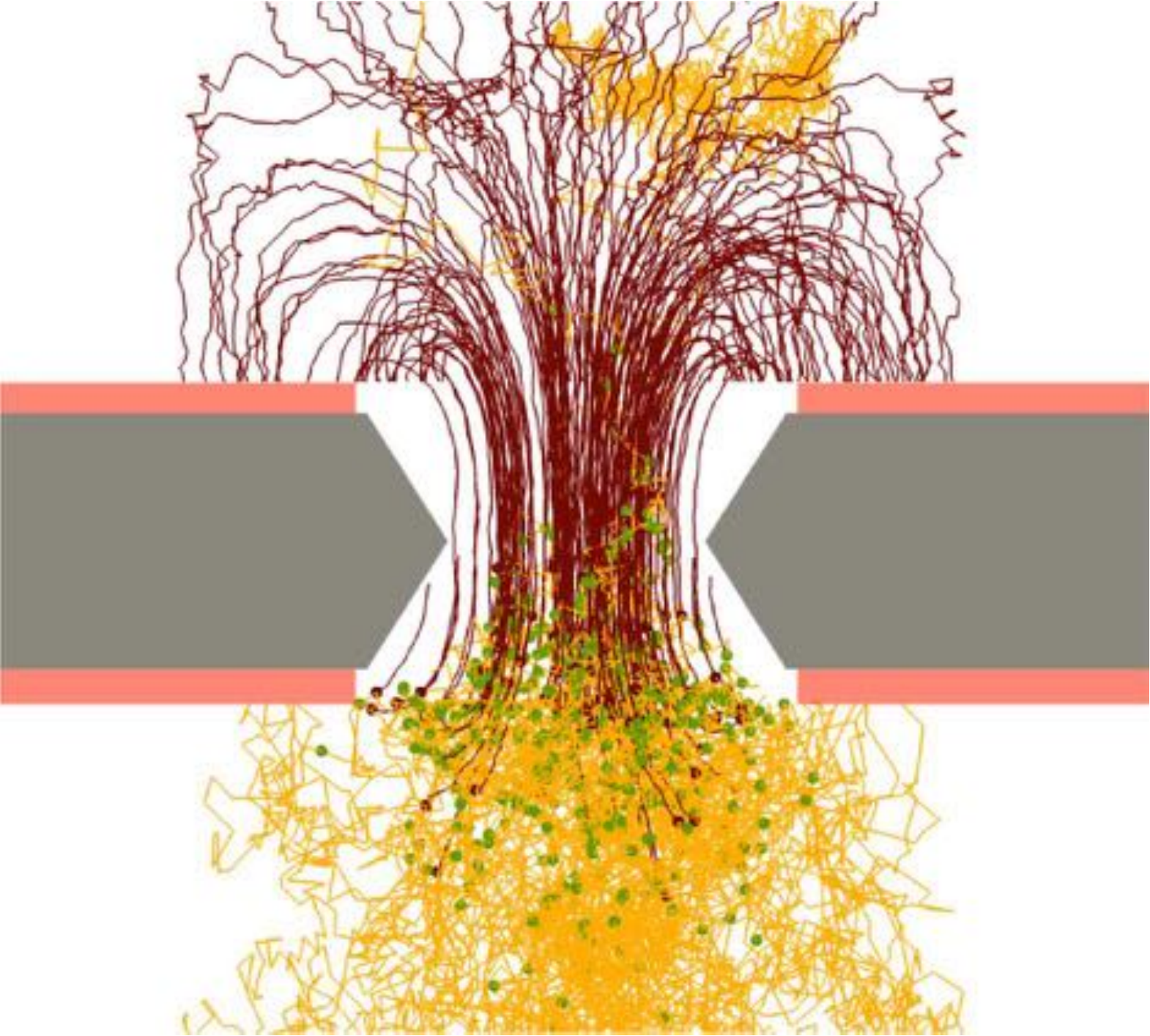}
\caption{Illustration of a microscopic GEM simulation by the ALICE experiment \cite{ALICE}. Brown and yellow lines correspond to ion and electron trajectories, respectively. Green dots represent the positions where an ionization took place.}
\label{Garfieldpp}
\end{figure}

\subsection{Microphysics modelling and simulation tools} \label{Micro}

There are numerous physics simulators that can describe the primary interaction process inside TPCs. Aimed at covering
the physics described in this work we will mention Geant4 \cite{geant}, SRIM \cite{SRIM} and NEST \cite{NEST}. Once
the primary ionization has been computed, it can be transported to the multiplication plane by using eq. \ref{THESOL3} together with experimental or Magboltz-based swarm parameters at the corresponding field $E_d$. But Magboltz has been recently empowered in various ways, enabling purely microscopic electron transport in several situations of interest to TPCs.

One example of these new developments is Garfield++ \cite{Garfield++}, that interfaces the Magboltz database with ROOT \cite{ROOT} and with several electric field solvers (Gmsh+Elmer \cite{Gmsh,Elmer}, neBem \cite{neBem}, CST \cite{CST} and COMSOL \cite{COMSOL}), allowing microscopic (i.e., cross section based) 3D transport in arbitrary fields and geometries. This allows in particular the computation of transmission through meshes and holes, where the high field gradients restrict the applicability of the hydrodynamic approximation \cite{refTranspRob, Kuger}.
The capabilities of Garfield++ have been increased with the extraction of Penning transfer coefficients \cite{NePen,ArPen, XePen}, that allow now an improved estimate of the gain process in several neon, argon and xenon-based mixtures. Although these coefficients are approximate (see, e.g., eq. \ref{Jesse}) and, as such, they might be geometry and field dependent, a good agreement has been shown if using the Penning transfer coefficients obtained for single wire counters (in Ar/CH$_4$ and Ar/CO$_2$), for the simulation of the multiplication process in thick GEMs \cite{AzevedoThickGEM}. A similar level of agreement has been found between single-wire and microbulk Micromegas operated under Xe/TMA admixtures \cite{XePen}. Clearly, given the up to 2 orders of magnitude discrepancy with a gain calculation based purely on elementary electron cross sections, this represents a major step forward. Recently, the computation of scintillation in pure noble gases \cite{oliveira} and some binary mixtures \cite{Mua} has been made available, too. An example aimed at illustrating the capabilities of Garfield++ is given in Fig. \ref{Garfieldpp}, showing the charge propagation during multiplication in a GEM hole. The ALICE collaboration has resorted to this kind of simulations in order to understand better what could be a feasible compromise in terms of energy resolution and ion back-flow in view of the intended 4-GEM upgrade of their TPC \cite{ALICE}.

\begin{figure}[h!!!]
\centering
\includegraphics*[width=6.5cm]{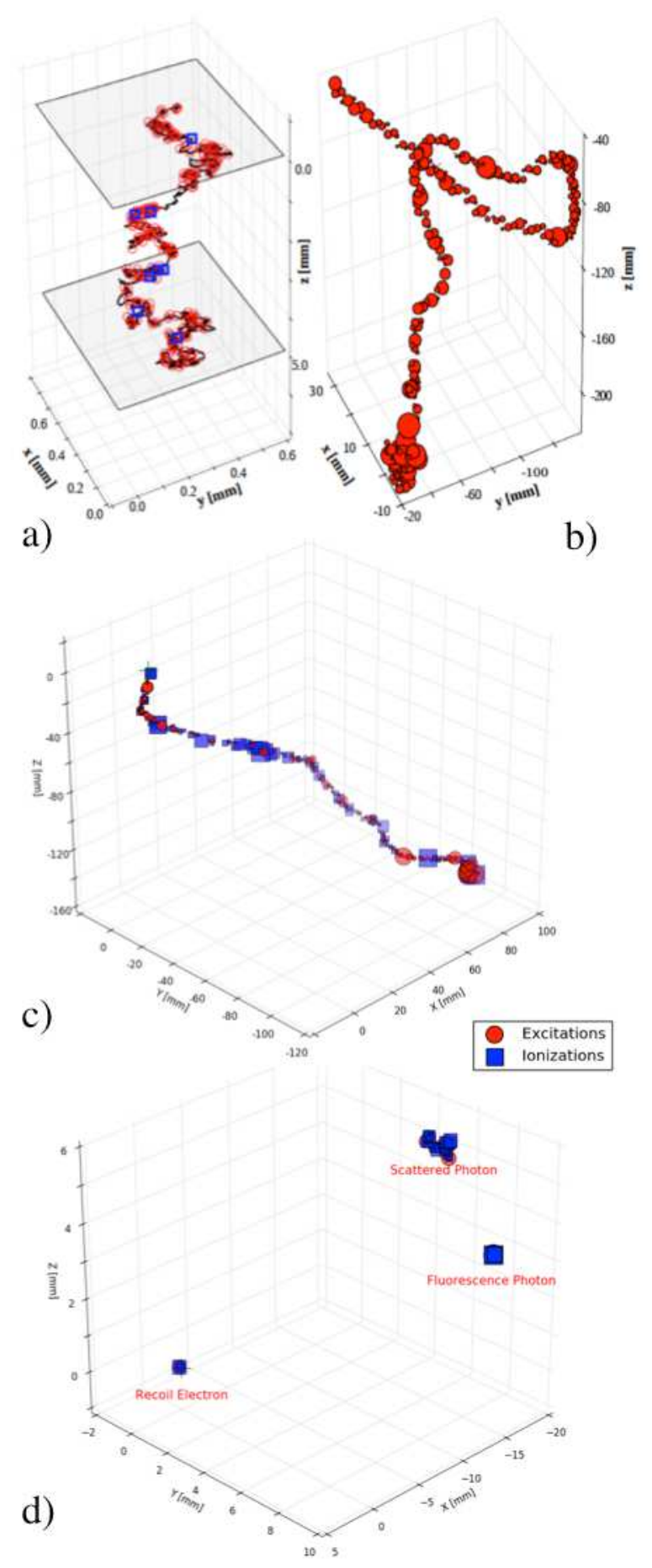}
\caption{a) Electron drifting under electroluminescence conditions, obtained with Garfield++ (circles indicate the positions of the promoted singlet and triplet states, and squares are higher lying states); b) 2.4\,MeV electron in xenon at 10\,bar (both figures taken from \cite{Mua}); c) a 50\,keV electron in CF$_4$ at 50\,mbar; d) a Compton interaction by a 140\,keV x-ray at 10\,bar. (Figures courtesy of C. Azevedo)}
\label{FigsMicroAll}
\end{figure}

Magboltz has been recently empowered through Degrad, too \cite{Magboltz}, that allows computation of primary electron and x-ray interactions in gases up to 1-2\,MeV. Degrad appplies the Born approximation to the elementary cross sections stored in Magboltz and includes, upon the release of one or several electrons in each ionizing encounter, the associated Auger electron, Coster-Kronig decay, x-ray fluorescence and shake-off processes. Examples of results obtained with Garfield++ and with Degrad are given in Fig. \ref{FigsMicroAll}.

\subsection{HV and field} \label{HVsec}

\begin{figure}[htb]
  \centering
  \includegraphics[width=8cm]{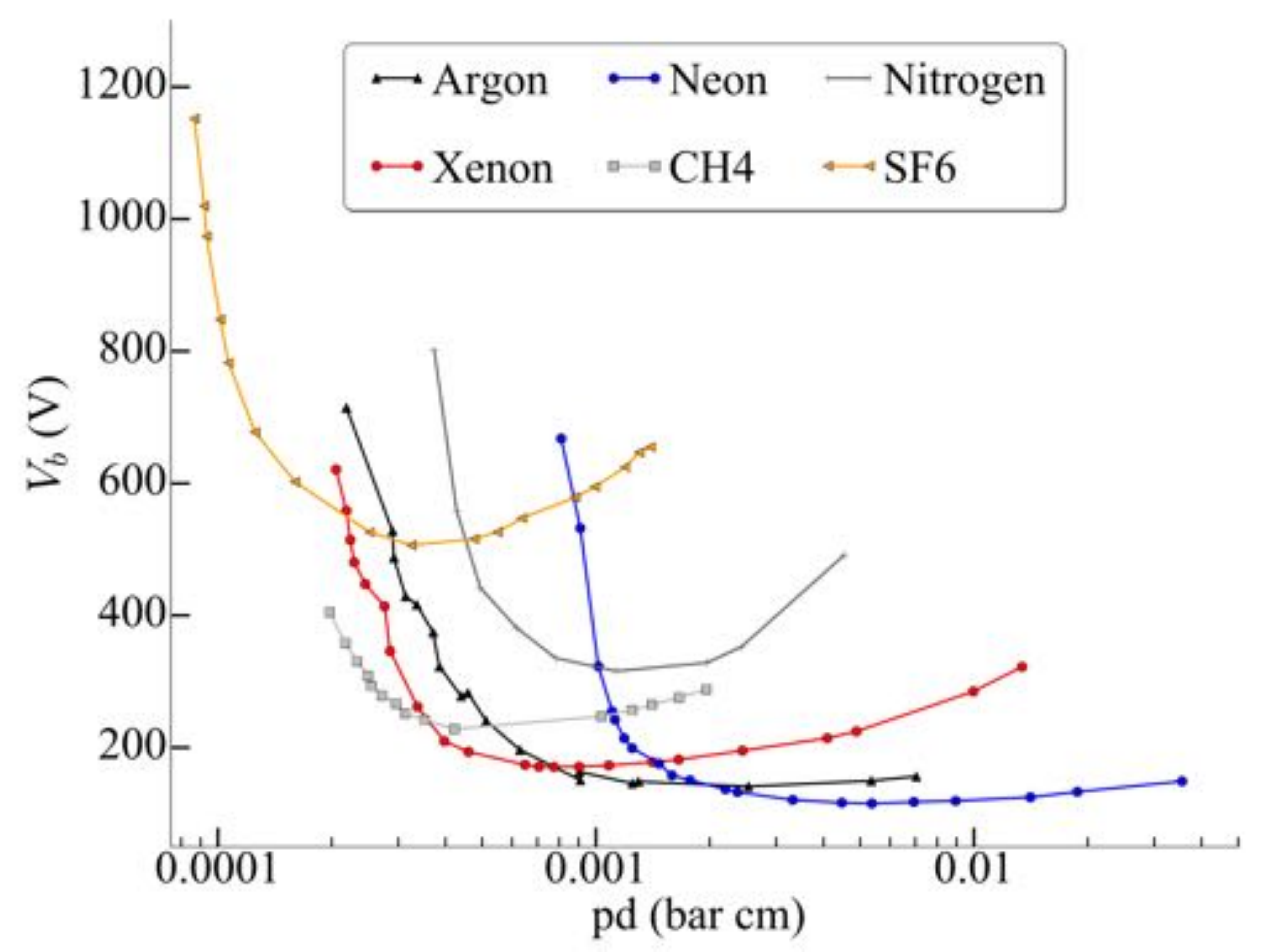}
  \caption{A recent experimental survey of Paschen curves obtained for different pure gases \cite{Maric}. The $x$-axis represents the product of the operating pressure ($P$) and   inter-electrode distance, or `gap' ($d$). The breakdown voltage $V_b$ is given on the $y$-axis.}
    \label{fig:Paschen}
\end{figure}

Given the broad range of operating pressures, and by virtue of the $E_d/P$ scalings in gas (table \ref{tab:Vb}),
the range of drift fields in modern TPCs is situated in the range $E_d=[0.01$-$1]$\,kV/cm. Although in extreme
situations, and given the large drift distances, the potential at the electrodes can approach the MV scale \cite{DeBonis:1692375}
most experiments do not exceed $\sim50$\,kV. The problem of reaching such HV levels is further complicated in some experiments
by the use of pure noble gases, given their low breakdown voltage V$_b$ (Fig. \ref{fig:Paschen}).
We briefly discuss here this particular situation, provided it is generally more demanding than the case of molecular
admixtures,\footnote{Except for some strong Penning mixtures (see \cite{Penning} for illustration).} and refer the reader to a recent
review work for additional details \cite{Rebel:2014uia}.

\begin{table}[h]
  \centering
\begin{tabular}{ |c|c|c|c| }
 \hline
Gas & RES &~E$_b$ [kV/cm]~& Ref.\\\hline
SF$_6$  &  1 & 91.8  & \cite{ChristophorouHV} \\
CO$_2$  & 0.3-0.35 & 29.3-32.3 & \cite{ChristophorouHV,VIJH1985287} \\
CH$_4$  & 0.4 & 36.7  & \cite{VIJH1985287} \\
CF$_4$  & 0.39-0.4 & 35.8-37.1 & \cite{ChristophorouHV,VIJH1985287} \\
N$_2$O  & 0.44-0.64 & 42.9-58.8  & \cite{ChristophorouHV,VIJH1985287} \\
air  & 0.28 & 26.1  & \cite{ChristophorouHV} \\
N$_2$  & 0.36-0.4 &35.0-36.7  & \cite{ChristophorouHV,VIJH1985287} \\
neon  & 0.1-0.11 &9.2-10.1  & \cite{VIJH1985287,Xiao_Xenon}  \\
argon  & 0.07-0.2 &6.47-18.4  & \cite{ChristophorouHV,VIJH1985287} \\
xenon  &~0.28-0.32~& 25.7-29.4 &~\cite{VIJH1985287,Xiao_Xenon}~\\ \hline \hline
~liquid argon~~& -- &40, 1000 & \cite{Swan1,Bay:2014jwa} \\
~liquid xenon~~& -- &400 & \cite{PhysRevA.9.2582}\\
 \hline
\end{tabular}

\caption{Values of the relative electric strength (RES) of different gases, taking SF$_6$ as 1 by convention. In the third column the absolute value (dielectric strength) has been calculated at 1\,bar pressure assuming the value for SF$_6$ from \cite{ChristophorouHV}. The dielectric strength for liquid xenon and argon is also shown.}
\label{tab:Vb}
\end{table}

\subsubsection{Gas phase}

The HV and pressure conditions that are of interest to TPCs are frequently outside the region where Paschen behaviour
approximately holds and the product  $P\! \cdot \! d$  represents an useful magnitude \cite{Meek}. This is because Paschen
behaviour is characteristic of avalanche-driven feedback processes stemming from the cathode, and the variable that
regulates the phenomenon in those conditions is the product $P \cdot  d$ (see \cite{Raizer}, for instance). For high values (note that the product $P \cdot  d$ is proportional to the number of molecules per unit area), self-propagation of the ionization front without the need of auxiliary processes at the cathode becomes the dominant mechanism: i.e., germinal electrons progress into an avalanche as they feel the field, but the fact that there is physically a cathode somewhere becomes irrelevant for spark development. Some authors take as an approximate criteria the value $P\! \cdot \! d \simeq 1\,\tn{cm} ~ \tn{bar}$, as the point where Paschen mechanism is no longer valid (e.g. \cite{Raizer}).

\begin{figure}[htb]
  \centering
  \includegraphics[width=\linewidth]{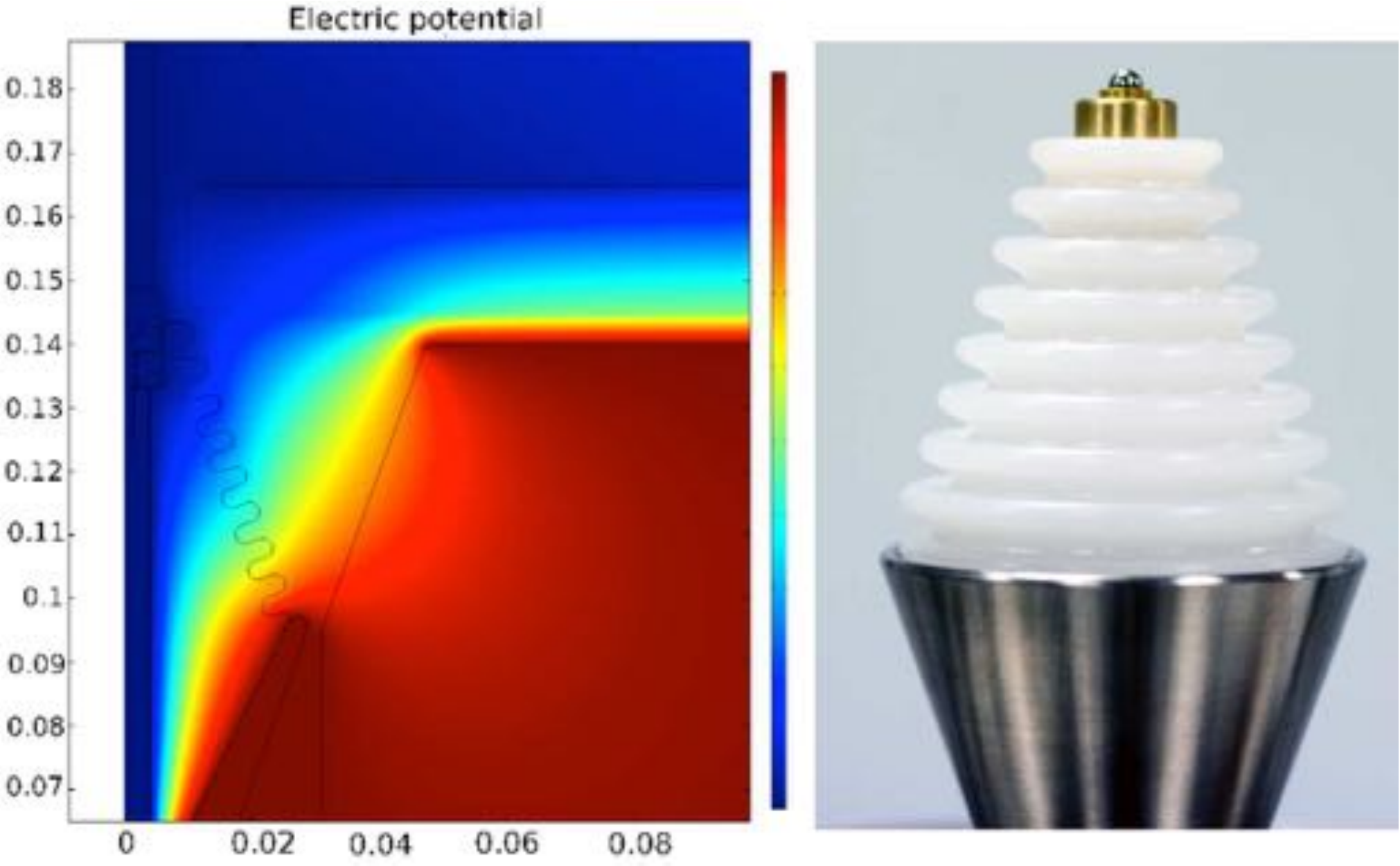}
  \caption{Left: electric potential in the region of the high voltage feedthrough of the NEXT TPC (as a reference, note that the feedthrough's central pin is situated to the left-side of the figure). The very compact design (courtesy of Hanguo Wang) stems from the necessity to minimize space in order to save the precious $^{136}$Xe content, and to shield against radioactive backgrounds. The right structure in red is the inner copper shielding (ground) and, above it, the field cage can be seen. The feedthrough tip touches the TPC cathode at about 2/3 of the figure's height (lengths are in meters). Right: a photo of the finalized device. This feedthrough is currently working at around 28\,kV under 7\,bar of pressure in NEXT-NEW, without any sign of sparking for several weeks.}
    \label{fig:conicHVFT}
\end{figure}

Therefore in practice, for gas gaps around 1\,cm and pressure around atmospheric, Paschen systematics must be abandoned and the relative electric strength is often used instead (table \ref{tab:Vb}). According to it, a 1\,cm column of argon gas can hold around 10\,kV at 1\,bar, and xenon will hold twice that. In practice, much lower values are found, and a conventional 2\,cm-long feedthrough can hardly prevent the formation of an arc between tip and cage at around 5\,kV in pure argon, in the absence of a proper design. Even a 10\,cm-long feedthrough will show issues when operated at 50\,kV in xenon at 10\,bar \cite{Rebel:2014uia}. One difficulty is related to the fact that the concept of dielectric strength ($E_b$) applies to plano-parallel electrodes at near-atmospheric pressure, whereas the situation depends in general on the electrode geometry and gas conditions. Another one, that is little understood and that represents a frequent practical limitation, is the propagation of surface discharges along insulators (\cite{Sobota_surf}, for instance). Even in the absence of a quantitative description of this latter process, electrostatic simulations can help at finding a field configuration that minimizes the accumulation of charges at the dielectric, partly alleviating the situation (Fig.\ref{fig:conicHVFT}-left). Besides modelling and design, systematic experimental measurements (e.g. \cite{Lockwitz}) together with the use of large safety factors ($\times2$-$4$) seems mandatory \cite{LZ}.

It must be noted that both HV insulation and detector performance depend critically on the purity of the noble gas: e.g. a high scintillation throughput (the most common goal behind the use of noble gases) will be connected to a low breakdown voltage, and viceversa. Therefore, the purity conditions during any HV test need to approach the ones expected in the final experiment. Besides avoiding HV breakdown, the electric field should be everywhere (except perhaps in the anode region) below the threshold for electroluminescence ($\simeq$ 1\,kV/cm/bar), and corona effect should be similarly avoided not to blind the optical sensors.

\subsubsection{Dual-phase}

\begin{figure}[htb]
  \centering
  \includegraphics[width=7cm]{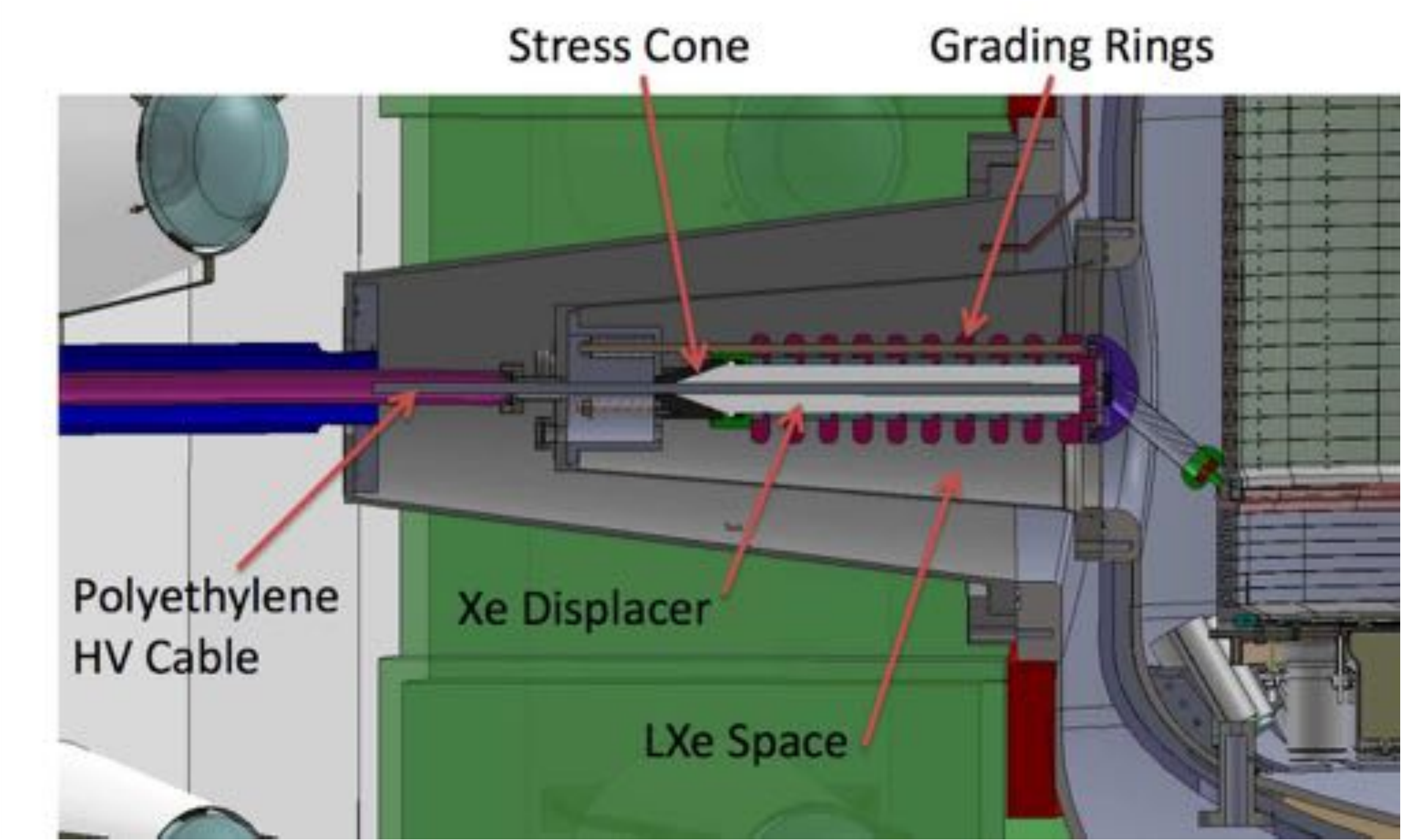}
  \caption{Schematic of the HV feedthrough of the LZ experiment \cite{LZ}, and connection to the cathode grid (in liquid phase). The field grading structures allow to maintain the field below 50\,kV/cm.}
    \label{fig:LZ_HVT}
\end{figure}

Despite the dielectric strength is much higher in liquid phase than in gas (table \ref{tab:Vb}), small amounts of locally dissipated heat can be sufficient to create bubbles, reducing for instance the dielectric strength from 1.4\,MV/cm \cite{Swan1} to around 40\,kV/cm \cite{Bay:2014jwa,Blatter} in the case of liquid argon. Other difficulties are related to the need of materials with compatible thermal expansion coefficients, low outgassing, as well as a low radioactivity content in experiments where a low background is needed (e.g. dark matter detection).\footnote{An example of the latter type is high density polyethylene, HDPE.} Similar to the gas phase, a gentle grading of the electric field between tip and ground improves the situation (Fig. \ref{fig:LZ_HVT}).

An extreme situation is that of the dual-phase DUNE far detector, that requires several 100's of kV to 1\,MV at the cathode. The feedthrough is about 2\,m-long since it has to traverse the $\sim$1\,m-thick walls of the cryostat and part of the gas phase. The design is based on the
one developed for ICARUS \cite{Amerio:2004ze}, where the feedthrough follows a coaxial configuration and the insulator is manufactured from a single block of cryogenic polyethylene. It has reached the 300\,kV milestone very recently \cite{1748-0221-12-03-P03021}.

\begin{figure}[htb]
  \centering
  \includegraphics[width=\linewidth]{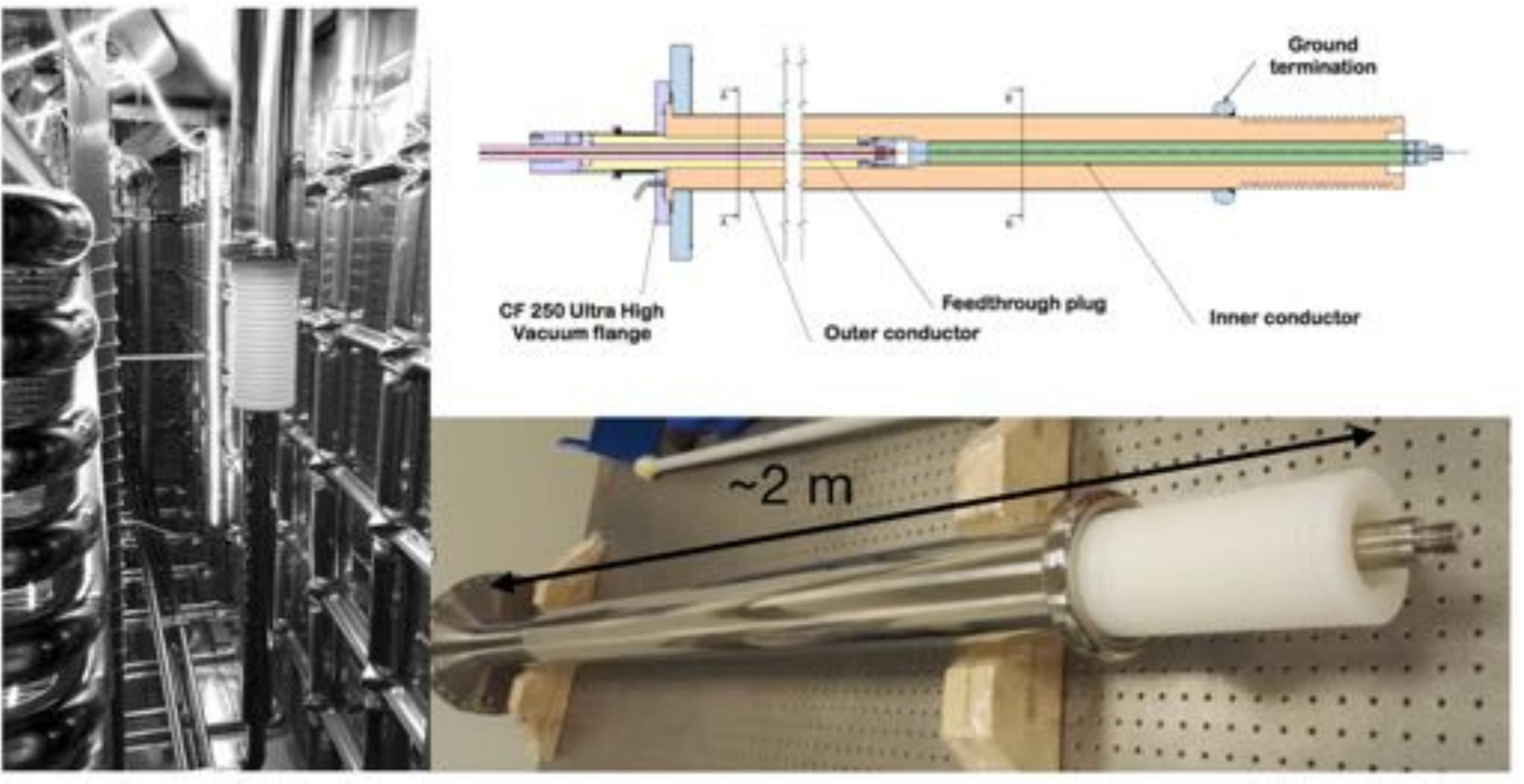}
  \caption{The 300\,kV-rated high voltage feedthrough developed for DUNE's dual-phase far detector (as an intermediate step towards the envisaged 600\,kV), \cite{1748-0221-12-03-P03021}. The left picture shows the feedthrough inside a $3\times1\times1$ m$^3$ prototype.}
    \label{fig:CRP-sketch}
\end{figure}

\subsection{Cryogenics}

Condensed phases of noble gases ($T_{boil}=87$\,K(argon) and $T_{boil}=165$\,K(xenon) at $P=1$\,bar) can be achieved by means of liquid nitrogen cooling. Generally, the cryostat is made of stainless steel or titanium, in a way that its interior allows operation under vacuum conditions as well as a slight pressurization (up to around 3-4\,bar). This allows on the one hand to achieve ultra-high purity conditions through pumping and bake-out; on the other hand, vapor pressure can be tuned to enhance performance (for details, see for instance \cite{LZ}).

\begin{figure}[htb]
  \centering
  \includegraphics[width=\linewidth]{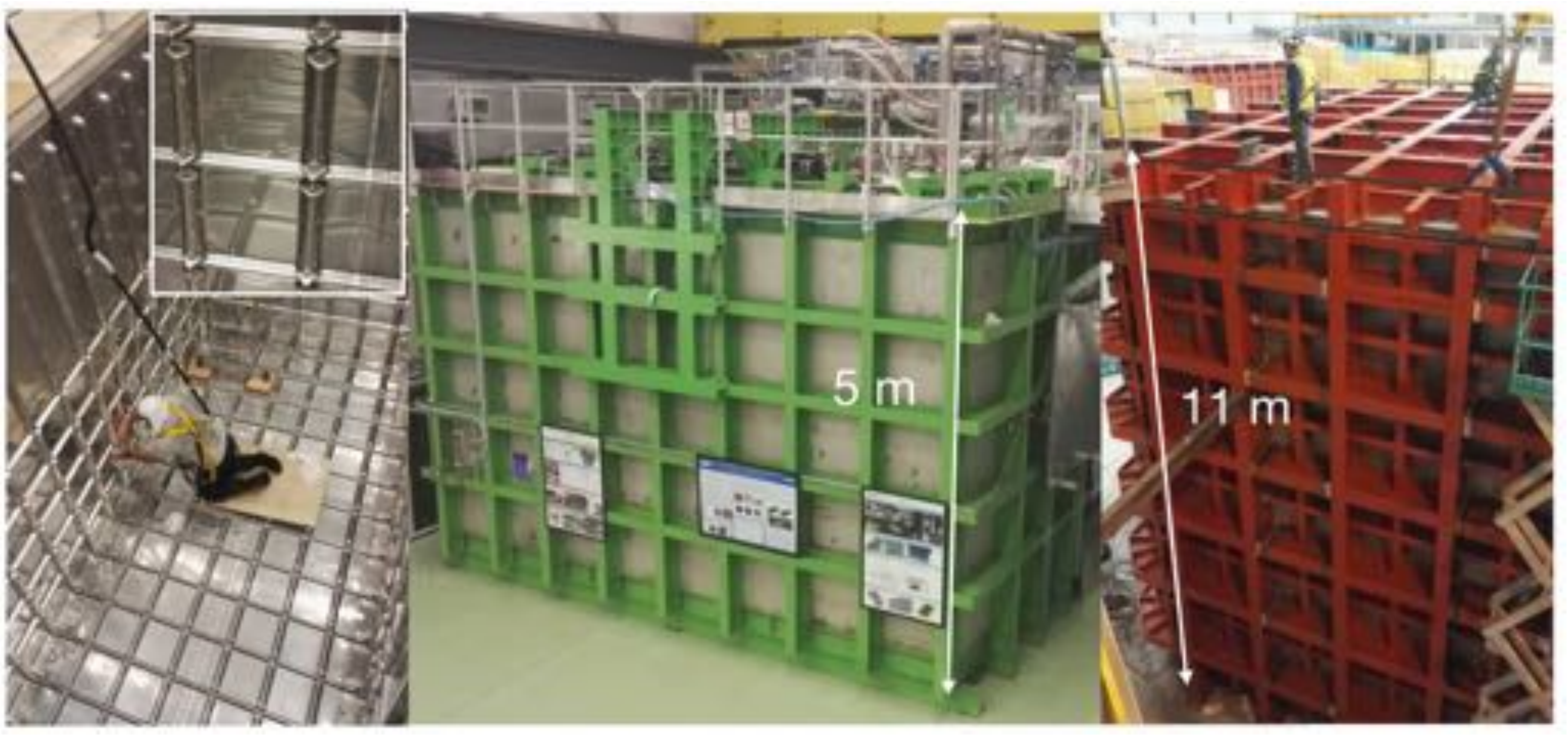}
  \caption{Pictures of the membrane cryostats installed at CERN. Left, middle: internal and external views of the cryostat hosting a $3\times1\times1$\,m$^3$ dual-phase TPC prototype \cite{Murphy:2016ged}. Right: the cryostat of the proto-DUNE $6\times6\times6$\,m$^3$ dual-phase demonstrator under construction in the CERN test beam area \cite{DeBonis:1692375}.}
    \label{fig:WA105-cryostat}
\end{figure}

Again, the case of the far-detector of DUNE is somewhat an exception, provided it cannot be neither pumped nor pressurized in a practical way, given its dimensions. Another scale-related problem is the fact that conventional `rigid' cryostats cannot absorb the thermal stress easily, so they have been replaced by corrugated stainless steel `membrane' cryostats, commonly used for the industrial transport of liquified natural gas (Fig. \ref{fig:WA105-cryostat}). The heat input of a kt-scale membrane cryostat is typically in the 10\,kW range.

Achieving and maintaining a liquid argon purity better than 100\,O$_2$\,ppt (section \ref{ChargeColl}) requires the ability of recirculating and purifying both the gas and liquid phase, in a completely sealed and leak-free system. During operation, the system has to satisfy a leak rate $<10^{-9}$\,mbar l s$^{-1}$ and maintain the inner volume at a constant $\sim$1\,atm pressure by actively cooling the boil-off gas through liquid nitrogen heat exchangers. The purification is based on cartridges made from copper pellets to remove oxygen and molecular sieves to remove water by physical absorption. To maintain the high level of purity, both liquid and (re-condensed) gas are constantly circulated through the cartridges. Since the structure of the cryostat is not compatible with vacuum operation, the initial impurities must be removed prior to filling by flushing and recirculating pure argon gas inside the cryostat.

\section{Gaseous chambers} \label{classification1}

A list of the main TPCs used for the detection of rare processes, presented in the order in which they appear in text, can be found in table \ref{TableAll}.
We will discuss in this section those based on gaseous media, classified by field of application, and next section will be devoted to dual-phase.

\onecolumn
\begin{table}[h]
  \centering
  \begin{tabular}{|c|c|c|c|c|c|c|c|c|c|c|c|c|c|c|c|c|c|c|c|c|c|}
     \hline
     ~ TPC~        &~$E_d$\,[V/cm]~& $B\,[\tn{T}]$ & ~ $H(\times S)$\,[m $\times$ m$^2$] ~ & $P$\,[bar] & image plane  & layout & medium & Ref\\
     \hline
     \hline
     ACTAR     & flexible     & -		& $0.25\,(\times\,0.25^2)$                      & 0.1-3	    & MM\,(bulk)		      & 3D    & generic (H$_2$, He, Ar\ldots)	      & \cite{ACTAR}         \\
     AT-TPC    & flexible     &~up to 2~& $1\,(\times\pi\,0.3^2)$                       & 0.05-1    & MM\,(microbulk)	      & 3D    &~generic (H$_2$, He, Ar, CO$_2$\ldots)~& \cite{AT-TPC2}       \\
     Warsaw    & flexible     & -		& ~ $0.21\,(\times\,0.18\times\,31)$ ~          & 1	        & ~ 4-GEM + PM + CCD ~   &~2D+1D~& Ar/He/CH$_4$/N$_2$-based              &~\cite{OTPClast}~     \\
     TUNL      & flexible     & - 	    & $0.21\,(\times\,0.3\times\,0.3)$              &~0.13-0.18~& MSAC + PMs + CCD	      &~2D+1D~& CO$_2$/N$_2$	                      & \cite{ZimmermanPhD}  \\
     ~NEXT-NEW~& 	200-600   & -	    & $0.53\,(\times\,\pi\,0.21^2)$                 & 5-15		& mesh + SiPMs + PMs     &~3D    & $^{136}$Xe-enriched xenon             & -                    \\
     PandaX-III& up to 1000   & -	    & $(2\times)\,1\,(\times\,\pi\,0.75^2)$         & 10	    & MM\,(microbulk)	      &~2D+2D~& ~ $^{136}$Xe-enriched Xe/TMA ~	      & \cite{PandaX-last}   \\
     DRIFT 	   &	600-700   & -		& $(2\times)\,0.5\,(\times\,1\times\,1)$        & 0.055	    & MWPC				      &~2D+2D~& CS$_2,$O$_2$-based	                  & \cite{DRIFTlast}     \\
     DMTPC	   &	150-250   & -	    & $(4\times)\,0.275\,(\times\,1\times\,1)$      &0.04-0.1	& mesh + PMs + CCDs	  &~2D+1D~& CF$_4$ 	                              &~\cite{Leyton,DDMpeople}~\\
     NEWAGE	   & 80-300       & -	    & 0.41\,($\times\,0.3\times\,0.3$)	            &0.2	    & $\mu$-PIC + GEM	      &~2D+2D~& CF$_4$ 				                  &\cite{NEWAGE}	        \\
     MIMAC	   & 100		  & -	    &~$(2\times)\,0.25\,(\times\,0.1\times\,0.1)$~  &0.05	    & MM\,(bulk)		      &~2D+2D~& CF$_4$/CHF$_3$/i-C$_4$H$_{10}$ 	      &~\cite{MIMAC0,DDMpeople}~\\
     TREX-DM   & flexible	  & -	    &~$(2\times)\,0.25\,(\times\,0.25\times\,0.25)$~&1-10		& MM\,(microbulk)	      &~2D+2D~& Ne, Ar -based	                      &\cite{TREX-DM_last}		\\
     T2K-ND	   & 200-300	  & 0.18	&~$(2\times)\,1.25\,(\times\,1\times\,2.55)$~   & 1			& MM\,(bulk)		      & 3D    & Ar/CF$_4$/i-C$_4$H$_{10}$             &\cite{T2KNIM}	\\
     CAST 	   & $\sim\!100$  & -		& 0.03\,($\times\,0.06\times\,0.06$)			& 1.4	    & MM\,(microbulk), INGRID &~2D+2D~& Ar/i-C$_4$H$_{10}$ 		              &~\cite{TREX-DM_last,Garza}~\\
     MuCap	   & 2000		  & - 	    & 0.12\,($\times\,0.15\times\,0.3$)		        & 10	    & MWPC	                  &~2D+2D~& D-depleted H$_2$                      & \cite{MuCap0}			\\
     \hline
     \hline
     DUNE-FD     & 1000       & -		& $(4 \times)\,12\,(\times60\times12)$  		& 1			& ~ LEMs + PMs	~         & 2D+2D & argon                                 & \cite{DUNE_CDR}           \\
     LUX         & 181		  & -		& $0.48\,(\times\pi\,0.235^2)$		            & 1-2       & mesh + 2 PM planes	  & 3D    & xenon				                  & \cite{LUX1}\\
     XENON1T     & 120		  & -		& $1\,(\times\pi\,0.5^2)$	        		    & 1-2       & mesh + 2 PM planes	  & 3D    & xenon	               				  & \cite{1TonLimit}\\
     PandaX-II   & 393.5      & -	    & $0.6\,(\times\pi\,0.32^2)$			        & 1-2	    & mesh + 2 PM planes	  & 3D    & xenon			                      & \cite{PandaNew0}\\
     DarkSide-50 &  200		  & -		& $0.35\,(\times\pi\,0.178^2)$	 	 	        & 1			& mesh + 2 PM planes     & 3D    & $^{39}$Ar-depleted argon	          & \cite{DarkSide}\\
     WARP(100\,l)&   90-330	  & - 	    & $0.6\,(\times\pi\,0.25^2)$	  	 	        & 1			& mesh + PMs			  & 3D    & argon				                  & \cite{Zani:2014lea}		\\
     \hline
     \hline
     ALICE       & 400 		  & 0.5     & $(2\times)\,2.5\,(\times18)$                  & 1         & MWPC (GEMs)$^{*a}$ + pads & 3D    & Ne/CO$_2$/N$_2$                     & \cite{ALICE}	\\
     STAR      	 & 135 		  & 0-0.5 	& $(2\times)\,2.1\,(\times18)$				    & 1	        & MWPC + pads			    & 3D    &Ar/CH$_4$	 	                      & \cite{STAR}		\\
     \hline
   \end{tabular}

  \caption{Some technical parameters of the most representative TPCs used in the search of rare processes, both in gas (top block) and dual (middle block) phase. For reference, the lowest block includes two
   paradigmatic collider TPCs. The size of the active dimension along the electric field is dubbed $H$ and $S$ is the active area. For dual-phase, the electric field is given for the liquid phase and the pressure for the gas phase. The compilation is illustrative since several of the collaborations are already heading towards an upgrade, e.g., NEXT \cite{JJreview}, MIMAC \cite{DDMpeople}, T2K-ND \cite{T2K-NDup}, DarkSide \cite{1742-6596-718-4-042016} or LUX \cite{LZ}.} \label{TableAll}
  \begin{tablenotes}
    \item[1] $^{*a}$ the ALICE TPC will replace its MWPC plane by a 4-GEM one.
  \end{tablenotes}
  \end{table}

\twocolumn
\subsection {Low energy nuclear physics}

Three main experimental configurations exist, in the field of low energy nuclear reactions, all amenable to the TPC technique: `active' or `reaction' (nuclei from the gas mixture act as targets), `decay' (unstable nuclei are implanted
in the chamber) and `embedded-target' (a very thin target is inserted in the chamber).\footnote{A review detailing
the physics accessible through the first type of experiments has been recently published in \cite{Saul}.} Given the aim of this review,
we make a distinction between TPCs aimed at general purpose experiments and those designed with a specific process in mind, e.g.,
$2p$, $3p~\beta$-delayed, $\beta\beta2\nu$ or $\beta\beta0\nu$ decay, for instance. We start with the first class.

\subsubsection{Maya, ACTAR and other general purpose TPCs}

There is a surge of general-purpose TPCs aimed at low energy nuclear physics experiments, with easily some 20 different
configurations existing or planned \cite{Saul,Pollaco}. A selection of some of the most established ones includes ACTAR \cite{ACTAR}, ANASEN \cite{ANASEN}, CAT \cite{CAT}, MAIKO \cite{Maiko}, Maya \cite{Maya}, MINOS \cite{MINOS}, MSTPC \cite{MSTPC} and TACTIC \cite{TACTIC}. One of the defining characteristics of this category of TPCs is that the emphasis is
generally put on the accurate 3D-reconstruction of the event, that implies for instance in the case of ACTAR the use of pixels at a pitch as small as $\Delta{x(y)}=2$\,mm (16384 electronic channels for a 25\,cm $\times$ 25\,cm readout plane). ACTAR builds on the experience of the Maya TPC, that runs at the Grand Acc\'el\'erateur National d'Ions Lourds (GANIL) since 2003 \cite{Maya}.

\begin{figure}[h!!!]
\centering
\includegraphics*[width=\linewidth]{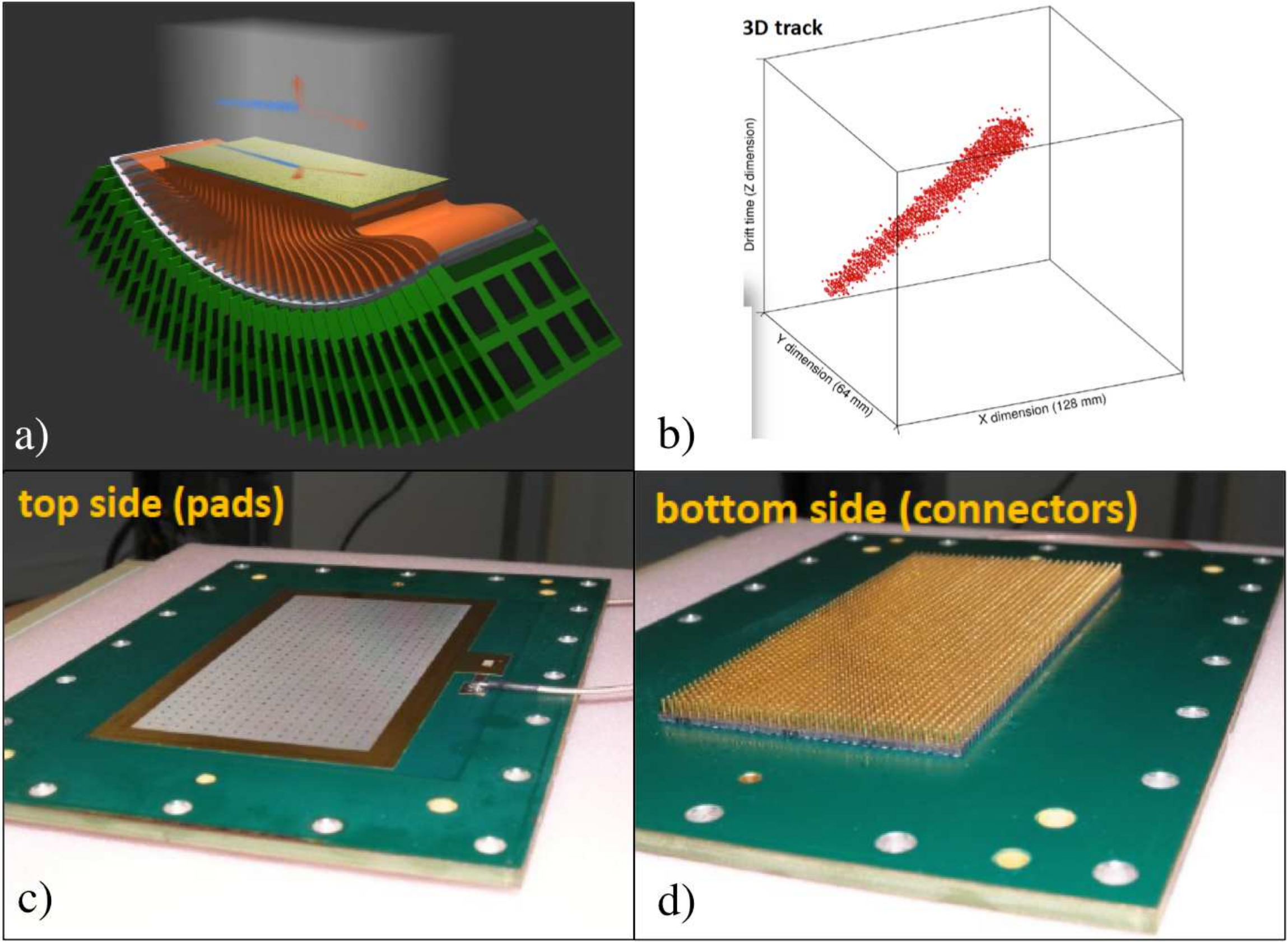}
\caption{The ACTAR TPC (figures courtesy of J. Giovinazzo): a) artistic design, highlighting the readout plane, flexible cables and front-end cards; b) a $\sim 12$\,cm-long $\alpha$-particle reconstructed with 2\,mm voxels in Ar/CH$_4$ (90/10) at 0.4\,bar; c) anode plane showing the induction pads (inner part); d) anode plane showing the connection pins (outer part).}
\label{ACTAR}
\end{figure}

This 3D-imaging philosophy is thus conceptually similar to what is usually done
in TPCs at colliders, but at a much smaller $(20$-$30)^3$\,cm$^3$ scale. Additionally, chamber gas and pressure (in the range 0.1-3\,bar) must be adjustable in order to ensure an adequate interaction probability and stopping power for a number of potentially relevant reactions. Being able to work reliably in environments involving very different gas and $d\varepsilon/dx$ conditions implies that a big part of the design effort must go into the charge amplification and readout: besides providing an adequate dynamic range, the electronics must be robust against micro-discharges, flexible in terms of settings (e.g. shaping time, amplification, time range and sampling time), preferably self-triggerable, and scalable. Similarly, a rugged design for the charge multiplication structure is desirable too. The Micromegas plane and the readout electronics of ACTAR, the latter based on the newly developed AGET chip \cite{aGET}, took several years of development. The high-density feedthrough structure, to which the Micromegas mesh is attached, must withstand -1/+3\,bar internal pressure, interfacing the induction pads with the exterior through gas-tight pin connections, in a reliable manner (Fig. \ref{ACTAR}). The anode plane is meant to allow assembly in two different chamber geometries with different aspect ratios (referred to as `reaction' and `decay' chamber). AGET is an enhanced version of AFTER, with updated components and including the capability of self-trigger.

\begin{figure}[h!!!]
\centering
\includegraphics*[width=7.5cm]{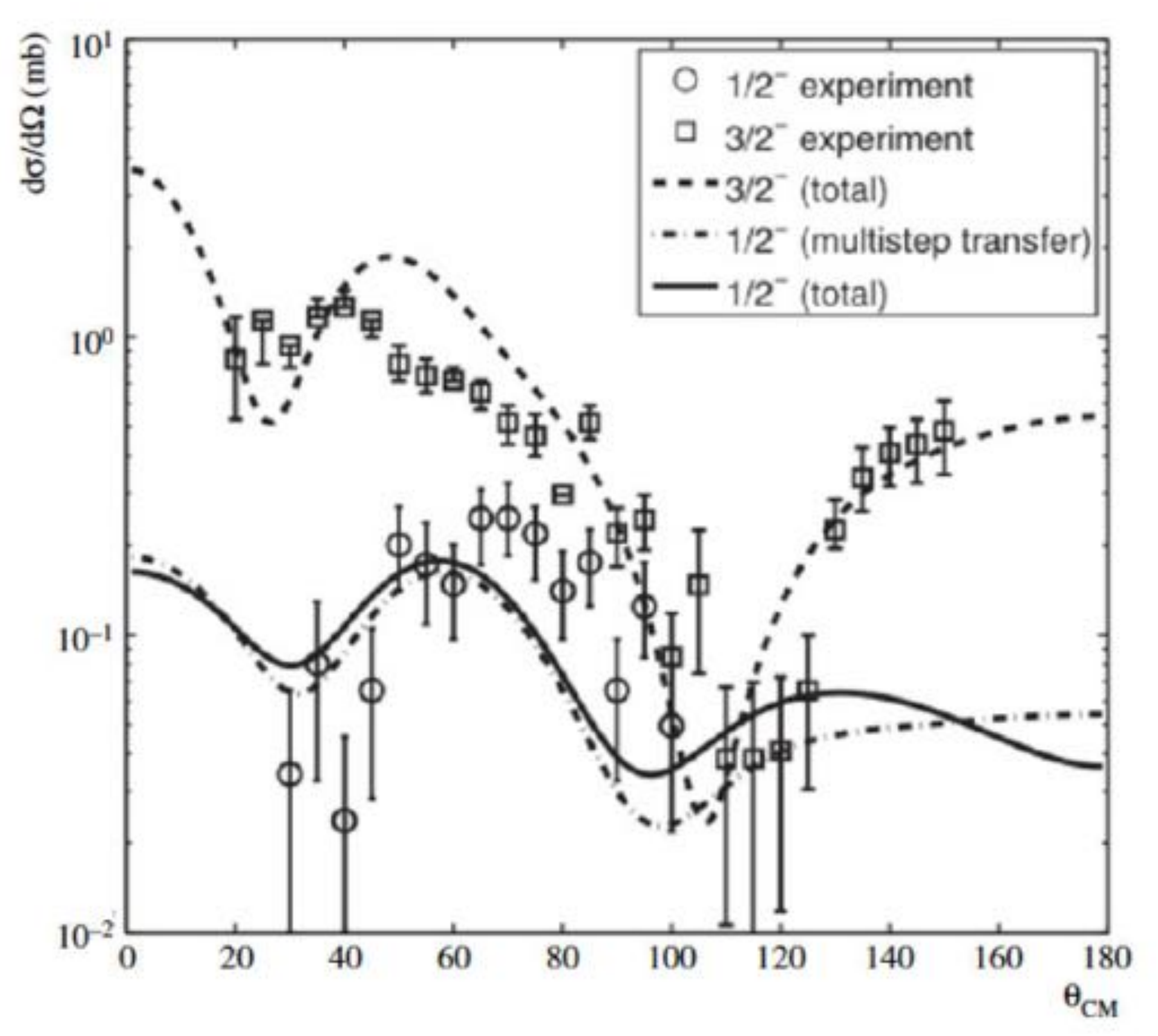}
\caption{Differential cross section for the reaction $^1$H$(^{11}$Li, $^{9}$Li)${^3}$H leading to the population of $3/2^-$ and $1/2^-$ states of ${^9}$Li, as measured with the Maya detector filled with hydrogen \cite{Maya1, Maya2}.}
\label{Maya}
\end{figure}

Operation of these chambers resorts normally to a main gas admixed with conventional VUV-quenchers (e.g., \mbox{i-C$_{4}$H$_{10}$}, CO$_2$)
at the minimum concentration compatible with a stable gain, in order to avoid spurious reactions.
In reaction studies, hydrogen or helium targets are widely used, since they are key to the study of nuclear reactions in inverse kinematics,
however for decay studies argon represents a more natural choice. The range of reactions of interest is indeed considerably broad \cite{Saul}. Fig. \ref{Maya} shows an example study of the 2 neutron halo of $^{11}$Li with a hydrogen target, one of the first experiments in Maya.

\subsubsection{The AT-TPC}

\begin{figure}[h!!!]
\centering
\includegraphics*[width=\linewidth]{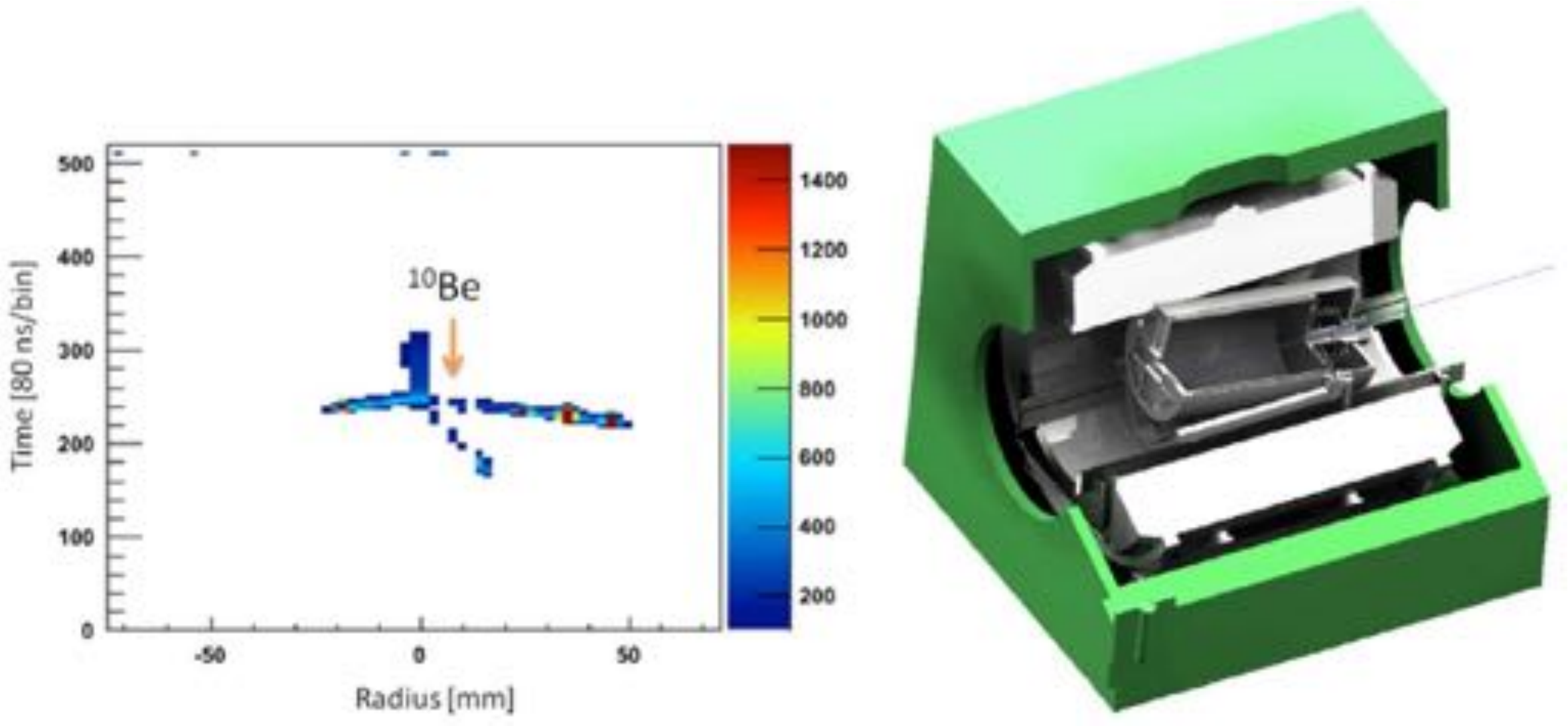}
\caption{Left: three body decay of $^{10}$Be observed in the AT-TPC prototype \cite{AT-TPC2}. Right: the AT-TPC final detector inside the bore of the 2\,T
super-conducting magnet.}
\label{ATTPC}
\end{figure}

The active target TPC at the Facility for Rare Isotope Beams (FRIB) is a general-purpose TPC that approaches, in size and complexity, other TPCs found at collider experiments. It resorts to magnetic field, as well, both to reduce transverse diffusion and to help at particle identification. A reduced
prototype, pAT-TPC, was commissioned with He/CO$_2$ (90/10) under $\alpha$-particles in \cite{AT-TPC} and realized a series of experiments
with light beams on helium (e.g. \cite{AT-TPC-results} and Fig. \ref{ATTPC}-left). Using microbulk Micromegas for charge amplification, a gain of around 200 was achieved in those conditions. The pAT-TPC detector has been shown for illustration earlier in this review (Fig. \ref{4TPCs}).

The final AT-TPC detector has been recently commissioned in \cite{AT-TPC2} and placed inside the magnet bore (Fig. \ref{ATTPC}-right).
It is a cylinder with about 1\,m in length and a readout plane with 0.6\,m diameter, covered by 10240 radial pads designed to provide a measurement of
 the $\varphi$ and $r$ coordinates. The TPC can be optionally tilted, in order to increase the angular resolution of particles scattered at low angles, as well as suppressing unreacted beam particles efficiently. A functionally similar TPC, dubbed S$\pi$RIT, read out with wires and with a box-shape instead of cylindrical has become available recently, too \cite{SPRIT}.

\subsubsection{The Warsaw and TUNL OTPCs} \label{2ppSec}

\begin{figure}[h!!!]
\centering
\includegraphics*[width=\linewidth]{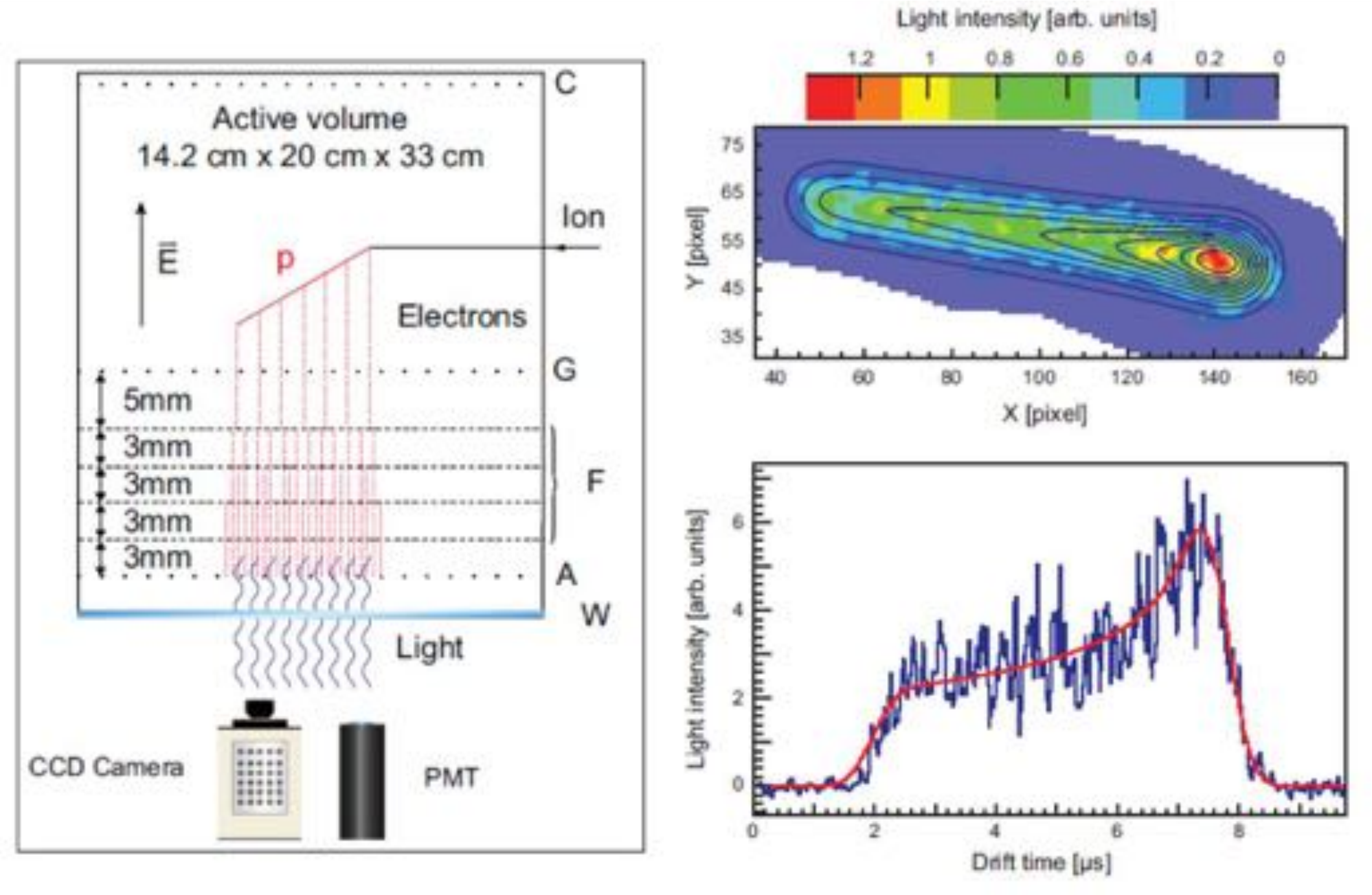}
\caption{Left: the Warsaw OTPC featuring a CCD camera, PM, 4-stage GEM and gating grid. Right: experimental results on the reconstruction of $\beta$-delayed proton emission of $^{48}$Ni. Top figure shows the $xy$ (CCD) projection, bottom one shows the $z$ (PM) projection. Contour plot (top) and red line profile (bottom) indicate a fit to the simulation package SRIM/TRIM. (Figs. from \cite{OTPClast})}
\label{Warsow TPC}
\end{figure}

The Warsaw TPC (Fig. \ref{Warsow TPC}) was designed for the measurement of the angular correlations in the 2-proton decay of neutron deficient nuclei but, as demonstrated recently by a team at TUNL, the technique is applicable to active targets as well. The concept was born at Charpak's group in the 80's \cite{Charpak}, and one of the original authors (W. Dominik) and colleagues proposed its application to the 2p problem. The idea is to use an optically-read TPC (OTPC) that images, thanks to a scientific CCD camera, the N$_2$ scintillation produced after several multiplication stages.\footnote{The term OTPC seems to have been coined after this work, despite it was not the first device of this kind.} Modern camera-read TPCs (based on either CMOS or CCD technology) offer a partially limited 2D view (through an $xy$ projection), due to the slow frame rate of low-noise cameras, up to around 1000\,fps. Instead, they provide a fine pixelization that easily allows imaging down to 50-200\,$\mu$m effective pixel size (i.e., defined at the object plane) on 100's of cm$^2$ areas. An additional projection along the $z$ axis can be obtained through an auxiliary PM, for instance. It should be noted that it is not well established to which extent this type of 2D$+$1D readout can be considered equivalent to a true 3D reconstruction (and, especially, under which experimental conditions). However, for low multiplicities and relatively straight tracks (as the ones studied in the Warsaw and TUNL experiments) 2D+1D readouts have shown to achieve full kinematic reconstruction (e.g. \cite{ZimmermanPRL}).

Conceptually (and with sufficient generality for present purposes) in this type of TPC and for this particular application one wishes to have a quaternary gas mixture with the following components:
\begin{enumerate}
\item A main (gain) gas, desirably cheap and readily available, for instance Ar.
\item A buffer gas that allows adjusting the track's stopping power and interferes only weakly with the gain and drift processes, e.g., He. In this way, the chamber can be operated at around atmospheric pressure.\footnote{If taking operation at atmospheric pressure as a constraint, the addition of He represents a (limited) analogy with a reduction of pressure, under which approximate $P$-scalings are expected to apply.}
\item A gas displaying a strong scintillation, stemming from direct excitation and/or mediated by atomic-molecular or photon transfers (`wavelength-shifting'), for instance N$_2$.
\item An electron cooler and VUV-quencher additive, thereby providing sufficient electron cooling and VUV-quenching to allow a fine tuning of both the drift and gain characteristics.
\end{enumerate}

\begin{figure}[h!!!]
\centering
\includegraphics*[width=8cm]{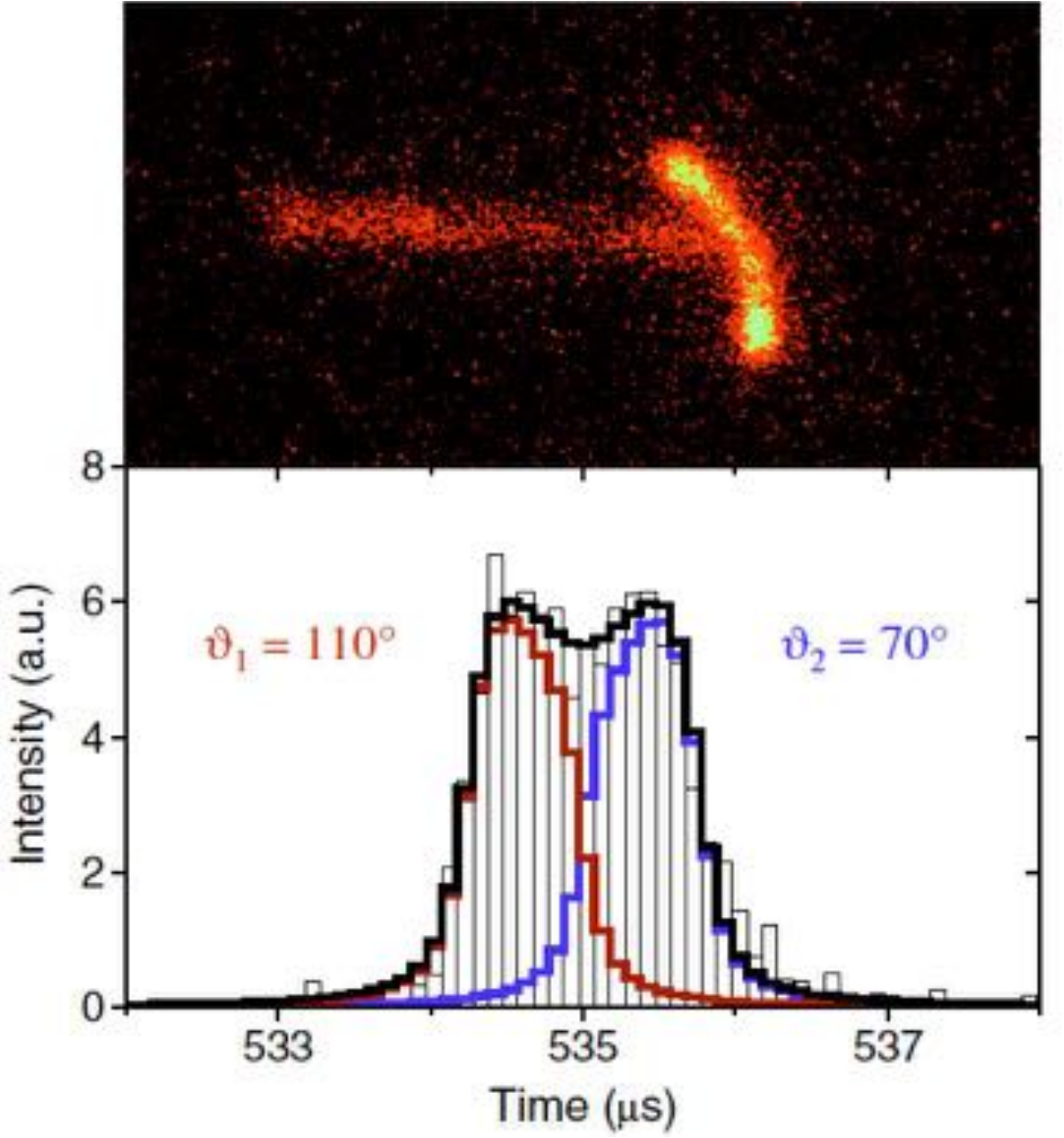}
\caption{The iconic image of the 2p decay of $^{45}$Fe obtained by the Warsaw group in 2007 \cite{OTPC2pp}. Top figure shows the reconstruction in the $xy$ plane through a CCD camera, and bottom one gives the $z$ projection, obtained with a PM.}
\label{2pIconic}
\end{figure}

Being designed for decay studies, the Warsaw TPC operates under fairly stable conditions concerning gas mixture, pressure, electric field
and scintillation structure, however some modifications were introduced over the years. As an example, the approach followed in the study of the 2p-decay was that of using a gas mixture consisting of Ar/He/CH$_4$/N$_2$ at $32/66/1/1$, a solid wave-length shifter
and multi-step avalanche configuration (MSAC) \cite{OTPCfirst}, allowing to image the process for the first time (Fig. \ref{2pIconic}). Successive versions used a simpler ternary mixture (Ar/He/N$_2$ at $49.5/49.5/1$) and a 4-stage GEM detector, after which no solid wavelength-shifter was needed. In a recent experiment,
He/N$_2$ at $98/2$ was used for studying the decay of $^6$He into $\alpha$ and D (deuterium) \cite{OTPClast2}. A description of one of the last experiments, together with an exhaustive list of previous ones can be found in a recent work \cite{OTPClast}. All these gas admixtures have a remarkable property: they can be flushed directly to the atmosphere for they are cheap, inert and environment friendly.

The Warsaw TPC possesses a remarkable technical feature, arising from the necessity to keep up with the extreme difference in $d\varepsilon/dx$ between (e.g.) Fe-nuclei and protons (nearly two hundred). In these conditions, when aiming for instance at 2p-decay in  $^{45}$Fe, the scintillation structure would be either too insensitive to the emitted protons or fully sensitive but just too unstable in the presence of the Fe-beam. A solution to this, based on the synchronous gating of an interposed grid, was implemented since the first prototypes.

The OTPC at the Triangle Universities Nuclear Lab (TUNL) was developed for the resonant photo-production of high lying Hoyle's states, meaning that a carbon target was needed. This suggested operation under CO$_2$ and a minimum amount of N$_2$ to ensure scintillation without introducing too many spurious reactions (80/20 was the final compromise). In order to optimize the stopping, the authors opted for a pressure reduction down to around 150\,mbar. Scintillation was produced through MSAC amplification and, in order to achieve enough light sensitivity, a gated image intensifier from the CHORUS experiment was used \cite{CHORUS}.

\begin{figure}[h!!!]
\centering
\includegraphics*[width=\linewidth]{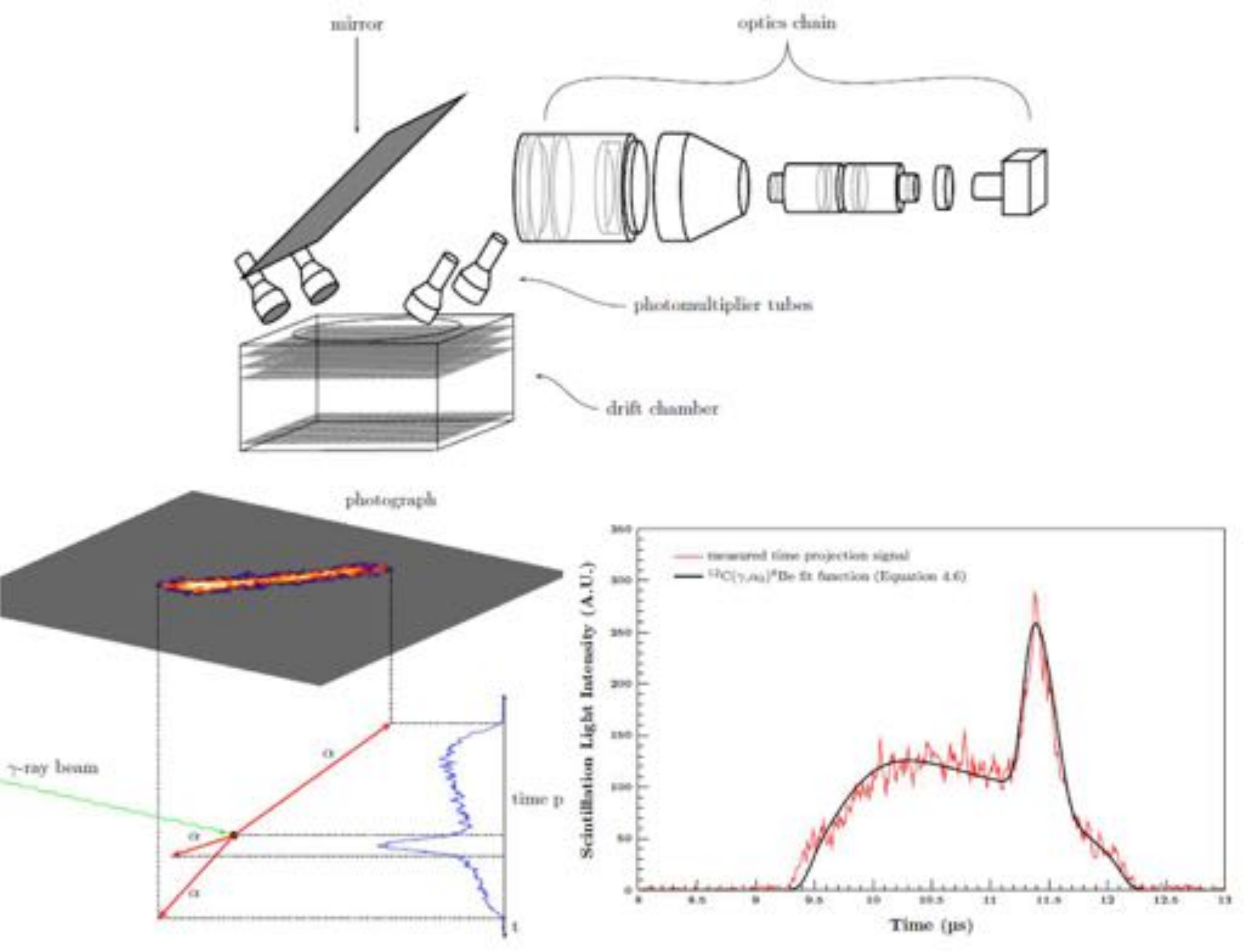}
\caption{Top: the TUNL OTPC coupled to the CHORUS optical system. Bottom: reconstruction of a typical $^{12}$C($\gamma$,$\alpha$)$^8$Be event, illustrating the correspondence between the $xy$ and $z$ projections (note that there are three nuclei in the final state). In the bottom-right figure the continuous line shows the result from SRIM/TRIM simulations.}
\label{TUNL}
\end{figure}

With the help of the SRIM simulation package, the authors showed that they could reconstruct the full reaction kinematics even in the presence of three tracks in the final state (e.g., Fig. \ref{TUNL}-bottom), and obtained the associated E1 and E2 cross sections for the $^{12}$C($\gamma$,$\alpha$)$^8$Be reaction in the neighborhood of the $J^\pi=2^+$ excited Hoyle state \cite{ZimmermanPRL}. For a detailed description of the system the reader is referred to \cite{ZimmermanPhD}.

\subsection{Neutrino-less double beta decay ($\beta\beta0\nu$)}

Neutrino-less double beta decay is a postulated (and exceedingly rare, were it to exist) nuclear process whose study represents a category on its own. Given its long half-life $T^{0\nu}_{1/2}\gtrsim 10^{26}$\cite{Kamland} it is genuinely a low-background experiment, unlike those discussed
in previous section. It targets a well known problem in the neutrino sector of the standard model of particle physics: the neutrino's nature and its
 mass. For this decay should proceed with a half life in an inverse relation with the square of the effective neutrino mass \cite{bbreview}.

Of all viable nuclei that can provide a measurable $\beta\beta0\nu$ lifetime, only the $^{136}$Xe
isotope is a gas at ambient temperature, making it amenable to the TPC technique. The most competitive experiments to date, on the other hand, are based on germanium \cite{Gerda} and liquid xenon, either TPC-based \cite{EXO} or diluted in liquid scintillator \cite{Kamland}. If the above techniques can not be made background-free, however, a positive claim may be challenged on a purely methodological basis.
In the event of a positive claim, the imaging capability provided by the high pressure phase offers a plausible solution to this experimental dilemma (the other one being to measure the process for several isotopes), and it is highly competitive when assuming measured figures of the achievable topological rejection factor \cite{Paola}. A recent compilation of the discovery potential of upcoming $\beta\beta0\nu$ experiments can be found in \cite{BB0new}.

\begin{figure}[htb]
  \centering
  \includegraphics*[width=7cm]{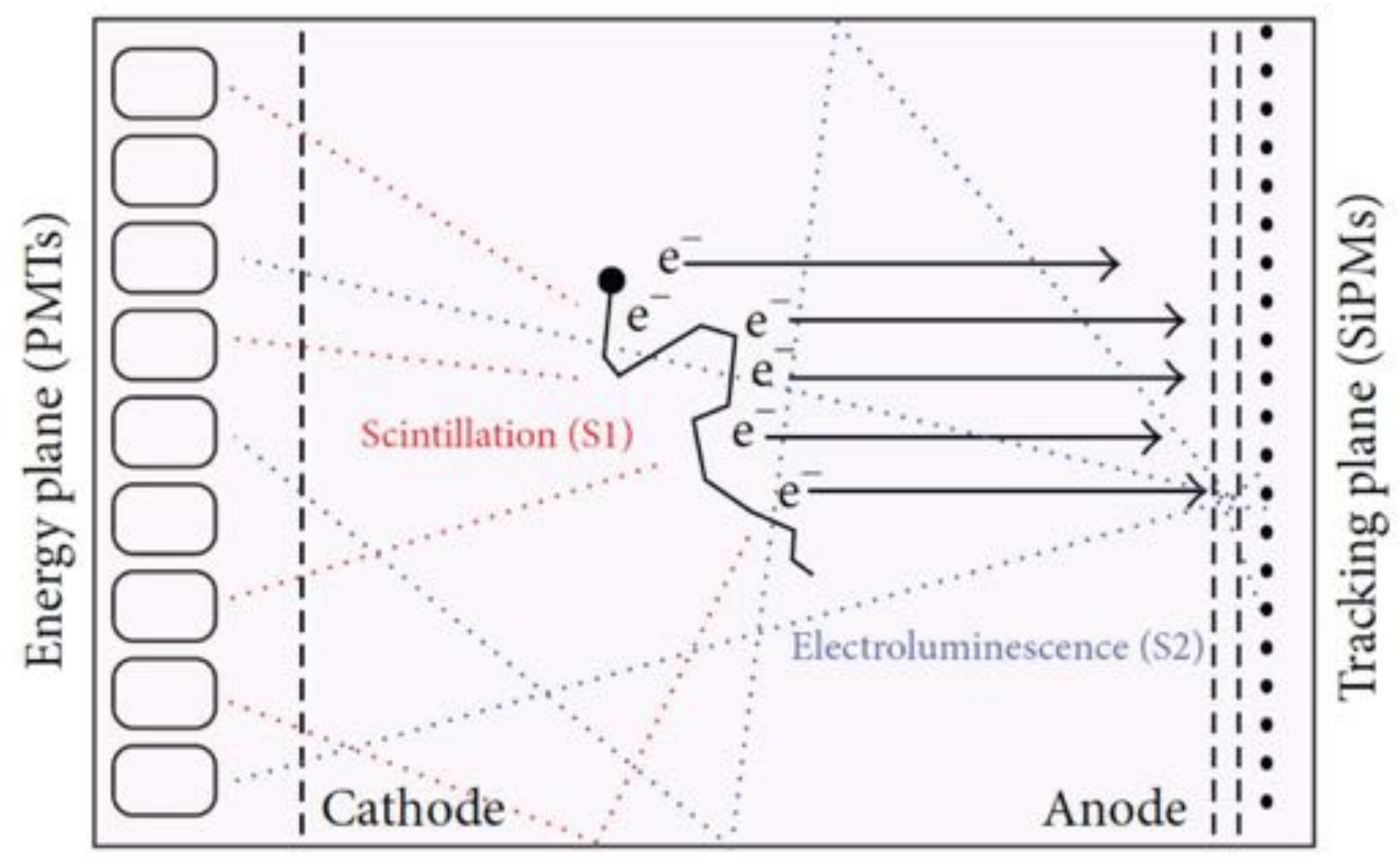}
  \includegraphics*[width=7cm]{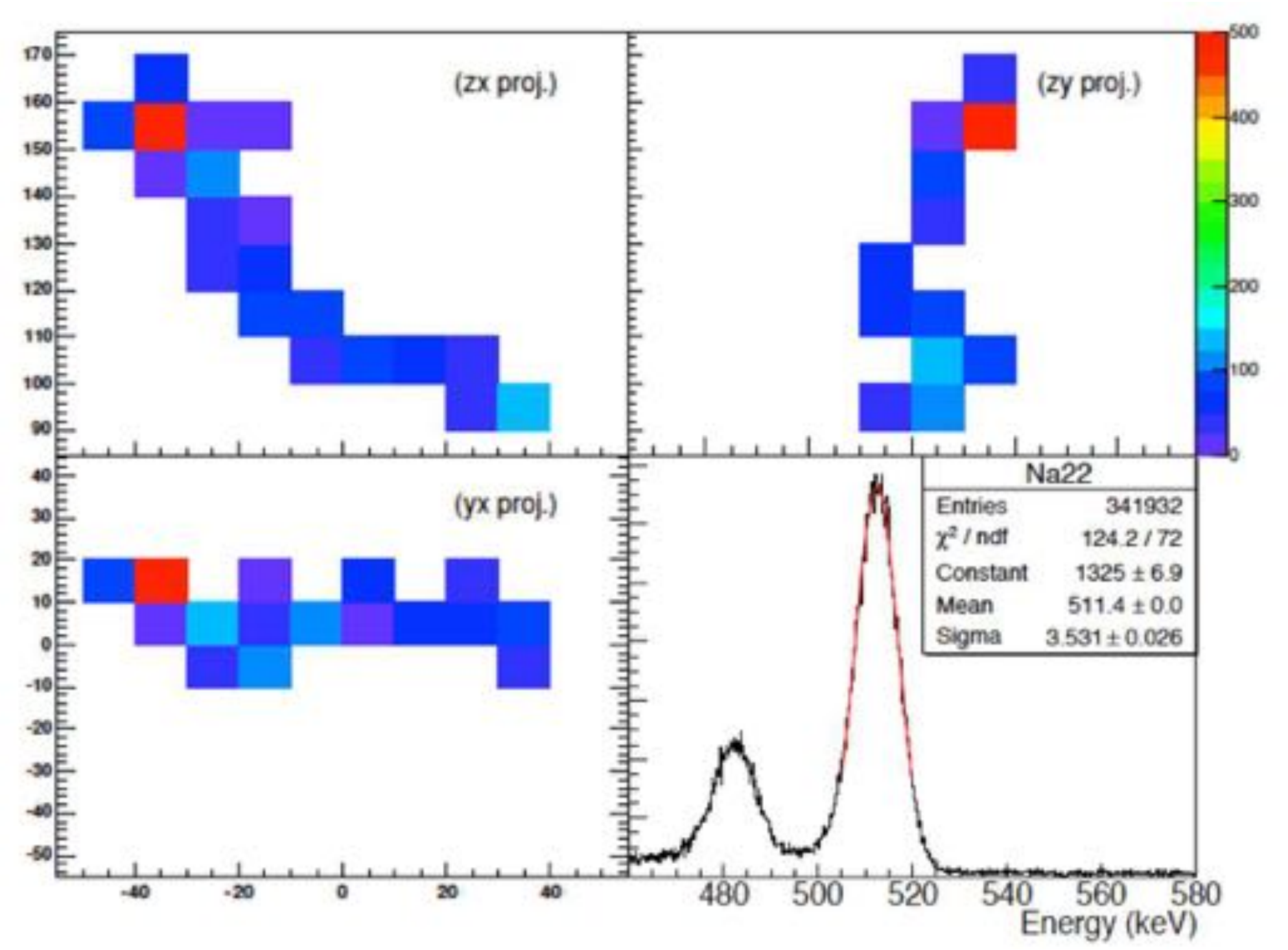}
  \caption{Top: sketch of the NEXT experiment \cite{JJreview}, showing the calorimetry (PM) plane on the left and the position-sensitive (SiPM) plane on the right. Bottom: cartesian projections derived from the 3D-reconstruction (1\,cm$^3$ voxel size) of a 1.275\,MeV photoelectron stemming from a $^{22}$Na source \cite{Paola}. The energy spectrum for the main annihilation peak obtained with the same source is also shown \cite{Lorca}.}
    \label{NEXT-results}
\end{figure}

\subsubsection{NEXT electroluminescence TPC} \label{NEXT_sec}

The most advanced $\beta\beta0\nu$ gaseous TPC, that represents a particularly elegant version of an optical TPC, has been built by the NEXT collaboration. The ground for the experiment was laid in \cite{Davefirst}, and a detailed description can be found in the experiment technical design report \cite{NEXT_TDR}. NEXT, unlike other optical TPCs discussed in this review, resorts to the electroluminescence process in order to approach the intrinsic energy resolution of the gas medium (i.e., the Fano factor $F_e$), so as to assure immunity against the irreducible $\beta\beta2\nu$ backgrounds as well as suppressing the $^{208}$Tl and $^{214}$Bi ones stemming from contamination. At 10\,bar the detector works at an optical gain around $m_{\gamma}=1000$ and this ensures, for a light collection efficiency of $\Omega_{EL}=0.03$ and a quantum efficiency around $30\%$, $\sim10$ collected photoelectrons per primary electron. Such values for $\Omega_{EL}$ are achievable through a PM plane placed behind a cathode mesh, with the help of a highly reflective teflon tube, that is placed field-cage inwards. As argued extensively early on \cite{Davefirst, NEXT_CDR}, and as shown in the 1\,kg-scale technological demonstrators \cite{DEMO,NEXTBNL}, a precise calorimetric measurement (`near-intrinsic') can be performed with these rather natural settings. The achieved energy resolution (FWHM) is just a factor of two away from the one expected for the Fano factor (Fig.\ref{NEXT-results}, bottom-right), i.e., 5.69\% at 30\,keV and 1.62\% at 511\,keV. A $1/\sqrt{\varepsilon}$-extrapolation yields 0.62-0.73\% at the $Q_{\beta\beta}$ of $^{136}$Xe \cite{Lorca}, ($Q_{\beta\beta}=2.45$\,MeV). At the same time, the $T_0$ of the event is naturally obtained in this configuration from the primary scintillation signal, allowing fiducialization in $z$.

\begin{figure}[htb]
  \centering
  \includegraphics[width=\linewidth]{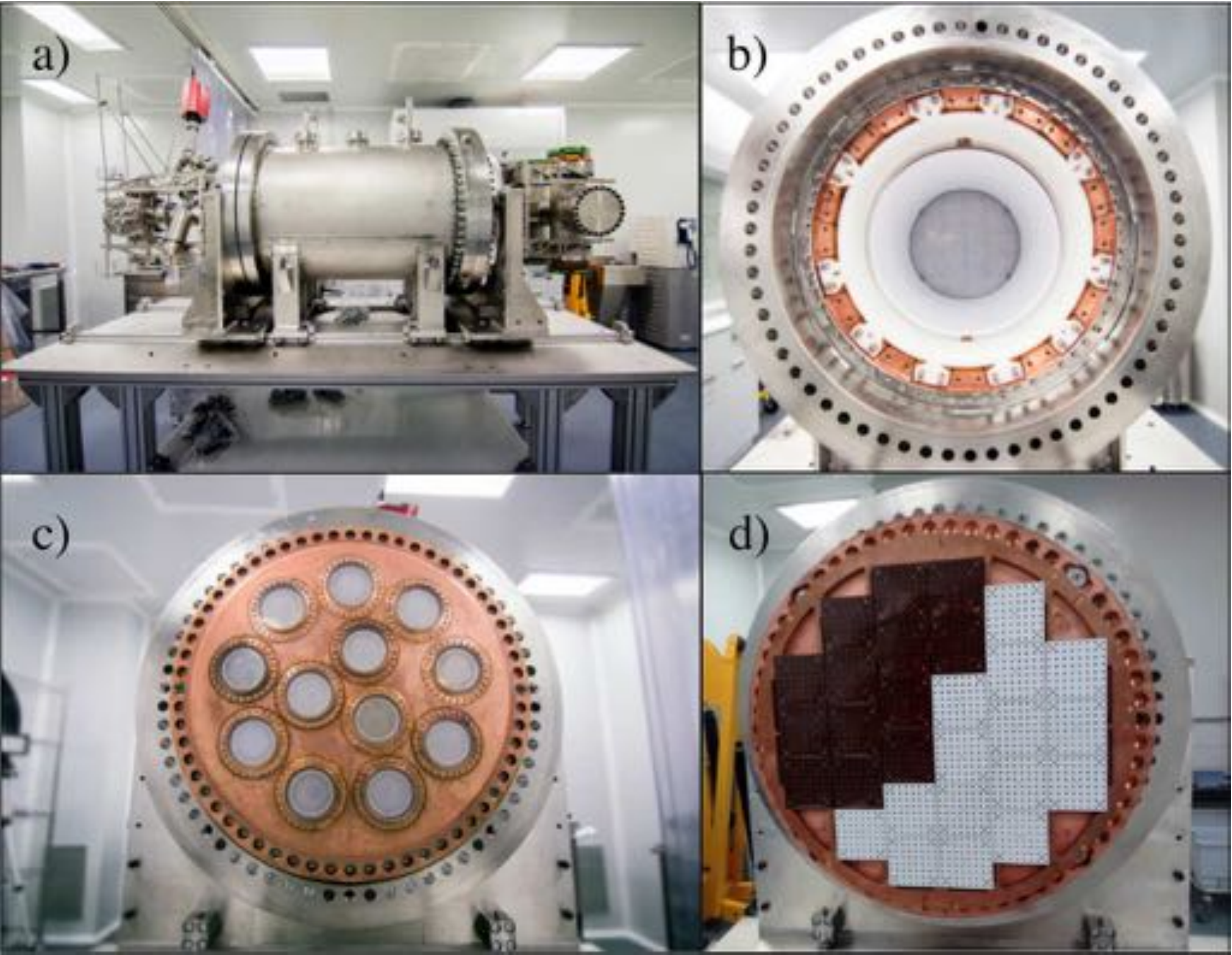}
  \caption{The NEXT-NEW detector, currently taking data at LSC: a) vessel; b) open view of the TPC showing (from outwards to inwards) the vessel, inner copper shielding, polyethylene/copper based field cage and teflon reflector tube (TPB-coated); c) PM plane; d) SiPM plane with (right) and without (left) teflon reflectors (TPB-coated as well).}
    \label{fig:NEXT-NEW}
\end{figure}

In order to achieve a faithful representation of the ionization footprint, a plane of SiPMs at 10\,mm-pitch is placed immediately behind the EL-region (Fig. \ref{NEXT-results}-top). Conventional (i.e., not VUV-grade) SiPMs coated with a solid wavelength-shifter (tetraphenyl butadiene, TPB) are used. In this configuration, the NEXT technological demonstrator has shown the availability of the golden `1 vs 2-blob signature' necessary for suppressing $\gamma$-backgrounds from $^{208}$Tl and $^{214}$Bi sources \cite{Paola}. The analysis was based on the study of the monochromatic double-escape peak of $^{208}$Tl at around 1.6\,MeV (mimicking a $\beta\beta0\nu$ event with about half the $Q_{\beta\beta}$ energy) and the 1.275\,MeV photoelectron from $^{22}$Na (mimicking background events in the same energy region, Fig. \ref{NEXT-results}). The efficiency and background suppression obtained for these reference cases could be reproduced with a full simulation of the detector, allowing to safely extrapolate competitive numbers for the experiment's final conditions \cite{Justo}.

Phase-I of the experiment (dubbed NEXT-NEW), is currently taking data at Laboratorio Subterr\'aneo de Canfranc (LSC). Besides measuring the normal $\beta\beta2\nu$ decay mode, its aim is to consolidate the technology's construction procedures in view of future experiment stages (100\,kg and 1\,ton), and to perform a detailed evaluation of the background sources, energy resolution and topological rejection factors. With its present imaging capabilities (benchmarked in the $\sim 1$\,kg technological demonstrators) NEXT is expected to achieve about 75\% $\beta\beta0\nu$ efficiency for 10\% efficiency to $^{208}$Tl and $^{214}$Bi photoelectrons in its fiducial region around the $Q_{\beta\beta}$ value (i.e., a factor 10 background suppression) according to \cite{JoshNN,Paola}. With the use of low-diffusion mixtures \cite{CO2Henriques}, this figure could be potentially enhanced to 85\% signal efficiency at a factor 20 background suppression, or about a factor 30 for the same signal efficiency \cite{JoshNN}.

\subsubsection{PandaX-III and NEXT-MM}

The first gaseous TPC to set a competitive lower limit to the $\beta\beta0\nu$ lifetime in xenon was the Gotthard TPC: $T^{0\nu}_{1/2} > 4.4 \times 10^{23}$\,yr \cite{Gotthard}. The collaboration resorted to a low diffusion mixture (Xe-CH$_4$, 96.1/3.9) and reconstructed the $xz$ and $yz$ event projections with the help of a MWPC readout. The energy resolution was limited to around 5\%(FWHM) at the $Q_{\beta\beta}$ of $^{136}$Xe and this, together with the radioactive contamination coming from the electrodes and the absence of $T_0$, were at the time very strong handicaps against the rapid rise of germanium detectors, that went on to dominate the field, till recently.

\begin{figure}[htb]
  \centering
  \includegraphics[width=\linewidth]{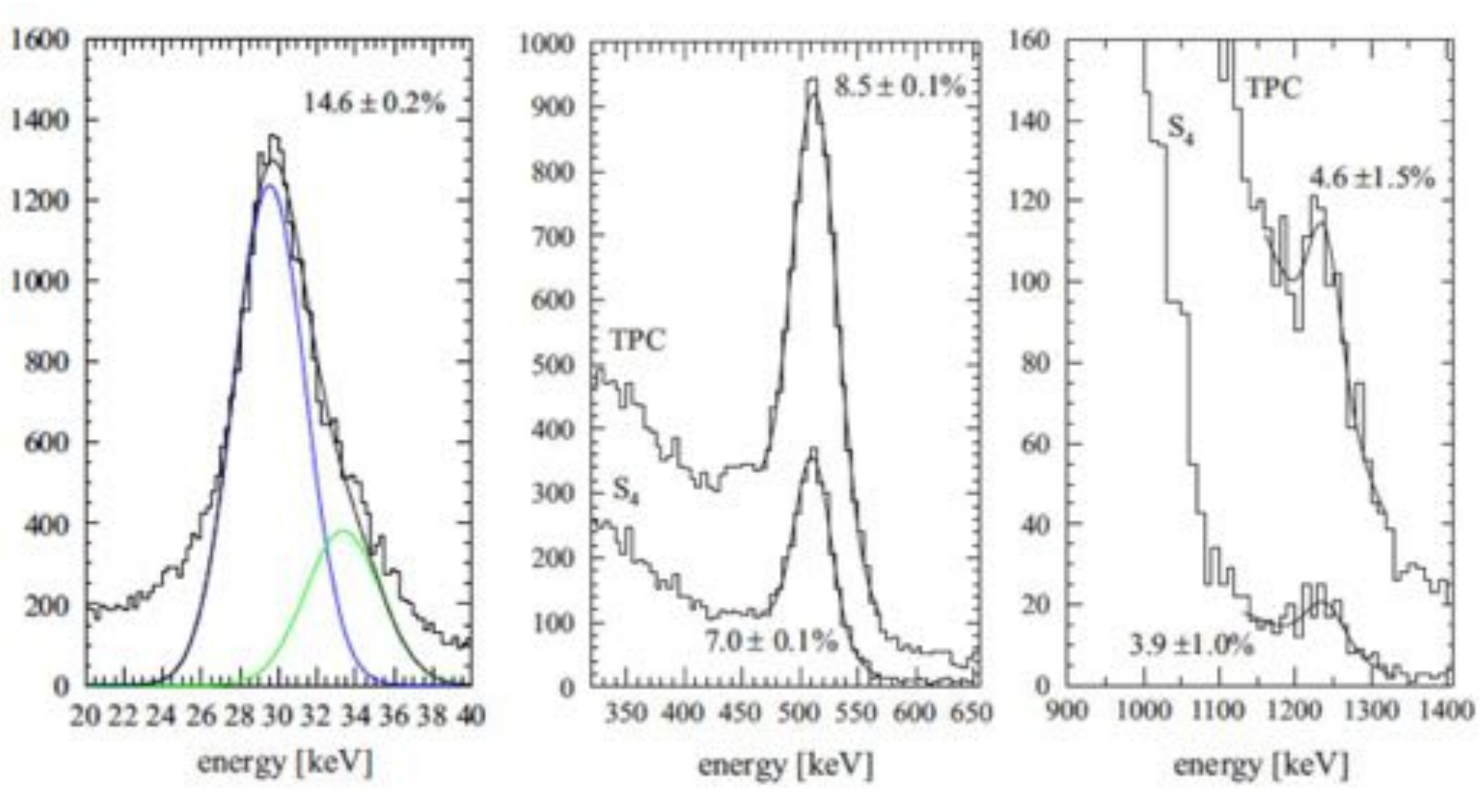}
  \caption{Energy resolution (FWHM) obtained for various photoelectron energies with the (1\,kg-fiducial mass) NEXT demonstrator for the Micromegas technology \cite{NEXTMM}. A direct $1/\sqrt{\varepsilon}$ extrapolation from the 0.511\,MeV and 1.275\,MeV energies yields an energy resolution around 3-4\% (FWHM) at the $Q_{\beta\beta}$ of $^{136}$Xe. An extrapolation from the 30\,keV results would yield 1.6\%.}
    \label{resMM}
\end{figure}

The NEXT collaboration developed a technological demonstrator to re-evaluate the charge readout concept, this time using the recently introduced \mbox{microbulk} Micromegas technology, in a 1\,kg-fiducial detector (dubbed NEXT-MM), \cite{Hector1, Diego1}. The proposal was to use a Xe/TMA Penning-fluorescent mixture ($\sim\!99/1$), that allows a reduced operating voltage in the multiplication region and some residual (potentially usable) scintillation in the TMA band \cite{TMAYasu}. The addition of TMA resulted in a charge gain of $m=200$ at 10\,bar, greatly improving on previous results obtained in pure xenon \cite{Balan}.\footnote{For small Micromegas ($\simeq 10$\,cm$^2$) manufactured with the microbulk technique, the maximum gains obtained for 5-20\,keV x-rays at 10\,bar were in the range of $m=400$-$500$ in Xe/TMA mixtures \cite{DianaTMA}, to be compared with $m\simeq100$ in pure xenon \cite{Balan}. For the large Micromegas detectors used in NEXT-MM the maximum operating gain decreased by around a factor of two.} This provided sufficient signal to noise ratio for a calorimetric measurement of $30$\,keV x-rays (Fig. \ref{resMM}-left): a $\mathcal{S}/\mathcal{N}\simeq 8$ was achieved on the ($8$\,mm$ \times 8$\,mm) pads when coupled to the FEE electronics, based on the AFTER chip. Energy resolutions obtained for 0.511, 1.275\,MeV $\gamma$-rays over the full TPC extrapolate to around 3\%(FWHM) at $Q_{\beta\beta}$ (Fig. \ref{resMM}).\footnote{An example of an event reconstructed in NEXT-MM has been given earlier, in Fig. \ref{Images}-d.} However, in the absence of a strong $T_0$-signal and due to the limited energy resolution, this option was discarded by the collaboration.

Lately, the PandaX-III collaboration (established at the China JinPing underground Laboratory, CJPL) has adopted this technological solution as the experiment's workhorse \cite{PandaX-last,GalanPandaX}. The idea is to benefit from a relatively simple technical implementation, with Micromegas assembled in tiles and read out through $x$-$y$ strips at 3\,mm pitch. The absence of PMs (a sizeable contribution to the experiment's radioactive content) may turn an advantage if earlier concerns on the imperious need of fiducialization would turn out to be unfounded,\footnote{The progeny from the ubiquitous radon, stemming from either emanation or gas contamination will unavoidably accumulate at the cathode. In particular, electrons from the $\beta$-decay of $^{214}$Bi (with a maximum energy of 3.27\,MeV), will thus become an additional background source (see \cite{Gotthard}, for instance).} and if the electron lifetime can be kept high enough to avoid attachment corrections. On the other hand, the departure of the energy resolution from the $1/\sqrt{\varepsilon}$-scaling, observed in the range [30-1200]\,keV (Fig. \ref{resMM}), was attributed partly to $\mathcal{S}/\mathcal{N}$ issues and partly to a form of inter-pad cross-talk \cite{NEXTMM}. If true, a more judicious selection of operating pressure and readout layout may improve the situation down to the energy resolution levels obtained for 30\,keV x-rays (extrapolating to 1.6\%(FWHM) at Q$_{\beta\beta}$). This is likely the practical limit of the microbulk technology in the experiment conditions, given today's knowledge.\footnote{In order to achieve this, the possibility of calibrating the detector gain on a large area by using $x$-$y$ strips (down to $3\%$ at $Q_{\beta\beta}$) will need to be demonstrated, though.}

\begin{figure}[htb]
  \centering
  \includegraphics[width=\linewidth]{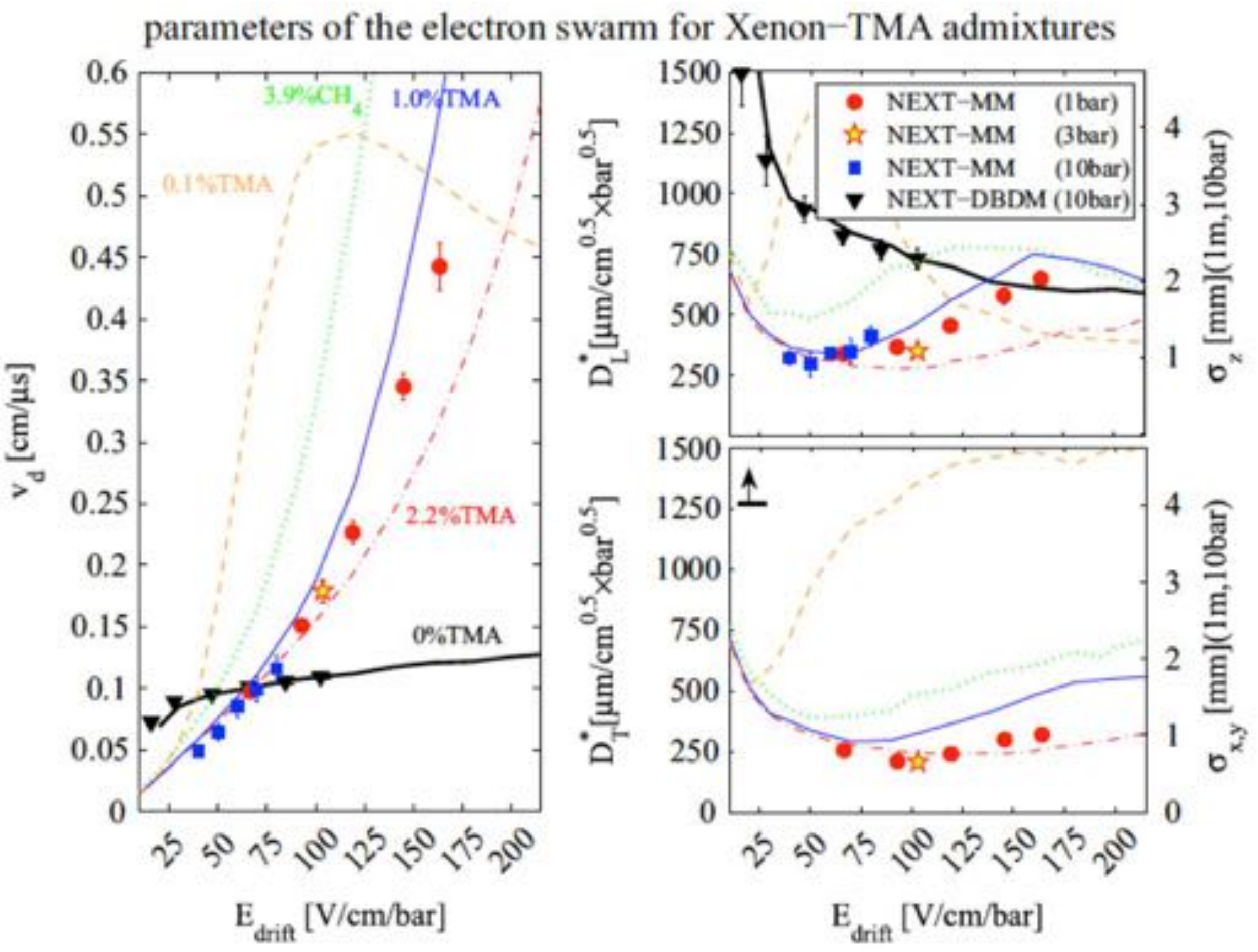}
  \caption{Drift velocity (left) and diffusion coefficients (right) obtained under xenon and xenon/TMA mixtures after \cite{NEXTMM}. Note the slightly different definition of the diffusion coefficients compared to the one used in text ($D_{L,T}^*|_{fig} \equiv D_{L,T}^*|_{text}\cdot\sqrt{P}$). For the sake of clarity, the corresponding diffusion expected for 1\,m drift at 10\,bar is also given.}
    \label{DiffCoeff}
\end{figure}

A particularly important asset are the low diffusion levels, slightly below $D_T^*=1$\,mm$/\sqrt{\tn{m}}$, that have been demonstrated in the proposed detector configuration (Fig. \ref{DiffCoeff}). Diffusion suppression relative to operation in pure xenon is known to contribute significantly to the background rejection factor, by an additional factor of two to three \cite{JoshNN, IguazLowD}. Given the performance measured in NEXT-MM and assuming background levels around $b=10^{-4}$\,c/keV/kg/y in a 200\,kg-xenon chamber, a sensitivity to half-lives around $10^{26}$\,yr was estimated in \cite{PandaX-last}, after three years of operation. Under the simple $\sqrt{Mt/b}$  scaling law assumed by the authors, the old 3.3\,kg Gotthard TPC would have to run for $4\times10^4$ years in order to set a comparable limit. Clearly, suppressing backgrounds down to the required levels in the absence of complete volume fiducialization represents the experiment's major challenge.

\subsection{WIMP dark matter}

Conventionally, TPCs based on gaseous media that are used for direct dark matter detection operate well below atmospheric pressure ($\sim20$-$130$\,mbar).
Contrary to condensed phases, these conditions allow to image the trajectory of nuclei recoiling upon elastic scattering
with gravitationally trapped WIMPs,\footnote{WIMP: weakly interacting massive particle. A generic category of particles commonly situated in the 100\,GeV-1\,TeV/c$^2$ mass, that is the range motivated by theories based on super-symmetry.} a scenario that is strongly motivated theoretically \cite{Feng}.
In particular, information about the direction of the nucleus (and even the sense of its momentum vector, in some conditions) can be obtained down to some 10's of keV$_r$,\footnote{Provided the energy deposited by a recoiling nucleus results in a reduced ionization by virtue of the ionization quenching factor $\mathcal{Q}$ (and maybe some charge recombination, too), sub-indexes $_r$ and $_{ee}$ are used to specify the primary particle (nucleus or electron, respectively).} potentially allowing to extract the apparent WIMP `wind' direction due to Earth's motion. If a recoiling nucleus is measured by these means, the reconstructed momentum vector of the WIMP should be compatible with the direction of motion of the Earth relative to the Galactic rest frame. If it is not, a WIMP-origin can be discarded and a background-origin established, thereby enhancing the detector sensitivity.

An extensive review on this topic has been written recently in \cite{DDMpeople} and the reader is referred to it for additional details on these and other `directional detection' techniques. At the same time, some recent proposals involving operation at near-atmospheric or even high pressure can be found later in text.

\begin{figure}[h!!!]
\centering
\includegraphics*[width=\linewidth]{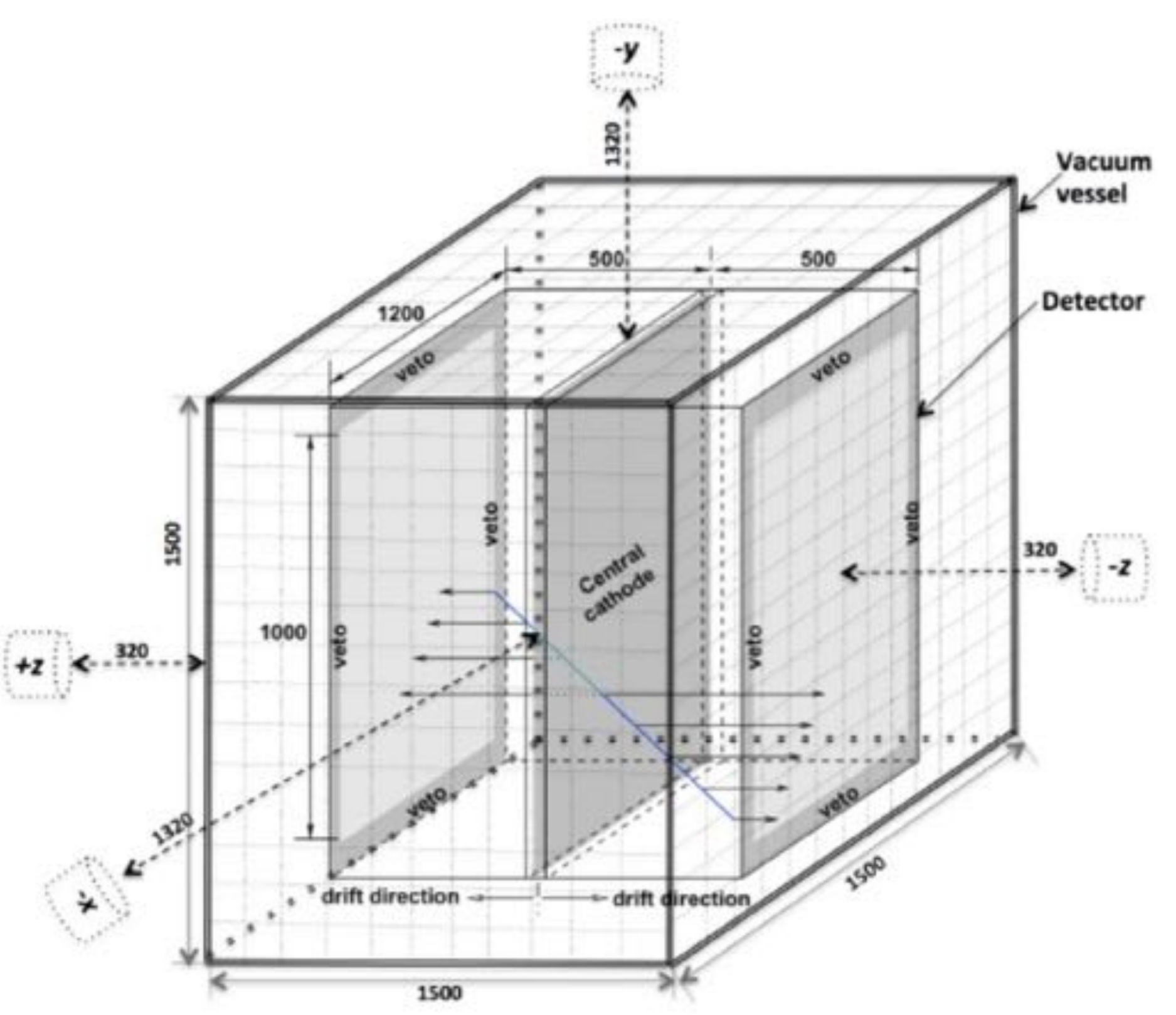}
\caption{The DRIFT-IIb chamber. The dashed cylinders indicate different positions of a $^{252}$Cf neutron source used during a calibration run.}
\label{DRIFTchamber}
\end{figure}

\subsubsection{DRIFT}

The DRIFT experiment, located at the STFC Boulby Underground Science Facility, is a low pressure MWPC-based TPC
with an active volume around 1\,m$^3$, that makes use of the negative ion technique. The wire plane at the MWPC's entrance (or `gate')
provides a $y$-measurement and the MWPC's central anode plane, set in crisscross at 90\,deg, provides an $x$-measurement.
The collaboration has been running m$^3$-scale detectors at Boulby since 2001.
In its latest incarnation, DRIFT uses a multi-compound CS$_2$/CF$_4$/O$_2$ (59/39/2) gas mixture in the range 50-70\,mbar
in a back-to-back TPC with a central cathode (Fig. \ref{DRIFTchamber}). Here CF$_4$ is the main target gas
and CS$_2$ an electronegative species aimed at capturing the ionization electrons and thus providing diffusion values close
to the thermal limit. A remarkable level down to $D^*_{L,T}\simeq1.5$\,mm$/\sqrt{\tn{m}}$ at a reduced field $E_d^*=4$\,kV/cm/bar was demonstrated in \cite{Snowden}. The addition of O$_2$ allows an estimate of the event's $T_0$ due to the presence of additional negatively-charged species with different drift velocities \cite{SnowdenII}, enabling fiducialization in the $z$-coordinate.

\begin{figure}[h!!!]
\centering
\includegraphics*[width=\linewidth]{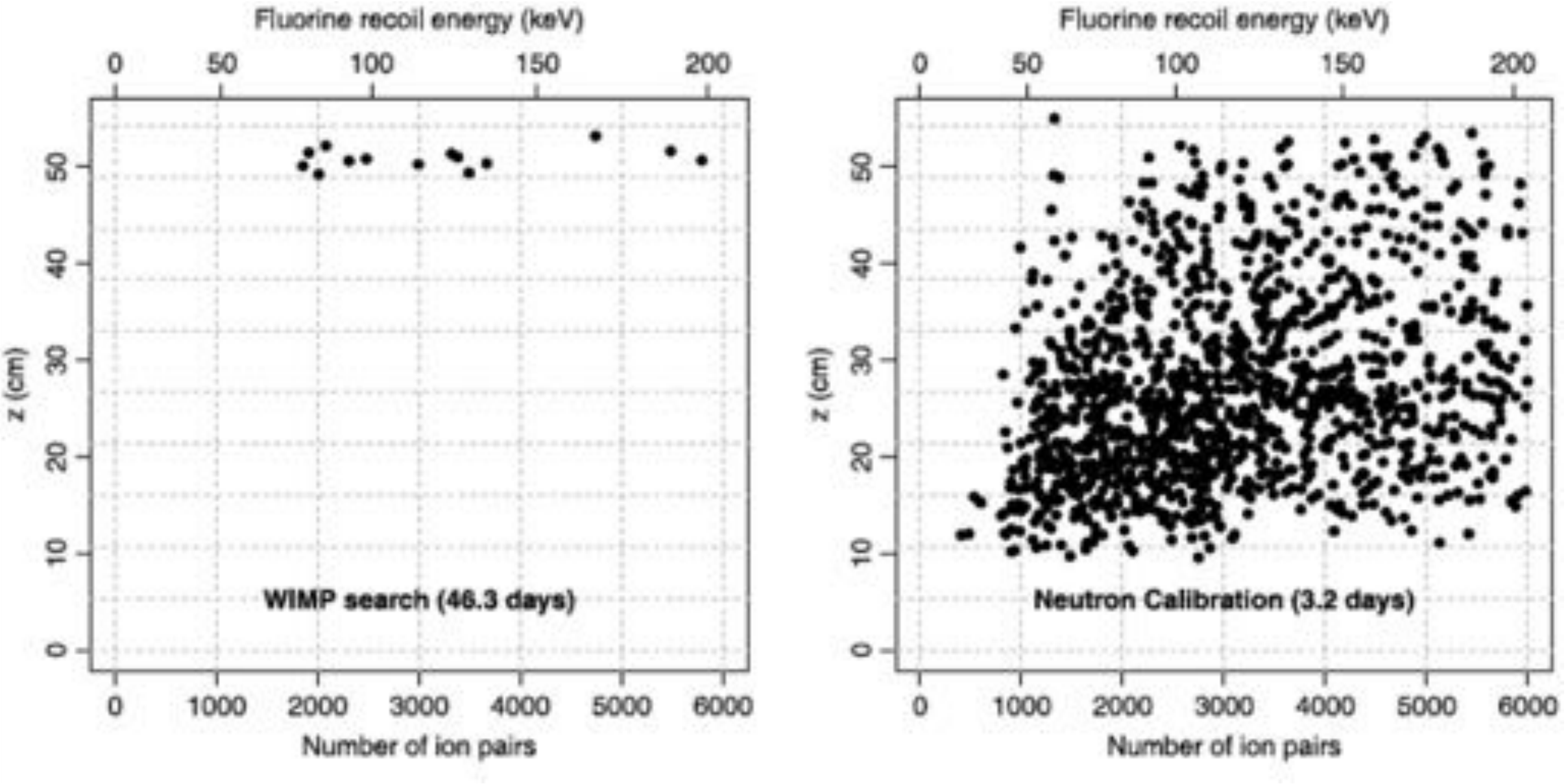}
\caption{Left: DRIFT results from a nuclear recoil search performed during 46.3\,days. Plotted is the number of events found at each $z$-position as a function of their ionization.
Right: same plot from a calibration run with neutron-induced nuclear recoils. After fiducializing off the cathode region at around $z=48.4$\,cm, no WIMP candidate survived the selection cuts. (After \cite{DRIFTlast})}
\label{DRIFTbackgroundFree}
\end{figure}

The drift velocity for the majority charge-carriers is $v_{d} = 5$\,cm/ms that, at a nominal sampling rate of 1\,MHz (1\,$\mu$s time bin), leads to 50\,$\mu$m voxel size in the $z$-dimension. Despite the relatively coarse wire separation in $x$, $y$ (2\,mm) the collaboration managed to suppress
all electrons and neutron recoils from their sample, through material selection, topological rejection and fiducialization. This has recently allowed
a zero-background measurement after 46.3 live-days of operation (Fig.\ref{DRIFTbackgroundFree}).

\begin{figure}[h!!!]
\centering
\includegraphics*[width=7cm]{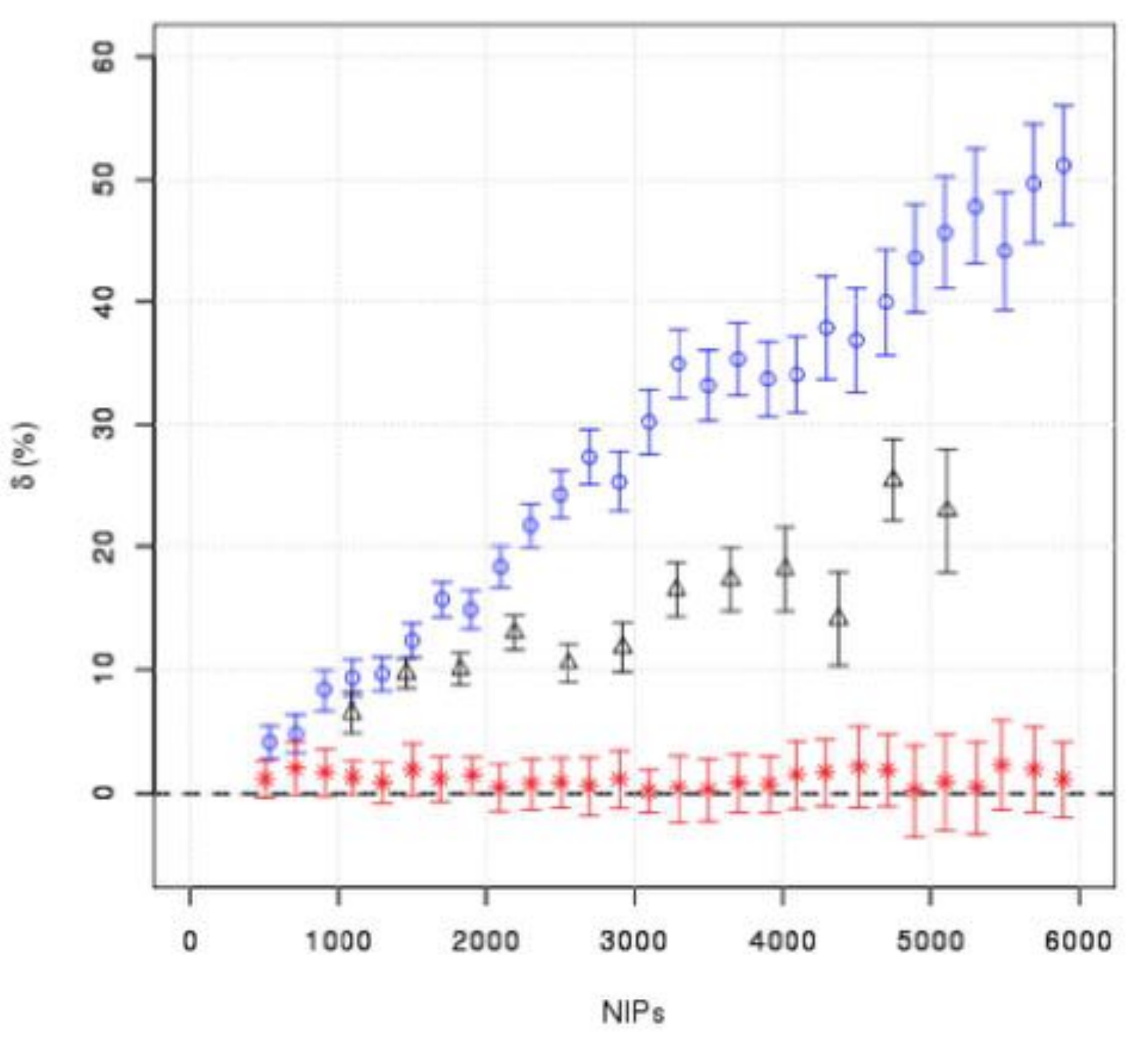}
\caption{Head-tail (i.e., directional) sensitivity obtained by the DRIFT collaboration through the asymmetry parameter $\delta_{h-t}$ ($\delta$ in figure). Blue circles correspond to measurements along the $\pm z$ axis, where the reconstruction precision is higher. Red asterisks correspond to measurements in one of the orthogonal directions (where the reconstruction is non-optimal). Black triangles correspond to a previous run. The $x$-axis indicates the estimated number of ion pairs (NIP), with 750 corresponding to about 40\,keV$_r$.}
\label{DRIFTHeadTail}
\end{figure}

Additionally, and thanks to the very fine $z$-voxelization, the capability to perform some head-tail identification above 40\,keV$_r$ could
be demonstrated, by resorting to interactions from a $^{252}$Cf neutron source at various positions \cite{headtail}. The experimental procedure followed
these steps: i) placing the source such that the most likely direction of the nuclear recoil is along the axis $\pm z$, $\pm y$, $\pm x$, ii) defining a simple head-tail
discriminant as $\alpha_{h-t}=\eta_1/\eta_2$ where $\eta_1$ and $\eta_2$ are the fraction of charge in each of the two identical segments in which the
track projection (in $x$,$y$ or $z$) is divided, and iii) studying the relative difference between the means of the $\alpha_{h-t}$ distributions for recoils coming from the $+$ axis (`R') and $-$ axis (`L') as:
\beq
\delta_{h-t} = 2 \frac{|\bar{\alpha}_R - \bar{\alpha}_L|}{\bar{\alpha}_R + \bar{\alpha}_L}
\eeq
Clearly, this discriminant will be 0 in the absence of a directional signature, something that is observed for the more coarsely estimated coordinates ($x$, $y$), and that is shown by asterisks in Fig. \ref{DRIFTHeadTail}. Along the $z$ dimension, however, a clear sensitivity exists as indicated by the blue circles, and down to about 40\,keV$_r$ (corresponding to 750 NIP -number of ion pairs - in the figure). Evaluating the actual power of this signature for the detection of dark matter is under investigation.

\begin{figure}[h!!!]
\centering
\includegraphics*[width=7cm]{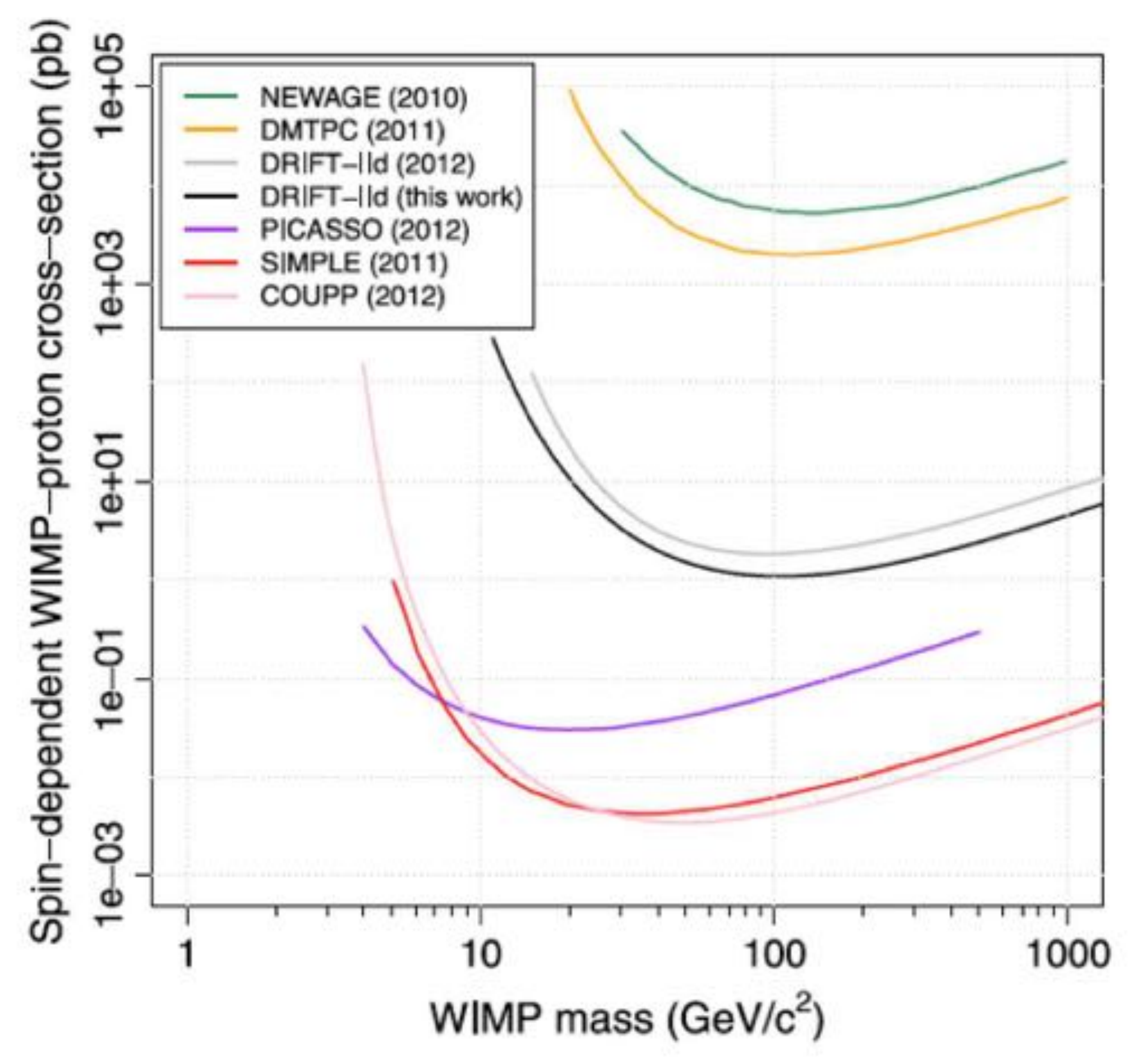}
\caption{Upper limits on the spin-dependent WIMP cross section (obtained circa 2015), together with some early limits obtained with other gaseous TPCs discussed in this work (DRIFT, DMTPC, NEWAGE), \cite{DRIFTlast}. (Added note: the latest NEWAGE limit is about one order of magnitude lower than the one shown here \cite{NEWAGE})}
\label{limits_SD}
\end{figure}

DRIFT currently sets the lowest spin-dependent cross section limit of a dark matter detector with directional sensitivity \cite{DRIFTlast}, about two orders
of magnitude above the absolute lowest limits (Fig. \ref{limits_SD}). Provided the TPC is already background free at the scale of a month, a $\gtrsim 10$ times enlarged detector and/or an array of them will be needed in the future in order to explore uncharted territories during realistic experiment live times.

\subsubsection{DMTPC and related developments}

TPCs imaged by CCD cameras were proposed by Buckland et al. in 1994 \cite{Buckland1}, for the directional detection of dark matter. The original proposal contemplated operation at a reduced electric field around $E_d^*$=500\,V/cm/bar inside a 0.35\,T magnetic field and under a CH$_4$-based mixture. TEA was proposed as wavelength-shifter, its UV scintillation registered at the CCD with the help of an ancillary image-intensifier, in similar fashion to \cite{Charpak}. Buckland and colleagues noticed that the $E\!\times\!{B}$-effect of Nygren (eq. \ref{TransDiff}) could bring a great advantage for the reconstruction of events especially at low pressure, and managed to obtain a transverse diffusion down to $D_T^*=0.56\tn{\,mm}/\sqrt{\tn{m}}$ at 25\,mbar, for the aforementioned operating fields. In the absence of magnetic field, however, the diffusion coefficient was nearly a factor of ten larger. Therefore, when the necessity of a readily scalable concept made the magnet design unviable \cite{MartoffReview} the original idea was abandoned.

\begin{figure}[h!!!]
\centering
\includegraphics*[width=7cm]{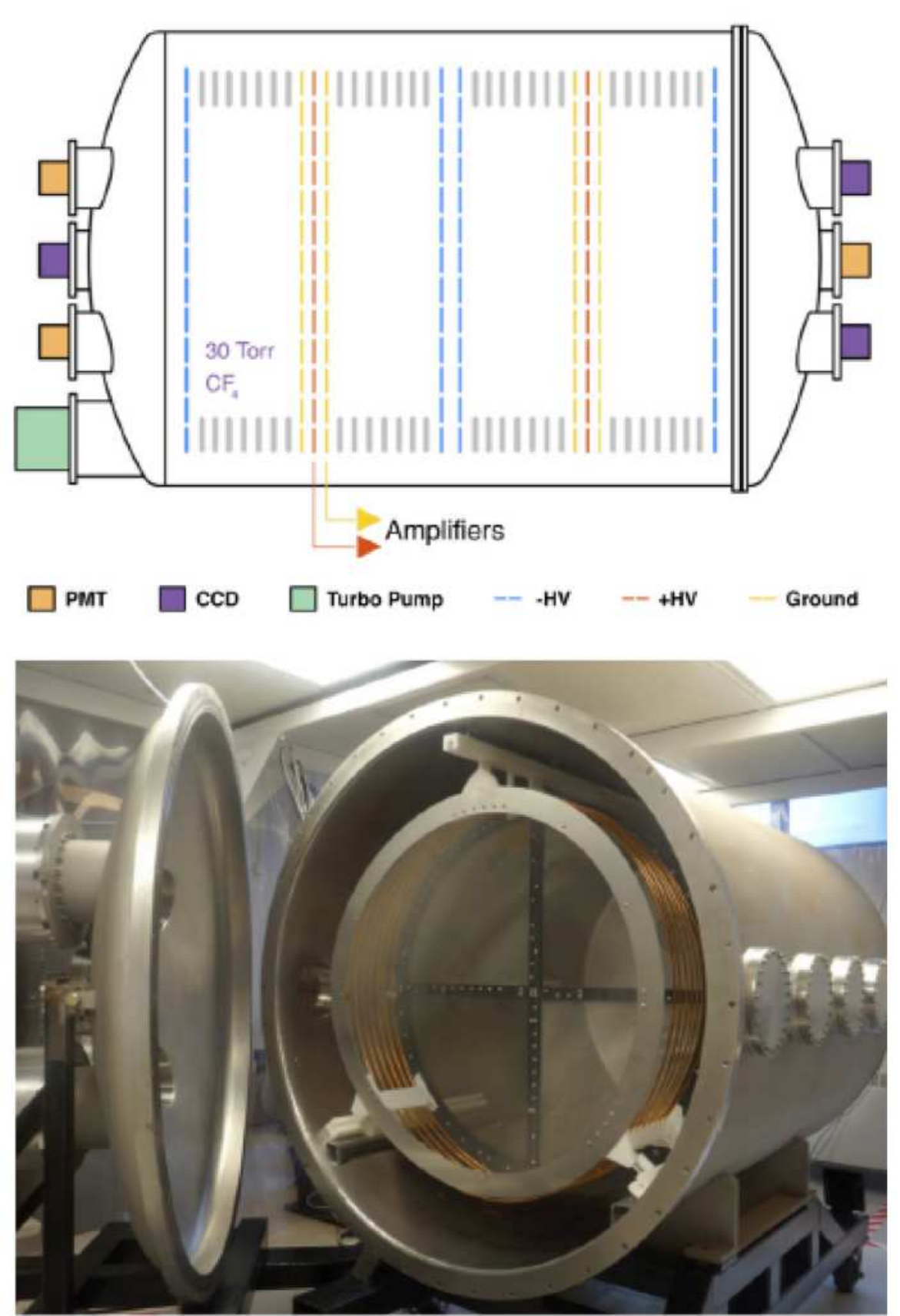}
\caption{Top: sketch of the latest DMTPC detector, highlighting its doubly-mirrored configuration that effectively divides the active volume in four sections.
Bottom: open view, showing the end-cap region and one of the four field cage sections.}
\label{DMTPCfig}
\end{figure}

Some other difficulties of the original Buckland's TPC were connected to the manipulation of TEA and its (non-optimal) UV-emission. A seemingly convenient alternative was introduced soon after in 1995 \cite{Pansky}
and its scintillation demonstrated under practical detector conditions: CF$_4$. The new substance turned out not only to have a much higher vapor pressure at ambient temperature and being more inert (and less toxic) than TEA: one of the CF$_4$ fragments released during the ionization process (CF$_3^*$) scintillated strongly in the visible range \cite{ArCF4Fraga}. These facts, coupled to the rapid rise and increased affordability of low-noise CCD cameras, gave some people a reason to stick to the optical concept, yet without magnetic field.
The doubly mirrored configuration of the contemporary version of Buckland's TPC (dubbed DMTPC) helps at minimizing diffusion down to tolerable $\sim2$\,mm levels, in a 1\,m$^3$-sized active volume (Fig. \ref{DMTPCfig}). In this case, the cathodes are placed in the middle of the volume and at the two end caps, dividing the active region in four sections, thereby reducing diffusion by a factor of two compared
to a seamless TPC of the same dimensions (see Fig. \ref{DMTPCfig}-top).

\begin{figure}[h!!!]
\centering
\includegraphics*[width=7cm]{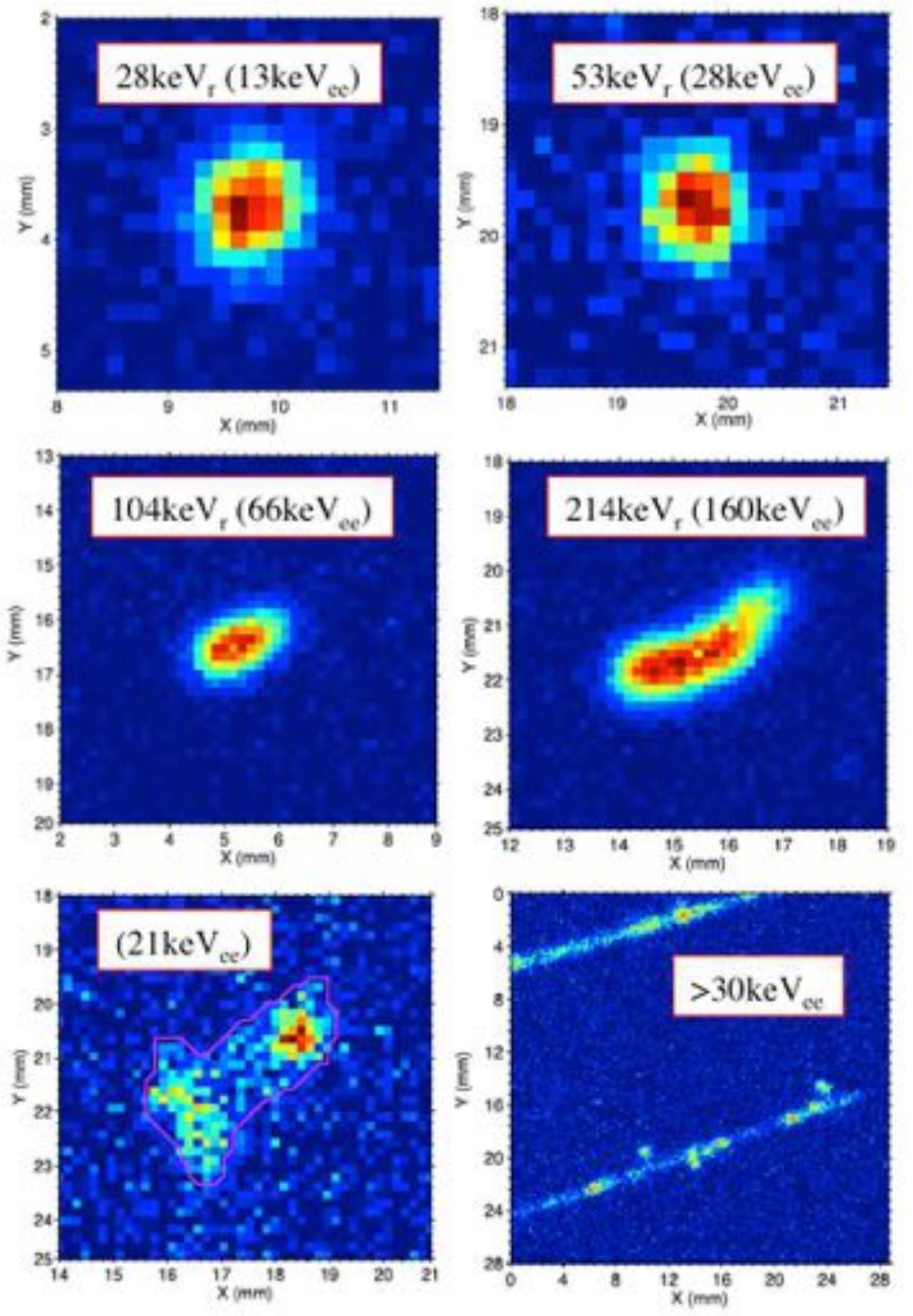}
\caption{Top and middle: recoiling nuclei obtained in CF$_4$ at around 130\,mbar with an optical readout based on a CCD camera. Bottom: electron events. The figure aims at qualitatively illustrating the discrimination (20\,keV$_r$) and directional (40\,keV$_r$) thresholds achieved by the authors. Figures adapted from \cite{Phan}, were a more quantitative discussion can be found.}
\label{Opticalevents}
\end{figure}

DMTPC uses parallel meshes for the generation of light, that is produced abundantly under an avalanche gain of $2\times10^5$ \cite{Leyton}. Promising results have been recently obtained with triple-GEM structures in \cite{Phan}, too, both in terms of discrimination power (ability to recognize electrons from nuclei of the same energy) and directionality (ability to recognize the nuclear recoil's direction), see Fig. \ref{Opticalevents}. In this latest work it was possible to demonstrate energy thresholds of 20\,keV$_r$ and 40\,keV$_r$, respectively, by imaging tracks with an equivalent pixel size at the object plane of $\Delta{x(y)}={165}$ $\mu$m. The CF$_4$ pressure was chosen to be around 130\,mbar.

Pure CF$_4$ provides a scintillation strength in the range of 0.15-0.3\,$\gamma$/e$^-$ \cite{MeVienna, CF4Fraga}, resulting in optical gains around $m_{\gamma}=7\times10^4$ for the cases discussed above. Assuming this performance and the light collection efficiency given by eq. \ref{OmegaCCD}, it can be expected that a 10 Megapixel sensor of size $A\!\!\sim\!5 \tn{\,cm} \times 5 \tn{\,cm}$, with a noise/pixel around $1\,\gamma$, could image a $1\tn{\,m} \times 1\tn{\,m}$ plane with an equivalent $200$\,$\mu$m pixel size at the object plane, and yet keeping a $\mathcal{S}/\mathcal{N}|_\gamma$ of 10. DMTPC currently foresees two asymmetric end-caps, equipped with PMs and two types of 10 Megapixel CCD cameras: a ProLine 9000 and a (UV-sensitive) back-illuminated from Spectral Instruments (1100S). DMPTC is expected to run at SNOLab during 2017. In order to reach the femtobarn landmark in the spin-dependent cross section sensitivity, it is estimated that the optical technique needs at present (imposing) exposures around 500\,m$^3$y \cite{Leyton}.

\subsubsection{NEWAGE and MIMAC}

\begin{figure}[h!!!]
\centering
\includegraphics*[width=\linewidth]{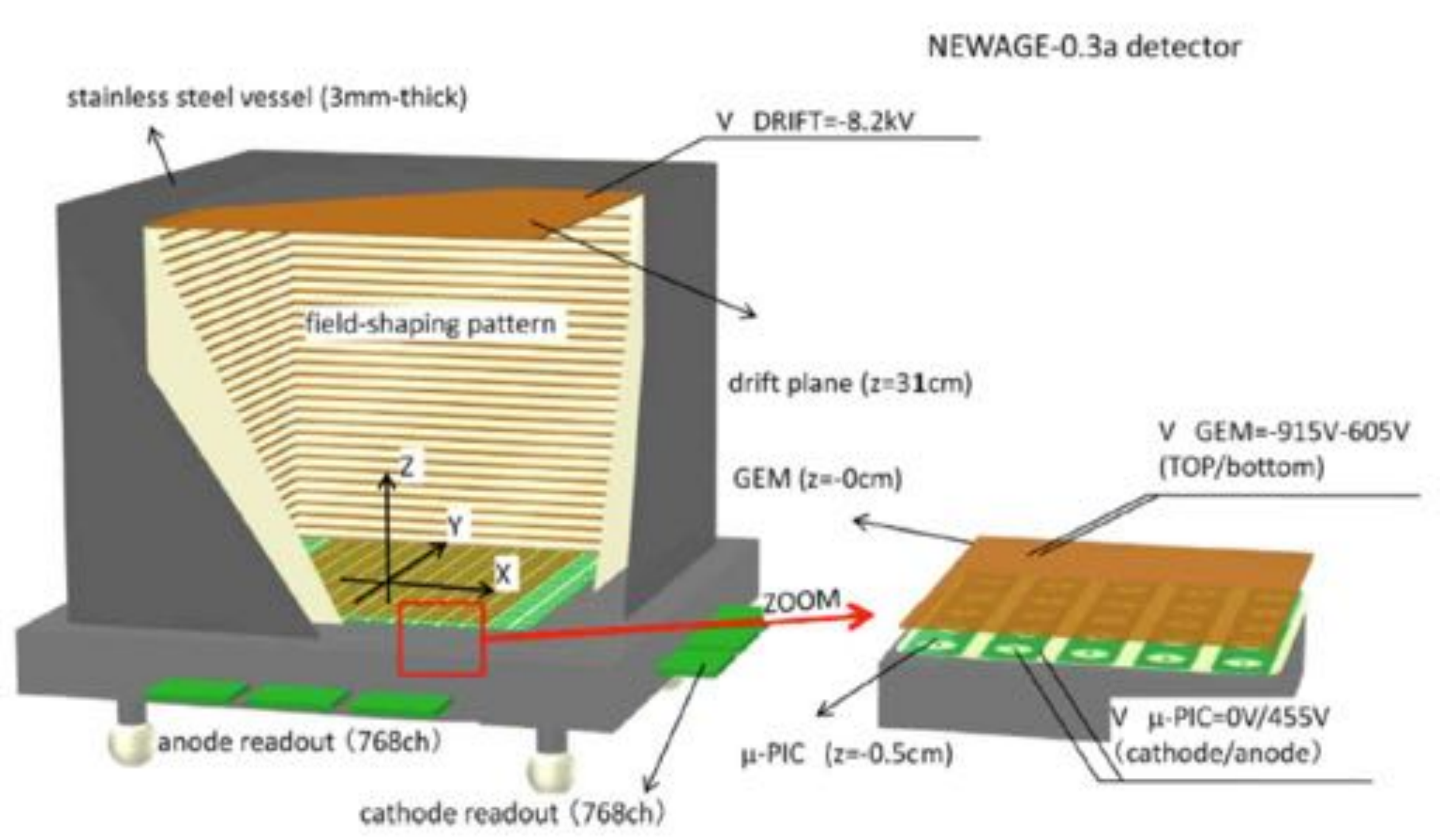}
\caption{The NEWAGE detector, version 0.3a, (left) and a close-up view (right) showing the tandem $\mu$-PIC + GEM equipping its multiplication region. NEWAGE (current version: 0.3b') operates under pure CF$_4$ at 100\,mbar.}
\label{NEWAGEFig}
\end{figure}

Conventional charge-readout TPCs based on MPGDs have been also developed for directional dark matter searches, under the names of NEWAGE \cite{NEWAGE} and MIMAC \cite{MIMAC1}. In this case, the main idea is to exploit the very fine segmentation and good scalability prospects that MPGDs naturally provide. Both systems read the $xz$ and $yz$ event projections through orthogonal pickup strips, at a pitch around $\Delta{x(y)} = 400$\,$\mu$m. NEWAGE (Fig. \ref{NEWAGEFig}) is currently deployed at the Kamioka underground lab, and consists of a $30 \tn{\,cm} \times 30 \tn{\,cm} \times 41 \tn{\,cm}$(drift) active volume, filled with 100\,mbar of CF$_4$. It makes use of the $\mu$-dot/$\mu$-PIC technology for charge amplification \cite{MicroDotSteve}, although in its latest design (version 0.3b') it uses an additional LCP-GEM plane in order to reach sufficient gain.\footnote{LCP: liquid crystal polymer.}

The lowest spin-dependent cross section limit of NEWAGE is modest (Fig. \ref{limits_SD}), but it shows that the readout technology is mature. Compared to operation in pure CF$_4$, MIMAC uses a slightly modified gas mixture consisting of CF$_4$/CHF$_3$/C$_4$H$_{10}$ (70/28/2) in order to fine tune the drift velocity. It is running at the Laboratoire Souterrain de Modane using conventional Micromegas with $128$-$256$\,$\mu$m gap \cite{MIMAC1}.

\subsubsection{TREX-DM}

\begin{figure}[h!!!]
\centering
\includegraphics*[width=\linewidth]{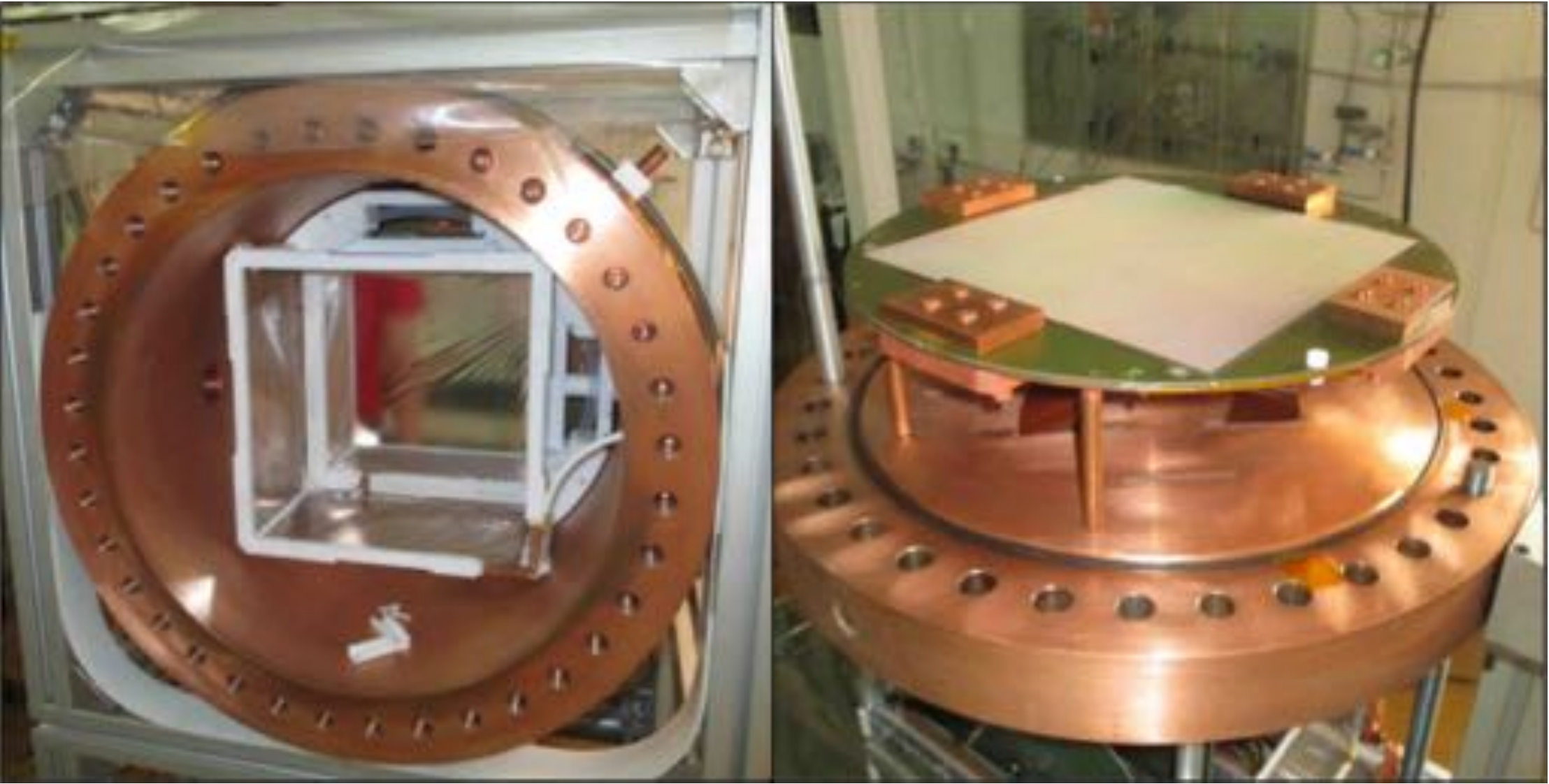}
\includegraphics*[width=\linewidth]{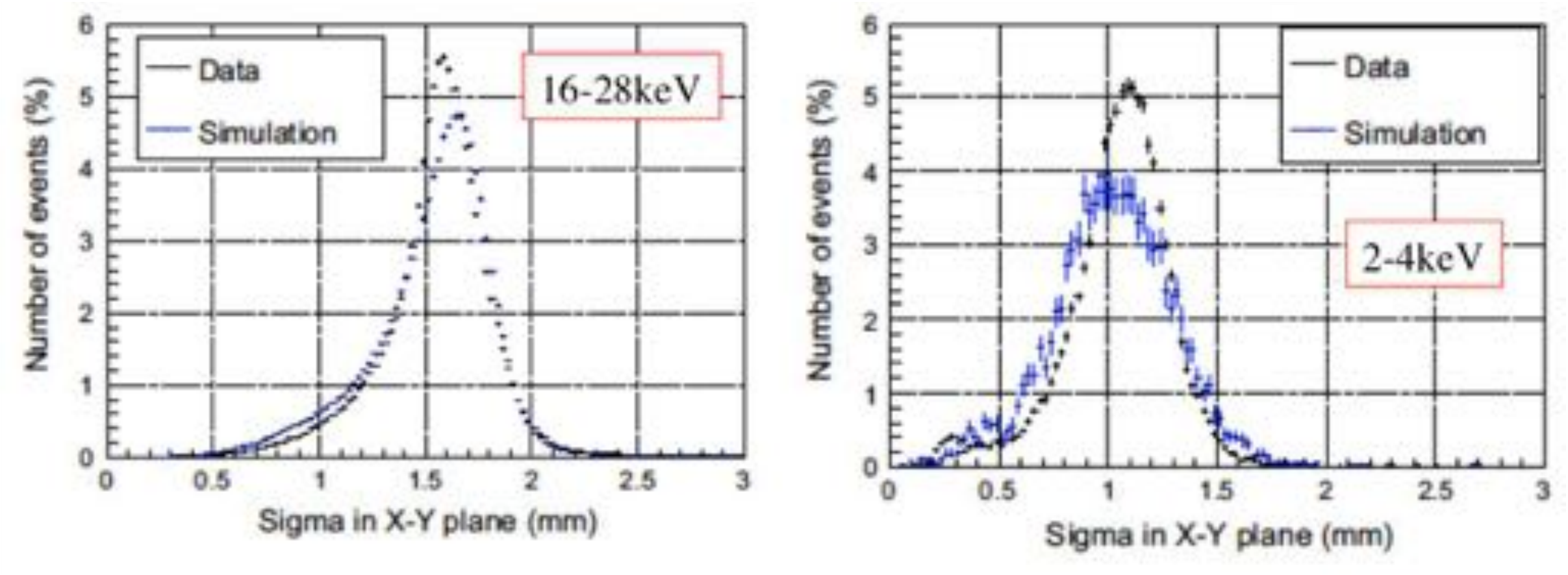}
\caption{Top-left: figure of the all-copper TREX-DM vessel and field cage. Top-right: the end-cap region, showing the Micromegas detector.
Bottom: transverse spread of electrons released after x-ray interactions at different energies, in a 1.2\,bar Ar/i-C$_4$H$_{10}$
mixture (95/5). The figure aims at illustrating the different track topologies, that depend chiefly on the event position within the chamber (the event itself is largely featureless), and are determined by diffusion and threshold effects. For details on the (particularly subtle) impact of cuts based on topological criteria in this situation, the reader is referred to \cite{TREX-DM_last2}.}
\label{TREXDMsetupAndFigs}
\end{figure}

The TREX-DM collaboration aims as well at making some (limited, in this case) use of topological information to eliminate x-ray events, although in light noble gases (Ne, Ar) and under pressurization \cite{TREX-DM_last}. The authors center the analysis on the identification of `point-like events', through which they could suppress keV-energy x-ray events (especially when they come from the electrodes), at near-atmospheric pressure (1-2\,bar). The vessel itself, on the other hand, has been designed to withstand 10\,bar, that is the targeted operating pressure after installation and commissioning (Fig. \ref{TREXDMsetupAndFigs}-top).

It is important to note that the observed topological features (Fig. \ref{TREXDMsetupAndFigs}-bottom) are presently dominated by diffusion and threshold effects, since events at keV-energies are largely featureless at near-atmospheric pressure, for practical purposes \cite{TREX-DM_last2}. Given that, the unavoidable reduction in diffusion (and further reduction of the track size), and the presumable decrease in working gain and electron lifetime with the intended $\times 5$-$8$ pressure increase could change the situation significantly. Based on the figures obtained at 1-2\,bar, under optimistic background assumptions \cite{TREX-DM_last_Err}, and taking a demonstrated 500\,eV energy threshold, the collaboration estimates that they can measure a WIMP with a mass in the 1-10\,GeV/c$^2$ range after one year of data, if its spin-independent cross section is in the attobarn region.

TREX-DM is largely a scaled-up version of the CAST experiment (section \ref{CAST}), with an $x,y$ strip pitch around $\Delta{x(y)} = 600$\,$\mu$m.
Contrary to experiments discussed earlier in this section, it does not anticipate any directional information. The final detector will rely on the microbulk Micromegas technology, that has shown competitive radiopurity levels down to 0.4\,Bq/m$^2$ \cite{RadioMM, RadioIras}. It will be installed at LSC during forthcoming months \cite{Irastorza}.

\subsection{Neutrino oscillations}

A gaseous TPC is a priori not a viable technology for the detection of neutrinos, in view of its low interaction probability as compared
to condensed phases. This is in particular the case for the far detectors (FD) in neutrino oscillation experiments. TPCs can be used as
particle trackers, though, or if placed close enough to the neutrino source, and conveniently pressurized, a TPC may constitute the
main detection volume of an experiment's near detector (ND).

\begin{figure}[h!!!]
\centering
\includegraphics*[width=\linewidth]{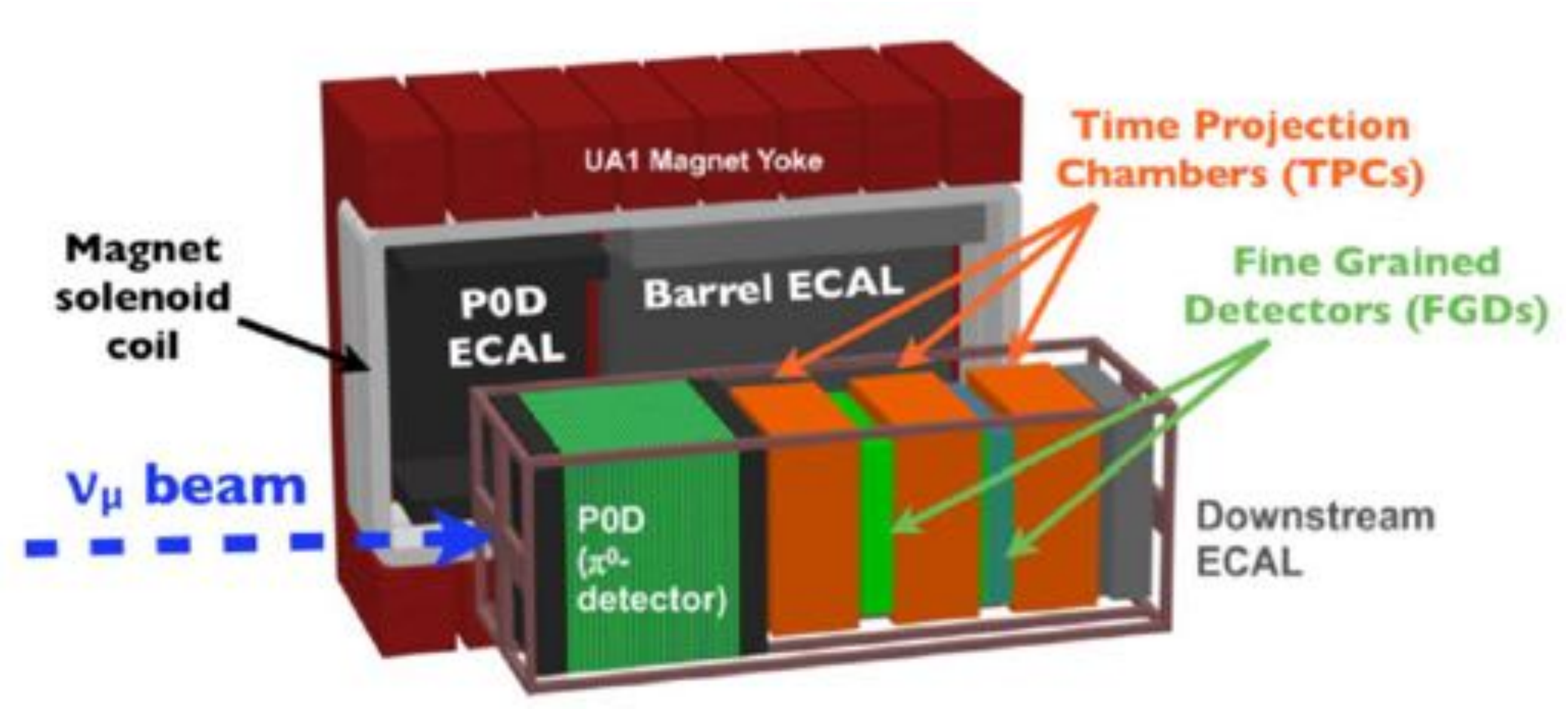}
\caption{View of the ND-spectrometer of the T2K neutrino experiment, showing the magnet, the $\pi_0$ detector, calorimeters, scintillator trackers and the 3-fold segmented TPC (from \cite{T2Kthesis}).}
\label{T2Kall}
\end{figure}

\subsubsection{The TPC of the T2K near detector }

The Tokai to Kamioka (T2K) experiment is a long baseline neutrino oscillation experiment that resorts to neutrinos produced at the J-PARC facility at Tokai and measures their survival probability in transit to the Super-Kamiokande detector at about 295\,km distance. The T2K-ND, placed at about 280\,m from the neutrino beam, is a particle spectrometer with a segmented TPC at its core (Fig. \ref{T2Kall}). As such, it is fundamentally different from most examples described in this review, that are aimed at performing stand-alone seamless particle imaging. The TPC acts mainly as a particle tracker, whose tasks are to identify and classify the event topology, and to perform particle identification (PID) through a combined $d\varepsilon/dx$ and momentum measurement (Fig. \ref{ElossT2K}). Surrounding detectors provide additional tracking, calorimetric and PID capabilities.

\begin{figure}[h!!!]
\centering
\includegraphics*[width=\linewidth]{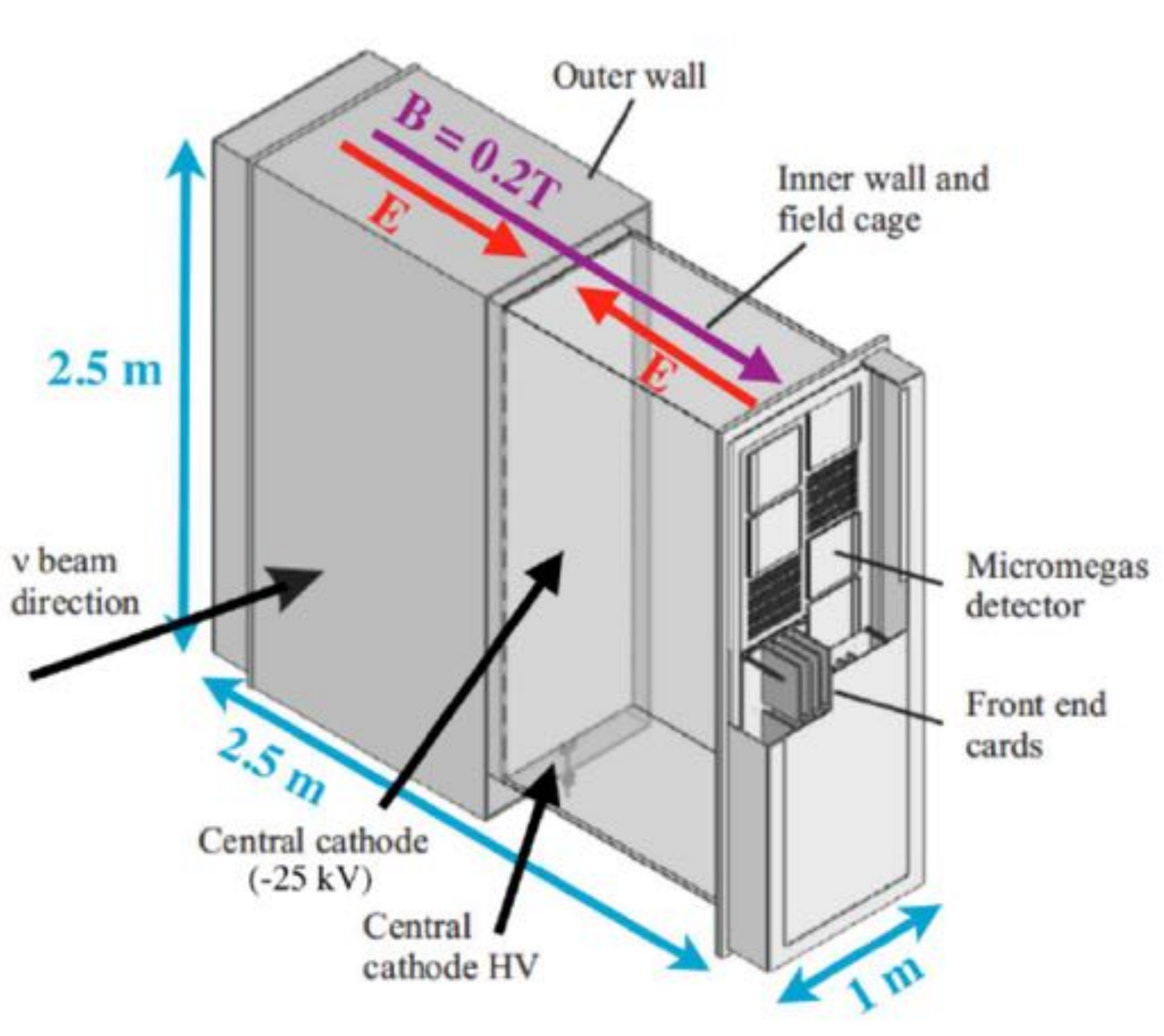}
\caption{One of the three TPC modules constituting the T2K near detector (from \cite{T2Kthesis}).}
\label{T2Kall2}
\end{figure}

T2K TPCs are particularly well known for including the first implementation of a Micromegas readout in a large particle physics experiment. They operate under a ternary gas mixture of Ar/CF$_4$/i-C$_4$H$_{10}$ (95/3/2) at atmospheric pressure, with i-C$_{4}$H$_{10}$ aimed at providing stable gain and CF$_4$ used for adjusting the drift velocity and transverse diffusion. Each TPC has a symmetric configuration with the cathode placed at its center ($1.25$\,m drift distance), a $2.5$\,m$\times 1$\,m readout plane and a pad size around 1\,cm$^2$. Signals induced in the pads are recorded with electronics based on the AFTER chip \cite{AFTER}.

\begin{figure}[h!!!]
\centering
\includegraphics*[width=\linewidth]{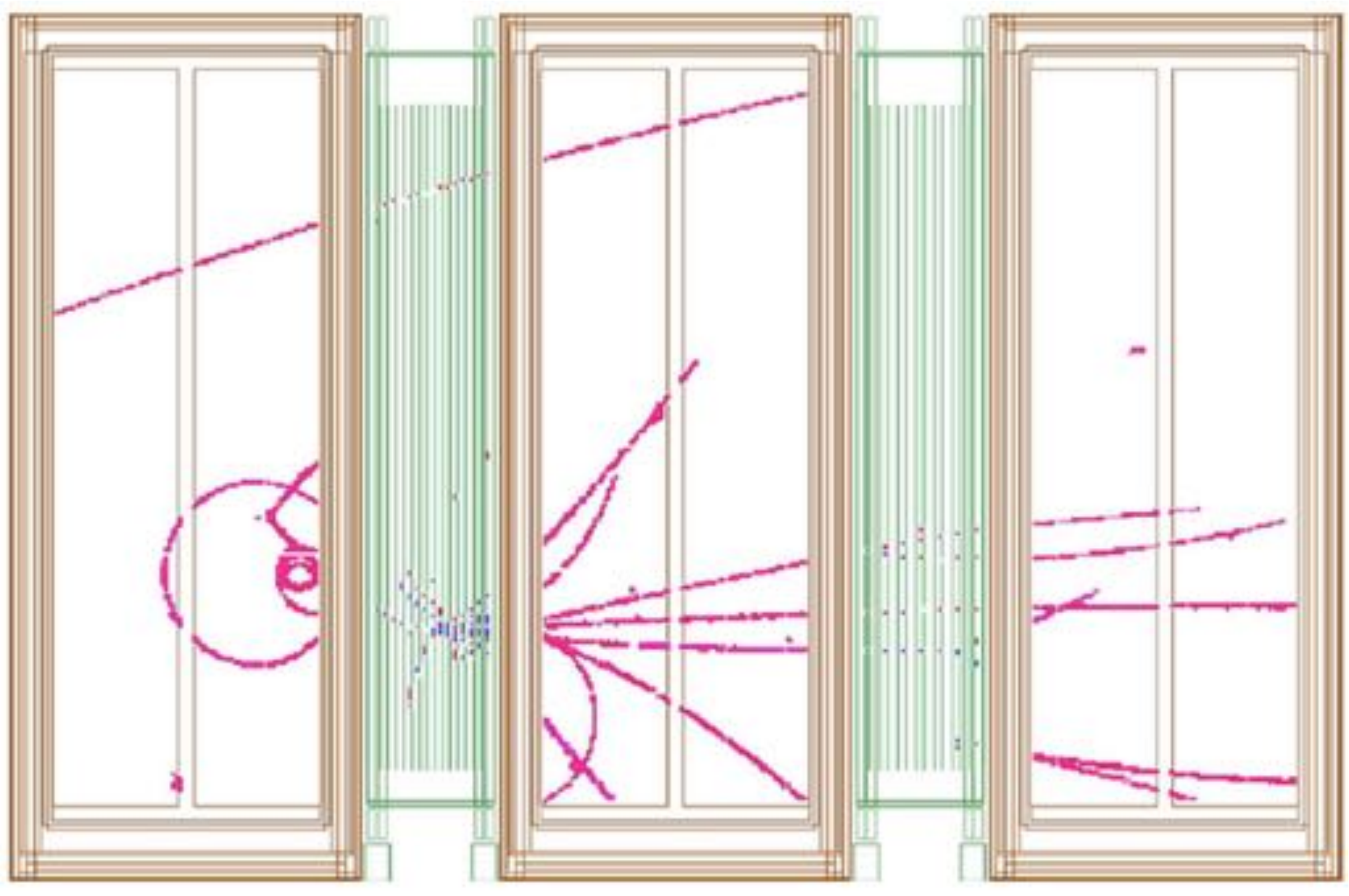}
\caption{An unusual event at the T2K spectrometer, featuring two simultaneous neutrino interactions immediately before the first and in between the first and second TPC modules \cite{T2KNIM}.}
\label{T2Kevt}
\end{figure}

The effect of the nominal magnetic field on the transverse diffusion is relatively minor in T2K operating conditions (and pad size does not allow presently to fully exploit this feature) but a value can be extracted through a special analysis, yielding: $D_T^* = 3.4$\,mm/$\sqrt{\tn{m}}$ ($B=0$\,T) and $D_T^* = 2.9$\,mm/$\sqrt{\tn{m}}$ ($B=0.18$\,T), \cite{T2KNIM}. It is worth noting that extrapolations to $B=5$\,T for future TPCs (e.g., at the ILC) hint at an extraordinary $D_T^* = 0.19$\,mm/$\sqrt{\tn{m}}$ landmark value by resorting to this gas mixture \cite{T2KDixit}. The T2K experiment has been running stably since 2010, providing a large wealth of results (e.g., \cite{T2Klast1}, \cite{T2Klast2} and references therein). An exemplary event, illustrative of the imaging capabilities of the T2K-TPC, is shown in Fig. \ref{T2Kevt}.

\subsubsection{Other proposals}

The DUNE collaboration (for a detailed description see section \ref{secWA105}) is exploring argon-based detectors (among other technologies) for its ND, that will be placed close to a new MW-scale neutrino beam line at Fermilab. This detector must precisely characterize the neutrino beam energy and composition and measure to unprecedented precision the cross sections and particle yields of the various neutrino scattering processes. At present, a pressurized argon-based TPC is under consideration, although the development is still in a conceptual phase \cite{JustoDUNE}.

\subsection{Solar axions}

\subsubsection{CAST}\label{CAST}

\begin{figure}[h!!!]
\centering
\includegraphics*[width=\linewidth]{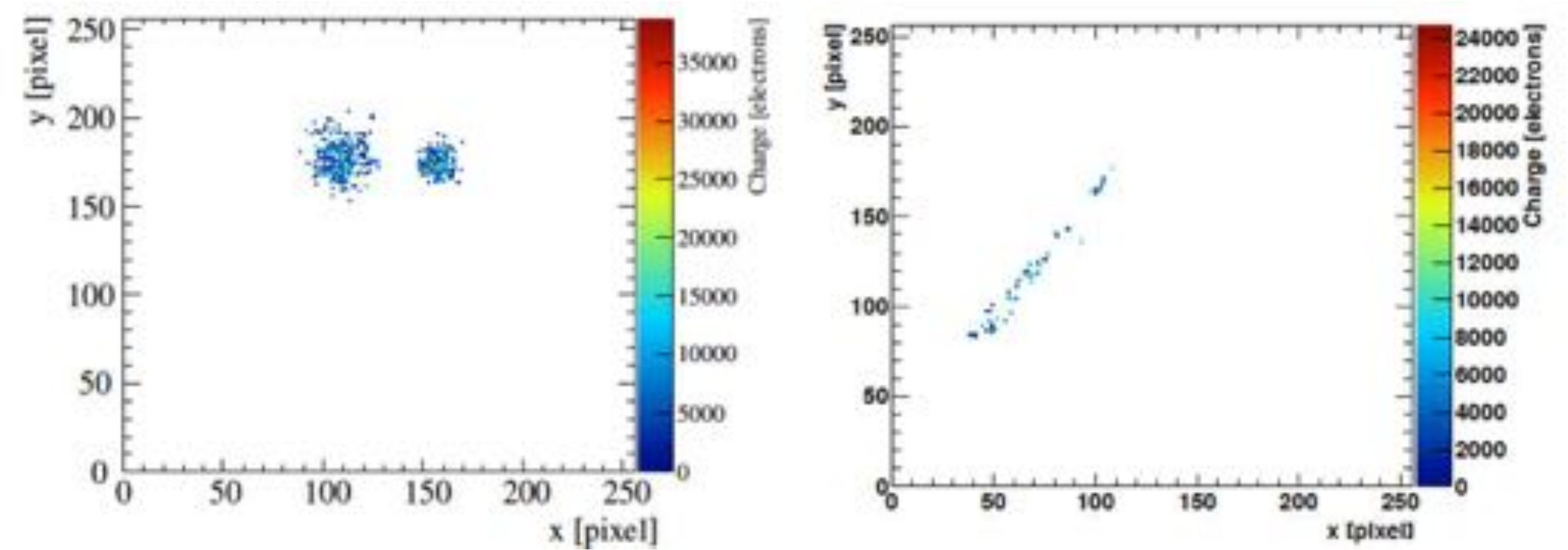}
\caption{Left: multi-site x-ray clusters observed in one of CAST's mini-\mbox{TPCs} when coupled to an InGrid readout (for this particular application a comparable performance can be achieved with microbulk Micromegas). Right: a cosmic ray event. (Figs. from \cite{INGRID_last_CAST})}
\label{CASTimage}
\end{figure}

\begin{figure}[h!!!]
\centering
\includegraphics*[width=\linewidth]{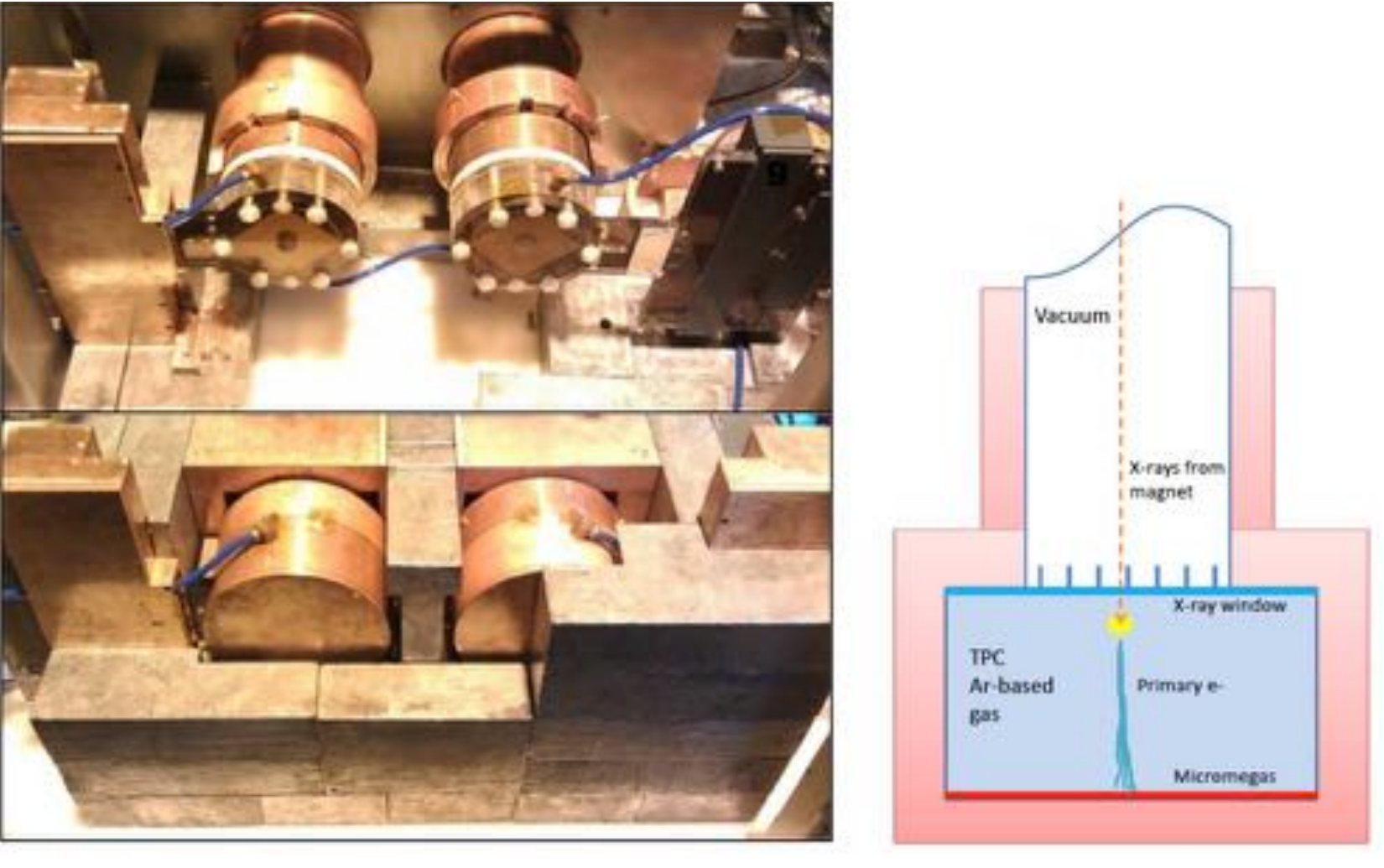}
\caption{Left: two CAST mini-TPCs placed at the beam-pipe bores of a decommissioned LHC dipole, before and after shielding (images from \cite{TREX-DM_last}).
Right: CAST working principle. X-rays from axion conversion in the magnetic field (above) traverse an ultra-thin window (defining the TPC cathode) and
interact in the gas, thereby producing small ionization clusters.}
\label{CastShielding}
\end{figure}

Mini-TPC arrays (readout surface $60\!\times\!60$\,mm$^2$) are presently used as the sensitive part of axion telescopes \cite{CAST1}.
The energy distribution of solar axions at Earth can be computed from solar models and peaks at around $\varepsilon=3$\,keV \cite{Dafni}. With the
assistance of an external magnetic field, axions can be converted to x-rays of the same energy (Primakoff effect) and detected in a suitable device.
CAST tracks the sun as it follows its apparent orbit on the sky.
Its TPC is based on a conventional Ar/i-C$_{4}$H$_{10}$ mixture (98/2) at around 1.5\,bar, to maximize conversion efficiency.\footnote{With a 3\,cm-long
drift region, these settings yield a conversion efficiency in excess of 70\% for x-ray energies below 8\,keV.} In one of the configurations,
a microbulk detector with an $x$-$y$ strip segmentation ($\Delta{x(y)}=500$\,$\mu$m) is used, while another one uses an InGrid device. The outstanding segmentation ($\Delta{x(y)}=50$\,$\mu$m) and single-electron sensitivity provided by InGrid allows to easily identify multi-site clusters as well as cosmic ray events that would otherwise create confusion during x-ray reconstruction in the keV energy region (Fig. \ref{CASTimage}). CAST operates at surface,
despite it is genuinely a low-background experiment, hence it uses passive shielding (Fig. \ref{CastShielding}-left,down). In this way, a background as low as $10^{-6}$\,keV$^{-1}$ cm$^{-2}$ s$^{-1}$ was recently achieved in the keV region \cite{Garza}.

\subsubsection{Other proposals}

\begin{figure}[h!!!]
\centering
\includegraphics*[width=\linewidth]{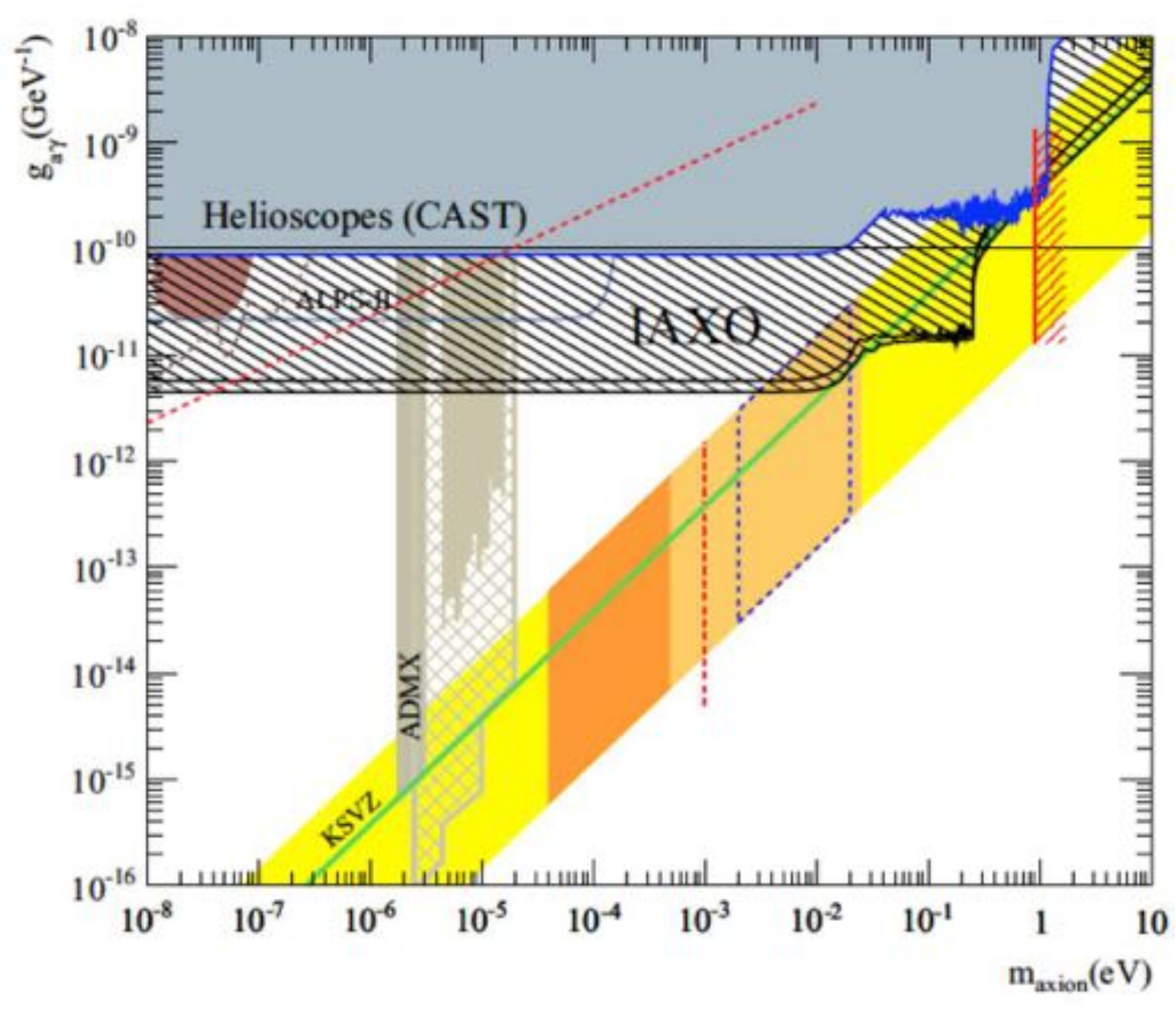}
\caption{Current limits to the axion properties (axion-photon coupling and axion mass) obtained with solar (CAST) and halo (ADMX)
detectors \cite{Dafni}. The limits projected for the IAXO solar telescope are also shown. The diagonal band comprises the
theoretically motivated region of parameters for a physical axion. (Added note: a slightly improved limit has been made
available recently in \cite{NatureCAST})}
\label{IAXO_limits}
\end{figure}

CAST has excluded an axion-photon coupling larger than $g_{a \gamma}=10^{-10}$\,GeV$^{-1}$, that implies axion masses $m_a\lesssim0.1$-$1$\,eV/$c^2$,
depending on the model (Fig. \ref{IAXO_limits}). The future IAXO observatory \cite{IAXO}, currently a consortium of around 40 institutes \cite{IAXOlett}, could improve the limit on the coupling constant by more than one order of magnitude. IAXO intends to use x-ray focusing through grazing angle lenses in order to reduce the sensor size (and backgrounds) even further.

Solar axions are amenable to detection on big TPC volumes, too, an idea that has been recently resurrected in
\cite{Galan}. In the proposed configuration, the TPC is operated inside the magnetic field
region, and the axion-photon conversion detected locally. In order to scan a competitive axion
parameter-space, the original proposal comprises a pressure and gas composition scan. With a seemingly
practical choice of pressures in the range 20-160\,mbar, and xenon as the main gas, a $\sim 1$\,m$^3$ active volume
TPC could outperform CAST.

At last, a low pressure TPC to study theories in extra-dimensions was proposed in \cite{KKaxions}. The idea in that case is that
a Kaluza-Klein solar axion with a mass in the keV/$c^2$-range would decay into two low energy photons, that could be detected as separated
clusters.

\section{Dual-phase chambers for neutrino and dark matter physics} \label{D-P}

Since the detection of rare processes often requires a dense target and good background discrimination, the idea to combine the properties of a gaseous argon TPC as a fully active tracking media and those of liquid argon for calorimetry \cite{Willis:1974gi} was introduced in 1977 and led to the concept of the liquid argon TPC \cite{Rubbia:1977zz}. Besides increasing the interaction probability compared to a classical argon+quencher gas TPC, this approach provides a very high resolution tracking device with good calorimetric response, and a $T_0$ signal for trigger (or for fiducialization, depending on the application) through primary scintillation. The enhanced ionization density due to the use of a condensed phase competes however with the availability of any stable form of charge amplification \cite{Bressi:1991yj, Breskin:2013ifa}. In a dual-phase TPC on the other hand, the liquid is kept close to its boiling point and charge multiplication can be obtained in the coexisting gas phase. 

The combination of the aforementioned properties makes dual-phase detectors based on charge multiplication perfectly suited to detect and fully characterize complex topologies such as those stemming from neutrino interactions, especially for giant multi-kton TPCs \cite{Rubbia:2004tz}. For the readout of smaller (m$^2$-scale) areas, a yet higher sensitivity has been achieved through electroluminescence, a technique capable of single-electron identification \cite{Aprile:2013blg}. This fact, together with the availability of a nuclear recoil signature, high density and self-shielding against superficial backgrounds makes dual-phase TPCs the leading technology for direct WIMP dark matter detection.

The key of the dual-phase concept lies on the ability to extract electrons from the liquid phase into the gas, something that has been demonstrated already decades ago \cite{1970JETPL..11..351D, Gushchin:1982}. This task can be performed with the help of auxiliary grids embedded in the liquid and an electric field of $E_{ext} \geq 2$\,kV/cm, allowing 100\% extraction efficiency. Once in the gas phase, the temperature conditions corresponding to a vapor pressure around 1\,bar ($T \gtrsim 80\,K$) are such that amplification can easily proceed there, under a mild reduction by a factor $2$-$3$ in the gas number density ($N$), relative to operation at NTP conditions.

\subsection{WA105 and DUNE} \label{secWA105}

\begin{figure}[htb]
  \centering
  \includegraphics[width=\linewidth]{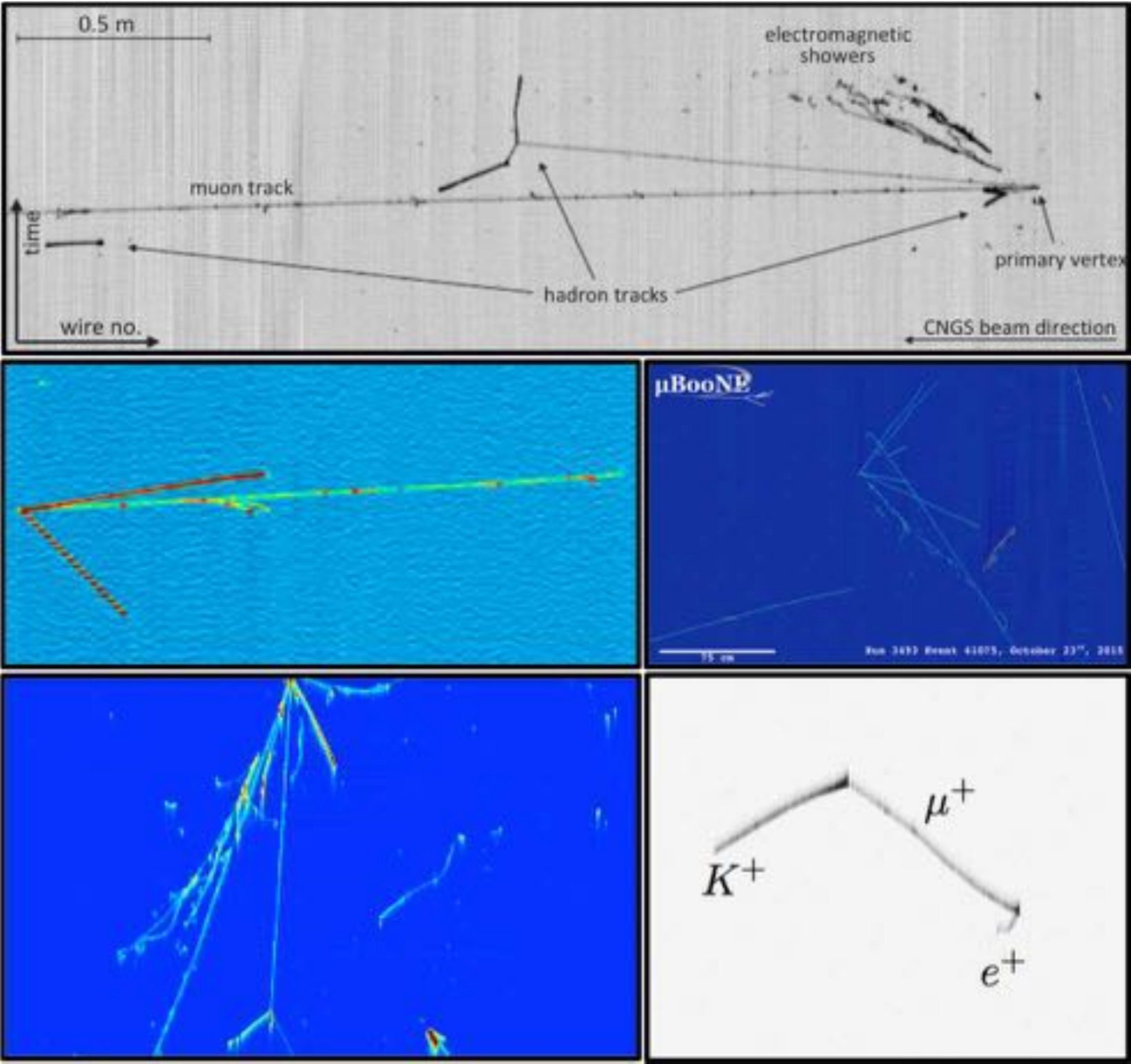}
  \caption{Typical event patterns from liquid argon TPCs. Top: a CNGS $\numu$~charged-current interaction in the ICARUS T600 detector \cite{Antonello:2012hu}. Middle-left: a $\numu$~charged-current interaction in ArgoNeuT \cite{Anderson:2011ce}. Middle-right: a neutrino interaction candidate in MicroBooNE \cite{MiniBoone}. Bottom-left: a cosmic ray -induced nuclear breakup in a dual-phase 5\,ton chamber \cite{Murphy:2016ged}. Bottom-right: a simulated proton decay in the $p\rightarrow K^+ + ~\bar{\nu}$ channel and subsequent decay chain, in GLACIER \cite{Stahl:2012exa}.}
    \label{Events1}
\end{figure}

The Deep Underground Neutrino Experiment (DUNE) (\cite{DUNE_CDR} and references therein) is an international effort to build a next-generation long-baseline oscillation experiment between Fermilab (Illinois, USA), where a new megawatt-scale neutrino beam-line will be built, and a far detector (FD) installed at the Sanford Underground Research Facility (SURF), at about 1300\,km distance. The FD will consist of $40$\,ktons of argon, and both single and dual-phase prototypes are being developed for the task. The anticipated imaging capabilities can be appreciated from previous single-phase liquid argon TPCs used in neutrino physics: the 600 ton ICARUS detector \cite{ICARUS}, ArgoNeut \cite{Argoneut} and MiniBoone \cite{MiniBoone}. They reconstruct the event projections ($xz$, $yz$) through crisscrossed wires immersed in the liquid. Notably, only one wire plane actually collects charge (dubbed `collection plane'), while the others register the bipolar transient signals induced as charges move through (`induction planes'). The latter are more prone to $\mathcal{S}/\mathcal{N}$ limitations and require of a special deconvolution analysis. At a $\mathcal{S}/\mathcal{N} > 10$, the collection plane projections for these experiments display a high level of detail, allowing to clearly identify a number of very diverse event topologies (Fig. \ref{Events1}). All these detectors are self-triggerable, thanks to the primary argon scintillation being detected by sparse PM arrays.

It has been shown that a $\mathcal{S}/\mathcal{N}\geq10$ in all planes is indeed necessary to provide adequate track reconstruction and particle identification via $d{\varepsilon}/dx$ \cite{Antonello:2012hu, DUNE_CDR}. However, maintaining such a performance in the next generation neutrino experiments like DUNE is challenging: without charge amplification, a similar imaging quality can only be preserved by segmenting the active volume in multiple sub-detectors of around 3\,m drift, as well as increasing the readout pitch to 5\,mm. From that perspective, a dual-phase detector provides an elegant alternative, allowing to construct fully homogeneous detectors of up to 12\,m drift with fewer readout channels (table \ref{TableArgon}).

\begin{figure}[htb]
  \centering
  \includegraphics[width=\linewidth]{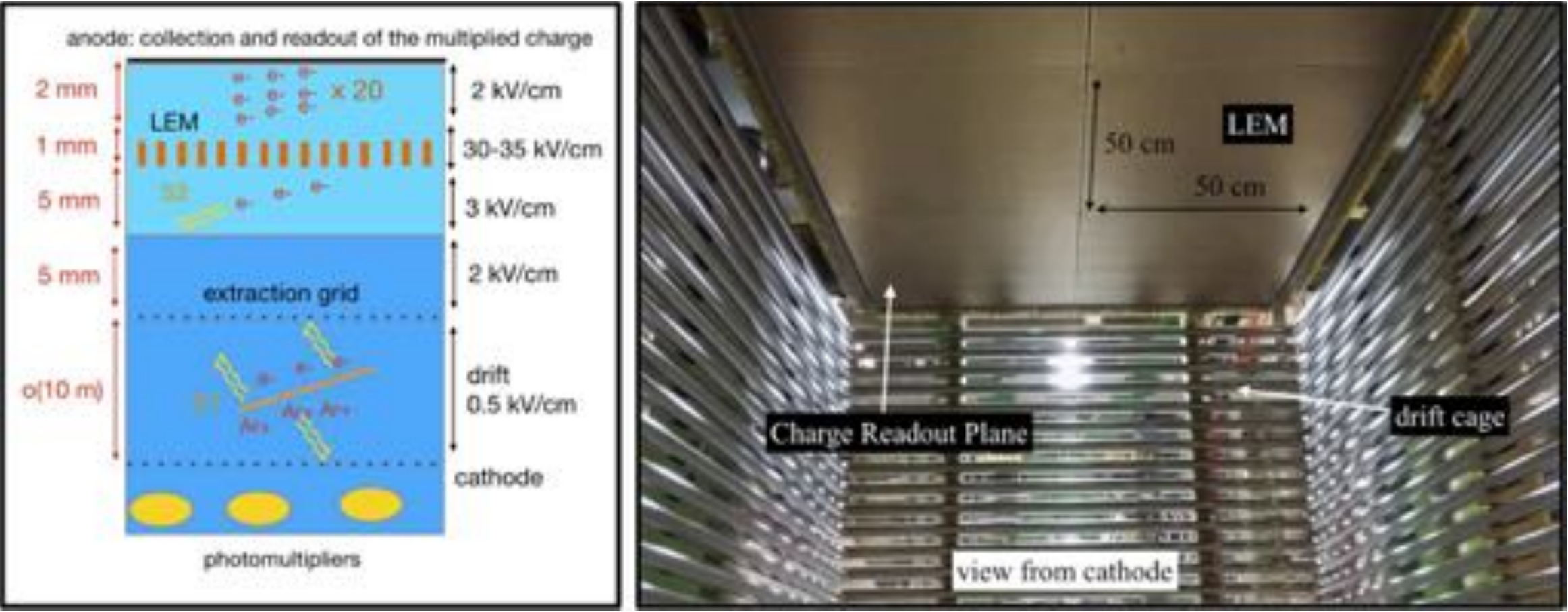}
  \caption{Left: operating principle of a dual-phase charge multiplication -based TPC. Right: inner view of the 5\,ton WA105 demonstrator (3$\times$1$\times$1\,m$^3$ readout area), currently under commissioning at CERN. The $50$\,cm$\times50$\,cm LEM tiles are shown at the top.}
    \label{DP_sketch}
\end{figure}

ETH-Zurich (and later the WA105 collaboration) have pioneered the use of large area thick GEMs (LEMs) coupled to a specially designed low-capacitance $x$-$y$ induction plane based on strip-like pickup structures, with 500\,pF/3m at a 3\,mm pitch (allowing about $1500\,$e$^-$ noise \cite{Badertscher:2013wm} - see Fig. \ref{2Dlayouts}-f). A sketch of the detector and an inner view of the 3$\times$1$\times$1\,m$^3$ (5\,ton) prototype currently under commissioning at CERN is shown in Fig. \ref{DP_sketch}. The choice of LEMs as the multiplication structure is based on multiple factors and is well documented \cite{Bondar:2008yw,Badertscher:2008rf}. They are robust, stable even in cryogenic conditions and can be economically manufactured by the printed circuit board industry. Effective gains of about 20-40 were reached on a dual-phase TPC equipped with a $10\times10$\,cm$^2$ readout in \cite{Cantini:2014xza}. A 250\,liter TPC was operated under stable gains of about 20 \cite{Badertscher:2013wm, Cantini:2013yba}. All measurements were performed at around 1\,bar vapor pressure, as intended in the final system. A field above 30\,kV/cm is required across the LEM in order to obtain charge multiplication, and a 2-5\,kV/cm field is applied in the induction region just in front of the $x$-$y$ pickup strips. The present LEM design consists of $50$\,cm$\times50$\,cm copper clad glass-fiber epoxy plates, with a thickness of about one millimeter and with mechanically drilled holes of 500\,$\mu$m diameter at a 800\,$\mu$m pitch.

\begin{figure}[ht!!]
  \centering
  \includegraphics[width=\linewidth]{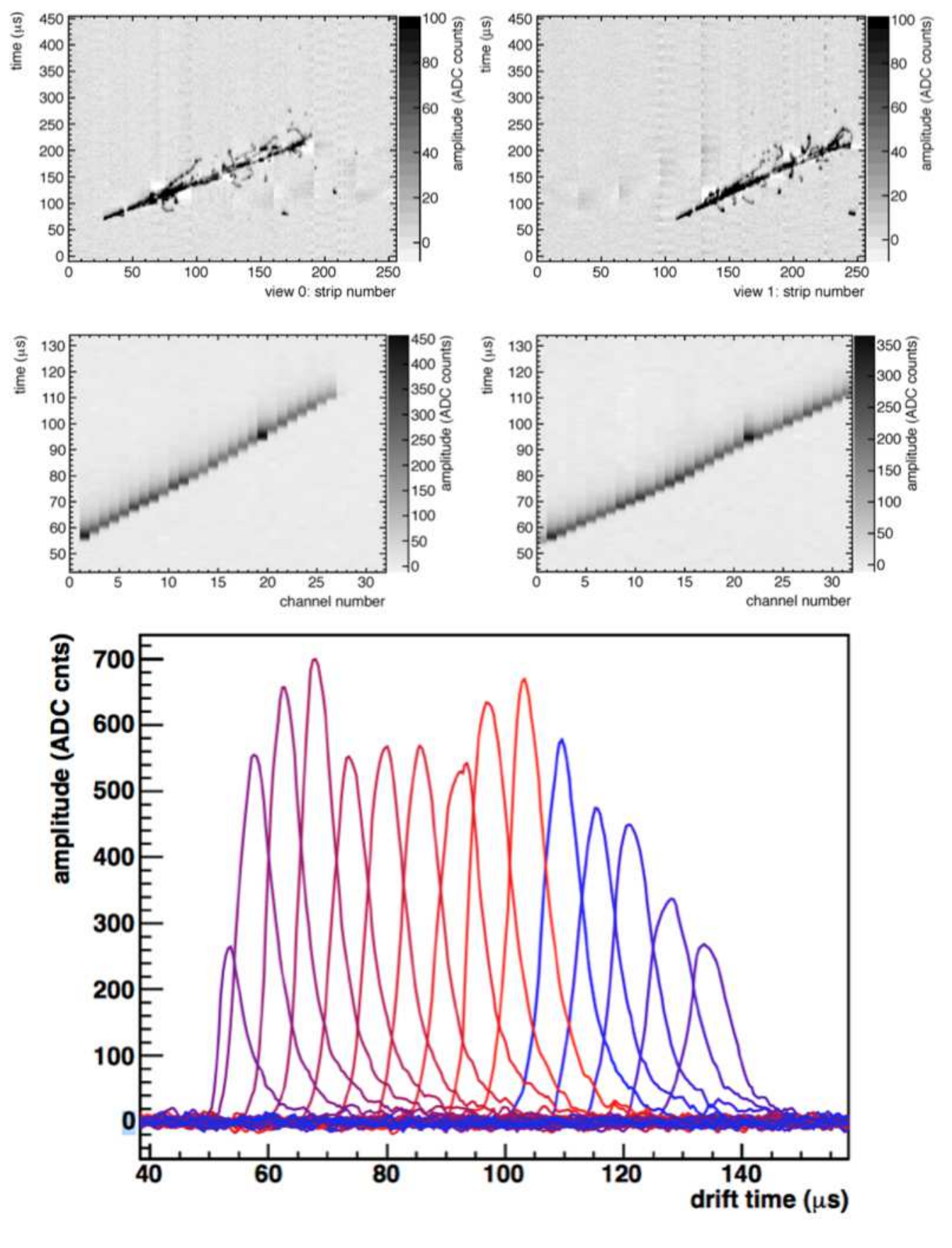}
  \caption{Reconstruction in dual-phase chambers. Top: $xz$ and $yz$ views of a cosmic ray shower collected in a 250\,liter argon TPC with $40$\,cm$\times$80\,cm anode plane. Middle: reconstructed muon in a 3\,liter TPC, with 10\,cm$\times$10\,cm anode plane. Bottom: typical signal from a muon event at an effective gain $m^*\simeq 20$.}
    \label{DP_wvf}
\end{figure}

\begin{figure}[htb]
  \centering
  \includegraphics[width=\linewidth]{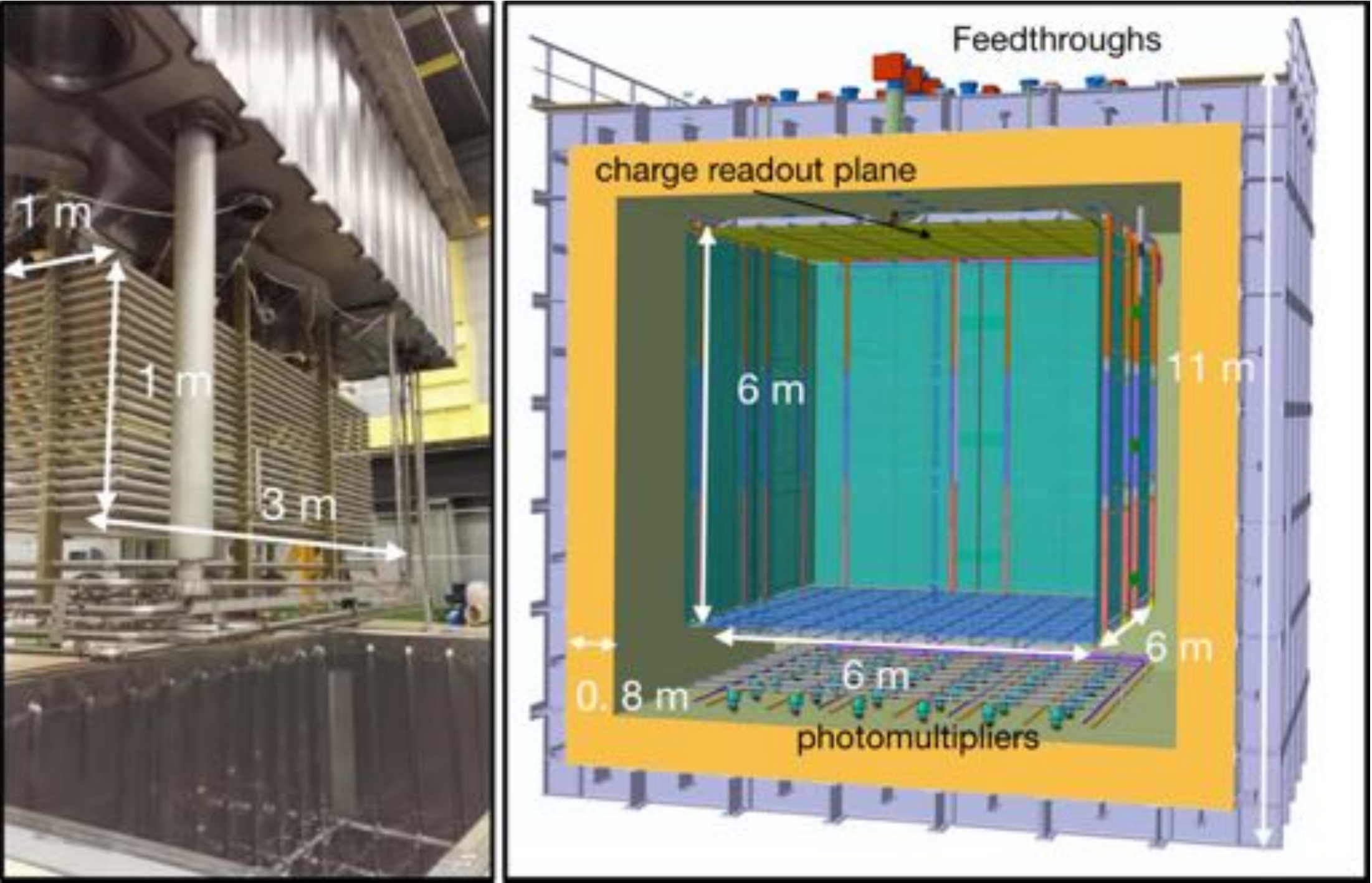}
  \caption{The WA105 dual-phase argon TPC prototypes at CERN. Left: the 3$\times$1$\times$1\,m$^3$ (5\,ton) active volume detector \cite{Murphy:2016ged} being inserted into the cryostat. Right: cut view of the 6$\times$6$\times$6\,m$^3$ (300\,ton) prototype \cite{DeBonis:1692375} already inside the cryostat.}
    \label{fig:WA105-proto}
\end{figure}

By considering an energy loss of $d\varepsilon/dx = 210$\,keV/mm for mips (30\% of which is recombined), a $W_I$-value of 23\,eV, an effective gain $m^*=20$ and a strip pitch $\Delta{x(y)} = 3$\,mm, a signal corresponding to $\bar{n}_{e,r} = 4 \times 10^5$ electrons can be estimated from eq. \ref{neFINAL!}. If assuming an equal amount of charge sharing between $x$ and $y$ collection strips, the sensor response function (eq.  \ref{SRF}) may be approximated through an additional reduction by a factor of two. With noise levels around $\sim1500\,$e$^-$, a $\mathcal{S}/\mathcal{N}$ of about 130 can be anticipated. After conservatively accounting for attachment along a 12\,m drift, the $\mathcal{S}/\mathcal{N}$ still reaches 10 for an O$_2$ concentration of 100\,ppt, and $\mathcal{S}/\mathcal{N}$=60 if 30\,ppt is finally achieved. As expected from these considerations, reconstruction of mips in small ($\simeq 4$\,kg) systems indicates indeed an outstanding image quality (Fig. \ref{DP_wvf}). By the time of publication of this work, nonetheless, the 5\,ton prototype (Fig. \ref{fig:WA105-proto}-left) has been in operation for about one month, displaying a comparable image quality (Fig. \ref{Events1} bottom-left). Although preliminary, these results already illustrate the potential of the dual-phase technology at the ton scale and m$^2$-readouts.

\begin{table}[htb]
\centering
\begin{tabular}{|c|c|}
    \hline
  property & value \\
    \hline
    \hline
  ~ boiling point at 1\,atm &87.303\,K\\
 ~  liquid density at boiling point                 ~                               & ~ 1.369\,g/cm$^3$~              \\
 ~  ionization energy $W_I$ \cite{PhysRevA.9.1438}  ~                               & ~ $23.6 ~ \pm ~ 0.5$\,eV ~      \\
 ~  electron drift velocity $v_d$ \cite{Li:2015rqa} ~                               &  1.648 (2.261)\,mm/$\mu$s      \\
 ~  ion drift velocity ($E_d\lesssim1$\,kV/cm) \cite{lar-prop,palestini}            &  8\,mm/s                        \\
 ~  mean $d\varepsilon/dx$ for mips                 ~                               & ~ 2.1\,MeV/cm ~                 \\
 ~  $\mathcal{R}$ for mips                          ~                               & ~ 0.29 (0.25) ~                \\
 ~  $\sigma_{_{L}}$(12\,m) \cite{Li:2015rqa}        ~                               & ~ 3.3\,mm ~                     \\
 ~  $\sigma_{_{T}}$(12\,m) \cite{Li:2015rqa}        ~                               & ~ 4.2\,mm ~                     \\
 ~  $\mathcal{A}$(30\,ppt O$_2$, 12\,m drift)       ~                               & ~ 0.53 (0.42)  ~               \\
 ~  active volume  (10\,kt-module)                  ~                               & ~ $60$\,m$\times12$\,m$\times12$\,m  \\
 ~  number of LEMs (10\,kt-module)                  ~                               & ~ $2880$                       \\
 ~  readout channels (10\,kt-module)                ~                               & ~ $153600$                     \\
 ~  cathode HV (10\,kt-module)                      ~                               & ~ $600$\,kV                      \\
 ~ $\mathcal{S}/\mathcal{N}$(100\,ppt O$_2$, 12\,m drift) ~                         & ~ $10$                         \\
 ~ $\mathcal{S}/\mathcal{N}$(30\,ppt O$_2$, 12\,m drift) ~                          & ~ $60$                         \\
    \hline
       \end{tabular}

     \caption{\label{tab:lar-prop} Some relevant properties of liquid argon and in particular those related to a future dual-phase DUNE-FD. For field-dependent magnitudes the first number refers to 500\,V/cm, the second one (in brackets) to 1\,kV/cm. For definitions of the attachment, $\mathcal{A}$, and recombination, $\mathcal{R}$, factors the reader is referred to eq. \ref{Reco}, \ref{Eq_hydro}.}
     \label{TableArgon}
   \end{table}

After a thorough system characterization, and with almost immediate character, the collaboration goals will be to demonstrate the scalability of the technology at workable purity levels that are suitable for the envisaged 10\,kton ($\times 4$) final modules, as well as evaluating the impact of the ion space charge on the detector performance. The phenomenon, due to the ions released during the ionization and multiplication processes is a common guest at collider TPCs (stemming from the ion back-flow from the multiplication region, or IBF). A priori, space charge is an unusual nuisance in this situation given the fact that the detector dose is overwhelmingly dominated by `just' background ionization from cosmic rays ($\mathcal{O}(100)$\,muons/m$^2$/s) and, to a lower extent, by $^{39}$Ar decays. However, the relatively low drift fields (translating into an ion drift velocity of less than 1\,cm per second -table \ref{TableArgon}) coupled to the large detector volume, yield an imposing charge density of the order of 10$^6$\,ions/cm$^3$ \cite{DeBonis:1692375}. This space charge may distort the electric field as well as potentially resulting in charge loss due to electron-ion recombination \cite{Bueno:2007um}. Although these effects are of particular concern for TPCs operating near ground level, the presence of $^{39}$Ar decays at rates of about $\sim$1\,Bq/kg should be taken into account in the evaluation of space charge for underground TPCs too. The consequences of space charge have been simulated for large liquid argon detectors with long drifts of up to 20\,m with varying results, some more pessimistic \cite{Romero:2016tla} than others \cite{Bueno:2007um,DeBonis:1692375}. Such simulations do not take into account however the motion of the liquid arising from temperature gradients and the recirculation system, and that result in characteristic velocities that are larger than the ion velocity along the field lines \cite{palestini}. This will increase the probability of the ions becoming neutralized on conductive surfaces \cite{DeBonis:1692375}. Detailed studies are currently ongoing with data from MicroBoone \cite{Mooney:2015kke} and ICARUS \cite{Torti:2016wub}.

The final detector foresees that LEMs, anode and grid will be assembled together in frames of $3\times3$\,m$^2$. Each of these structures functions as an independent system with its own feedthrough and is suspended by means of three adjustable cables that can precisely align the readout to the liquid level. Together with the space charge and purity levels, this concept will be tested on the 6$\times$6$\times$6\,m$^3$ (300\,ton) prototype currently under construction, and that aims at taking data from a charged particle beam in 2018 (Fig. \ref{fig:WA105-proto}-right). The physics program is compelling, given that never before has such a large liquid argon TPC been exposed to a charged particle beam. The events will be fully contained, which allows a detailed evaluation of the detector's calorimetric performance. It will also provide a wealth of cross section measurements in the energy of interest for the DUNE long baseline neutrino oscillation program.

\subsubsection{Some considerations about signals and backgrounds}

We give in the following some illustrative examples (the selection is not exhaustive) of the impact of accurate imaging for signal/background identification in DUNE.

\paragraph{Electron neutrino appearance.}

A high precision measurement of the $\nu_e$~appearance signal in an almost pure $\nu_\mu$~beam is essential for studies of CP-violation in the neutrino sector. At the typical beam energies for long baseline neutrino experiments ($\sim$ 1 GeV), reactions are mainly from charged-current interactions of the neutrino on the argon nucleus $\nu_l + n \rightarrow p + l$. The outgoing lepton provides a clear signature on the flavor of the interacting neutrino: electrons will develop into showers whereas muons will yield long straight tracks. One particular source of background is the neutral-current (NC) interaction of a muon neutrino with production of a $\pi^0$ in the final state ($\nu_\mu+\tn{Ar}\rightarrow\nu_\mu+\pi^0+X$). This NC-$\pi^0$ background is of particular importance to neutrino oscillation experiments as it can be experimentally misidentified as a $\nu_e$ appearance signal from charged-current interaction. The $\pi^0$ decays preferentially into pairs of energetic $\gamma$'s which will lead to pair production. In most neutrino detectors, $\gamma$-events will thus appear almost identical to electron signals, especially if both $\gamma$-induced showers overlap. The possibility to identify the one shower (electron signal) versus two showers ($\gamma$ background) events is an unique advantage of the fine spatial and calorimetric sampling of the liquid argon technology.
First, the 14\,cm interaction length compared to the fine 3\,mm-pitch of the readout means that the $\gamma$ pairs can leave a visible gap between its origin and the beginning of the shower. Second, the energy deposit at the beginning of the shower will be different for a single ionizing particle from the electron signal or the two ionizing particles of each electron/positron pair. Hence exploiting both topological and $d\varepsilon/dx$ information provides a very powerful tool for NC-$\pi^0$ background subtraction even if both $\gamma$ showers overlap spatially.

\paragraph{Proton decay.}

Nucleon decay signals are characterized by their topology and their kinematics. The total energy of the event should be close to the nucleon rest mass.
Here again the full event imaging and good particle identification capabilities of a liquid argon TPC combined with its low energy detection threshold give the technology a significant improvement in sensitivity  over water Cherenkov detectors in several nucleon decay modes. Multi-prong events like $p\rightarrow e^+ + ~ \pi^+ + ~ \pi^-$ (when all daughters are reconstructed) and channels involving kaons such as $p\rightarrow K^+ + ~\bar{\nu}$ or $p\rightarrow \mu^+ + ~K^0$ are particularly suitable. They offer typically one order of magnitude improvement over water Cherenkov detectors in similar background conditions.
In the latter, the emerging kaon is below its Cherenkov threshold in water ($\sim$500\,MeV) and therefore can only be detected via its decay products. On the contrary, a liquid argon TPC can detect and reconstruct such kaons with an extremely high efficiency as well as tag many of its decay products. Kaon backgrounds essentially originate from interactions of atmospheric neutrinos and neutral hadrons produced by cosmic muons or surrounding material (e.g rock) outside of the TPC. For instance, neutral kaons produced outside of the fiducial volume may undergo inelastic scattering (primarily charge exchange) in the TPC, resulting in a single positive kaon.

\paragraph{Low energy supernovae neutrinos.}

In a liquid argon TPC, neutrinos originating from supernovae ($\varepsilon <$100\,MeV) can be detected through three different channels: i) charged-current interactions ($\nu_e + ~ ^{40}\tn{Ar}\rightarrow e^- + ~ ^{40}\tn{K}^{*}$, $\bar{\nu}_e + ~ ^{40}\tn{Ar}\rightarrow e^+ + ~ ^{40}\tn{Cl}^{*}$), ii) neutral-current interactions ($\barparen{\nu} + ~ ^{40}\tn{Ar}\rightarrow \barparen{\nu} +~ ^{40}\tn{Ar}^{*}$) and iii) elastic scattering on orbital electrons ($\barparen{\nu} + e^-\rightarrow \barparen{\nu} + e^-$). All have thresholds at the MeV scale. Events at such low energies are rather dull from a topological point of view, consisting of a short electron track ($\mathcal{O}$(10cm)) that can be accompanied by a multitude of smaller hits arising from $\gamma$'s released during nuclear and atomic de-excitation. Those hits are often referred to as `sparkles'. Triggering and reconstructing such event topologies represent great challenges and, certainly, achieving an excellent signal to noise ratio constitutes a fundamental prerequisite in this area. The radioactive isotope $^{39}$Ar (that is naturally present in the TPC) also produces similar sparkles, which adds to the complexity of reconstructing those low energy neutrino events.

\subsection{LUX, PandaX and XENON} \label{DMDP}

\begin{figure}[htb]
 \centering
 \includegraphics[width=\linewidth]{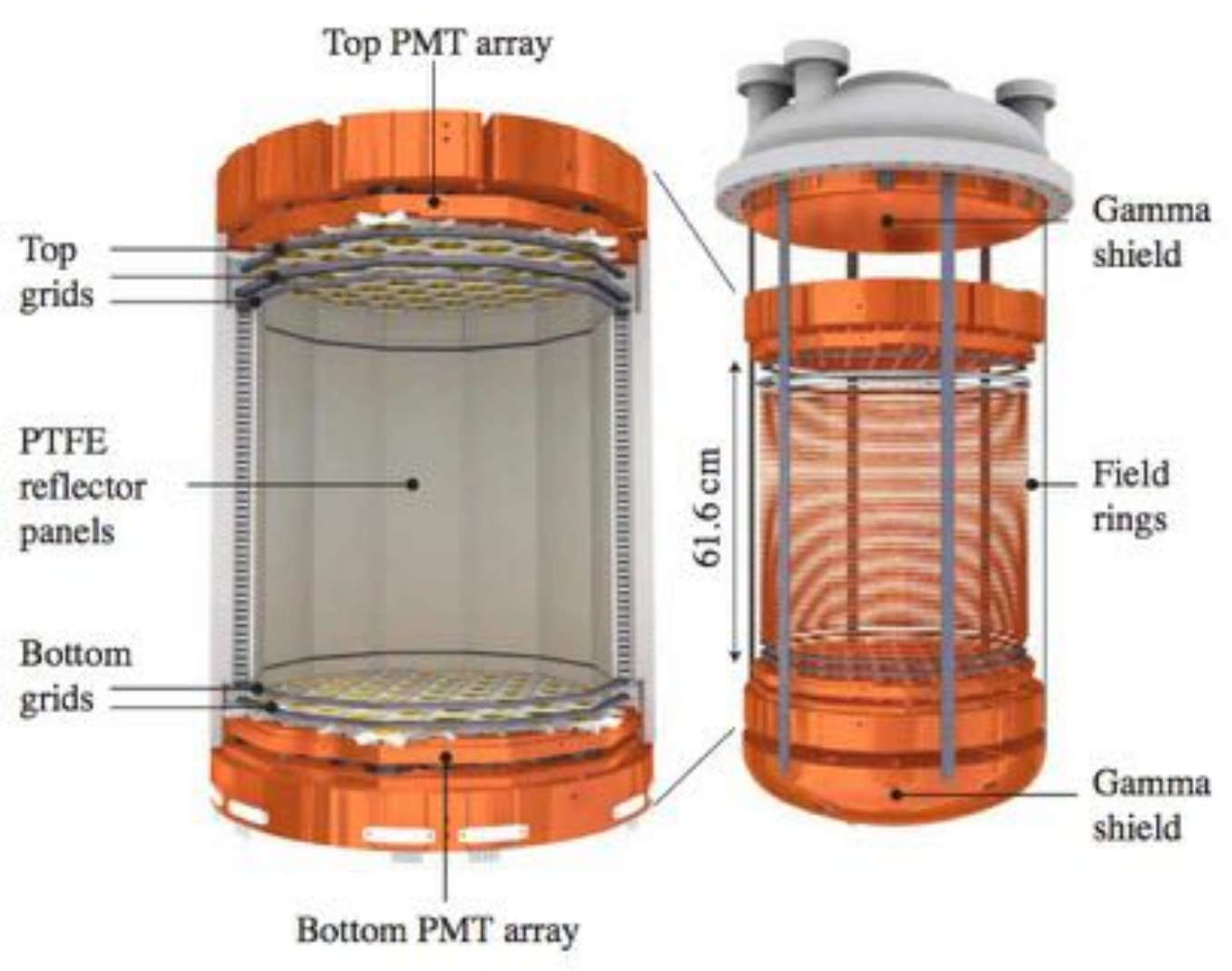}
\caption{Sketch of the basic elements of the LUX experiment \cite{LUX}.}
\label{sketchLUX}
\end{figure}

\begin{figure}[htb]
  \centering
  \includegraphics[width=0.95\linewidth]{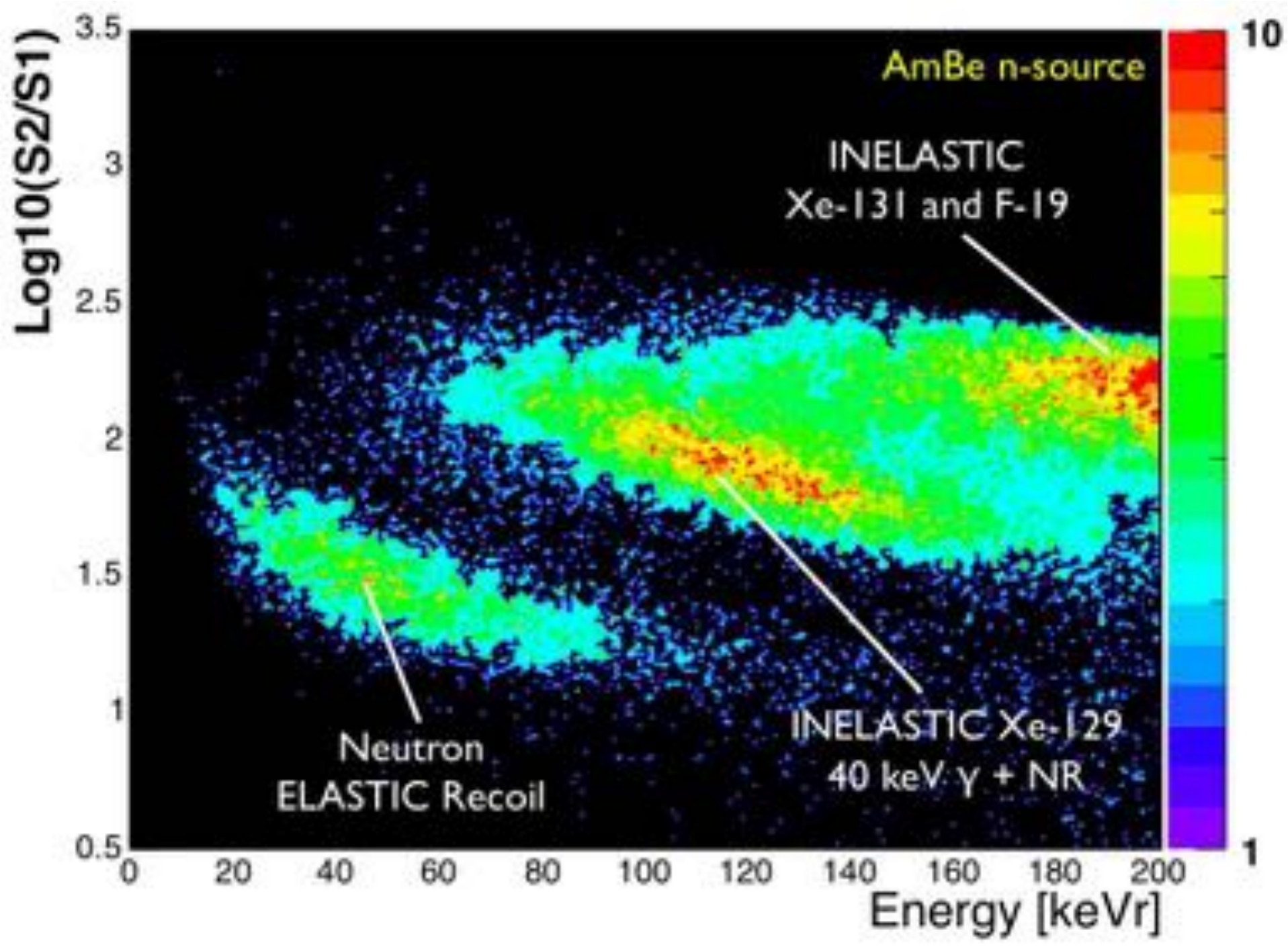}
  \caption{Ratio of ionization (S2) to scintillation (S1) yields as a function of the energy of the nucleus, depending on the type of interaction \cite{Aprile:2006kx}. There is a clear separation between neutron-induced elastic nuclear recoils and the rest of interactions.}
    \label{fig:nuc_recoil}
\end{figure}

The dual-phase approach to the detection of WIMP dark matter was laid out in the early 90's \cite{Barabash, Benetti} and is best realized, nowadays, through xenon TPCs (Fig. \ref{sketchLUX}). Natural xenon has different isotopes with masses around 130\,GeV/$c^2$, an ideal value for studying nuclei recoiling upon its elastic interaction with super-symmetric particles having masses around the electro-weak scale (100\,GeV/$c^2$), and that naturally explain the relic dark matter abundance \cite{Feng}. The technique benefits from key facts like: i) the shielding capability of the condensed phase to external backgrounds (self-shielding); ii) the high sensitivity to recoiling nuclei (down to 10\,keV$_r$); iii) the peculiar ionization and scintillation yields anticipated for WIMP-induced nuclear recoils (section \ref{image}). This latter feature is often expressed in a S2/S1 (ionization/scintillation) vs S2 representation; a selection of nuclear recoil events based on this criterion allows a background suppression of $10^2$-$10^3$ \cite{Agnes:2015lwe,Szydagis:2014xog}, (Fig. \ref{fig:nuc_recoil}).

\begin{figure}[htb]
 \centering
 \includegraphics[width=\linewidth]{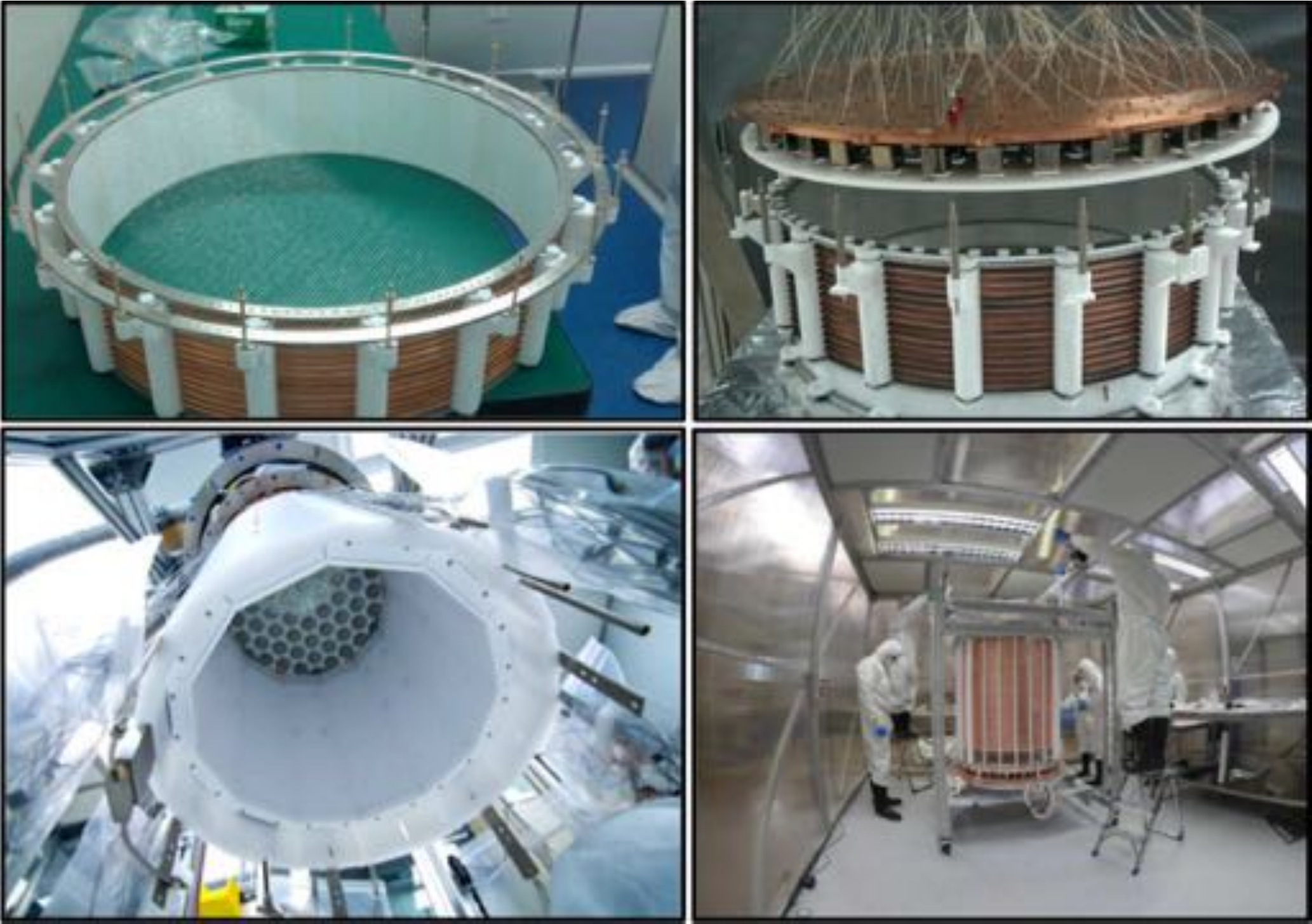}
\caption{Top: field cage of the PandaX detector before and after assembling one of the PM planes. Bottom-left: field cage and top PM plane of the LUX detector. Bottom-right: XENON1T (the existing detector with the best projected sensitivity) during assembly. (Photos from the collaborations' web sites \cite{CollPage})}
\label{DM_Xenon}
\end{figure}

\begin{figure}[htb]
  \centering
  \includegraphics[width=0.95\linewidth]{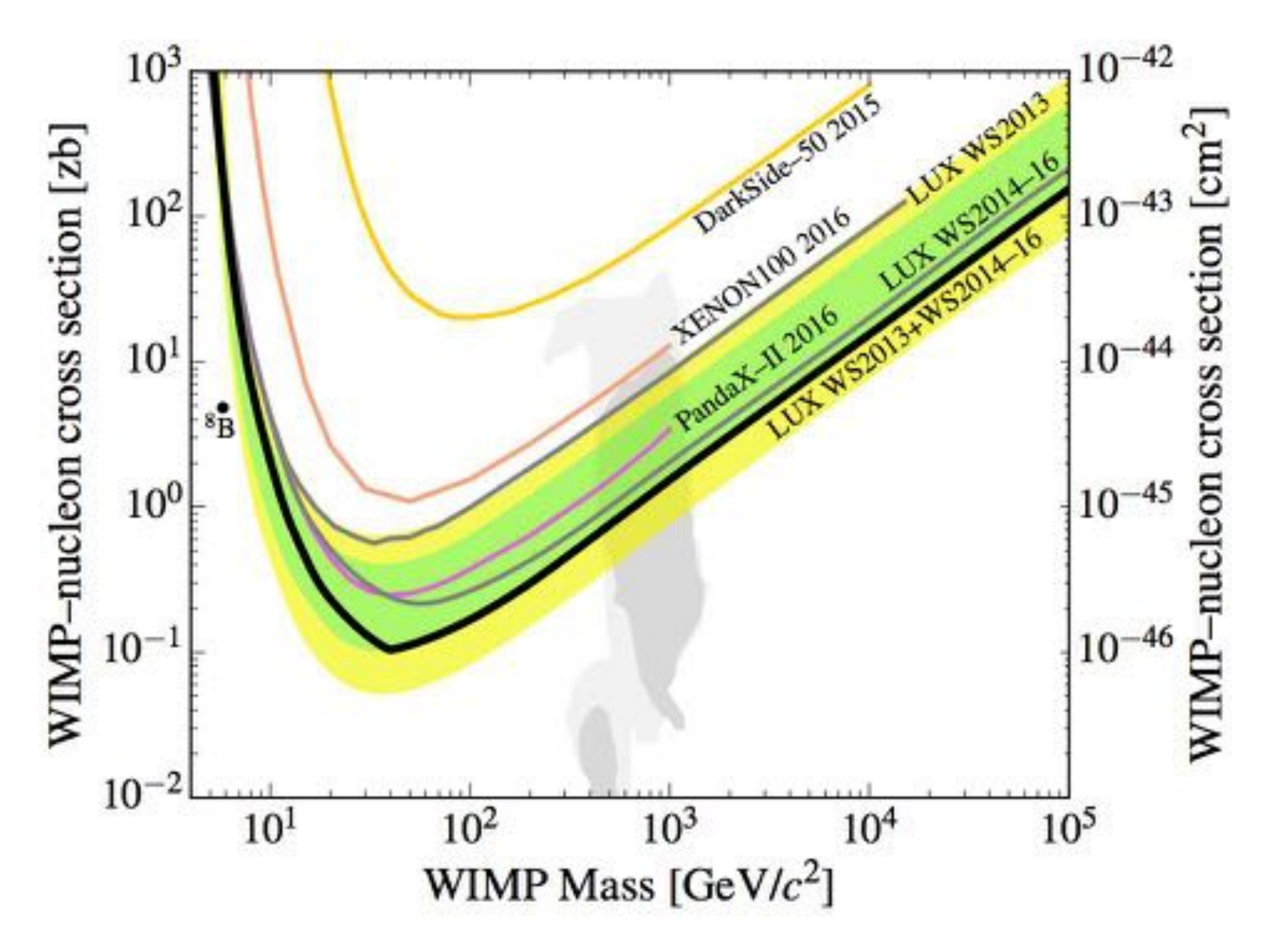}
  \caption{Exclusion plots for direct dark matter searches from different experiments. At a WIMP mass of 50\,GeV/$c^{2}$,
the LUX experiment excludes a cross section larger than $2.2 \times 10^{-46}$\,cm$^2$ at 90\% confidence level (re-printed from \cite{Akerib:2016vxi}).}
\label{fig:DM_results}
\end{figure}

Dual-phase xenon technology is mature, and several TPCs have been built over the last 15 years \cite{Cao:2014jsa,Alner:2005wt, Angle:2009xb, Aprile:2011dd, Majewski:2011st} (Fig. \ref{DM_Xenon}). Moreover, they currently set the lowest experimental bounds to the WIMP dark matter cross section (Fig. \ref{fig:DM_results}). Since the technical solutions adopted by the leading experiments are all very similar, we'll focus on the one that had the lowest published limits at the time of writing, the LUX experiment \cite{Akerib:2016vxi} (Fig. \ref{sketchLUX}).\footnote{Meanwhile, this limit has been superseded both by PANDAX and XENON1T collaborations (!) \cite{PandaNew,1TonLimit}.}

The large underground xenon (LUX) TPC contains a fiducial mass of 250 kg of LXe in a cylinder of 55\,cm height by 24\,cm radius. The main TPC volume is isolated from the external radioactivity by passive shields and active vetos. The active volume is confined by high-reflectivity teflon panels (95\% reflectivity to xenon VUV light \cite{Yamashita2004692}), leading to a light collection yield of $8$\,phe/keV at the center of the detector (corresponding to a collection efficiency of $\Omega\simeq 50\%$). Grids are constructed from stainless steel, that has shown to be 57\% reflective to xenon light \cite{BRICOLA2007260}. Two arrays of 61 PMs placed at the anode and cathode regions are responsible for detecting both the primary and secondary scintillation signals. The detector is subdivided in 4 regions (drift, electroluminescence/EL, and two buffer regions) using five meshes with optical transparency above 96\% except for the entrance to the EL region (88\%), \cite{LUX}.

\begin{figure}[htb]
 \centering
 \includegraphics[width=0.95\linewidth]{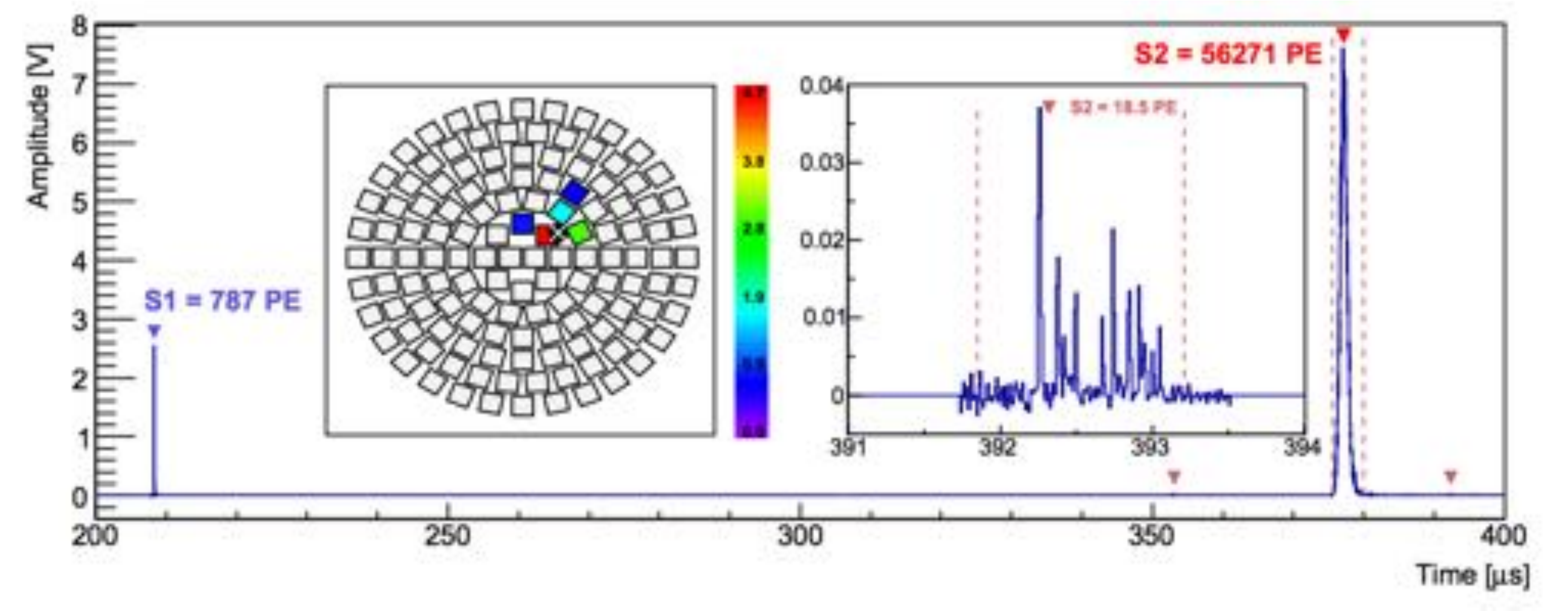}
 \includegraphics[width=0.45\linewidth]{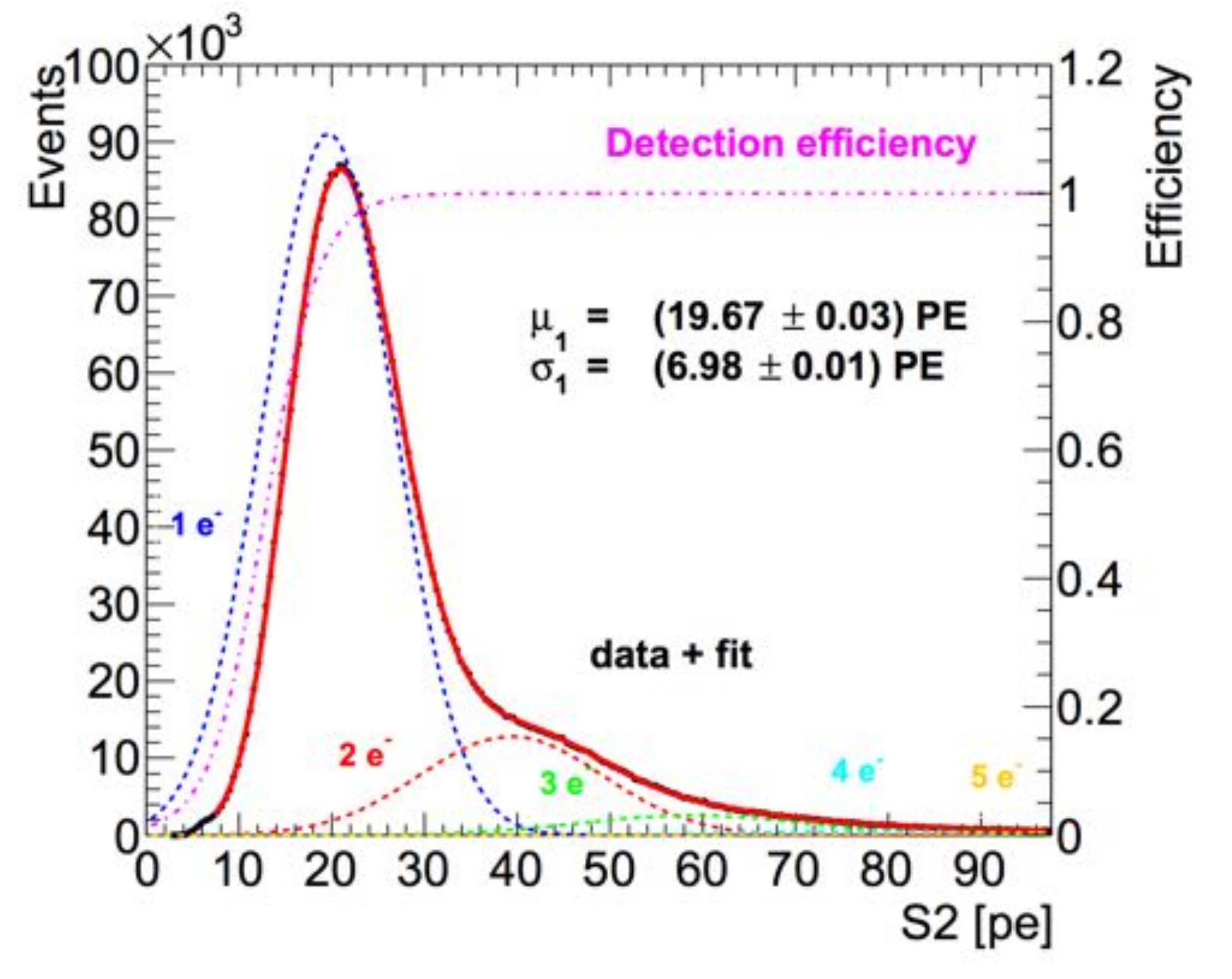}
 \includegraphics[width=0.45\linewidth]{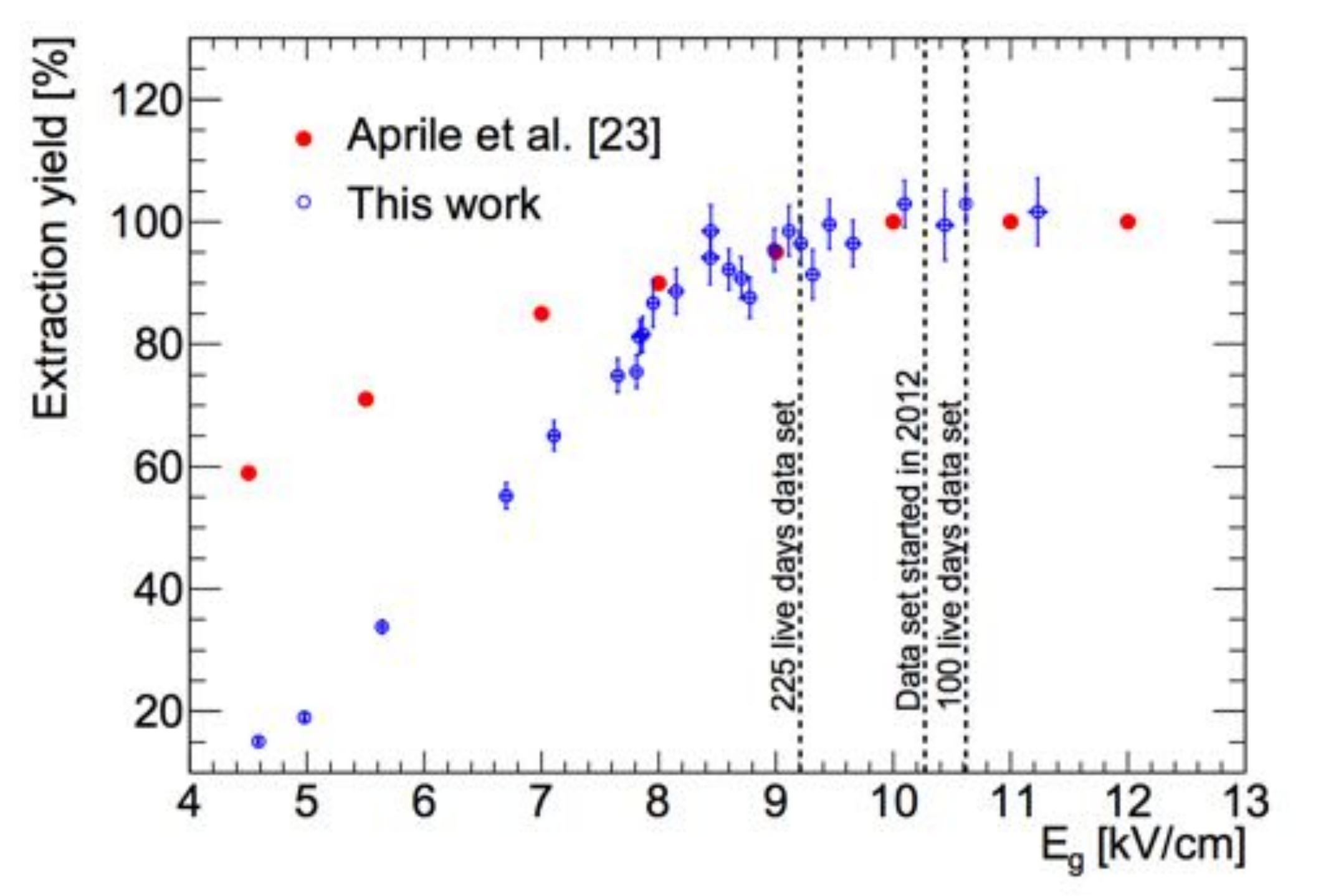}
\caption{Top: example of a XENON-100 event, indicating the primary (S1) and secondary (S2) scintillation signals. Two small S2 signals below 150\,photoelectrons (pe in figure) are observed and indicated by red triangles, $145$\,$\mu$s after S1 and $17$\,$\mu$s after the main S2. The second one is shown in the inset, together with its top array PM pattern, revealing a localised signal compatible with a single electron. Bottom-left: example of the low energy spectrum, comprising a sum of one to five electron's S2 signals, together with the efficiency function. Bottom-right: electron extraction efficiency. (Figures taken from \cite{Aprile:2013blg})}
\label{fig:single_e}
\end{figure}

While these detectors can collect at the PM planes nearly half of the primary scintillation released, no less impressive is their performance in terms of sensitivity to the primary ionization. As an example, the XENON collaboration recognizes single ionization electrons as light flashes emitted during their transit along the EL region
(Fig. \ref{fig:single_e}-top). These signals turn out to be a powerful tool to understand the technical performance of the detector:
e.g. allowing to obtain gain maps of the EL region, the efficiency of the S2-peak finder algorithm (Fig. \ref{fig:single_e} bottom-left), or the electron extraction efficiency (Fig. \ref{fig:single_e} bottom-right). Similar to the giant TPC of the DUNE collaboration, gas purity is an essential issue here too, however the smaller scale makes the problem much more benign.

\subsection{DarkSide}

In spite of its higher abundance and affordability when compared to xenon, the presence of the (unstable) $^{39}$Ar isotope in atmospheric argon (AAr) creates a background activity of nearly 1\,Bq/kg, thereby limiting the detector sensitivity for dark matter searches.
Since $^{39}$Ar is produced by cosmic ray interactions in the upper atmosphere, principally via the
$^{40}$Ar(n,2n)$^{39}$Ar reaction, argon gas from underground (UAr) may be used instead. Extracting $^{39}$Ar-depleted UAr has been recently accomplished thanks to underground CO$_2$ wells streams, that contain argon at concentrations of 400-600\,ppm, depleted to the $6\times10^{-3}$ \,Bq/kg level \cite{39ArLevel}.

\begin{figure}[htb]
  \centering
  \includegraphics[width=\linewidth]{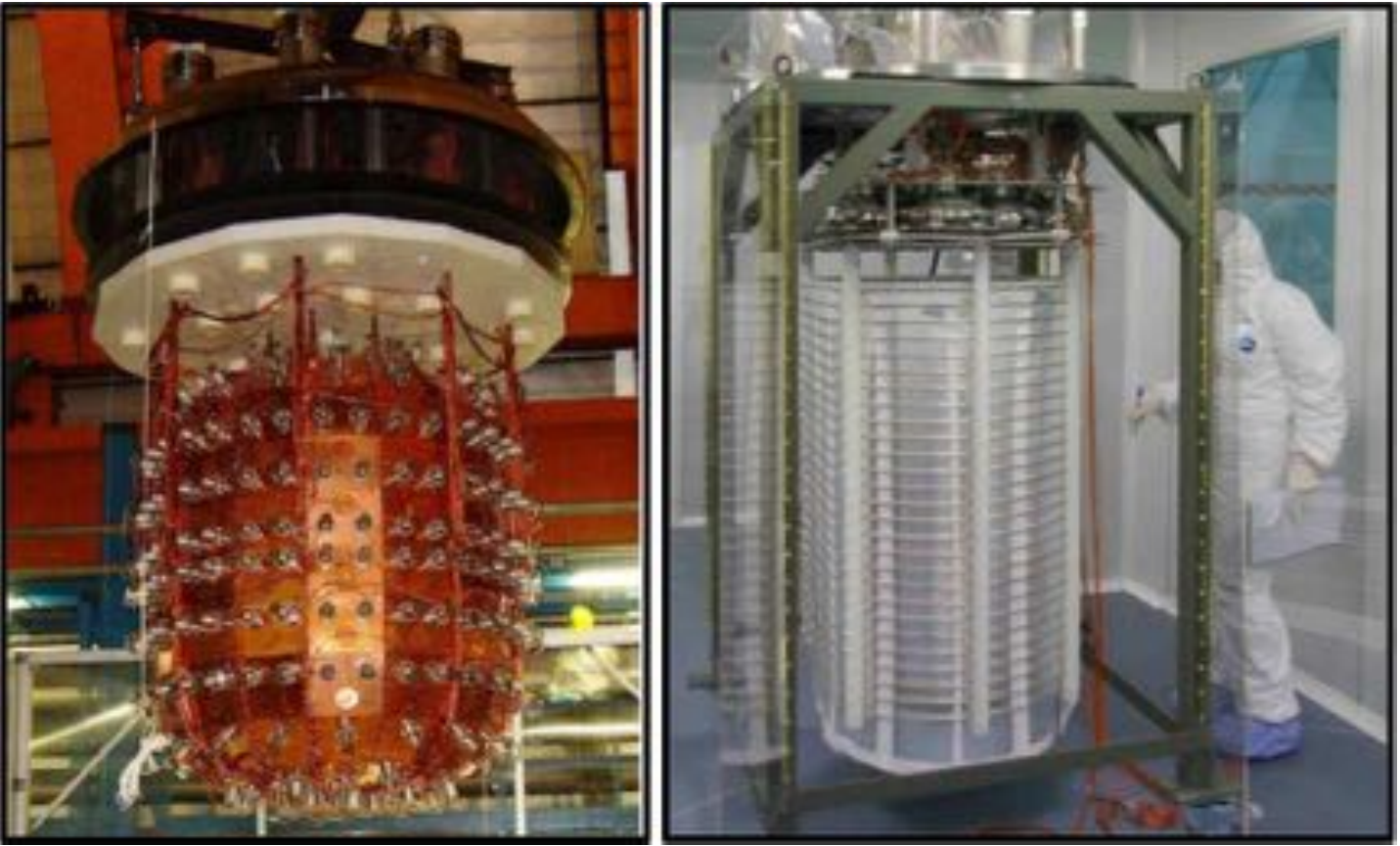}
  \caption{The WArP (left) and ArDM (right) experiments (figures from \cite{Zani:2014lea} and \cite{1475-7516-2017-03-003}, respectively).}
    \label{fig:DP-LAR-DM-detectors}
\end{figure}

\begin{figure}[htb]
 \centering
\includegraphics[width=\linewidth]{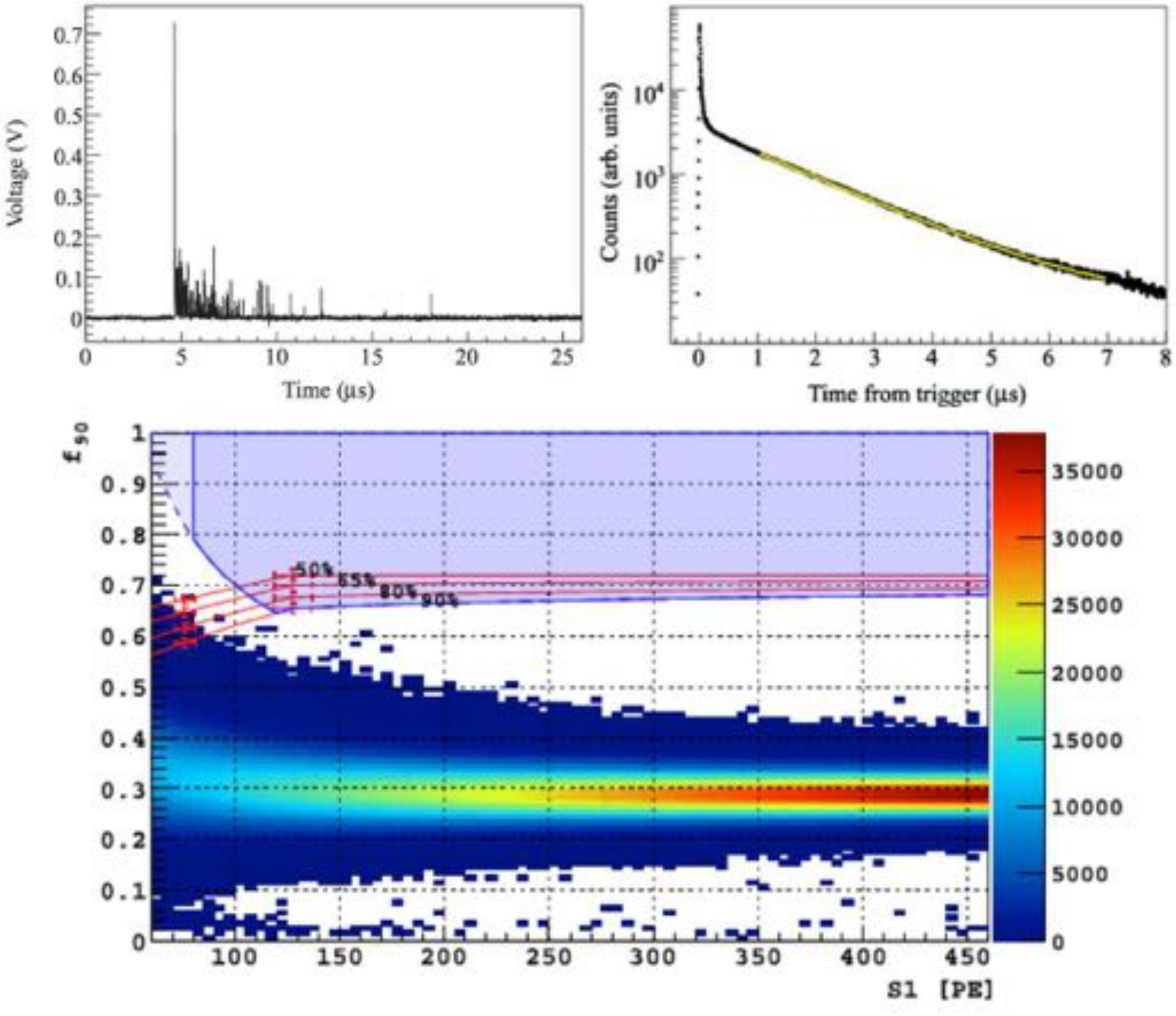}
\caption{Top: image of a typical waveform displaying primary scintillation (S1) in liquid argon (left), and a multiple-waveform average \cite{PhysRevC.78.035801}, (right). Bottom: 2D distribution of the main signal characteristics in an $f_{90}$-S1 representation as observed by the DarkSide-50 experiment (dominated by $^{39}$Ar $\beta$-decays). $f_{90}$ represents the fraction of S1 signal recorded within the first 90\,ns, and is typically 0.3 for electrons and 0.7 for nuclear recoils. The shaded area represents the dark matter search box. Percentages indicate different acceptance cuts (after \cite{DarkSide}).}
\label{fig:DS_PSD}
\end{figure}

Dual-phase argon detectors for dark matter searches were developed by WArP(2.3\,l) and later WArP(100\,l) in the first decade of 2000 \cite{Zani:2014lea} (Fig. \ref{fig:DP-LAR-DM-detectors}-left). Although the collaboration did not publish competing physics results and the 100\,l chamber suffered technical difficulties, it paved the road to contemporary experiments such as ArDM \cite{1475-7516-2017-03-003} and DarkSide-50 \cite{1748-0221-9-01-C01034}. There are two significant technological differences between argon and xenon dual-phase TPCs:
 \begin{enumerate}
 \item Unlike for xenon, reflectors and PM surfaces need to be coated with a thin layer of wavelength shifter, to convert the argon scintillation (peaked at $\sim$127\,nm) to around 430\,nm and enhance reflectivity as well as photo-sensor response. A widely used and well-characterized wavelength shifting material is the organic chemical 1,1,4,4-Tetraphenyl Butadiene (TPB) \cite{MCKINSEY1997351}.
 \item In addition to the background rejection factor due mainly to the quenching effect observed for nuclei (that is similar to the one observed for xenon \cite{Cap}) argon displays a distinct feature relative to xenon, which is the large $\sim 1\mu$s time constant of the triplet state (Fig. \ref{fig:DS_PSD}). The fact that (contrary to x-rays and electrons) low energy nuclei populate preferentially the singlet state when exciting argon, allows a background reduction of $7.6 \times 10^{-7}$ between ~10 and 110\,keV at about 50\% acceptance \cite{PhysRevC.78.035801}.
\end{enumerate}

\begin{figure}[htb]
 \centering
 \includegraphics[width=\linewidth]{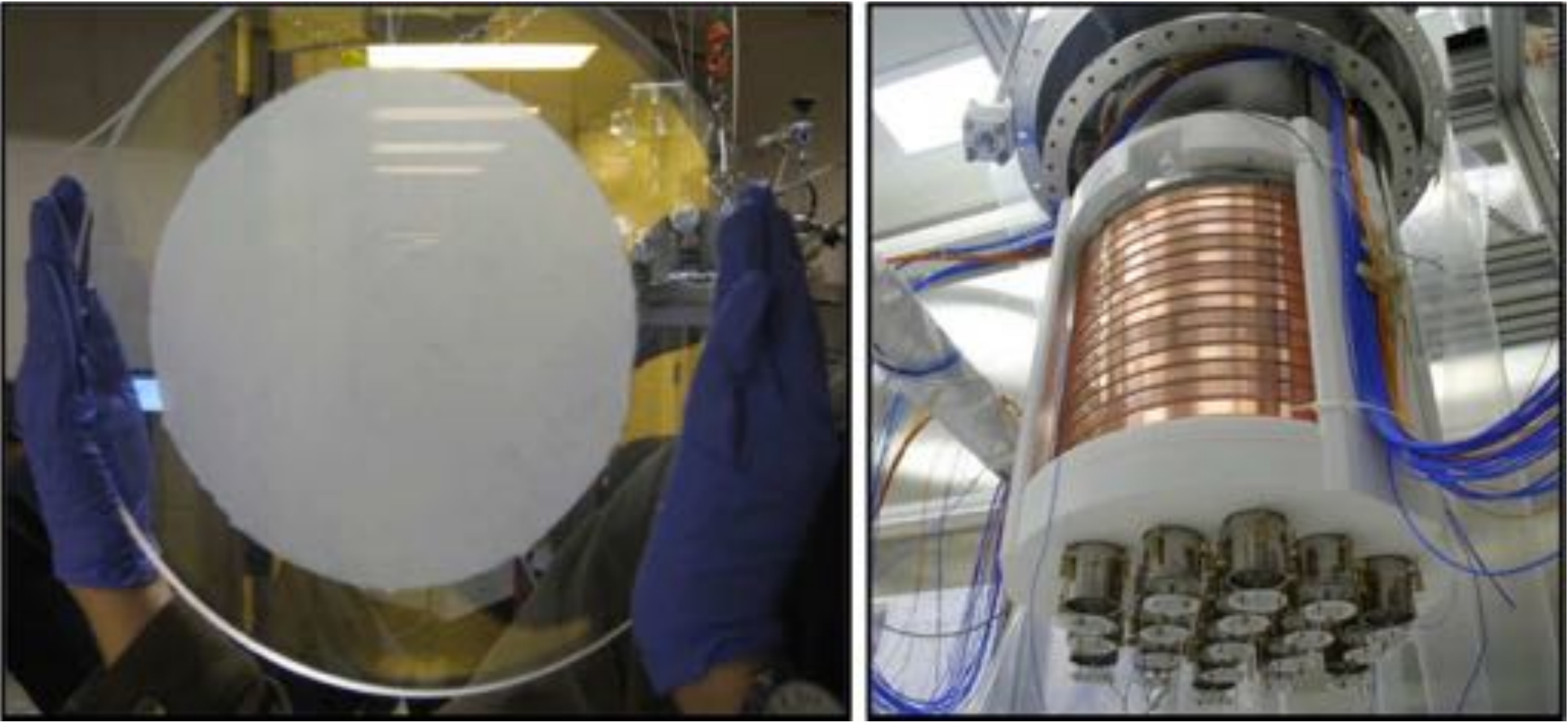}
\caption{Some of the assets of the DarkSide collaboration. Left: image of the fused silica plate coated with ITO and $2$\,$\mu$m-thick TPB to shift the argon VUV light. Right: field cage. (Figs. from \cite{DS_figs})}
\label{fig:DS_TPC}
\end{figure}

In 2015, the 1\,ton argon dark matter (ArDM) TPC (Fig. \ref{fig:DP-LAR-DM-detectors}-right) was operated at LSC in single-phase mode, that is, without drift field and hence no S2 signal. Good detector performance has been achieved and the results are encouraging for upcoming operation in dual-phase \cite{1475-7516-2017-03-003}. Also recently, the DarkSide collaboration has built and operated a 50\,kg detector (DarkSide-50), and has published results from a global analysis using data obtained with UAr extracted from a mine in Colorado \cite{Back:2012pg}. The cross section upper limit of $6.1 \times 10^{-44}$\,cm$^2$ at 100\,GeV/c$^2$ \cite{Agnes:2014bvk} has clearly made the case for the use of UAr in future larger argon dark matter experiments. DarkSide-50 is a cylindrical TPC with 35.6\,cm in height and 35.6\,cm in diameter. It has two arrays of 19 PMs both above and below the active volume. Besides TPB-coating the teflon reflectors, DarkSide has implemented an elegant solution for the end-caps, by resorting to ancillary quartz plates combined with an ITO+TPB coating (Fig. \ref{fig:DS_TPC}-left).\footnote{ITO: indium tin oxide, a high conductivity compound transparent in the visible range.}

\subsection{Future projects}

The ultimate goal of future experiments is to probe the cross sections down to or beyond the so-called `neutrino  floor', where the experiment becomes sensitive to coherent neutrino scattering on the target nuclei \cite{Baudis:2014naa}. The latter is the objective of the envisaged Argo project (300\,tons of UAr) and Darwin (40\,tons of xenon \cite{Albers}). The planned DarkSide-20k experiment, with a target mass of 20\,tons of UAr, will reach a sensitivity to the WIMP-nucleon cross sections of $10^{-47}$\,cm$^2$  for a WIMP of 1\,TeV/$c^2$ mass after 5\,years of running \cite{1742-6596-718-4-042016}.
Procurement of the necessary quantity of low radioactivity UAr for DarkSide-20k is the critical technical challenge for the experiment, and will be addressed within the framework of the \textit{Urania} and \textit{Aria} projects. The Urania project plans to extract 100\,kg per day of UAr from the same mine in Colorado that provided the UAr for DarkSide 50, and the Aria project will provide a cryogenic distillation plant capable of reducing the residual $^{39}$Ar in the UAr by a factor of 10 per pass, at a rate of 150\,kg per day \cite{1742-6596-718-4-042016}.

\section{Other TPCs and ideas} \label{classification6}

Other TPC types and some fashionable (far from established) ideas, are discussed in this section.

\subsection{Other TPC types}

\begin{figure}[htb]
 \centering
 \includegraphics[width=\linewidth]{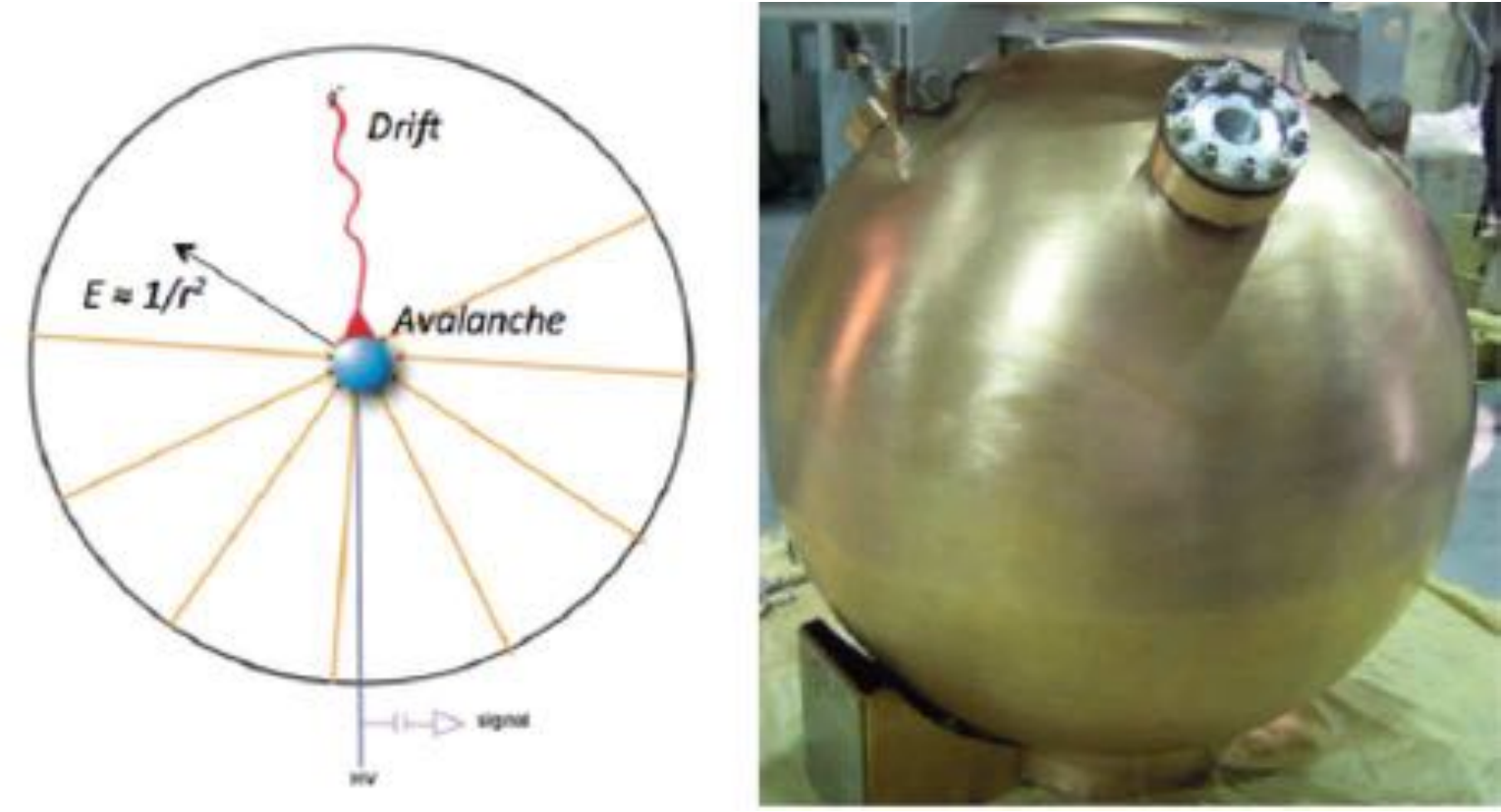}
\caption{Left: the basic principle of the spherical TPC, namely, using a single channel for reading a large pressurized volume thanks to its symmetric design. Right: a 60\,cm-diameter copper prototype by the NEWS collaboration.}
\label{SphericalTPC}
\end{figure}

Arguably, one of the most unconventional TPCs to date is the so-called `spherical TPC' by Giomataris \cite{SphGioma}, whose
working principle is sketched in Fig. \ref{SphericalTPC}. By using a single
channel and a large spherical volume (compatible with pressurization) this detector can provide a high sensitivity to rare processes
(both in interaction probability and event energy), while using the radial track extension (in the form of temporal spread) to identify and
separate extended tracks (e.g. cosmic rays). It has found application in a number of environments, indeed: with N$_2$ filling
at 0.5\,bar, it has been used for fast-neutron spectroscopy \cite{NeuDet}, with light-nucleus filling such as H$_2$, He and Ne it seems as if it could improve on present dark matter limits at masses below 6\,GeV/$c^2$ \cite{Gerbier}, and when operated at 10\,bar of xenon it provides an estimated number of neutrino events between 300 and 500 for a SuperNova explosion within our galaxy \cite{Vergados}.
Several projects are exploring these directions, chiefly, NEWS \cite{Gerbier}. NEWS has shown operation at an energy threshold as low as 17\,eV, theoretically enabling single-electron counting if not limited by the detector dark rate. Recently, the collaboration has published a new upper limit for low WIMP masses around 500\,MeV/c$^2$ \cite{NEWS_last}, by applying an analysis threshold of 0.76\,keV$_r$ (36\,eV trigger threshold).

Another remarkable TPC is the one of the MuCap experiment that, based on a 10\,bar deuterium-depleted H$_2$-TPC \cite{MuCap0, MuCap}, was devised for the measurement of the muon capture rate:
\beq
\mu^{-} + p \rightarrow n + \nu_{\mu}
\eeq

\begin{figure}[htb]
 \centering
 \includegraphics[width=6cm]{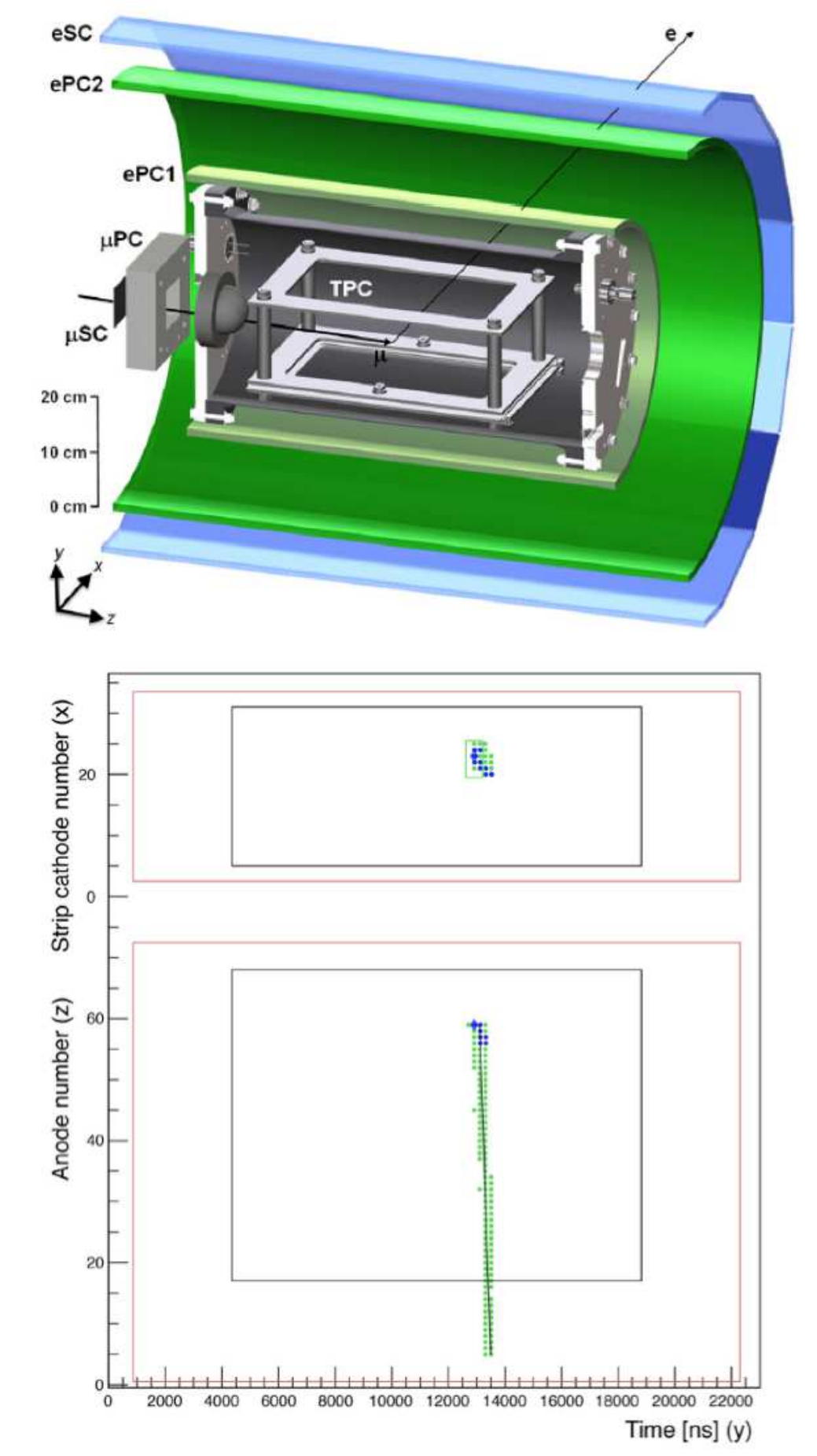}
\caption{Top: design of the MuCap experiment, showing the high pressure H$_2$ TPC at its core. Bottom: reconstruction of a fully stopped muon, displaying a characteristic end-blob (from \cite{MuCap}).}
\label{MuCap}
\end{figure}

This was finally achieved at the unprecedented level of precision of $\Lambda_S = 714.9\pm5.4_{stat}\pm5.1_{syst}$ s$^{-1}$ \cite{MuCapMeas}. The chamber's dimensions were $15\times12\times 30$\,cm$^3$ and it operated at the Paul Scherrer Institute (PSI) from 2004-2007. Read out with crisscrossing $x$-$y$ wires at $1$-$4$\,mm pitch, it achieved a working gain of 125 (notable, given the fact that no VUV-quenching gas was used), allowing to reconstruct fully stopping muons inside the chamber (Fig. \ref{MuCap}). The collaboration has developed a new TPC as part of the MuSun experiment, operated under deuterium at 30\,K \cite{MuSunProp}.

Some other recent TPC proposals devoted to the imaging of rare processes, at different stages of development, involve i) the triple GEM-based HypTPC, for searches of the H-dibaryon at high incoming beam rates (around 10$^5$-10$^6$\,Hz/cm$^2$) at J-PARC \cite{Sako}; ii) the thick GEM-based TPC for the study of deeply-bound kaonic clusters (also at J-PARC) \cite{E15}; iii) the $\beta\beta0\nu$ xenon TPC of the AXEL experiment \cite{AXEL}, that relies on a very thick (5-10\,mm) GEM-like teflon structure coupled to VUV-sensitive SiPMs; iv) the recently proposed active target TPC for $\gamma$-spectroscopy at ELI \cite{GammaReactions}; and v) the ALPHA-g experiment, that aims at a measurement of the gravitational force over anti-hydrogen, by continuously monitoring its decay to pions under free-fall conditions \cite{ALPHA}. Recent imaging TPCs whose subject of study can be hardly categorized as `rare process' include those devoted to precision measurements of the neutron lifetime \cite{NeutronLifetime}, study of fission \cite{fissionTPC}, and low energy nuclear reactions \cite{Fidias}. Other TPCs that emphasize image reconstruction, in astronomy and applied science, involve x-ray polarimetry through the accurate reconstruction of the photoelectron (e.g. \cite{BellaXray, PrieskornXray, LiXray, Iwakiri}), $\gamma$-ray polarimetry through pair production \cite{HARPO}, Compton Cameras \cite{ComptonCamera} and thermal neutron detectors \cite{DoroMuTPC}. They are particularly synergetic with many TPCs described in this review.

Given the recent surge of interest in the study of neutrino coherent scattering with nuclei, several dual-phase TPCs (either argon or xenon -based) have been proposed over the last years \cite{IEE_NCS,Bulu,ZepNCS}. One of the most active current efforts is led by the RED collaboration \cite{RED}, with a design conceptually similar to the ones described in the dark matter section (\ref{DMDP}).

At last, this brief TPC overview would not be complete without mentioning the new generation of TPCs used as trackers at various spectrometers like for instance Crystal Ball \cite{CrysBall}, FOPI \cite{FOPI}, super-FRS \cite{FRSTPC}, NICA \cite{NICATPC} and certainly those for future colliders (e.g. ILC \cite{Attie} and CPEC \cite{CPEC}). Although with smaller drift distances, radial/curved TPCs have been popularized recently, too (e.g., \cite{cyl1,cyl2}).

\subsection{Novel ideas}

\subsubsection{Barium tagging in $\beta\beta0\nu$ experiments}

\begin{figure}[htb]
 \centering
 \includegraphics[width=\linewidth]{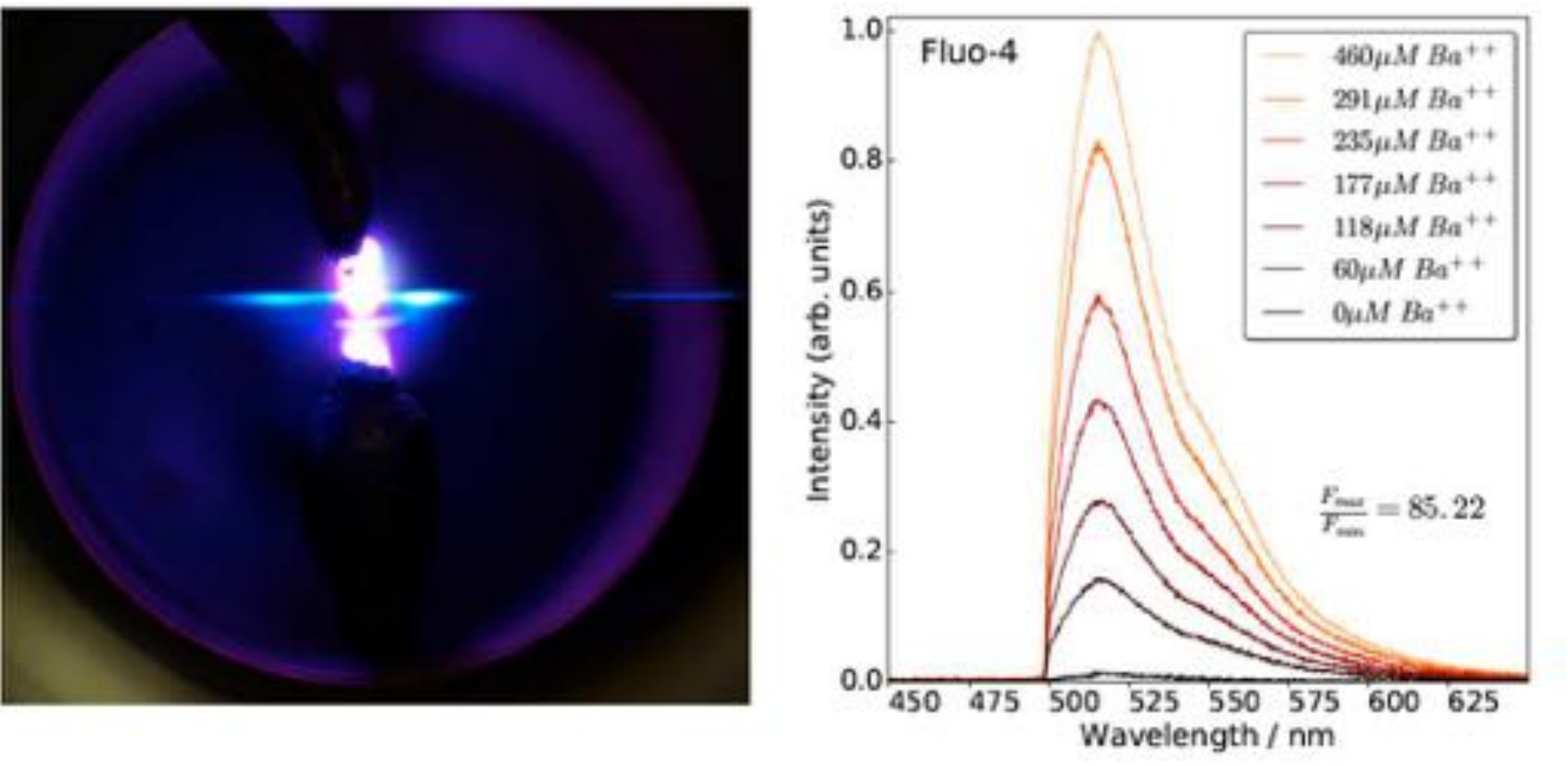}
\caption{Two promising techniques under development for barium tagging in xenon. Left: scintillation in blue (laser-interrogation) and red (response) for a Ba$^{+}$-rich electrical arc as measured by the EXO collaboration, after \cite{Rollin:2011gla}. Right: fluorescent response of the Fluo-4 molecule upon chelation (achieving a maximum at a concentration of roughly $460\,\mu$M of Ba$^{++}$) and comparison to the un-chelated one (by definition at $0\,\mu$M).}
\label{BaTaFig}
\end{figure}

Identification of the $^{136}$Ba daughter from $^{136}$Xe $\beta\beta0\nu$ decay, proposed by Moe in \cite{Moe}, has been pursued in the context of the EXO collaboration, both in liquid and gas phase, since approximately year 2000 \cite{Danilov}. It is a daunting idea, implying for a typical 1\,ton experiment a selectivity of 1 part per $10^{27}$. In its original proposal, this was achieved through the interrogation of the Ba$^+$ ion at 493.54\,nm (blue) by measuring its fluorescence at 649.87\,nm (red), \cite{Rollin:2011gla} (see, e.g., Fig. \ref{BaTaFig}-left). The process requires neutralization of Ba$^{++}$ into Ba$^{+}$ (something that is not energetically viable in a perfect xenon gas at room temperature, but may be be achieved in the presence of molecular clusters or molecular additives with low ionization potential) as well as an efficient transfer to the interrogation station (e.g. \cite{Brunner:2014sfa}). Possibilities for in-situ tagging in liquid/gas \cite{In-situ} and in solid \cite{SolXe} have been explored, too. While partial success has been obtained in the interrogation, neutralization and ion-transfer stages, a complete concept is still to be realized (for a recent proposal, see \cite{BaLast}).

Alternatively, the NEXT collaboration aims at achieving the selectivity needed through single molecule fluorescent imaging (SMFI), \cite{BaTa2}. The basic idea here is to resort to a coating consisting of a molecule that changes its fluorescent properties after trapping a Ba$^{++}$ ion (a process dubbed `chelation'), and so interrogation can proceed directly on the cathode region.
The Arlington group is currently pursuing an implementation of the complete concept, having achieved a signal to background ratio of 85 upon interrogation of the Ba$^{++}$-chelated Fluo-4 molecule \cite{2016 JINST 11 P12011}.

\subsubsection{Positive ion TPC}

\begin{figure}[htb]
 \centering
 \includegraphics[width=\linewidth]{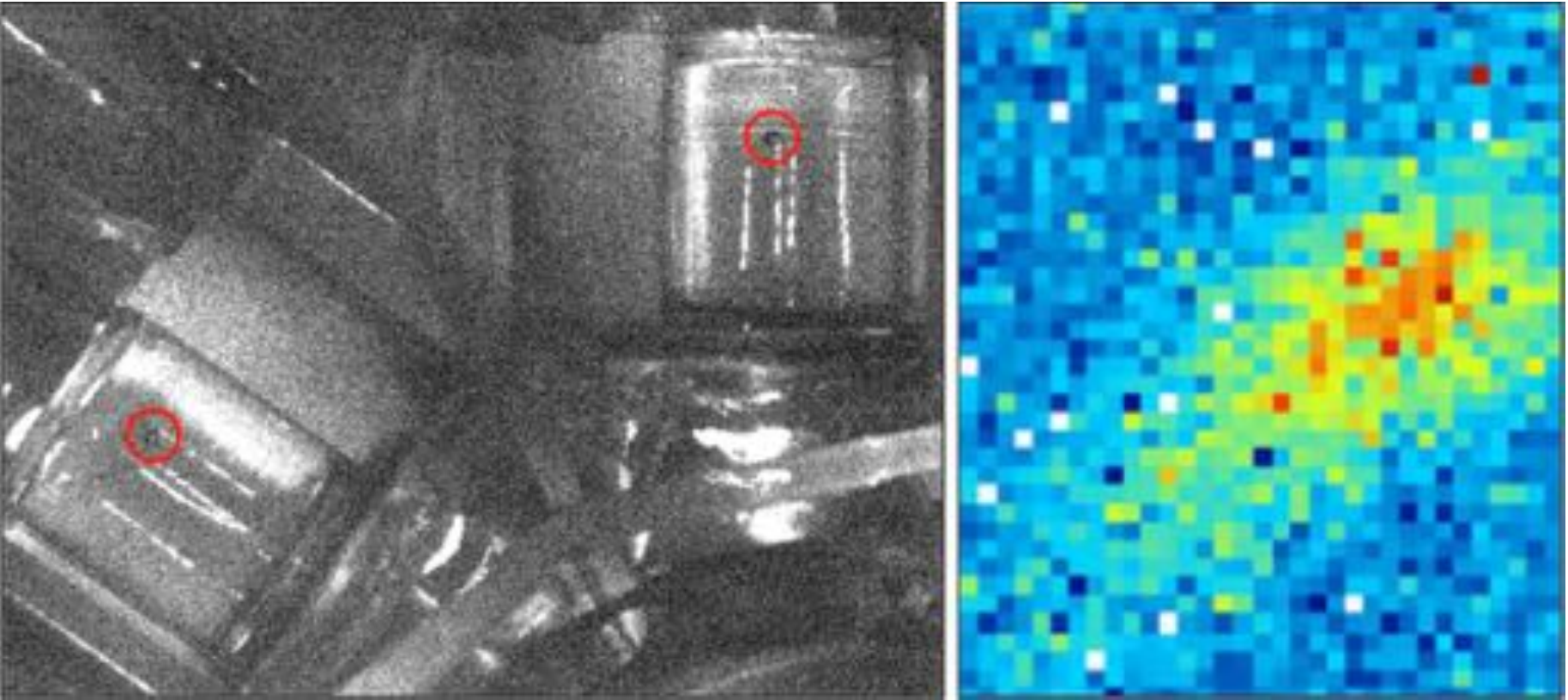}
\caption{Left: two CCD views of a nuclear recoil from $^{252}$Cf, reconstructed in a xenon bubble chamber (after \cite{arXiv:1702.08861v1}). Bubbles formed upon the passage of the nucleus are encircled. Right: an $\alpha$-particle from a $^{241}$Am source, detected with the TopMetal ASIC in air, without resorting to auxiliary multiplication \cite{TopMetal}.}
\label{BubbleAndPos}
\end{figure}

A new and radically different approach towards an ion-TPC consists on using the \emph{positive} ions produced during the ionization \cite{PositiveIon}. Indeed, positive ions can ionize the gas or even the cathode with high efficiency, if a sufficiently high kinetic energy ($\mathcal{O}$(10's of eV)) can be gained in the electric field. In practice, this requires pressures as low as 1-5\,mbar \cite{Fabi}, however. But, even at high pressures, secondary ionization still takes place (residually) upon ion neutralization at the cathode, provided an Auger electron is emitted and it can overcome the gas barrier so as to diffuse into the main gas volume.\footnote{Ejected electrons are highly non-thermal and can interact elastically in the gas, eventually bouncing back to the cathode. The probability to overcome this process (dubbed electron extraction efficiency) represents and important contribution, especially in pure noble gases.} This leads to the well-known phenomenon of ion feedback \cite{Raether}, usually characterized by a secondary emission probability $\gamma_i$. Since the probability depends on the difference between the ionization potential of the ion and twice the work function of the surface material, a suitable material selection, handling, and treatment will be needed. The author in \cite{PositiveIon} aims at $\gamma_i=2\times 10^{-3}$, a value compatible with the values obtained in vacuum by Hagstrum in \cite{Hagstrum} after including electron extraction efficiencies in gas obtained from \cite{CoelhoExtr}. Positive ions could hence make available a $T_o$-signal for energy deposits of the order 10-20\,keV, as well as precise topological information for the case of extended tracks capable of creating a sufficiently high ionization density. In the detection scheme proposed in \cite{PositiveIon}, electrons released from the cathode need to be locally amplified, something that can be for instance achieved with a microstrip or MWPC configuration. At the same time, the possibility of a direct measurement of the ions charge without amplification has been demonstrated for $\alpha$-particles, with the new TopMetal ASIC (15\,e$^-$ noise per $83\,\mu$m pixel) in \cite{TopMetal}, (Fig.\ref{BubbleAndPos}-right). Quite certainly, work in this direction will continue.

\subsubsection{Bubble chamber TPC}

Bubble chambers play an important role in dark matter searches thanks to their capability to tune the energy threshold so as to become insensitive to background electrons. Combining this feature with a standard identification based on the ionization and scintillation in liquid xenon holds the promise for a yet stronger background suppression factor in the identification of nuclear recoils, and has been recently proposed in \cite{Xenon bubble chamber}. Although the joint operation with the three measurement modes enabled remains unproven, the formation of bubbles in liquid xenon upon the passage of nuclei has been meanwhile demonstrated in \cite{arXiv:1702.08861v1}, (Fig. \ref{BubbleAndPos}-left).

\subsubsection{Columnar recombination as a directional marker}

In \cite{ColReco}, the possibility of using columnar recombination as a directional handle in dark matter searches, was proposed. The basic idea is to enable
the extraction of some topological information (in this case, the track's direction) from an image that is in fact structure-less. The reasoning
goes along the lines of the earlier Jaff\'e theory for charge recombination and subsequent measurements (e.g. \cite{Ramsey, Kanne, Jaffe}), however the proposal focused on Xe/TMA admixtures, that at the time held a strong promise. Since electrons and ions along the ionization trail have a higher chance to encounter each other when they are produced along the direction of the electric field, a higher recombination would ensue, thereby producing a directional marker. Moreover,
such a directional marker has been measured for $\alpha$-tracks and shown to be enhanced at high pressure \cite{Diana}. If columnar recombination becomes a viable
technique, it could eliminate the low exposure (low $Mt$ product) problem of traditional (low pressure) directional dark matter detectors. The original idea seems to be very much alive, since columnar recombination is indeed expected in pure noble gases and it has been identified earlier in \cite{Ramsey}, for instance. Additionally, recombination light has been measured in that case as well as in CF$_4$ at high pressure \cite{Morozov}. Unfortunately, although there are by now several theoretical evaluations of the impact of a directional signal based on this technique \cite{arXiv:1503.03937v3, arXiv:1704.03741v1, JinLi}, there is, to date, no quantitative information about the effect for low energy nuclei.

\subsubsection{Low-diffusion electroluminescent TPC}

For gaseous TPCs relying on the scintillation of pure noble gases, diffusion can be relatively large ($\mathcal{O}(10$\,mm$/\sqrt{\tn{m}}$)), thus limiting their imaging capabilities. Historically, the inclusion of molecular additives in order to improve this situation has been considered to be problematic, by virtue of the enhancement of processes like attachment, photon attenuation, VUV-quenching or charge recombination.

\begin{figure}[htb]
 \centering
 \includegraphics[width=\linewidth]{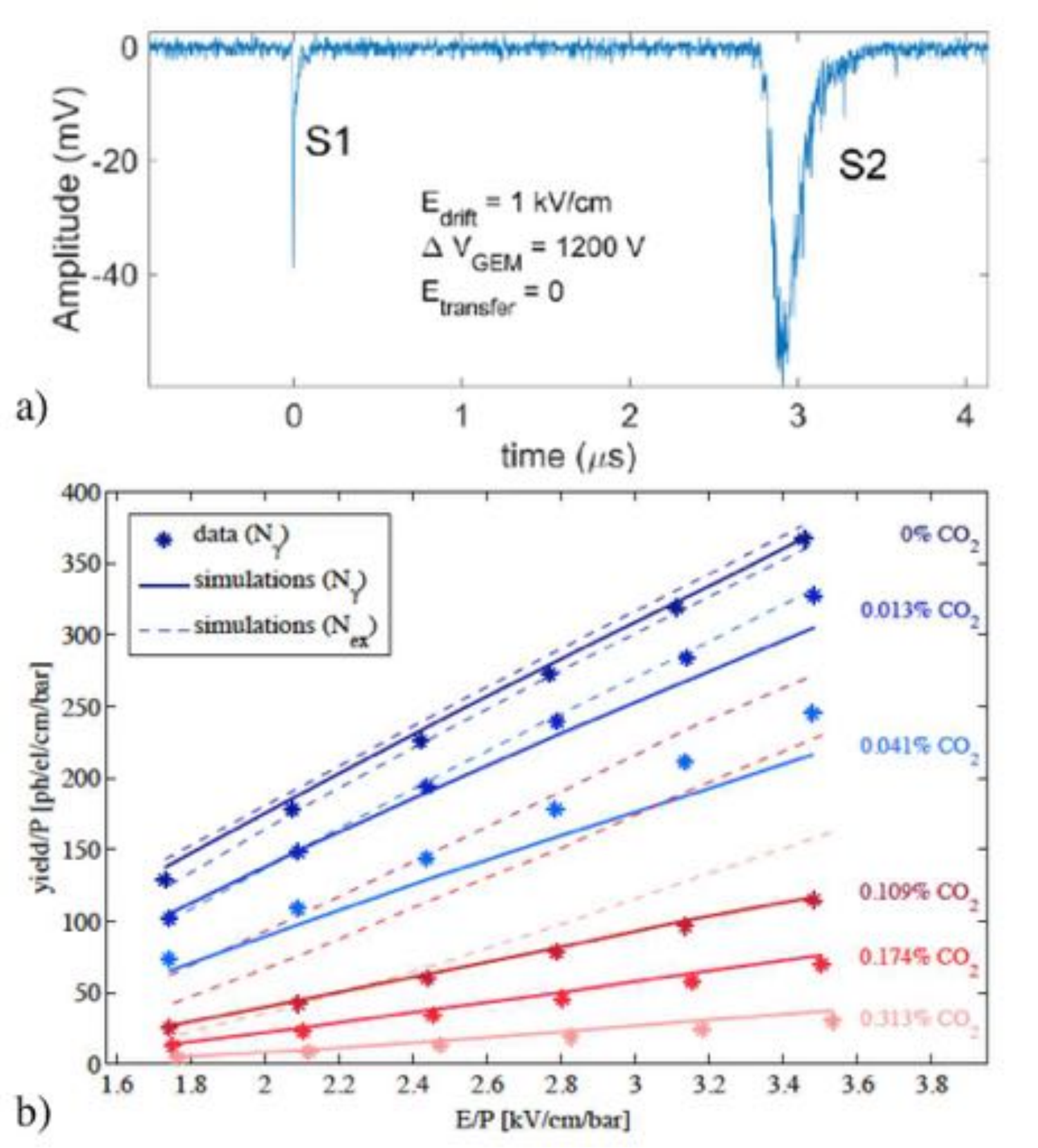}
\caption{Top (a): simultaneous measurement of scintillation and ionization in a liquid xenon TPC, through a bubble-assisted LHM \cite{ErdalLast}. Bottom (b): measurements and simulations of the electroluminescence yield in Xe/CO$_2$ mixtures, after \cite{Mua}.}
\label{LowDiff}
\end{figure}

It has been recently shown that additives like CH$_4$, CO$_2$ or CF$_4$ maintain indeed a sizeable amount of secondary scintillation under additive concentrations that are (expectedly) capable of providing sufficient electron cooling, for the case of xenon \cite{CO2Henriques}. The key aspect is the relatively short time constant of the triplet state of xenon (100\,ns), compared to the excimer quenching rates at the concentrations of interest (0.01\%-0.5\%). With the help of simulations \cite{Mua, homeo}, a reduction of the transverse diffusion by a factor nearing four can be anticipated in conditions where the available scintillation (suppressed by a factor five at 10\,bar) allows reconstruction of both primary and secondary scintillation, at sufficiently good energy resolution. Moreover, the situation can be interpreted through a fully microscopic simulation, that has been recently developed for the task \cite{Mua} (Fig. \ref{LowDiff}-bottom). The NEXT collaboration is currently working towards the demonstration of the achievable diffusion and energy resolution levels in a real-size demonstrator at pressures around 10\,bar. Emphasis is put on the handling of the minute additive concentration, getter compatibility and stability.

\subsubsection{Scintillating-negative ion TPC}

Following recent seminal work in \cite{Phan}, the CYGNUS\_RD collaboration is actively pursuing a simplified version of a negative ion TPC through the addition of strongly scintillating species, chiefly CF$_4$, read out in optical mode. Indeed, the collaboration has achieved an intermediate milestone by successfully operating a SF$_6$-based TPC at near atmospheric pressure (0.8\,bar) in \cite{NITEC}.
By combining the low diffusion of a negative ion TPC with the high segmentation and high sensitivity of a CMOS camera \cite{Marafini:2016kzd},
the collaboration can expect to approach the angular resolution limit coming from the angular spread of the elastic interaction, which is
used for directional WIMP detection. Similar to the case of other common gaseous wavelength-shifters (e.g., \cite{BreskinTeaSeminal}) one may expect some fractional energy transfer to the CF$_3^*$ scintillation precursor during the multiplication process, however it remains to be demonstrated to which extent this is the case.

\subsubsection{Charge multiplication in the liquid}

Achieving amplification directly in the liquid phase would avoid many of the complications inherent to dual-phase chambers, resulting in a higher construction and operational simplicity. A new concept was recently proposed in \cite{arXiv:1303.4365}, based on scintillation-induced amplification through cascaded GEMs/THGEMs coated with CsI. Each stage would produce some electroluminescence but no avalanche multiplication, since the latter is hardly achievable in liquid. During the exploration of this idea, the authors realized that they could indeed detect both scintillation and ionization directly in the liquid, if conditions were such that a steady bubble would be formed underneath the amplification structure  \cite{ErdalLast, JINST 10 (2015) P08015}, Fig. \ref{LowDiff}-top. They named this device `bubble-assisted liquid hole multiplier (LHM)'. Although a realistic concept (involving auxiliary heat sources) exists towards the implementation of this technique in liquid TPCs \cite{JINST 10 (2015) P08015}, this step has not been taken yet.

\section{Conclusions}\label{conc}

More than 40 years after its invention, time projection chambers continue to represent the leading technique for the reconstruction of rare physical processes, and there is no sign of fatigue nor any shortage of ideas about how to further enhance performance in the most competitive research fields. TPC development keeps building on the increasingly sophisticated understanding of a number of instrumental microscopic processes, micro-fabrication techniques, the development of low-noise, robust and configurable electronics, the mastering of high pressure as well as vacuum and cryogenic conditions, and the surge in position-sensitive photon detectors.

\ack
The authors want to thank Fabio Sauli for his encouragement, Ben Jones and Biagio Rossi for essential support on the liquid argon section, Carlos Azevedo (especially concerning the technicalities of Degrad and Garfield++ simulations), Davide Pinci and Filippo Resnati (for many discussions on the optical readout), Justo Martin-Albo (for help with the DUNE-ND section), Beatriz Fernandez and Manuel Caama\~no (for discussions on the nuclear reaction section), Jingbo Wang (for help with the interpretation of the hydrodynamic framework) and Andrew Laing (for careful reading of the manuscript). The five-peer review process enabled by NIM, as well as contributions from the referees is greatly appreciated. The support of NEXT and WA105 collaborations, RD51 collaboration and the CERN Neutrino platform, as well as discussions with their members over the years, is greatly acknowledged, too. DGD is supported by MINECO (Spain) under the Ramon y Cajal program.

\end{document}